\documentclass[%
 reprint,
 superscriptaddress,
 amsmath,amssymb,
 aps,
 prx,
 longbibliography,
 floatfix
]{revtex4-2}

\usepackage[dvipdfmx]{graphicx}
\usepackage{dcolumn}
\usepackage{multirow}
\usepackage{bm}
\usepackage{ulem}
\usepackage{txfonts}
\usepackage{hyperref}
\usepackage{color} 
\usepackage{xcolor}
\usepackage{listings}
\DeclareMathAlphabet{\mathlcal}{U}{dutchcal}{m}{n}

\newcommand{\bs}   {\boldsymbol}

\newcommand{\e}{{\rm e}}
\newcommand{\imag}{{\rm i}}
\newcommand{\dd}{{\rm d}}

\hypersetup{
  colorlinks=true,
  linkcolor=[rgb]{0.60,0.00,0.00},
  citecolor=[rgb]{0.00,0.00,0.60},
  urlcolor=[rgb]{0.00,0.00,0.60},
  setpagesize=false
}

\definecolor{codegreen}{rgb}{0,0.6,0}
\definecolor{codegray}{rgb}{0.5,0.5,0.5}
\definecolor{codepurple}{rgb}{0.58,0,0.82}
\definecolor{backcolour}{rgb}{0.95,0.95,0.92}

\lstdefinestyle{mystyle}{
    backgroundcolor=\color{backcolour},   
    commentstyle=\color{codegreen},
    keywordstyle=\color{magenta},
    numberstyle=\tiny\color{codegray},
    stringstyle=\color{codepurple},
    basicstyle=\ttfamily\footnotesize,
    breakatwhitespace=false,         
    breaklines=true,                 
    captionpos=b,                    
    keepspaces=true,                 
    numbers=left,                    
    numbersep=5pt,                  
    showspaces=false,                
    showstringspaces=false,
    showtabs=false,                  
    tabsize=2
}

\begin{document}

\title{  
  Quantum Power Method by a Superposition of Time-Evolved States
}

\author{Kazuhiro~Seki}
\affiliation{Computational Quantum Matter Research Team, RIKEN Center for Emergent Matter Science (CEMS), Saitama 351-0198, Japan}

\author{Seiji~Yunoki}
\affiliation{Computational Quantum Matter Research Team, RIKEN Center for Emergent Matter Science (CEMS), Saitama 351-0198, Japan}
\affiliation{Computational Materials Science Research Team, RIKEN Center for Computational Science (R-CCS),  Hyogo 650-0047,  Japan}
\affiliation{Computational Condensed Matter Physics Laboratory, RIKEN Cluster for Pioneering Research (CPR), Saitama 351-0198, Japan}

\begin{abstract}
  We propose a quantum-classical hybrid algorithm of the power method,
  here dubbed as quantum power method, to evaluate $\hat{\cal H}^n |\psi\rangle$ with quantum
  computers, where $n$ is a nonnegative integer, $\hat{\cal H}$ is a time-independent
  Hamiltonian of interest, and $|\psi \rangle$ is a quantum state. We show that the
  number of gates required for approximating $\hat{\cal H}^n$
  scales linearly in the
  power and the number of qubits, making it a promising application for
  near term quantum computers. Using numerical simulation, we show that
  the power method can control systematic errors in approximating the
  Hamiltonian power ${\hat{\cal H}^n}$ for $n$ as large as 100.
  As an application, we
  combine our method with a multireference Krylov-subspace-diagonalization scheme
  to show how one can improve the estimation of
  ground-state energies and the ground-state fidelities found using a
  variational-quantum-eigensolver scheme. Finally, we outline other
  applications of the quantum power method, including several
  moment-based methods. We numerically demonstrate the connected-moment
  expansion for the imaginary-time evolution and compare the results
  with the multireference Krylov-subspace diagonalization.
\end{abstract}

\date{\today}

\maketitle

\section{Introduction}
Numerically solving quantum many-body systems 
is one of the most
useful approaches for yet challenging issues in
condensed-matter physics
and quantum chemistry~\cite{LeBlanc2015,Motta2017,motta2019groundstate,Eriksen2020}. 
With classical computers,  
a repeated multiplication of 
a Hamiltonian $\hat{\mathcal{H}}$ of interest
to a properly chosen state, 
i.e., the power iteration,  
is an essential element of
various practical and advanced numerical techniques 
such as Krylov-subspace methods~\cite{Liesen} 
including the Lanczos method~\cite{Dagotto1994,Jaklic1994,Jaklic2000,Weisse,Prelovsek,Koch2019}, and
polynomial-expansion methods~\cite{Weisse_book}.
Such methods allow for calculating 
not only ground states 
but also dynamics~\cite{TalEzer1984,Park1986,Vijay2002,Iitaka2003,Mohankumar2006}
of quantum many-body systems. 
A major obstacle in these methods is, however, 
the exponential growth of the dimension of the Hilbert space 
with its system size $N$.
The Lanczos method has been implemented 
also with the variational Monte Carlo technique 
to systematically improve variational 
states towards the exact ground state~\cite{Sorella2001}.  
While the variational Monte Carlo method 
allows for substantially larger $N$ than 
the full-Hilbert space approaches, 
an affordable number of the Lanczos iterations is
practically limited to a few due to the $O(N^n)$ number
of terms constituting $\hat{\mathcal{H}}^n$.  

Recently, 
simulating quantum many-body systems with 
quantum computers~\cite{Feynman1982,Aspuru-Guzik2005,Wecker2015,McArdle2020,bauer2020quantum}
attracts great interest due to 
experimental realizations of and advances on quantum devices 
~\cite{Nakamura1999,optical_RMP2007,Ladd2010,RevModPhys.85.623,Barends2014,Chow2014,Kelly2015,Riste2015,Arute2019,Asavanant2019}.
Quantum computers will allow for a rather more direct access
to quantum states defined in a Hilbert space of potentially huge dimensions 
that cannot be treated with classical computers. 
At present, 
quantum computers are prone to noises 
and computations have to be accomplished 
with a small number of gates. 
In this regard, 
the variational-quantum-eigensolver (VQE) scheme 
has been proposed to simulate quantum many-body
systems using noisy intermediate-scale quantum devices~\cite{Preskill2018} 
and classical computers in a hybrid manner~\cite{Peruzzo2014,Wecker2015vqe,O'Malley2016,McClean2016,Kandala2017,Li2017,Mazzola2019}. 
VQE calculations with noisy quantum devices 
are now becoming affordable for fairly larger systems~\cite{arute2020hartreefock}
than in the earlier studies. 
While the majority of VQE schemes 
is devoted for gate reduction at the expense of the increased number of measurements,  
a measure-and-reuse technique has been proposed
for reducing the number of qubits~\cite{Liu2019}. 
Such a qubit-reuse technique 
has been demonstrated by evaluating
the ground-state energy of the one-dimensional Heisenberg model 
accurately only with a few trapped-ion qubits~\cite{fossfeig2020holographic}.

Moreover, to bypass
variational parameter optimization and 
ansatz-state cultivation inherent in the VQE scheme,  
several versions of Krylov-subspace methods  
have been proposed. 
The quantum Lanczos (QLanczos) method~\cite{Motta2019} generates
a Krylov subspace by evolving a reference state 
with an approximate imaginary-time evolution
~\cite{Yeter-Aydeniz2020,Nishi2020,Gomes2020,yeteraydeniz2020scattering}. 
The multireference-selected quantum Krylov (MRSQK)
algorithm generates a set of states spanning a Krylov subspace
by evolving selected reference states in real time~\cite{Stair2020}. 
As a related method, a quantum version of the filter-diagonalization (QFD)
method has been developed~\cite{parrish2019quantum}. 
A version of the inverse-iteration method
suitable for quantum computers~\cite{Kyriienko_2020}
makes use of an integral representation
of the inverse of the Hamiltonian~\cite{Childs2017}. 
Recently, an implementation of
the exact imaginary-time evolution 
with the help of ancillary qubits and Grover's search algorithm
has been proposed~\cite{Liu2020}.
Subspace-diagonalization schemes,
with subspaces
not restricted to a Krylov subspace
but intended to approximate a particular 
set of eigenspaces of the Hamiltonian of interest,  
have been implemented 
for calculating not only the ground state but also excited states of 
correlated quantum-chemistry systems 
~\cite{McClean2017,Colless2018,Parrish2019,nakanishi2018subspacesearch,heya2019subspace,huggins2019nonorthogonal}. 

In this paper, we propose a quantum power method,
a version of the power method
suitable for quantum-classical hybrid computing 
of quantum many-body systems. 
The method is based on
a time-discretized form of 
the higher-order derivative  
$\hat{\mathcal{H}}^{n}=\imag^n \dd^n\hat{U}(t)/\dd t^n|_{t=0}$ 
of the time-evolution operator $\hat{U}(t)=\e^{-\imag \hat{\mathcal{H}}t}$,  
by which the Hamiltonian power $\hat{\mathcal{H}}^{n}$ is represented
as a linear combination of $\hat{U}(t)$ at different time ($t$) variables 
close to $t=0$. 
The approximated Hamiltonian power retains its Hermiticity 
by engaging the time-discretized formalism with 
a central-finite-difference scheme for the time derivatives and
the symmetric Suzuki-Trotter decomposition of the time-evolution operators. 
Assuming 
a $\mathlcal{k}$-local Hamiltonian $\hat{\mathcal{H}}$ composed of $O(N)$ terms, 
the number of the gates
required for approximating $\hat{\mathcal{H}}^n$ in the quantum power method is $O(n\mathlcal{k}N)$, 
where $N$ is the system size (i.e., the number of qubits).  
We numerically demonstrate that the quantum power method can control the systematic errors, 
due to the finite-difference scheme for the time derivatives and the Suzuki-Trotter decomposition 
of the time-evolution operators, in approximating
the Hamiltonian power $\hat{\mathcal{H}}^{n}$ 
with $n$ as large as 100 for $N$ up to 24.
We apply the quantum power method to generate a Krylov subspace 
and perform, using noiseless numerical simulations, 
the multireference Krylov-subspace diagonalization for a one-dimensional spin-$1/2$ Heisenberg model with 
various reference states including those obtained by the VQE scheme. 
We find that the estimated ground-state energy as well as the ground-state fidelity 
are significantly improved with increasing the power $n$, 
thus providing a way to systematically improve the VQE scheme.  
We also apply the Krylov-subspace diagonalization combined with the quantum power method 
to a Fermi-Hubbard model to demonstrate that the quantum power method remains effective even when the Hamiltonian in the qubit representation is not local.
Furthermore, we briefly outline other applications of the quantum power method.

The rest of the paper is organized as follows.
In Sec.~\ref{sec:main},
we first summarize the main ideas and formulas of the quantum power method and
list in a table major symbols used throughout the paper. 
In Sec.~\ref{sec:form},
we provide the derivations of the main formulas and the technical details of
the quantum power method,
which includes the description of 
the central-finite-difference scheme for the time derivatives,
basic properties of the approximated Hamiltonian power, 
and the Suzuki-Trotter decomposition of the time-evolution operators. 
In Sec.~\ref{sec:ksd}, 
we review the Krylov-subspace-diagonalization scheme  
for an application of the quantum power method. 
In Sec.~\ref{sec:result},
we numerically demonstrate 
the quantum power method by considering the spin-$1/2$ Heisenberg model
on a one-dimensional periodic chain as an example. 
After defining the spin-$1/2$ Heisenberg model, we first
numerically show that the systematic errors in the quantum power method 
are well controlled to be essentially exact.  
We then present numerical results of
the Krylov-subspace diagonalization 
combined with the quantum power method.
The paper is summarized with discussions 
in Sec.~\ref{sec:conclusion}.
The Krylov-subspace-diagonalization scheme based on the quantum power is compared 
with other algorithms reported recently and the distinctions
of our method are highlighted in Appendix~\ref{app:ksd}.
The Krylov-subspace diagonalization
combined with the quantum power method 
is also numerically demonstrated for the Fermi-Hubbard model 
in Appendix~\ref{app:Hubbard}.
Explicit forms of 
the higher-order symmetric Suzuki-Trotter decompositions
generalized for multipartitioned Hamiltonians 
and their error analysis are provided in 
Appendix~\ref{app:ST}. 
An alternative formalism of approximating the Hamiltonian power
is discussed in Appendix~\ref{app:HST}.
For other applications of the quantum power method, 
some properties of the moments and cumulants are discussed in the context of the 
quantum power method, and the connected-moment
expansion (CMX) for the imaginary-time evolution 
is demonstrated by numerical simulations in Appendix~\ref{app:moment}.  
The Lanczos method with an emphasis on 
its connection to the moments 
is also described in Appendix~\ref{app:lanczos}.  
Throughout the paper, we set the reduced Planck constant $\hbar=1$.

\begin{center}
  \begin{figure*}
    \includegraphics[width=2.0\columnwidth]{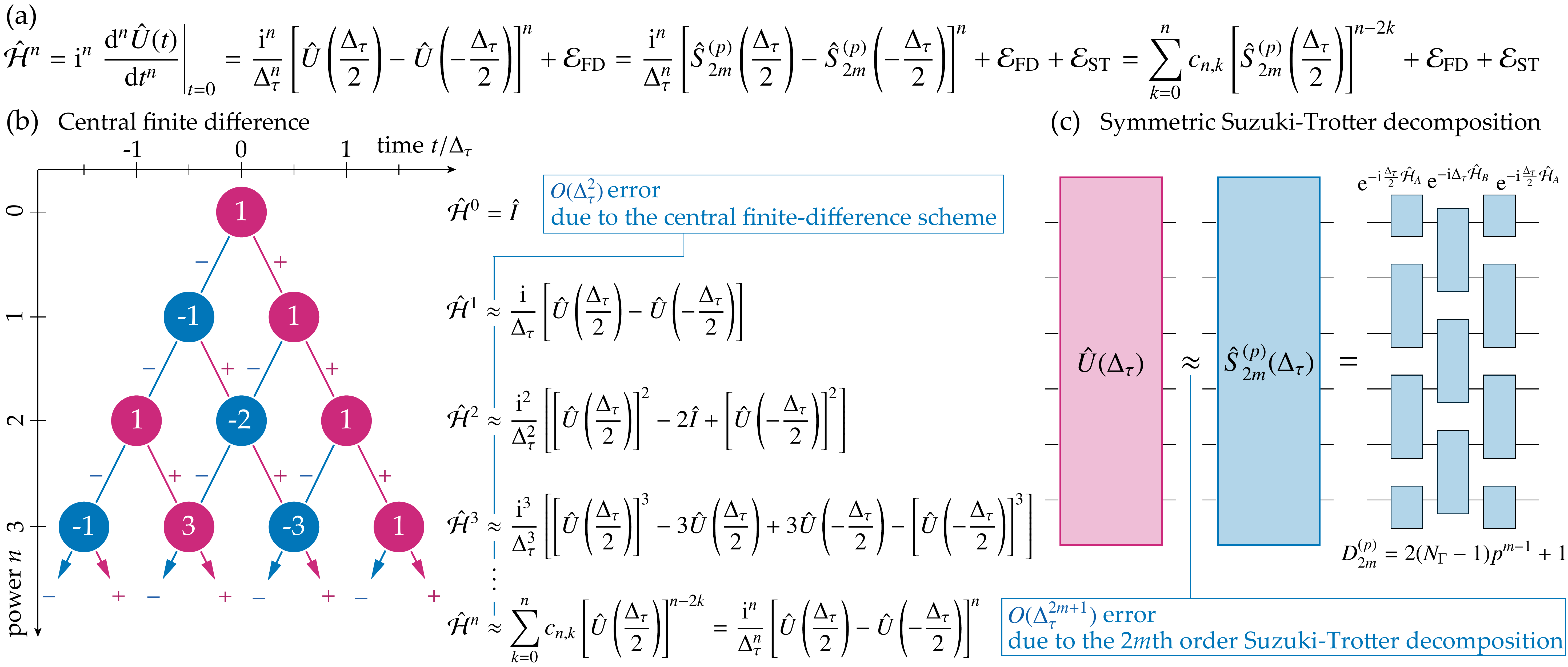}
    \caption{
    Overview of the quantum power method proposed here. 
    (a)
    The Hamiltonian power
    $\hat{\mathcal{H}}^n$
    is approximated 
    as a linear combination of the time-evolution operators
    $[\hat{U}(\Delta_\tau/2)]^{n-2k}$
    for $k=0,1,\dots,n$, 
    in which each
    $\hat{U}(\Delta_\tau/2)$
    is further decomposed 
    into $\hat{S}_{2m}^{(p)}(\Delta_\tau/2)$
    using the symmetric Suzuki-Trotter decomposition. 
    Here, $\Delta_\tau$ is a small time interval, and thus real positive number. 
    $\mathcal{E}_{\rm FD}$ and $\mathcal{E}_{\rm ST}$ 
    denote systematic errors due to 
    the finite-difference scheme for the time derivatives and
    the symmetric Suzuki-Trotter decomposition of the time-evolution operators, respectively.
    (b)
    An illustration of the central-finite-difference scheme  
    for the $n$th-order derivative of the time-evolution operator $\hat{U}(t)$ at $t=0$.
    Pascal's triangle with an alternating sign in time $t$ and power $n$ 
    provides coefficients $c_{n,k}$ of a linear combination
    of the time-evolution operators that approximates the Hamiltonian power $\hat{\mathcal{H}}^n$. 
    The systematic error due to the finite-difference scheme is 
    $\mathcal{E}_{\rm FD}\sim O(\Delta_\tau^{2})$.
    (c)
    A quantum circuit for the $2m$th-order symmetric 
    Suzuki-Trotter decomposition $\hat{S}_{2m}^{(p)}(\Delta_\tau)$ 
    of the time-evolution operator
    $\hat{U}(\Delta_\tau)=\e^{-\imag \hat{\mathcal{H}} \Delta_\tau}$ 
    with the systematic error of
    $O(\Delta_\tau^{2m+1})$.
    The systematic error $\mathcal{E}_{\rm ST}$ due to 
    the Suzuki-Trotter decomposition 
    for approximating the Hamiltonian power $\hat{\mathcal{H}}^n$ in (a) 
    is $O(\Delta_\tau^{2m})$ because of the factor 
    $1/\Delta_\tau^n$ in $c_{n,k}$.  
    $D_{2m}^{(p)}\,[=2(N_\Gamma-1)p^{m-1}+1]$ is the circuit depth of a single $\hat{S}_{2m}^{(p)}(\Delta_\tau)$ 
    for a Hamiltonian $\hat{\mathcal{H}}$ subdivided into $N_\Gamma$ parts,
    and $p$ is typically an $O(1)$ integer parameter for
    the symmetric Suzuki-Trotter decomposition,
    independent of the number $N$ of qubits. 
    The figure refers to $m=1$, $p=3$, and $N=6$
    for a 2-local Hamiltonian with $N_\Gamma=2$. 
    The $r$th-order Richardson extrapolation improves systematically the systematic errors as 
    $\mathcal{E}_{\rm FD}\sim O(\Delta_\tau^{2r+2})$ and $\mathcal{E}_{\rm ST}\sim O(\Delta_\tau^{2m+2r})$ 
    at the expense of increasing the number $(r+1)(n+1)$ of terms in the linear combination. 
    This implies that the lowest-order symmetric Suzuki-Trotter decomposition with $m=1$ is adequate  
    to control these systematic errors consistently. 
    The number of gates, 
    each of which is indicated by a small blue rectangle in (c), required 
    to approximately represent the Hamiltonian power $\mathcal{H}^n$ 
    scales as
    $O(n \mathlcal{k} N)$ for a $\mathlcal{k}$-local Hamiltonian with a prefactor $D_{2m}^{(p)}$.
    }     
    \label{overview}
  \end{figure*} 
\end{center}

\begin{table*}
  \caption{
    Major symbols used in this paper.
    \label{table.symbols} }
  \begin{tabular}{llll}
    \hline
    \hline
    Symbol &  Type & Description  & Defining equation(s)\\ \hline
    $N$ & integer & number of qubits & -- \\
    $t$ & real number & time  & Eq.~(\ref{unitary}) \\
    $\Delta_\tau$ & real number & time interval & Eqs.~(\ref{QPW}) and (\ref{powerH}) \\
    $\hat{U}$ & unitary operator & exact time-evolution operator & Eq.~(\ref{unitary})\\
    $\hat{S}_{2m}^{(p)}$ & unitary operator & $2m$th-order symmetric Suzuki-Trotter decomposition of $\hat{U}$ &
    Eqs.~(\ref{Suzuki}), (\ref{ST2}), and (\ref{ST2mp}) \\
    $p$ & integer & parameter in $\hat{S}_{2m}^{(p)}$ determining its accuracy for given $m$, $p\geqslant 3$ and odd
    &   Eq.~(\ref{ST2mp}) \\ 
    $\mathcal{\hat{H}}$ & Hermitian operator & time-independent Hamiltonian &
    Eqs.~(\ref{unitary}), (\ref{Ham_SWAP}), (\ref{Ham_Hubbard}), and (\ref{JWHub})\\
    $\mathcal{\hat{H}}^n$ & Hermitian operator & exact Hamiltonian power & Eq.~(\ref{powerH_exact}) \\
    $\mathcal{\hat{H}}_{\rm ST}^n $ & Hermitian operator & approximated Hamiltonian power & Eqs.~(\ref{QPW}) and (\ref{HST}) \\
    $\mathcal{\hat{H}}_{\rm ST(r)}^n $ & Hermitian operator & approximated Hamiltonian power with $r$th-order Richardson extrapolation& Eq.~(\ref{eq:Richardson}) \\
    $\{c_{n,k}\}_{k=0}^n$ & complex number & coefficients appearing in  $\mathcal{\hat{H}}_{\rm ST}^n$ & Eq.~(\ref{eq:cs}) \\
    $N_\Gamma$ & integer & number of noncommuting parts in $\hat{\mathcal{H}}$ & Eq.~(\ref{H_Gamma})\\
    $D_{2m}^{(p)}$ & integer & number of noncommuting exponentials in $\hat{S}_{2m}^{(p)}$ & Eq.~(\ref{depth})\\
    $\{s_i\}_{i=1}^{D_{2m}^{(p)}}$ & real number & coefficients appearing in exponents of $\hat{S}_{2m}^{(p)}$,
    available via program in Listing~1 & Eq.~(\ref{si}) \\
    $\mathcal{K}$ & vector subspace & (block) Krylov subspace & Eq.~(\ref{eq:ksub}) \\
    $M_{\rm B}$ & integer & block size (i.e., number of reference states) in block Krylov subspace $\mathcal{K}$ & Eq.~(\ref{eq:ksub}) \\
    $\bs{H}$ & Hermitian matrix & matrix representation of $\hat{\mathcal{H}}$ in (block) Krylov subspace $\mathcal{K}$ & Eq.~(\ref{Hsubspace}) \\
    $\bs{S}$ & Hermitian matrix & overlap matrix in (block) Krylov subspace $\mathcal{K}$ & Eq.~(\ref{Ssubspace}) \\
    $|\Psi_0\rangle$  & quantum state & exact ground state of $\hat{\mathcal{H}}$ with the exact ground-sate energy $E_0$ & Eq.~(\ref{eq:scheq}) \\
    $|\Psi_{\rm KS}\rangle$ & quantum state & approximated ground state of $\hat{\mathcal{H}}$ spanned in $\mathcal{K}$ with the corresponding energy $E_{\rm KS}$ & Eq.~(\ref{Psi_subspace}) \\
    $d$ & real number & operator distance, $0 \leqslant d \leqslant 1$ & Eq.~(\ref{distance}) \\
    $J$ & real number & exchange interaction in Heisenberg model / transfer integral in Hubbard model & Eq.~(\ref{Ham_SWAP}) / Eq.~(\ref{Ham_Hubbard}) \\
    $U_{\rm H}$ & real number & on-site interaction in Hubbard model & Eq.~(\ref{Ham_Hubbard}) \\    
    \hline
    \hline
  \end{tabular}
\end{table*}

\section{Main formulas}\label{sec:main}

Here, we summarize the main ideas and formulas of the quantum power method. 
Figure~\ref{overview} illustrates an overview of the formalism for the quantum power method 
based on the higher-order derivative of the time-evolution operator $\hat{U}(t)$, 
which is decomposed approximately using the symmetric Suzuki-Trotter decomposition.
Table~\ref{table.symbols} lists major symbols used in this paper. 
The derivations of the main formulas and the technical details are described in Sec.~\ref{sec:form}.

\subsection{Main ideas}

Let $\hat{\mathcal{H}}$ be the time-independent Hamiltonian.  
Then, the time-evolution operator $\hat{U}(t)$ is given by 
\begin{equation}
  \hat{U}(t) = \e^{-\imag \hat{\mathcal{H}}t}
  =\sum_{n=0}^{\infty} \frac{(-\imag t)^n}{n!} \hat{\mathcal{H}}^n,  
  \label{unitary} 
\end{equation}
where time $t$ is real. 
The quantum power method proposed here is based simply by noticing that
the $n$th power of the Hamiltonian, $\hat{\mathcal{H}}^n$, is given 
by the $n$th derivative of the time-evolution operator at $t=0$, i.e., 
\begin{equation}
  \hat{\mathcal{H}}^n
  =\imag^n \left.\frac{\dd^n \hat{U}(t)}{\dd t^n}\right|_{t=0}.
  \label{powerH_exact}
\end{equation}

The main purpose of this paper is to formulate, 
on the basis of the relation in Eq.~(\ref{powerH_exact}), 
a quantum-classical hybrid scheme which scales polynomially in both
the number $N$ of qubits and the power $n$
for evaluating approximately the Hamiltonian power
$\hat{\mathcal{H}}^n$ in a controlled manner.
To this end, below we show that 
the Hamiltonian power $\hat{\mathcal{H}}^n$ can be approximated by 
a linear combination of unitaries under a controlled accuracy,
if the central finite-difference scheme
for the time derivative in Eq.~(\ref{powerH_exact}) and 
the symmetric Suzuki-Trotter decomposition
of the time-evolution operator $\hat{U}(t)$
are employed (see Fig.~\ref{overview}).

\subsection{Quantum power method}\label{sec:qpm}

\subsubsection{Main formulas}\label{sec:qpm_main}

By applying 
the central finite-difference scheme with a small time interval $\Delta_\tau$
for the time derivative in Eq.~(\ref{powerH_exact}) and
the symmetric Suzuki-Trotter decomposition
of the time-evolution operator, 
the Hamiltonian power $\mathcal{\hat{H}}^n$ is approximated as 
\begin{alignat}{1}
  \hat{\mathcal{H}}^n
  &=
  \hat{\mathcal{H}}^n_{{\rm ST}}(\Delta_\tau)
  + O(\Delta_\tau^{2})
  + O(\Delta_\tau^{2m}),
  \label{QPW}
\end{alignat}
where
\begin{alignat}{1}
  \hat{\mathcal{H}}_{\rm ST}^n(\Delta_\tau)
  &=
  \sum_{k=0}^n
  c_{n,k}
  \left[\hat{S}_{2m}^{(p)}
    \left(\frac{\Delta_\tau}{2}\right)\right]^{n-2k} \label{HSTLC} \\ 
  &=
  \frac{\imag^n}{\Delta_\tau^n}
  \left[
    \hat{S}_{2m}^{(p)}\left(\frac{\Delta_\tau}{2}\right)
    -\hat{S}_{2m}^{(p)}\left(-\frac{\Delta_\tau}{2}\right)      
    \right]^n 
  \label{HST}
\end{alignat}
with 
\begin{equation}
  c_{n,k}
  =  \frac{\imag^n}{\Delta_\tau^n}(-1)^k\binom{n}{k} 
  \label{eq:cs}
\end{equation}
being coefficients for the central finite-difference scheme 
[see  Fig.~\ref{overview}(b) for a graphical derivation of $c_{n,k}$]. 
Note that the coefficient $c_{n,k}$ depends on $\Delta_\tau$, and if its dependence is 
denoted explicitly as $c_{n,k}(\Delta_\tau)$, the coefficient satisfies that 
$c_{n,k}(\Delta_\tau)=(-1)^n c_{n,k}(-\Delta_\tau)=c_{n,k}^*(-\Delta_\tau)$.

In Eq.~(\ref{QPW}), $O(\Delta_\tau^{2})$ represents the systematic error $\mathcal{E}_{\rm FD}$ 
due to the finite-difference scheme for the time derivatives, and 
$O(\Delta_\tau^{2m})$ denotes the systematic error $\mathcal{E}_{\rm ST}$ 
due to the Suzuki-Trotter decomposition of the time-evolution operators. 
$\hat{S}_{2m}^{(p)}(\Delta_\tau)$ is
the $2m$th-order symmetric Suzuki-Trotter decomposition 
of $\hat{U}(\Delta_\tau)$, given in Eq.~(\ref{ST2mp}), and satisfies that 
\begin{equation}
  \hat{U}(\Delta_\tau)=\hat{S}_{2m}^{(p)}(\Delta_\tau)+O(\Delta_\tau^{2m+1})
  \label{Suzuki}
\end{equation}
and
\begin{equation}
  \left[\hat{S}_{2m}^{(p)}(\Delta_\tau)\right]^\dag
  =\left[\hat{S}_{2m}^{(p)}(\Delta_\tau)\right]^{-1} 
  =\hat{S}_{2m}^{(p)}(-\Delta_\tau).
  \label{unitarityS}
\end{equation}
The superscript $p$ in
$\hat{S}_{2m}^{(p)}\left(\Delta_\tau\right)$
is an odd-integer parameter with $p \geqslant 3$ that determines prefactors
of the residual terms in $O(\Delta_\tau^{2m+1})$ but
does not change the order of the approximation in $\Delta_\tau$
(for numerical demonstrations, see Appendix~\ref{app:ST}). 
The order $O(\Delta_\tau^{2m})$ of
the Suzuki-Trotter error $\mathcal{E}_{\rm ST}$ in Eq.~(\ref{QPW}) is
decreased by one from the naively expected order $O(\Delta_\tau^{2m+1})$
because of the factor $1/\Delta_\tau^n$ in $c_{n,k}$. 
$\hat{\mathcal{H}}^n_{{\rm ST}}(\Delta_\tau)$ is
the central quantity in the quantum power method that 
approximates the Hamiltonian power $\mathcal{\hat{H}}^n$. 

Three remarks are in order. 
First,
Eq.~(\ref{QPW}) already reveals a
remarkable advantage in the quantum power method: 
in order to control the systematic errors
$\mathcal{E}_{\rm FD}$ and $\mathcal{E}_{\rm ST}$ with the same order of accuracy, 
it is enough to 
adopt the lowest-order Suzuki-Trotter decomposition with $m=1$,
independently of the power $n$.
Second,
Eq.~(\ref{HSTLC}) indicates that 
the Hamiltonian power $\hat{\mathcal{H}}^n$ is 
approximated by a linear combination of the $n+1$
Suzuki-Trotter-decomposed time-evolution operators. 
Third, 
Eq.~(\ref{HST}) indicates that 
$\mathcal{\hat{H}}_{\rm ST}^n(\Delta_\tau)$ satisfies
the law of exponents 
\begin{equation}
  \hat{\mathcal{H}}_{\rm ST}^n(\Delta_\tau) =
  \left[\hat{\mathcal{H}}^1_{{\rm ST}}(\Delta_\tau)\right]^n.
  \label{HSTpower}
\end{equation}
Moreover, 
$\hat{\mathcal{H}}^n_{\rm ST}(\Delta_\tau)$ is 
Hermitian and an even function
of $\Delta_\tau$, i.e.,
\begin{equation}
  \hat{\mathcal{H}}^n_{\rm ST}(\Delta_\tau)
  =\left[\hat{\mathcal{H}}^n_{\rm ST}(\Delta_\tau)\right]^\dag
  =\hat{\mathcal{H}}^n_{\rm ST}(-\Delta_\tau), \label{eq:Hst}
\end{equation}
indicating that the systematic error 
in odd powers of $\Delta_\tau$ is
absent in Eq.~(\ref{QPW}).

\subsubsection{Richardson extrapolation}\label{sec:richard}
The systematic errors 
$\mathcal{E}_{\rm FD}$ and 
$\mathcal{E}_{\rm ST}$ in Eq.~(\ref{QPW})  
can be controlled by varying the time interval $\Delta_\tau$. 
However, it is often practically useful to reduce
the systematic errors by not taking too small $\Delta_\tau$ 
in the algorithmic level.
A better error estimate can be achieved
by systematically eliminating lower-order errors in Eq.~(\ref{QPW})
with the Richardson extrapolation.  

In the Richardson extrapolation, 
$\hat{\mathcal{H}}^n_{{\rm ST}}(\Delta_\tau)$ and 
$\hat{\mathcal{H}}^n_{{\rm ST}}(\Delta_\tau/h)$ with some real $h$  
(such that $0<h\not=1$) are used to eliminate the leading terms of 
the systematic errors $\mathcal{E}_{\rm FD}$ and $\mathcal{E}_{\rm ST}$ simultaneously 
in Eq.~(\ref{QPW}) as 
\begin{equation}
  \hat{\mathcal{H}}^n = \hat{\mathcal{H}}^n_{{\rm ST}(1)}(\Delta_\tau)
  + O(\Delta_\tau^4) + O(\Delta_\tau^{2m+2}), 
  \label{powerHDelta1}
\end{equation}
where
\begin{equation}
  \hat{\mathcal{H}}^n_{{\rm ST}(1)}(\Delta_\tau) =
  \frac{h^2 \hat{\mathcal{H}}^n_{\rm ST}(\Delta_\tau/h)-
    \hat{\mathcal{H}}^n_{\rm ST}(\Delta_\tau)}{h^2-1} 
\end{equation}
is the first-order Richardson extrapolation of
$\hat{\mathcal{H}}^n_{\rm ST}(\Delta_\tau)$. 
Since 
$\hat{\mathcal{H}}^n_{\rm ST}(\Delta_\tau)$ is an even function
of $\Delta_\tau$,  $\hat{\mathcal{H}}^n_{{\rm ST}(1)}(\Delta_\tau)$ is also 
an even function of $\Delta_\tau$ and thus 
the systematic errors $\mathcal{E}_{\rm FD}$ and $\mathcal{E}_{\rm ST}$ 
in odd powers of $\Delta_\tau$
are absent in Eq.~(\ref{powerHDelta1}).

We can use the Richardson extrapolation recursively to further eliminate 
the leading terms of the systematic errors in Eq.~(\ref{powerHDelta1}).  
Namely, 
the $r$th-order Richardson extrapolation $\hat{\mathcal{H}}^n_{{\rm ST} (r)}(\Delta_\tau)$ 
of the approximated Hamiltonian power  
can be obtained recursively as 
\begin{alignat}{1}
  \hat{\mathcal{H}}^n
  &=
  \hat{\mathcal{H}}^n_{{\rm ST}(r)}(\Delta_\tau)
  + O(\Delta_\tau^{2+2r})
  + O(\Delta_\tau^{2m+2r}),
  \label{QPWr}
\end{alignat}
where 
\begin{equation}
  \hat{\mathcal{H}}^n_{{\rm ST}(r)}(\Delta_\tau) =
  \frac{h^{2r} \hat{\mathcal{H}}^n_{{\rm ST}(r-1)}(\Delta_\tau/h)-
    \hat{\mathcal{H}}^n_{{\rm ST}(r-1)}(\Delta_\tau)}{h^{2r}-1}
  \label{eq:Richardson}
\end{equation}
with 
$\hat{\mathcal{H}}^n_{{\rm ST} (0)}(\Delta_\tau) 
\equiv\hat{\mathcal{H}}^n_{{\rm ST}}(\Delta_\tau)$ 
and therefore the systematic errors $\mathcal{E}_{\rm FD}$ and $\mathcal{E}_{\rm ST}$  
are reduced to $O(\Delta_\tau^{2+2r})$ and $O(\Delta_\tau^{2m+2r})$, respectively, 
after the $r$th-order Richardson extrapolation. 
One can easily show that 
\begin{equation}
\hat{\mathcal{H}}^n_{{\rm ST}(r)}(\Delta_\tau) =  \left[ \hat{\mathcal{H}}^n_{{\rm ST}(r)}(\Delta_\tau) \right]^\dag 
= \hat{\mathcal{H}}^n_{{\rm ST}(r)}(-\Delta_\tau)
\label{eq:hir}
\end{equation} 
because $\hat{\mathcal{H}}^n_{{\rm ST}(0)}(\Delta_\tau)$ is Hermitian and is an even function of $\Delta_\tau$, 
and therefore  
the systematic errors $\mathcal{E}_{\rm FD}$ and $\mathcal{E}_{\rm ST}$ in odd powers 
of $\Delta_\tau$ are absent in Eq.~(\ref{QPWr}). 
However, 
$\hat{\mathcal{H}}^n_{{\rm ST}(r)}(\Delta_\tau)$ is no longer the $n$th power of 
$\hat{\mathcal{H}}^{n=1}_{{\rm ST}(r)}(\Delta_\tau)$, 
i.e.,
$\hat{\mathcal{H}}^n_{{\rm ST}(r)}(\Delta_\tau) \ne
\left[\hat{\mathcal{H}}^{1}_{{\rm ST}(r)}(\Delta_\tau)\right]^n$, when $r\geqslant 1$, 
but obviously
$\hat{\mathcal{H}}^n_{{\rm ST}(r)}(\Delta_\tau) =
\left[ \hat{\mathcal{H}}^{1}_{{\rm ST}(r)}(\Delta_\tau)\right]^n + O(\Delta_\tau^{2+2r}) 
+ O(\Delta_\tau^{2m+2r})$.
In our numerical simulations, we set $h=2$ when the Richardson extrapolation is used.

Since $\hat{\mathcal{H}}^n_{{\rm ST} (0)}(\Delta_\tau)$ is
a linear combination of $n+1$ unitaries 
$\left\{[\hat{S}_{2m}^{(p)} (\frac{\Delta_\tau}{2} )]^{n-2k} \right\}_{k=0}^n$, 
$\hat{\mathcal{H}}^n_{{\rm ST}(r)}(\Delta_\tau)$ is 
a linear combination of $(r+1)(n+1)$ 
unitaries 
$\left\{\left\{[\hat{S}_{2m}^{(p)} (\frac{\Delta_\tau}{2h^l} )]^{n-2k} \right\}_{k=0}^n\right\}_{l=0}^r$. 
Equation~(\ref{QPWr}) hence reveals another significant feature of the quantum power method that 
the lowest-order symmetric Suzuki-Trotter decomposition with $m=1$ suffices to 
systematically and consistently eliminate  
the lower-order systematic errors in $\mathcal{E}_{\rm FD}$ and $\mathcal{E}_{\rm ST}$ 
with only a polynomial increase of computational complexity.
In Sec.~\ref{sec:error}, we will show by numerical simulations that these systematic errors 
in the approximated Hamiltonian power 
are well controlled with the time interval $\Delta_\tau$ 
for the power $n$ as large as 100.

For the application purpose of the quantum power method, it is important that the symmetry of the Hamiltonian 
$\mathcal{\hat{H}}$ is still respected in the approximated Hamiltonian power $\hat{\mathcal{H}}^n_{{\rm ST}(r)}(\Delta_\tau)$. 
This is indeed the case in the quantum power method because 
\begin{equation}
  \left[ \mathcal{\hat{H}}, \hat{\mathcal{H}}^n_{{\rm ST}(r)}(\Delta_\tau) \right] = O(\Delta_\tau^{2m+2r}). 
\label{eq:symmetry}
\end{equation}
Therefore, the symmetry of the Hamiltonian $\mathcal{\hat{H}}$ is preserved in the quantum power method 
within the systematic error $\mathcal{E}_{\rm ST}$ due to the Suzuki-Trotter decomposition that can be well controlled. 
Notice that there is no contribution from the systematic error $\mathcal{E}_{\rm FD}$ due to the finite-difference scheme 
for the time derivatives in the right-hand side of Eq.~(\ref{eq:symmetry}) 
because
$\left[ \mathcal{\hat{H}}, \hat{U}(\Delta_\tau) \right] =0$.

\subsubsection{Gate count}\label{sec:qp_gc}
In the quantum power method, the Hamiltonian power $\hat{\mathcal{H}}^n$ is approximated 
with $\hat{\mathcal{H}}^n_{{\rm ST}(r)}(\Delta_\tau)$, which is a linear combination of 
$(r+1)(n+1)$ unitaries 
$\left\{\left\{[\hat{S}_{2m}^{(p)} (\frac{\Delta_\tau}{2h^l} )]^{n-2k} \right\}_{k=0}^n\right\}_{l=0}^r$, 
i.e., Suzuki-Trotter-decomposed time-evolution operators, 
and each unitary is treated separately. Therefore, the gate count is determined by the number of gates 
required for describing $[\hat{S}_{2m}^{(p)} (\pm\frac{\Delta_\tau}{2h^l} )]^n$ in a quantum circuit 
because the number of gates required scales linearly with the power of $\hat{S}_{2m}^{(p)} (\frac{\Delta_\tau}{2h^l} )$ 
and is independent of the argument.

The number of the noncommuting exponentials in $\hat{S}_{2m}^{(p)}(\Delta_\tau)$ corresponds to 
the circuit depth of a quantum circuit for a single time-evolution operator $\hat{U}(\Delta_\tau)$
approximated by  $\hat{S}_{2m}^{(p)}(\Delta_\tau)$, and thus it gives a prefactor for the gate count.
Let us assume that the Hamiltonian $\hat{\mathcal{H}}$ can be 
divided into $N_\Gamma$ parts as  
\begin{equation}
  \hat{\mathcal{H}}=
  \underbrace{
    \hat{\mathcal{H}}_A+
    \hat{\mathcal{H}}_B+
    \hat{\mathcal{H}}_C+
    \cdots+
    \hat{\mathcal{H}}_Z
  }_{N_{\Gamma} \text{\ terms}},
  \label{H_Gamma}
\end{equation}
where generally 
$[\hat{\mathcal{H}}_\Gamma, \hat{\mathcal{H}}_{\Gamma^\prime}]\not=0$
if $\Gamma \not = \Gamma^\prime$
but terms within each $\hat{\mathcal{H}}_\Gamma$ commute to each other (here, $\Gamma,\Gamma^\prime=A,B,\cdots,Z$). 
As derived in Sec.~\ref{sec:depth}, 
the number $D^{(p)}_{2m}$ of noncommuting exponentials in  
$\hat{S}_{2m}^{(p)}(\Delta_\tau)$ 
is given by
\begin{equation}
  D^{(p)}_{2m} = 2(N_\Gamma-1)  p^{m-1}+1.
  \label{depth}
\end{equation}
As illustrated in Fig.~\ref{overview}(c), 
the simplest case with $m=1$, $p=3$, and $N_\Gamma=2$, 
for which $D^{(p)}_{2}=3$, a quantum circuit for a single $\hat{S}_{2m}^{(p)}(\Delta_\tau)$ 
has the circuit depth $D^{(p)}_{2m}$, and thus the circuit depth required for 
$\hat{\mathcal{H}}_{\rm ST}^n(\Delta_\tau)$ is at most $O(n)$ with a prefactor $D^{(p)}_{2m}$. 
It should be recalled here that,
as far as the quantum power method is concerned, 
the lowest-order symmetric Suzuki-Trotter decomposition (i.e., $m=1$)
is sufficient.

Let $N$ be the number of qubits. 
Assuming that a Hamiltonian $\hat{\mathcal{H}}$
is $\mathlcal{k}$ local and consists of $O(N)$ terms,
each of which is a Pauli string of length at most $\mathlcal{k}$,
the number of gates required for
$\hat{S}_{2}^{(p)}(\Delta_\tau)$ is
$O(\mathlcal{k} N)$~\cite{NielsenChuang}
with a prefactor $D^{(p)}_{2}$. 
Therefore,  
the number of gates required for
$\hat{\mathcal{H}}_{\rm ST}^n(\Delta_\tau)$ is
$O(n \mathlcal{k} N)$ with
a prefactor $D^{(p)}_{2} = 2 N_\Gamma-1 \sim O(1)$,
where $O(1)$ implies that the quantity is
independent of $n$ and $N$.
For example,
for the spin-1/2 Heisenberg model considered in Sec.~\ref{sec:result}, 
the locality of the Hamiltonian is independent of the
system size, i.e., $\mathlcal{k} \sim O(1)$, and
hence the gate count for
$\hat{\mathcal{H}}_{\rm ST}^n(\Delta_\tau)$ scales as $O(nN)$.
On the other hand,
when a fermionic Hamiltonian is considered,
the locality of the Hamiltonian may depend on
the system size $N$ due to a fermion-to-qubit mapping such as 
the Jordan-Wigner transformation~\cite{Jordan1928} or 
the Bravyi-Kitaev transformation~\cite{Bravyi2002}
(see also Refs.~\cite{Seeley2012,Tranter2013,Havlicek2017}). 
The Jordan-Wigner transformation represents 
a fermionic operator with an $O(N)$ number of Pauli operators,
i.e., $\mathlcal{k}\sim O(N)$,
and hence the gate count for 
$\hat{\mathcal{H}}_{\rm ST}^n(\Delta_\tau)$ 
scales asymptotically as $O(n N^2)$.  
The Bravyi-Kitaev transformation represents 
a fermionic operator with an $O(\log{N})$ number of Pauli operators, 
i.e., $\mathlcal{k}\sim O(\log{N})$,
and hence the gate count for
$\hat{\mathcal{H}}_{\rm ST}^n(\Delta_\tau)$
scales asymptotically as $O(n N \log{N})$.

As described above, the $r$th-order Richardson extrapolation does not alter  
the number of gates required, but 
the number of the Suzuki-Trotter-decomposed time-evolution operators 
in $\hat{\mathcal{H}}_{{\rm ST(r)}}^n(\Delta_\tau)$
increases as $(r+1)(n+1)$.
Therefore, for example, to evaluate the expectation value of 
$\hat{\mathcal{H}}_{{\rm ST}(r)}^n(\Delta_\tau)$ 
with respect to a given state $|\psi\rangle$, the 
$(r+1)(n+1)$ 
number of state overlaps such as 
$\langle \psi|[\hat{S}_{2m}^{(p)}(\frac{\Delta_\tau}{2h^l})]^{n-2k}|\psi\rangle$ 
have to be estimated.
However, these quantities 
can be evaluated on quantum computers separately in parallel
with respect to $k$ and $l$.

\subsubsection{Possible circuit structure
  for the linear combination of time-evolution operators}  

\begin{center}
  \begin{figure}[hbt!]
    \includegraphics[width=.95\columnwidth]{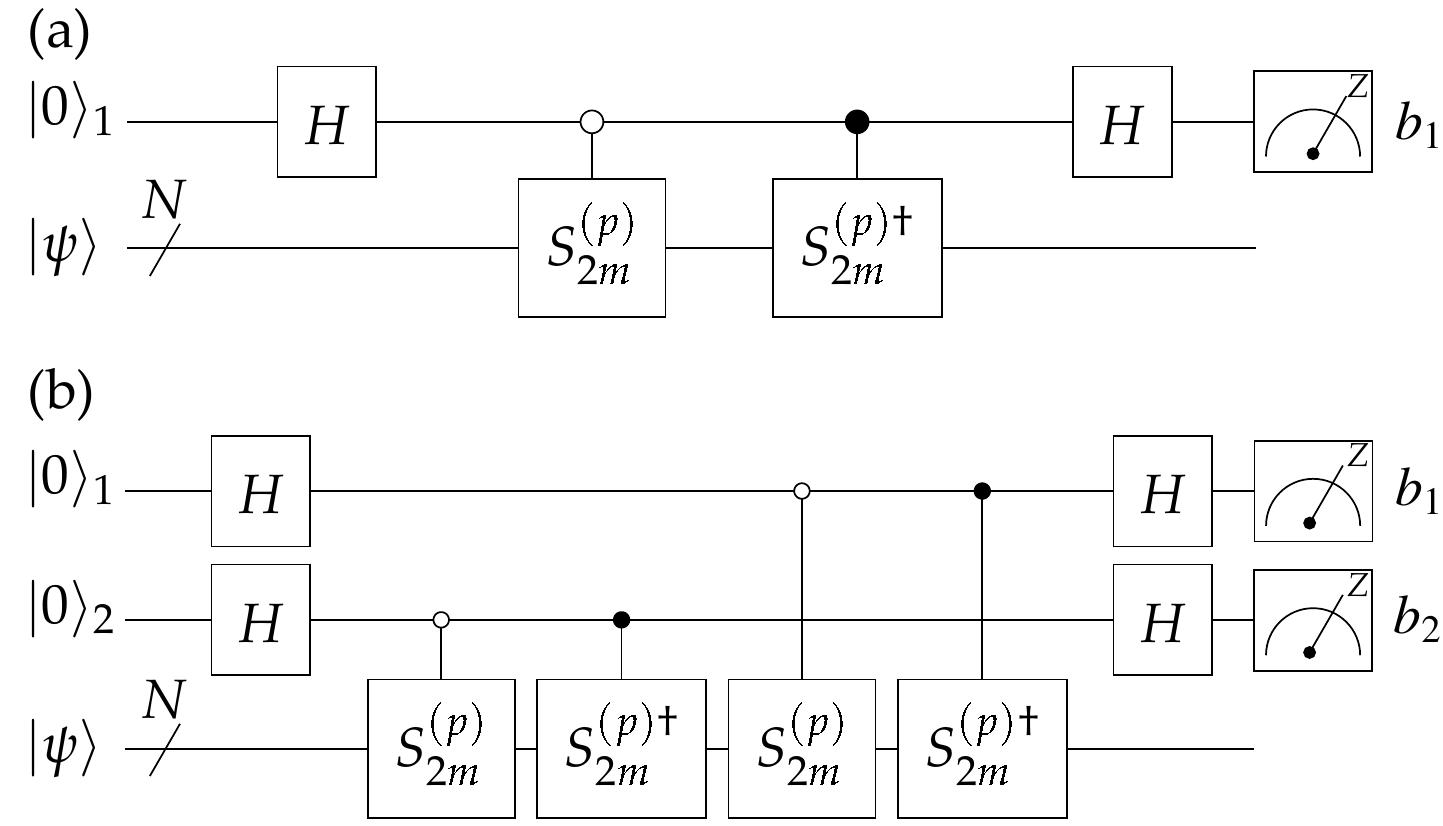}
    \caption{
      Circuit with 
      $N$ register qubits and $n$ ancilla qubits 
      for probabilistically generating the state 
      $\propto [\hat{S}_{2m}^{(p)}(\Delta_\tau/2)-
        \hat{S}_{2m}^{(p)}(-\Delta_\tau/2)]^n|\psi\rangle$ 
      in the register qubits       
      for (a) $n=1$ and (b) $n=2$.
      $H$, $S_{2m}^{(p)}$, and $S_{2m}^{(p)\dag}$ in the circuit
      denote the Hadamard gate,
      $\hat{S}_{2m}^{(p)}(\Delta_\tau/2)$, and
      $\hat{S}_{2m}^{(p)}(-\Delta_\tau/2)$, respectively. 
      A controlled-unitary gate with a solid (open) circle indicates that 
      the unitary gate is applied only if the control qubit is set to 1 (0).  
      The probability $P_{b_1 b_2 \cdots b_n}$
      for finding the bit string 
      $b_1 b_2 \cdots b_n = 1 1 \cdots 1$
      in the ancilla qubits 
      is given in Eq.~(\ref{prob}).
      \label{LC}
    } 
  \end{figure}
\end{center}

The form of the approximated Hamiltonian
power $\hat{\mathcal{H}}_{\rm ST}^{n}(\Delta_\tau)$ 
in Eq.~(\ref{HST}) suggests a direct treatment of
the linear combination of the Suzuki-Trotter-decomposed time-evolution 
operators with a single quantum circuit~\cite{Childs2012,kosugi2019construction,kosugi2019charge} 
that forms 
a simple recursive structure.    
Figure~\ref{LC} shows such a circuit structure for 
probabilistically generating the state 
$\propto
[\hat{S}_{2m}^{(p)}(\Delta_\tau/2)-
  \hat{S}_{2m}^{(p)}(-\Delta_\tau/2)]^n
|\psi\rangle$,
among $2^n$ superposed states, 
in the $N$ register qubits along with $n$ ancilla qubits.  
However,
the probability for finding the desired state in the
register qubits becomes exponentially small in general if $n$ is large.
Let us define $P_{b_1 b_2 \cdots b_n}$ as
the probability for finding a bit string
$b_1 b_2 \cdots b_n$ by measuring the $n$ ancilla qubits  
($b_k=0$ or $1$ for $1 \leqslant k \leqslant n$). 
Then the probability for 
finding the bit string $11\cdots 1$,
which is relevant for $\hat{\mathcal{H}}_{\rm ST}^{2n}(\Delta_\tau)$~\cite{odd_power}, 
is given by 
\begin{equation}
  P_{11 \cdots 1}
  =\frac{1}{4^{n}}
  (-1)^{n}\langle \psi|
  [\hat{S}_{2m}^{(p)}(\Delta_\tau/2)-
    \hat{S}_{2m}^{(p)}(-\Delta_\tau/2)]^{2n}
  |\psi\rangle.
  \label{prob}
\end{equation}
If $|\psi\rangle$ were an eigenstate of
$\hat{S}_{2m}^{(p)}(\Delta_\tau/2)$ with
an eigenvalue $\e^{\imag \lambda(\Delta_\tau)}$,
it oscillates as $P_{11\cdots 1}=[\sin{\lambda(\Delta_{\tau})}]^{2n}$, 
but otherwise it is exponentially small. 
Therefore,
as far as near-term applications
with a limited number of gates are concerned,
the linear combination of the Suzuki-Trotter-decomposed time-evolution
operators is better treated
with classical computers in the form of Eq.~(\ref{HSTLC}).
However, we anticipate that,
once a noiseless quantum computer is realized,
the product form of Eq.~(\ref{HST})
might have the advantage of 
robustness against loss of significance for small $\Delta_\tau$.

\subsubsection{Summary of the proposed method}

Figure~\ref{overview} summarizes the quantum power method. 
In the quantum power method, the Hamiltonian power $\hat{\mathcal{H}}^n$ is approximated 
to $\hat{\mathcal{H}}^n_{\rm ST}(\Delta_\tau)$ represented 
as a linear combination of
the $n+1$ Suzuki-Trotter-decomposed time-evolution operators
$\{[\hat{S}_{2m}^{(p)}(\frac{\Delta_\tau}{2})]^{n-2k}\}_{k=0}^{n}$.
The systematic error
$\mathcal{E}_{\rm FD}$ due to the finite-difference scheme for the time derivatives is 
$O(\Delta_\tau^2)$, and 
the systematic error 
$\mathcal{E}_{\rm ST}$ due to the Suzuki-Trotter decomposition of the time-evolution operators is
$O(\Delta_\tau^{2m})$. 
These systematic errors $\mathcal{E}_{\rm FD}$ and
$\mathcal{E}_{\rm ST}$ can be  
both improved systematically with the $r$th-order Richardson extrapolation to 
$O(\Delta_\tau^{2+2r})$ and 
$O(\Delta_\tau^{2m+2r})$, respectively, 
by approximating the Hamiltonian power $\hat{\mathcal{H}}^n$ 
with $\hat{\mathcal{H}}^n_{{\rm ST}(r)}(\Delta_\tau)$, which is given as 
a linear combination of the $(r+1)(n+1)$ Suzuki-Trotter-decomposed time-evolution operators
$\left\{\left\{[\hat{S}_{2m}^{(p)}(\frac{\Delta_\tau}{2h^l})]^{n-2k}\right\}_{k=0}^{n}\right\}_{l=0}^r$.
While the linear combination of the Suzuki-Trotter-decomposed
time-evolution operators is treated classically, 
each Suzuki-Trotter-decomposed time-evolution operator 
$[\hat{S}_{2m}^{(p)}(\frac{\Delta_\tau}{2h^l})]^{n-2k}$
is evaluated on quantum computers.

\subsection{Comparison with direct evaluation and classical computation}

The direct evaluation of 
$\langle \psi|\hat{\mathcal{H}}^n|\psi\rangle$ 
requires the expectation values of
$O({\rm min}(N^n,4^N))$ operators,
possibly containing long strings of Pauli operators,  
provided that the Hamiltonian 
$\hat{\mathcal{H}}$ consists of $O(N)$ terms.
Although
the depth of the circuits for these terms is $O(1)$, the
$O({\rm min}(N^n,4^N))$
measurements make the direct evaluation of 
$\langle \psi|\hat{\mathcal{H}}^n|\psi\rangle$
unfeasible as soon as the power $n$ and the number $N$ of qubits are large. 

In classical computation, the computational complexity scales as
$O(n N_{\rm D} )$ for the evaluation of $\mathcal{\hat{H}}^n|\psi\rangle$, 
when the Hamiltonian $\mathcal{\hat{H}}$ is local and thus the Hamiltonian matrix is sparse. 
Here, $N_{\rm D}$ is the dimension of the Hilbert space, e.g.,  
$N_{\rm D}=2^N$ for the spin-1/2 Heisenberg model.  
This implies that the computational complexity of the classical computation scales
exponentially in $N$.

In the quantum power method proposed here, 
the gate count for approximating the Hamiltonian power $\mathcal{\hat{H}}^n$ 
scales as $O(n \mathlcal{k} N)$ for a $\mathlcal{k}$-local
Hamiltonian composed of $O(N)$ terms.
In addition, the number of
state overlaps required to evaluate is 
$(r+1)(n+1)$, which is polynomial in $n$ and independent of $N$.
Therefore, 
although it is approximate, the quantum power method 
is a potentially promising application for
near-term quantum devices and would have
a quantum advantage over the classical counterpart of the power method.

\section{Derivations of main formulas}\label{sec:form}
Here, we provide the derivations of the main formulas in Sec.~\ref{sec:main} 
and describe technical details of the quantum power method.

\subsection{Hamiltonian power as a linear combination of unitary time-evolution operators}\label{sec:hp}
As shown in Eq.~(\ref{powerH_exact}), the Hamiltonian power $\hat{\mathcal{H}}^n$ is given by the 
$n$th derivative of the time-evolution operator $\hat{U}(t)$ at $t=0$. 
Here we show that, using the central finite-difference scheme for the time derivatives, the
Hamiltonian power can be approximated
by a linear combination of the time-evolution operators. 
 
By introducing a small time interval $\Delta_\tau$,
we replace the time derivative in Eq.~(\ref{powerH_exact})
with the central finite-difference as 
\begin{equation}
  \hat{\mathcal{H}}^{n}=\hat{\mathcal{H}}^n(\Delta_\tau) 
  + O(\Delta_\tau^2), 
  \label{powerH}
\end{equation}
where
\begin{equation}
  \hat{\mathcal{H}}^n(\Delta_\tau)
  =
  \sum_{k=0}^n
  c_{n,k}
  \hat{U}\left(\left(\frac{n}{2}-k\right)\Delta_\tau\right)  \label{naiveHFD}
\end{equation}
and $c_{n,k}$ is given in Eq.~(\ref{eq:cs}). 
The systematic error $O(\Delta_\tau^2)$ in Eq.~(\ref{powerH}) is due to the finite differentiation and 
this is the same systematic error $\mathcal{E}_{\rm FD}$ in Eq.~(\ref{QPW}).  
Equations~(\ref{powerH}) and (\ref{naiveHFD}) thus 
indicate that the $n$th power of the Hamiltonian, $\hat{\mathcal{H}}^n$,
can be approximated with a controlled accuracy as a linear combination of
the time-evolution operators evaluated at $n+1$ different time variables.

From the unitarity of the time-evolution operator
and its accordance with the time-reversed evolution,  
\begin{equation}
  \left[\hat{U}(t)\right]^{\dag}=\left[\hat{U}(t)\right]^{-1}=\hat{U}(-t), 
  \label{unitarity}
\end{equation}
it follows that the approximated Hamiltonian power $\hat{\mathcal{H}}^n(\Delta_\tau)$
is Hermitian and an even function of $\Delta_\tau $ i.e., 
\begin{equation}
  \hat{\mathcal{H}}^n(\Delta_\tau)
  =\left[\hat{\mathcal{H}}^n(\Delta_\tau)\right]^\dag
  =\hat{\mathcal{H}}^n(-\Delta_\tau).
  \label{Hermiticity}
\end{equation}
In the last equality, we have used that $c_{n,k}$ in Eq.~(\ref{eq:cs}) is an even (odd)  
function of $\Delta_\tau$ when $n$ is even (odd). 
Since $\hat{\mathcal{H}}^n(\Delta_\tau)$ is an even function
of $\Delta_\tau$, 
the systematic error $\mathcal{E}_{\rm FD}$ in odd powers of $\Delta_\tau$
is absent in Eq.~(\ref{powerH}).
Moreover, with the multiplication law of the time-evolution operator 
$\hat{U}\left(t\right) \hat{U}\left(t'\right)=\hat{U}\left(t+t'\right)$,
Eq.~(\ref{naiveHFD}) can be written as 
\begin{alignat}{1}
  \hat{\mathcal{H}}^n(\Delta_\tau)
  &=
  \sum_{k=0}^n
  c_{n,k}
  \left[\hat{U}\left(\frac{\Delta_\tau}{2}\right)\right]^{n-2k}
  \notag \\
  &=
  \sum_{k=0}^n
  c_{n,k}
  \left[\hat{U}\left(\frac{\Delta_\tau}{2}\right)\right]^{n-k}
  \left[\hat{U}\left(-\frac{\Delta_\tau}{2}\right)\right]^{k}
  \notag \\
  &=
  \frac{\imag^n}{\Delta_\tau^n}
  \left[
    \hat{U}\left(\frac{\Delta_\tau}{2}\right)
    -\hat{U}\left(-\frac{\Delta_\tau}{2}\right)
    \right]^n.
\label{powerHFD}
\end{alignat}
The last line in Eq.~(\ref{powerHFD}) indicates 
that the approximated Hamiltonian power
$\hat{\mathcal{H}}^{n}(\Delta_\tau)$ satisfies
a law of exponents
\begin{equation}
  \hat{\mathcal{H}}^{n}(\Delta_\tau)=
  \left[ \hat{\mathcal{H}}^{1}(\Delta_\tau)\right]^n. 
  \label{Hpower}
\end{equation}  
Namely, 
$\hat{\mathcal{H}}^n(\Delta_\tau)$ is exactly the $n$th power of 
$\hat{\mathcal{H}}^{n=1}(\Delta_\tau)$
for $n\geqslant 0$. 
In fact, Eq.~(\ref{powerHFD}) can be understood simply as 
\begin{alignat}{1}
  \mathcal{\hat{H}}^n = \left[ \imag \left.\frac{\dd \hat{U}(t)}{\dd t}\right|_{t=0} \right]^n = \left[ \hat{\mathcal{H}}^{1}(\Delta_\tau)\right]^n 
+ O(\Delta_\tau^2).
\end{alignat}

The systematic error in Eq.~(\ref{powerH}) can be systematically improved by the Richardson extrapolation. 
Following the same procedure described in Sec.~\ref{sec:richard}, the leading order of the systematic error 
can be eliminated recursively by the $r$th-order Richardson extrapolation as
\begin{equation}
  \hat{\mathcal{H}}^n = \hat{\mathcal{H}}^n_{(r)}(\Delta_\tau)
  + O(\Delta_\tau^{2+2r}),
  \label{powerHDeltar}
\end{equation}
where
\begin{equation}
  \hat{\mathcal{H}}^n_{(r)}(\Delta_\tau) =
  \frac{h^{2r} \hat{\mathcal{H}}_{(r-1)}^n(\Delta_\tau/h)-
    \hat{\mathcal{H}}_{(r-1)}^n(\Delta_\tau)}{h^{2r}-1}
  \label{Richardson}
\end{equation}
with $\hat{\mathcal{H}}^n_{(0)}(\Delta_\tau) \equiv \hat{\mathcal{H}}^n(\Delta_\tau)$.  
Because $\hat{\mathcal{H}}^n(\Delta_\tau)$ satisfies Eq.~(\ref{Hermiticity}), one can readily show that 
$ \hat{\mathcal{H}}^n_{(r)}(\Delta_\tau)$ is also Hermitian and is an even function of $\Delta_\tau$, 
and therefore the systematic error $\mathcal{E}_{\rm FD}$ in odd powers of $\Delta_\tau$
is absent in Eq.~(\ref{powerHDeltar}). 
Since $\hat{\mathcal{H}}^n_{(0)}(\Delta_\tau)$ is
a linear combination of the time-evolution operators at $n+1$ different times, 
$\hat{\mathcal{H}}^n_{(r)}(\Delta_\tau)$ is 
a linear combination of the time-evolution operators at $(r+1)(n+1)$ different times. 
Note also that 
$\hat{\mathcal{H}}^n_{(r)}(\Delta_\tau) \ne \left[ \hat{\mathcal{H}}^{1}_{(r)}(\Delta_\tau)\right]^n$ for $r\geqslant 1$, 
but obviously $\hat{\mathcal{H}}^n_{(r)}(\Delta_\tau) = \left[ \hat{\mathcal{H}}^{1}_{(r)}(\Delta_\tau)\right]^n + O(\Delta_\tau^{2+2r})$.

There are three additional remarks 
regarding the properties of the approximated Hamiltonian power 
$\hat{\mathcal{H}}^n(\Delta_\tau)$.
First,
if a forward or backward, instead of central, 
finite-difference scheme is employed in Eq.~(\ref{naiveHFD}), 
the Hermiticity and the even dependence on $\Delta_\tau$ 
of $\hat{\mathcal{H}}^n(\Delta_\tau)$  
in Eq.~(\ref{Hermiticity}) are both violated.
Therefore, the central finite-difference scheme
is a crucial choice.
Second,
when the time-evolution operator 
$\hat{U}(\Delta_\tau)$ is approximated 
by a Suzuki-Trotter decomposition,  
the corresponding Suzuki-Trotter error $\mathcal{E}_{\rm ST}$ 
appears in Eqs.~(\ref{naiveHFD}) and (\ref{powerHFD}).
Since the implementation of a higher-order Suzuki-Trotter decomposition 
on quantum computers requires many layers of gates, it is essential to control 
$\mathcal{E}_{\rm ST}$ with a lower-order Suzuki-Trotter decomposition. 
Third, 
if a symmetric Suzuki-Trotter decomposition, which retains
the equivalence between 
the inverse of the time evolution and 
the time-reversed evolution
[the right-most equality in Eq.~(\ref{unitarity})], 
is employed to decompose the time-evolution operators in
Eqs.~(\ref{naiveHFD}) and (\ref{powerHFD}), 
the resulting $\hat{\mathcal{H}}^n(\Delta_\tau)$ 
still satisfies the Hermiticity
and the even dependence on $\Delta_\tau$, as given in Eq.~(\ref{eq:Hst}). 
Therefore, it is important to adopt a symmetric Suzuki-Trotter decomposition 
(see Sec.~\ref{STunitarity} for details).

\subsection{Suzuki-Trotter decomposition}~\label{ST}
The formalism described above in Sec.~\ref{sec:hp}
is based on the exact time-evolution operator 
$\hat{U}(t)$ in Eq.~(\ref{unitary}).  
However, on quantum computers, 
the time-evolution operator with its exponent composed of
the sum of noncommuting operators usually has to be 
represented as a product of time-evolution operators
with each exponent composed of the sum of commuting operators. 
For this purpose, the Suzuki-Trotter decomposition is employed
to approximately decompose the time-evolution operator. 

In this regard,
we should emphasize that one of the crucial steps for the successful quantum power method is 
to determine properly in which stage the time-evolution operators 
in $\hat{\mathcal{H}}^n(\Delta_\tau)$ should be approximated by the Suzuki-Trotter decomposition, 
either in Eq.~(\ref{naiveHFD}) or in Eq.~(\ref{powerHFD}). 
Although Eqs.~(\ref{naiveHFD}) and (\ref{powerHFD}) are exactly the same
if the exact time-evolution operators are used, 
they are no longer the same in general once the time-evolution operators are approximated.  
Therefore, there are at least two routes to formulate the quantum power method.
As we shall discuss in details, these 
two approaches give us two different algorithms that scale differently in the power $n$. 
It turns out that when the power $n$ is larger than four, 
the algorithm formulated on the basis of Eq.~(\ref{powerHFD}) 
with the lowest-order symmetric Suzuki-Trotter decomposition
is preferable, otherwise the formalism based on 
Eq.~(\ref{naiveHFD}) with the higher-order symmetric
Suzuki-Trotter decompositions is favored in terms of the gate counts.

To understand the difference of these two approaches,
in this section, we briefly summarize a systematic construction of
the higher-order symmetric Suzuki-Trotter decompositions~\cite{Suzuki1990,Yoshida1990,Suzuki1991JMP}
for the quantum power method.

\subsubsection{Recursive construction of higher-order Suzuki-Trotter decompositions}\label{sec:rec_ST}

We now describe a systematic construction
of the symmetric Suzuki-Trotter decompositions. 
Let us define $x=-\imag \Delta_\tau$ to simplify the notation. 
The second-order symmetric decomposition $\hat{S}_2(\Delta_\tau)$
of the time-evolution operator $\hat{U}(\Delta_\tau)$
for the Hamiltonian $\hat{\mathcal{H}}$ of the form in Eq.~(\ref{H_Gamma})
is given by 
\begin{equation}
  \hat{U}(\Delta_\tau) =
  \e^{x \hat{\mathcal{H}}}
  = \hat{S}_2(\Delta_\tau)+O(\Delta_\tau^3),
  \label{S2}
\end{equation}
where
\begin{equation}
\hat{S}_{2}(\Delta_\tau)=
\underbrace{
  \overbrace{
    \e^{\frac{x}{2}\hat{\mathcal{H}}_A}
    \e^{\frac{x}{2}\hat{\mathcal{H}}_B}
    \e^{\frac{x}{2}\hat{\mathcal{H}}_C}
    \cdots}^{N_{\Gamma}-1 \text{\ exponentials}}
  \e^{x\hat{\mathcal{H}}_Z}
  \overbrace{
    \cdots
    \e^{\frac{x}{2}\hat{\mathcal{H}}_C}
    \e^{\frac{x}{2}\hat{\mathcal{H}}_B}
    \e^{\frac{x}{2}\hat{\mathcal{H}}_A}
  }^{N_{\Gamma}-1 \text{\ exponentials}}
}_{2N_{\Gamma}-1 \text{\ exponentials}}.
\label{ST2}
\end{equation}
Equation~(\ref{ST2}) can be derived 
by using the well-known decomposition 
$\e^{x(\hat{\mathcal{H}}_A+\hat{\mathcal{H}}_B)}
=
\e^{\frac{x}{2}\hat{\mathcal{H}}_A}
\e^{x\hat{\mathcal{H}}_B}
\e^{\frac{x}{2}\hat{\mathcal{H}}_A}
+O(\Delta_\tau^3)$
repeatedly,
e.g., 
$\e^{x(\hat{\mathcal{H}}_{A}+\hat{\mathcal{H}}_{B}+\hat{\mathcal{H}}_{C})}                               
=
\e^{\frac{x}{2}\hat{\mathcal{H}}_{A}}
\e^{      x   (\hat{\mathcal{H}}_{C} + \hat{\mathcal{H}}_{B} ) }                                       
\e^{\frac{x}{2}\hat{\mathcal{H}}_{A}} + O(\Delta_\tau^3)   
= \e^{\frac{x}{2}\hat{\mathcal{H}}_{A}} \e^{\frac{x}{2}\hat{\mathcal{H}}_{B}} 
  \e^{      x    \hat{\mathcal{H}}_{C}}                                                                   
  \e^{\frac{x}{2}\hat{\mathcal{H}}_{B}} \e^{\frac{x}{2}\hat{\mathcal{H}}_{A}}
  + O(\Delta_\tau^3)$.
The subscript ``$2$'' implies that $\hat{S}_{2}(\Delta_\tau)$ correctly represents 
$\hat{U}(\Delta_\tau) = \e^{x \hat{\mathcal{H}}}$ 
to $O(\Delta_\tau^2)$.
It is readily found that $\hat{S}_2(\Delta_\tau)$ satisfies 
\begin{equation}
\left[\hat{S}_2(\Delta)\right]^\dag = \hat{S}_2(-\Delta_\tau)
\end{equation}
and
\begin{equation}
  \hat{S}_2(\Delta_\tau)\hat{S}_2(-\Delta_\tau)=\hat{S}_2(-\Delta_\tau)\hat{S}_2(\Delta_\tau)=\hat{I},
  \label{Scommutes}
\end{equation}
where $\hat{I}$ is the identity operator, 
and therefore $\hat{S}_2(\Delta_2)$ is unitary.

It is noteworthy that 
if we write $\hat{S}_2(\Delta_\tau)$ in the form 
$\hat{S}_2(\Delta_\tau)=\exp\left[x \hat{\mathcal{H}}
  +x^2 \hat{R}_2 + x^3 \hat{R}_3 + \cdots \right]$,   
then the residual terms $\hat{R}_k$ with $k$
even are zero~\cite{Yoshida1990}. 
This can be confirmed as follows.   
Equation~(\ref{Scommutes}) indicates that  
$\hat{S}_2(\Delta_\tau)$ commutes with
$\hat{S}_2(-\Delta_\tau)=[\hat{S}_2(\Delta_\tau)]^{-1}$,
implying that
$\hat{I}
=\hat{S}_2(\Delta_\tau) \hat{S}_2(-\Delta_\tau)
=\exp\left[2(x^2 \hat{R}_2 + x^4 \hat{R}_4 + x^6 \hat{R}_6 + \cdots)\right]$
for arbitrary $x\,(=-\imag \Delta_\tau)$.  
We thus obtain that $\hat{R}_2=\hat{R}_4=\hat{R}_6=\cdots=0$.
This property holds for 
the higher-order symmetric decompositions 
described below, 
as they satisfy the relation corresponding to Eq.~(\ref{Scommutes}) 
by construction~\cite{Yoshida1990}.

Starting with $\hat{S}^{(p)}_2(\Delta_\tau) \equiv \hat{S}_2(\Delta_\tau)$, 
the higher-order decomposition 
$\hat{S}_{2m}^{(p)}(\Delta_\tau)$ for $m\geqslant 2$
that satisfies Eq.~(\ref{Suzuki}) 
can be constructed recursively as 
\begin{alignat}{1}
  \hat{S}_{2m}^{(p)}(\Delta_\tau)&=
  \left[\hat{S}_{2m-2}^{(p)}(k_m^{(p)} \Delta_\tau)\right]^{(p-1)/2}\notag \\
  &\times
  \hat{S}_{2m-2}^{(p)}(\tilde{k}_m^{(p)} \Delta_\tau)
  \left[\hat{S}_{2m-2}^{(p)}(k_m^{(p)} \Delta_\tau)\right]^{(p-1)/2}, 
  \label{ST2mp}
\end{alignat}
where 
$\tilde{k}_{m}^{(p)}=1-(p-1)k_{m}^{(p)}$, 
$k_{m}^{(p)}= [(p-1)-(p-1)^{1/(2m-1)}]^{-1}$, and  
$p$ is an odd integer with $p \geqslant 3$~\cite{Hatano2005}.  
The superscript ``$(p)$'' implies that
$\hat{S}_{2m}^{(p)}$
consists of a product of $p$ $\hat{S}_{2m-2}^{(p)}$'s. 
The parameter $k_{m}^{(p)}$ is determined so as to
eliminate the residual term $x^{2m-1} \hat{R}_{2m-1}$
in $\ln \hat{S}_{2m}^{(p)}(\Delta_\tau)$ and thus  
\begin{equation}
  \hat{S}_{2m}^{(p)}(\Delta_\tau)=
  \exp{\left[
      {x \hat{\mathcal{H}}}
      +x^{2m+1}\hat{R}_{2m+1}+\cdots
      \right]}.
  \label{residual}
\end{equation}
Namely, $k_{m}^{(p)}$ is the solution of
$(p-1)\left[k_{m}^{(p)}\right]^{2m-1}+
\left[\tilde{k}_{m}^{(p)}\right]^{2m-1}=0$
under the condition
$(p-1)k_{m}^{(p)}+\tilde{k}_{m}^{(p)}=1$. 
It is obvious that $\hat{S}_{2m}^{(p)}(\Delta_\tau)$ satisfies 
\begin{equation}
\left[ \hat{S}_{2m}^{(p)}(\Delta_\tau) \right]^\dag = \hat{S}_{2m}^{(p)}(-\Delta_\tau).
\end{equation}
Since 
$\hat{S}_{2m}^{(p)}(\Delta_\tau)$ also satisfies 
\begin{equation}
  \hat{S}_{2m}^{(p)}(\Delta_\tau)\hat{S}_{2m}^{(p)}(-\Delta_\tau)=
  \hat{S}_{2m}^{(p)}(-\Delta_\tau)\hat{S}_{2m}^{(p)}(\Delta_\tau)=\hat{I},
  \label{S2mcommutes}
\end{equation}
the residual terms of even power such as  $x^{2m} \hat{R}_{2m}$
are absent in the exponent of $\hat{S}_{2m}^{(p)}(\Delta_\tau)$ in Eq.~(\ref{residual}), 
shown by the same argument for $m=1$. 
Some of the higher-order symmetric Suzuki-Trotter decompositions
are explicitly provided in Appendix~\ref{app:hoST}. 
As shown in Appendix~\ref{ST_numerical}, 
the parameter $p$ affects the accuracy of
the decomposition for a given $m$.

\subsubsection{Unitarity and time-reversed evolution of $\hat{S}_{2m}^{(p)}(\Delta_\tau)$}\label{STunitarity}

As implied in Eq.~(\ref{S2mcommutes}), 
$\hat{S}_{2m}^{(p)}(\Delta_\tau)$ retains 
not only the unitarity but also
the equivalence between the inverse and time-reversed evolution,
as given in Eq.~(\ref{unitarityS}). 
Therefore, 
the Hermiticity and the even dependence on $\Delta_\tau$ of 
$\hat{\mathcal{H}}^n(\Delta_\tau)$ in Eq.~(\ref{Hermiticity})
are both retained even when
the exact time-evolution operators in Eqs.~(\ref{naiveHFD}) and (\ref{powerHFD})
are approximated by simply replacing them with $\hat{S}_{2m}^{(p)}$'s.
Indeed, the main formula of the quantum power method in Eq.~(\ref{QPW}) is obtained
by replacing $\hat{U}(\Delta_\tau/2)$ with $\hat{S}_{2m}^{(p)}(\Delta_\tau/2)$ in Eq.~(\ref{powerHFD}) 
and the approximated Hamiltonian power $\hat{\mathcal{H}}^n_{\rm ST}(\Delta_\tau)$ satisfies Eq.~(\ref{eq:Hst}). 
The same relations are also satisfied for $\hat{\mathcal{H}}^n_{{\rm ST}(r)}(\Delta_\tau)$ 
after the $r$th-order Richardson extrapolation, as given in Eq.~(\ref{eq:hir}).

In contrast to the symmetric Suzuki-Trotter decomposition,
an asymmetric Suzuki-Trotter decomposition $\hat{F}(\Delta_\tau)$, such as 
$\hat{F}(\Delta_\tau)=\e^{x\hat{\mathcal{H}}_A}\e^{x\hat{\mathcal{H}}_B}\cdots\e^{x\hat{\mathcal{H}}_Z}$,
results in 
\begin{equation}
  \left[\hat{F}(\Delta_\tau)\right]^\dag
  =\left[\hat{F}(\Delta_\tau)\right]^{-1} 
  \not=\hat{F}(-\Delta_\tau). 
  \label{unitarityA}
\end{equation}
Thus, 
$\hat{F}(\Delta_\tau)$ retains the unitarity but 
the inverse is no longer equivalent to the time-reversed evolution. 
In this case,
either the Hermiticity or the even dependence on $\Delta_\tau$ of 
$\hat{\mathcal{H}}^n(\Delta_\tau)$ in Eq.~(\ref{Hermiticity})
is violated if 
the exact time-evolution operators in Eqs.~(\ref{naiveHFD}) and (\ref{powerHFD})
are approximated by $\hat{F}$'s.
For example, if we consider an operator
$\hat{H}_{\rm H}(\Delta_\tau)=
\imag[\hat{F}(\Delta_\tau)-\hat{F}^\dag(\Delta_\tau)]/\Delta_\tau$ 
to approximate $\imag[\hat{U}(\Delta_\tau)-\hat{U}(-\Delta_\tau)]/\Delta_\tau$,
it satisfies the Hermiticity but is no longer an even function of $\Delta_\tau$ as 
$\hat{H}_{\rm H}(\Delta_\tau)
=[\hat{H}_{\rm H}(\Delta_\tau)]^\dag
\not=\hat{H}_{\rm H}(-\Delta_\tau)$.
On the other hand, an operator 
$\hat{H}_{\rm E}(\Delta_\tau)=
\imag[\hat{F}(\Delta_\tau)-\hat{F}(-\Delta_\tau)]/\Delta_\tau$ 
is an even function of $\Delta_\tau$ but no longer satisfies the Hermiticity as 
$\hat{H}_{\rm E}(\Delta_\tau)
=\hat{H}_{\rm E}(-\Delta_\tau)
\not=[\hat{H}_{\rm E}(\Delta_\tau)]^\dag$.
Therefore, 
the symmetric Suzuki-Trotter decomposition $\hat{S}_{2m}^{(p)}(\Delta_\tau)$ is
essential for the resulting Suzuki-Trotter approximated 
$\hat{\mathcal{H}}^n(\Delta_\tau)$ 
to retain both the Hermiticity and the even dependence on $\Delta_\tau$. 
Note that asymmetric Suzuki-Trotter decompositions
and their connection to symmetric ones 
have been studied in Ref.~\cite{Suzuki1992}.

\subsubsection{Circuit depth for a single time-evolution operator approximated by the Suzuki-Trotter decomposition}
\label{sec:depth}

We now consider the circuit depth $D^{(p)}_{2m}$
required for a single time-evolution operator $\hat{U}(\Delta_\tau)$
approximated by the symmetric Suzuki-Trotter decomposition
$\hat{S}_{2m}^{(p)}(\Delta_\tau)$, as in Eq.~(\ref{Suzuki}) 
[also see Fig.~\ref{overview}(c)].
We define $D^{(p)}_{2m}$ as the number of
noncommuting exponentials appearing in $\hat{S}^{(p)}_{2m}(\Delta_\tau)$. 
The depth of  $\hat{S}_{2}(\Delta_\tau)$ is thus given by 
$D^{(p)}_2=2N_\Gamma-1$, as explicitly shown in Eq.~(\ref{ST2}).
Since $\hat{S}^{(p)}_{2m}(\Delta_\tau)$ consists of
a product of $p$ $\hat{S}_{2m-2}^{(p)}$'s, 
the depth of $\hat{S}^{(p)}_{2m}(\Delta_\tau)$ 
without contracting commuting exponentials is 
$p D^{(p)}_{2m-2}$. 
However, since $\hat{S}_{2m}^{(p)}(\Delta_\tau)$ involves $p-1$ products
of two consecutive $\hat{S}_{2m-2}^{(p)}$'s, 
between which two commuting exponentials reside,
$p-1$ exponentials can be contracted.
We thus obtain that $D^{(p)}_{2m}=p D^{(p)}_{2m-2}-(p-1)$ 
or equivalently $D^{(p)}_{2m}-1=p [D^{(p)}_{2m-2}-1].$
By using this relation recursively, we can find that  
\begin{alignat}{1}
  D^{(p)}_{2m} - 1
  &= p \left[D^{(p)}_{2m-2} -1\right] \notag \\
  &= p^2 \left[D^{(p)}_{2m-4} -1\right]  \notag \\
  &= \cdots  \notag \\
  &= p^{m-1} \left[D^{(p)}_{2} -1\right].
  \label{Dm}
\end{alignat}
Substituting
$D^{(p)}_2=2N_\Gamma-1$
in Eq.~(\ref{Dm}) yields Eq.~(\ref{depth}). 
Recalling that $p$ is a typically $O(1)$ integer parameter, the depth increases exponentially with $m$ but is 
independent of the number $N$ of qubits. 
Therefore, the lower-order Suzuki-Trotter decomposition is 
highly desirable to shallow the depth of a quantum circuit.

\subsubsection{Two routes for quantum power method}\label{sec:tworoutes}

While the time-evolution operators satisfy the multiplication law
 $\hat{U}(\Delta_\tau)\hat{U}(\Delta_\tau^\prime)=\hat{U}(\Delta_\tau+\Delta_\tau^\prime)$, 
this is no longer correct when the time-evolution operators are approximated by the Suzuki-Trotter 
decomposition, i.e., 
$\hat{S}_{2m}^{(p)}(\Delta_\tau)\hat{S}_{2m}^{(p)}(\Delta_\tau^\prime) \ne \hat{S}_{2m}^{(p)}(\Delta_\tau+\Delta_\tau^\prime)$.  
Therefore, it is crucial to carefully consider when the time-evolution operators
in the approximated Hamiltonian power 
$\hat{\mathcal{H}}^n(\Delta_\tau)$ should be replaced with the symmetric
Suzuki-Trotter decomposition, either in Eq.~(\ref{naiveHFD}) 
or in Eq.~(\ref{powerHFD}). 
This implies that there exist two different routes
to formulate the quantum power method.  
Indeed, 
these two approaches provide 
two different algorithms of the quantum power method that differ  in
the scaling of complexity but control 
the systematic errors $\mathcal{E}_{\rm FD}$ and $\mathcal{E}_{\rm ST}$ with
essentially the same accuracy. 
The quantum power method formulated in Sec.~\ref{sec:qpm_main} is based on Eq.~(\ref{powerHFD}) 
that scales much better when the power $n$ is large. 
In Appendix~\ref{app:HST}, an alternative algorithm is formulated on the basis of Eq.~(\ref{naiveHFD}), 
which is favored when the power $n$ is small
(e.g., $n\leqslant4$ when $p=3$).

\section{Krylov-subspace diagonalization}\label{sec:ksd}

As an application of the quantum power method,
here we consider the Krylov-subspace diagonalization. 
We first 
define a block Krylov subspace and review
the subspace-diagonalization scheme~\cite{Chatelin}. 
We then describe how the quantum power method 
is combined with the Krylov-subspace diagonalization. 
Other applications of the quantum power method are outlined in 
Appendix~\ref{app:moment} and Appendix~\ref{app:lanczos}.

\subsection{Block Krylov subspace}

The block Krylov subspace of the Hamiltonian 
$\hat{\mathcal{H}}$ with reference states $\{|q_k\rangle\}_{k=1}^{M_{\rm B}}$
is given as
\begin{widetext}
  \begin{equation}
    \mathcal{K}_{n}\left(\hat{\mathcal{H}},\{|q_k\rangle\}_{k=1}^{M_{\rm B}}\right)=
            {\rm span}
            \left(
            |q_1\rangle, \cdots |q_{M_{\rm B}}\rangle, \  
            \hat{\mathcal{H}} |q_1\rangle, \cdots, \hat{\mathcal{H}}|q_{M_{\rm B}} \rangle,\
            \cdots, \ 
            \hat{\mathcal{H}}^{n-1} |q_1\rangle, \cdots, \hat{\mathcal{H}}^{n-1}|q_{M_{\rm B}} \rangle
            \right),  
            \label{eq:ksub}
  \end{equation}
\end{widetext}
where we call $M_{\rm B} \geqslant 1$ the block size. 
We should note that the reference states $\{|q_k\rangle\}_{k=1}^{M_{\rm B}}$ do not have to be orthogonal 
to each other but they are linearly independent.
If $M_{\rm B}=1$, $\mathcal{K}_{n}\left(\hat{\mathcal{H}},\{|q_k\rangle\}_{k=1}^{M_{\rm B}}\right)$
reduces to the conventional Krylov subspace. 
By defining 
\begin{equation}
  |u_{i}\rangle = \hat{\mathcal{H}}^{l-1}|q_k \rangle, 
  \label{uk}
\end{equation}
with $i=k+(l-1)M_{\rm B}$ and $l=1,2,\cdots,n$, 
the block Krylov subspace can be written simply as 
$
\mathcal{K}_{n}\left(\hat{\mathcal{H}},\{|q_k\rangle\}_{k=1}^{M_{\rm B}}\right)=
        {\rm span}
        \left(
        \{
        |u_i\rangle\}_{i=1}^{nM_{\rm B}}
        \right).
        $

\subsection{Rayleigh-Ritz technique}

Suppose that the ground state $|\Psi_0\rangle$ 
of the Hamiltonian $\hat{\mathcal{H}}$, satisfying 
\begin{equation}
  \hat{\mathcal{H}} |\Psi_0 \rangle = E_0 |\Psi_0  \rangle   
  \label{eq:scheq}
\end{equation}
with $E_0$ being the ground-state energy,  
should be approximated
with the (nonorthonormal) basis states 
$\{|u_{i}\rangle\}_{i=1}^{n M_{\rm B}}$ in 
$\mathcal{K}_n(\hat{\mathcal{H}},\{|q_k\rangle\}_{k=1}^{M_{\rm B}})$
as
\begin{equation}
  |\Psi_0 \rangle \approx
  |\Psi_{\rm KS}\rangle \equiv
  \sum_{i=1}^{nM_{\rm B}} v_{i} |u_i\rangle, 
  \label{Psi_subspace}
\end{equation}
where $\{v_{i}\}_{i=1}^{nM_{\rm B}}$ are
the expansion coefficients to be determined. 

The expansion coefficients $\{v_{i}\}_{i=1}^{nM_{\rm B}}$
can be determined by minimizing
the energy expectation value
$\langle \Psi_{\rm KS} | \hat{\mathcal{H}}|\Psi_{\rm KS} \rangle$ 
under the constraint 
$ \langle \Psi_{\rm KS} |\Psi_{\rm KS} \rangle= 1$.
To this end, let us define the following function:
\begin{alignat}{1}
  \mathcal{F}(\bs{v},\bs{v}^*) &=
  \langle \Psi_{\rm KS} | \hat{\mathcal{H}}|\Psi_{\rm KS} \rangle 
  -\epsilon \left(\langle \Psi_{\rm KS} |\Psi_{\rm KS} \rangle -1
  \right) \notag \\
  &=\bs{v}^\dag \bs{H} \bs{v} 
  -\epsilon\left(\bs{v}^\dag \bs{S} \bs{v} -1\right) \notag \\
  &=\sum_{ij} v_i^* \left(H_{ij}-\epsilon S_{ij}\right) v_j + \epsilon,
\end{alignat}
where $\epsilon$ is a Lagrange multiplier, 
$[\bs{v}]_i=v_{i}$, 
\begin{equation}
  [\bs{H}]_{ij}=H_{ij}=\langle u_i |\hat{\mathcal{H}}| u_j \rangle
  \label{Hsubspace}
\end{equation}
is the subspace Hamiltonian matrix, and 
\begin{equation}
  [\bs{S}]_{ij}=S_{ij}=\langle u_i | u_j \rangle
  \label{Ssubspace}
\end{equation}
is the subspace overlap matrix.
Then, the condition $\partial \mathcal{F}/\partial {v_{i}^*} =0$ 
for $1 \leqslant i\leqslant nM_{\rm B}$ 
yields a generalized eigenvalue problem 
\begin{equation}
  \bs{H} \bs{v} = \epsilon \bs{S} \bs{v}.
  \label{geneig}
\end{equation}
Since both $\bs{H}$ and $\bs{S}$ are Hermitian, 
the condition 
$\partial \mathcal{F}/\partial {v_{i}} =0$
for $1 \leqslant i\leqslant nM_{\rm B}$
yields the same equation.
The lowest eigenvalue $\epsilon$ and 
the corresponding eigenvector $\bs{v}$ in Eq.~(\ref{geneig})
provide an approximation to the ground-state energy $E_0$ and 
the expansion coefficients $\{v_{i}\}_{i=1}^{nM_{\rm B}}$ in Eq.~(\ref{Psi_subspace}), 
respectively. 
Note that when $M_{\rm B}=1$, the matrices ${\bs{H}}$ and ${\bs{S}}$ correspond to the 
Hankel matrices $\bs{\mathcal{M}}_{n-1}$ and $\bs{\mathcal{L}}_{n-1}$, respectively,
defined in Eqs.~(\ref{eq:hankel_m}) and (\ref{eq:hankel_l}).

Since $\bs{S}$ is a Hermitian matrix,
it can be diagonalized by a unitary matrix 
$\bs{V}$ as 
\begin{equation}
  \bs{V}^\dag \bs{S} \bs{V} = \bs{s}, 
  \label{eq:ed}
\end{equation}
where $\bs{s}$ is the diagonal matrix that contains
the eigenvalues of $\bs{S}$.
Note that $\bs{s} > 0$ because $\bs{S}$ is 
a Gram matrix and hence is positive definite.
By using a matrix 
\begin{equation}
  \bs{W}=\bs{V}\bs{s}^{-1/2},  
\end{equation}
Eq.~(\ref{geneig}) can be transformed 
to a standard Hermitian eigenvalue problem of the form 
\begin{equation}
  \bs{T}\bs{q}=\epsilon \bs{q},
  \label{standardeig}
\end{equation}
where 
\begin{equation}
  \bs{T}\equiv\bs{W}^\dag \bs{H} \bs{W}
  \label{Tmat}
\end{equation}
and $\bs{q}=\bs{W}^{-1}\bs{v}$.
Thus, by solving the eigenvalue problem of Eq.~(\ref{standardeig}),
one can obtain $\epsilon$ and $\bs{v}=\bs{W}\bs{q}$. 
The eigenvector $\bs{v}$ with the lowest eigenvalue $\epsilon$ 
provides the coefficients in the approximate ground state
$|\Psi_{\rm KS}\rangle$ [see Eq.~(\ref{Psi_subspace})] 
with its energy $E_{\rm KS}$ of the 
Hamiltonian $\hat{\mathcal{H}}$  
in the Krylov subspace 
$\mathcal{K}_{n}\left(\hat{\mathcal{H}},\{|q_k\rangle\}_{k=1}^{M_{\rm B}}\right)$. 

We note that if we use the Cholesky decomposition $\bs{S}=\bs{R}^\dag\bs{R}$ with $\bs{R}$ being an upper-triangular matrix, 
instead of the eigen decomposition in Eq.~(\ref{eq:ed}), $\bs{T}$ reduces 
to the tridiagonal matrix in the Lanczos method when $M_{\rm B}=1$~\cite{Chatelin}.

\subsection{Quantum-classical-hybrid Krylov-subspace method} \label{sec:ksm}

Considering the Rayleigh-Ritz technique in 
a quantum-classical-hybrid computation,
it is suited for quantum hardware to evaluate the matrix elements of
$\bs{H}$ in Eq.~(\ref{Hsubspace}) and 
$\bs{S}$ in Eq.~(\ref{Ssubspace}), 
because the states $\{|u_i\rangle\}_{i=1}^{nM_{\rm B}}$ 
are defined on the Hilbert space of $N_{\rm D}=2^N$ dimensions,
for example, for the spin-$1/2$ Heisenberg model. 
On the other hand,
the eigenvalue problem in the $n M_{\rm B}$-dimensional block Krylov subspace 
given in Eq.~(\ref{geneig}) or Eq.~(\ref{standardeig}) 
can be solved on classical computers,
assuming that the Krylov subspace 
approximates reasonably well
the eigenspace of the ground state 
with relatively small $n$ and $M_{\rm B}$, 
despite that the dimension $N_{\rm D}$ of the full Hilbert space could be much larger than $nM_{\rm B}$. 
This feature is shared with 
other quantum-classical-hybrid subspace-diagonalization 
schemes reported previously~\cite{McClean2017,Colless2018,Parrish2019,nakanishi2018subspacesearch,heya2019subspace,huggins2019nonorthogonal}.

We can now approximate the Hamiltonian power
$\hat{\mathcal{H}}^{l-1}$ appearing in the Krylov-subspace basis 
$|u_{i} \rangle$ given in Eq.~(\ref{uk}) as 
\begin{equation}
|u_{i}\rangle = |\tilde{u}_{i}\rangle + O(\Delta_\tau^{2+2r}) + O(\Delta_\tau^{2m+2r}), 
\label{eq:ui0}
\end{equation}
where 
\begin{equation}
  |\tilde{u}_{i}\rangle = \hat{\mathcal{H}}^{l-1}_{{\rm ST}(r)}(\Delta_\tau)|q_k \rangle  
  \label{eq:ui}
\end{equation}
with $i=k+(l-1)M_{\rm B}$ 
for $1 \leqslant k \leqslant M_{\rm B}$ and $1 \leqslant l \leqslant n$.
Note that the systematic errors in Eq.~(\ref{eq:ui0}) are absent when $l=1$. 
As described in Sec.~\ref{sec:qpm}, to approximate the Hamiltonian power $\hat{\mathcal{H}}^{l-1}$ 
by $\hat{\mathcal{H}}^{l-1}_{{\rm ST}(r)}(\Delta_\tau)$ as in Eq.~(\ref{eq:ui}),
the Suzuki-Trotter-decomposed time-evolution operators $\hat{S}_{2m}^{(p)}(\pm\Delta_\tau/2)$ have to be applied 
at most $l-1$ times to a state $|q_k \rangle$. 
This implies that the circuit depth required
for constructing the block Krylov subspace
$\mathcal{K}_n\left(\hat{\mathcal{H}}_{{\rm ST}(r)}(\Delta_\tau),\{|q_k\rangle\}_{k=1}^{M_{\rm B}}\right)$ 
is at most $O(n)$ with a prefactor of $D_{2m}^{(p)}$.
The circuit depth does not depend on the order $r$ of the Richardson extrapolation.  

With the basis states defined in Eq.~(\ref{eq:ui0}),
the subspace Hamiltonian matrix and the overlap matrix are approximated respectively as 
\begin{equation}
  {H}_{ij} = \tilde{H}_{ij} + O(\Delta_\tau^{2+2r}) + O(\Delta_\tau^{2m+2r}) 
\end{equation}
and 
\begin{equation}
  S_{ij} = \tilde{S}_{ij} + O(\Delta_\tau^{2+2r}) + O(\Delta_\tau^{2m+2r}),
\end{equation}
where 
\begin{equation}
  [\tilde{\bs{H}}]_{ij}=
  \tilde{H}_{ij}=
  \langle \tilde{u}_i |\hat{\mathcal{H}}| \tilde{u}_j \rangle
  = \langle q_k|\hat{\mathcal{H}}_{{\rm ST}(r)}^{l-1}(\Delta_\tau)\,  \hat{\mathcal{H}}\,  \hat{\mathcal{H}}_{{\rm ST}(r)}^{l'-1}(\Delta_\tau) |q_{k'}\rangle
  \label{Hsubspace1}
\end{equation}
and
\begin{equation}
  [\tilde{\bs{S}}]_{ij}=
  \tilde{S}_{ij}=\langle \tilde{u}_i | \tilde{u}_j \rangle 
  = \langle q_k|\hat{\mathcal{H}}_{{\rm ST}(r)}^{l-1}(\Delta_\tau) \, \hat{\mathcal{H}}_{{\rm ST}(r)}^{l'-1}(\Delta_\tau) |q_{k'}\rangle
  \label{Ssubspace1}
\end{equation}  
with $i=k+(l-1)M_{\rm B}$ and $j=k'+(l'-1)M_{\rm B}$ for 
$1 \leqslant k,k' \leqslant M_{\rm B}$ and $1 \leqslant l,l' \leqslant n$ 
in the block Krylov subspace 
$\mathcal{K}_n\left(\hat{\mathcal{H}}_{{\rm ST}(r)}(\Delta_\tau),\{|q_k\rangle\}_{k=1}^{M_{\rm B}}\right)$. 
Here, the Hermiticity of the approximated Hamiltonian power $\hat{\mathcal{H}}_{{\rm ST}(r)}^{l-1}(\Delta_\tau)$ 
in  Eq.~(\ref{eq:hir}) is used.
Note also that 
$\hat{\mathcal{H}}_{{\rm ST}(r)}^{l-1}(\Delta_\tau)  \hat{\mathcal{H}}_{{\rm ST}(r)}^{l'-1}(\Delta_\tau)\ne 
\hat{\mathcal{H}}_{{\rm ST}(r)}^{l+l'-2}(\Delta_\tau)$ for $r\geqslant 1$, 
but this equation is satisfied when $r=0$.

More specifically, 
${\tilde{H}}_{ij}$ and ${\tilde{S}}_{ij}$ 
in terms of $\hat{S}_{2m}^{(p)}(\pm\Delta_\tau/2)$ without the Richardson extrapolation are given respectively as 
\begin{alignat}{1}
  \tilde{H}_{ij}
  &=
  \sum_{\nu=0}^{l-1}
  \sum_{\nu'=0}^{l'-1}
  c_{l-1,\nu}^*
  c_{l'-1,\nu'}
  \notag \\
  &
  \times
  \langle q_k |
  \left[\hat{S}_{2m}^{(p)}\left(-\frac{\Delta_\tau}{2}\right)\right]^{l-1-2\nu}
  \hat{\mathcal{H}}
  \left[\hat{S}_{2m}^{(p)}\left(\frac{\Delta_\tau}{2}\right)\right]^{l'-1-2\nu'}
  |q_{k'}\rangle
  \label{HsubspaceS}
\end{alignat}
and 
\begin{alignat}{1}
  \tilde{S}_{ij}
  &=
  \sum_{\nu=0}^{l-1}
  \sum_{\nu'=0}^{l'-1}
  c_{l-1,\nu}^*
  c_{l'-1,\nu'}
  \notag \\
  &
  \times
  \langle q_k |
  \left[\hat{S}_{2m}^{(p)}\left(-\frac{\Delta_\tau}{2}\right)\right]^{l-1-2\nu}
  \left[\hat{S}_{2m}^{(p)}\left(\frac{\Delta_\tau}{2}\right)\right]^{l'-1-2\nu'}
  | q_{k'} \rangle.
  \label{SsubspaceS}
\end{alignat}
Note that the Suzuki-Trotter-decomposed time-evolution operators 
$[\hat{S}_{2m}^{(p)}(-\Delta_\tau/2]^{l-1-2\nu}$ and $[\hat{S}_{2m}^{(p)}(\Delta_\tau/2]^{l'-1-2\nu'}$
in Eq.~(\ref{SsubspaceS}) can be combined exactly as in the form shown in Eq.~(\ref{Ssubspace3}) 
when the Richardson extrapolation is not used. 
However, here we deliberately do not combine these two terms because it is helpful when the extension for the 
$r$th-order Richardson extrapolation is considered. 
Assuming that $\mathcal{\hat{H}}$ consists of $O(N)$ local terms, 
the number of state overlaps required 
for constructing all matrix elements of $\tilde{\bs{H}}$ and $\tilde{\bs{S}}$ 
is $O(n^2 M_{\rm B}^2 N)$ and $O(n^2 M_{\rm B}^2)$, respectively. 
If the $r$th-order Richardson extrapolation
is employed, the number of state overlaps
to be evaluated is increased by a factor of $(r+1)^2$. 
The state overlaps in 
Eqs.~(\ref{HsubspaceS}) and (\ref{SsubspaceS}) 
can be evaluated with an Hadamard-test-like circuit, 
for example~\cite{Romero2018,Dallaire-Demers2019,McArdle2019,Stair2020}.

However, for the purpose of solving the generalized eigenvalue problem in Eq.~(\ref{geneig})
or the corresponding 
standard eigenvalue problem in Eq.~(\ref{standardeig}), one could
evaluate the matrix elements in Eqs.~(\ref{Hsubspace}) and (\ref{Ssubspace}) 
more directly as 
\begin{equation}
  {H}_{ij} = \tilde{H}_{ij}^\prime + O(\Delta_\tau^{2+2r}) + O(\Delta_\tau^{2m+2r}) 
\end{equation}
and 
\begin{equation}
  S_{ij} = \tilde{S}_{ij}^\prime + O(\Delta_\tau^{2+2r}) + O(\Delta_\tau^{2m+2r}),
\end{equation}
where 
\begin{equation}
    [\tilde{\bs{H}}^\prime]_{ij}=
  \tilde{H}_{ij}^\prime= \langle q_k|\hat{\mathcal{H}}^{l+l'-1}_{{\rm ST}(r)}(\Delta_\tau) |q_{k'}\rangle  \label{Hsubspace2}
\end{equation}
and 
\begin{equation}
    [\tilde{\bs{S}}^\prime]_{ij}=
  \tilde{S}_{ij}^\prime =\langle q_k | \hat{\mathcal{H}}^{l+l'-2}_{{\rm ST}(r)}(\Delta_\tau) |q_{k'}\rangle  
  \label{Ssubspace2}
\end{equation}
with $i=k+(l-1)M_{\rm B}$ and $j=k'+(l'-1)M_{\rm B}$ for
$1 \leqslant k,k' \leqslant M_{\rm B}$ and
$1 \leqslant l,l' \leqslant n$  
in the block Krylov subspace
$\mathcal{K}_n\left(\hat{\mathcal{H}},\{|q_k\rangle\}_{k=1}^{M_{\rm B}}\right)$.  
To be more specific, the matrix elements of $\tilde{\bs{H}}^\prime$ 
and $\tilde{\bs{S}}^\prime$
for $r=0$,
i.e., without the Richardson extrapolation, are given as
\begin{alignat}{1}
  \tilde{H}_{ij}^\prime
  &=
  \sum_{\nu=0}^{l+l'-1}
  c_{l+l'-1,\nu}
  \langle q_k |
  \left[\hat{S}_{2m}^{(p)}\left(\Delta_\tau/2\right)\right]^{l+l'-1-2\nu}
  |q_{k'}\rangle
  \label{Hsubspace3}
\end{alignat}
and 
\begin{alignat}{1}
  \tilde{S}_{ij}^\prime
  =
  \sum_{\nu=0}^{l+l'-2}
  c_{l+l'-2,\nu}
  \langle q_k |
  \left[\hat{S}_{2m}^{(p)}\left(\Delta_\tau/2\right)\right]^{l+l'-2-2\nu}
  |q_{k'}\rangle.
  \label{Ssubspace3}
\end{alignat}
The number of state overlaps required 
for constructing all matrix elements of
both $\tilde{\bs{H}}^\prime$ and $\tilde{\bs{S}}^\prime$ 
is thus $O(n M_{\rm B}^2)$. 
If the $r$th-order Richardson extrapolation
is employed, the number of state overlaps
to be evaluated is increased by a factor of $(r+1)$.

Therefore, the approach based on Eqs.~(\ref{Hsubspace2}) and (\ref{Ssubspace2}) 
is better than that based on Eqs.~(\ref{Hsubspace1}) and (\ref{Ssubspace1}) in the sense that  
fewer state overlaps are
required to approximately solve the Krylov-subspace diagonalization. 
However, although these two approaches are equivalent within the systematic errors, the approach 
based on Eqs.~(\ref{Hsubspace2}) and (\ref{Ssubspace2}) loses the exact meaning of
the variational principle for the ground state obtained by solving the (generalized) eigenvalue problem 
(also see Ref.~\cite{parrish2019quantum}).
This is because 
the approach based on Eqs.~(\ref{Hsubspace1}) and (\ref{Ssubspace1}) respects the subspace structure, 
which is generated by the Krylov 
subspace $\mathcal{K}_n\left(\hat{\mathcal{H}}_{{\rm ST}(r)}(\Delta_\tau),\{|q_k\rangle\}_{k=1}^{M_{\rm B}}\right)$, 
as opposed to the other approach, 
and thus the lowest eigenstate of the the (generalized) eigenvalue problem with the matrix elements 
in Eqs.~(\ref{Hsubspace1}) and (\ref{Ssubspace1}) satisfies exactly the variational principle. 
In addition, we find that the approach based on Eqs.~(\ref{Hsubspace1}) and (\ref{Ssubspace1}) is more stable for 
numerical simulations. Therefore, we adopt the approach based on Eqs.~(\ref{Hsubspace1}) and (\ref{Ssubspace1}) 
in our numerical simulations shown in this paper
unless otherwise stated. 
In Appendix~\ref{app:ksd}, our method described here is compared with other algorithms for the Krylov-subspace diagonalization.

\section{Numerical demonstration}\label{sec:result}

In this section,
we demonstrate the quantum power method 
by numerically simulating a spin-$1/2$ Heisenberg model. 
We first define the Hamiltonian of the Heisenberg model,
and then show how the quantum power method can control the 
systematic errors in approximating the Hamiltonian power $\mathcal{\hat{H}}^n$.
Next, as an application of the quantum power method,
we show the numerical results of the multireference Krylov-subspace
diagonalization combined with the quantum power method for the Heisenberg model. 
The numerical results of the multireference Krylov-subspace diagonalization 
combined with the quantum power method 
for a Fermi-Hubbard model, which involves 
more technical details, 
are also provided in Appendix~\ref{app:Hubbard}.

\subsection{Heisenberg model}\label{sec:model}

The spin-1/2 Heisenberg model is described by the following Hamiltonian:  
\begin{eqnarray}
  \hat{\mathcal{H}} =
  \frac{J}{4}
  \sum_{\langle i,j \rangle} 
  \left(
  \hat{I}_i \hat{I}_j +
  \hat{X}_i \hat{X}_j +
  \hat{Y}_i \hat{Y}_j +
  \hat{Z}_i \hat{Z}_j  
  \right)
  =
  \frac{J}{2}
  \sum_{\langle i,j \rangle}
  \hat{\mathcal{P}}_{ij},
  \label{Ham_SWAP}
\end{eqnarray}
where
$J>0$ is the antiferromagnetic exchange interaction, 
$\langle i,j \rangle$
runs over all nearest-neighbor pairs of qubits $i$ and $j$ 
connected with the exchange interaction $J$, and 
$\{ \hat{X}_{i}, \hat{Y}_{i}, \hat{Z}_{i} \}$ and $\hat{I}_i$
are the Pauli operators and the identity operator
acting on the $i$th qubit. 
$\hat{\mathcal{P}}_{ij}$
is the {\sc swap} operator which 
acts on the $i$th and $j$th qubits as 
$\hat{\mathcal{P}}_{ij} |a\rangle_i |b \rangle_j =|b\rangle_i |a \rangle_j $.
In the Hamiltonian in Eq.~(\ref{Ham_SWAP}), 
the constant (identity) term $\hat{I}_i \hat{I}_j$ is added to
the conventional Heisenberg Hamiltonian 
and thus $\hat{\mathcal{H}}$ is simply a sum of {\sc swap} operators.
Indeed, the second equality in Eq.~(\ref{Ham_SWAP}) follows
from the identity
\begin{equation}
  \hat{I}_i \hat{I}_j +
  \hat{X}_i \hat{X}_j +
  \hat{Y}_i \hat{Y}_j +
  \hat{Z}_i \hat{Z}_j  
  =
  2
  \hat{\mathcal{P}}_{ij}
  \label{SWAP}
\end{equation}
for $i\not=j$.

We consider the Hamiltonian $\hat{\mathcal{H}}$ on
a one-dimensional periodic chain with $N$ sites (i.e., $N$ qubits),
and assume that $N$ is even. 
Then, the Hamiltonian can be written as
\begin{equation}
  \hat{\mathcal{H}} =\frac{J}{2}\sum_{i=1}^{N} \hat{\mathcal{P}}_{i,i+1}, 
  \label{Ham_SWAP1D}
\end{equation}
where $i+1$ in the subscript should be read as $1$ if $i=N$
due to the periodic boundary conditions.
For the use in the Suzuki-Trotter decomposition of the time-evolution operator, 
we divide the Hamiltonian into two parts (i.e., $N_\Gamma=2$) as
\begin{equation}
  \hat{\mathcal{H}}=\hat{\mathcal{H}}_{A}+\hat{\mathcal{H}}_{B}
  \label{eq:hamAB}
\end{equation}
with 
\begin{equation}
  \hat{\mathcal{H}}_{A}=\frac{J}{2}\sum_{i=1}^{N/2} \hat{\mathcal{P}}_{2i,2i+1}
\end{equation}
and
\begin{equation}
  \hat{\mathcal{H}}_{B}=\frac{J}{2}\sum_{i=1}^{N/2} \hat{\mathcal{P}}_{2i-1,2i}.
\end{equation}
Notice that
$[\hat{\mathcal{P}}_{2i,2i+1}, \hat{\mathcal{P}}_{2j,2j+1}]=[\hat{\mathcal{P}}_{2i-1,2i},  \hat{\mathcal{P}}_{2j-1,2j}]=0$, 
where $[\hat{A},\hat{B}]=\hat{A}\hat{B}-\hat{B}\hat{A}$ is the commutator of two operators $\hat{A}$ and $\hat{B}$.

For the one-dimensional spin-1/2 Heisenberg model 
$\mathcal{\hat{H}}=\mathcal{\hat{H}}_A+\mathcal{\hat{H}}_B$ 
given in Eqs.~(\ref{Ham_SWAP}) and (\ref{eq:hamAB}), 
the time-evolution operator $\hat{U}(t) = \e^{-\imag \hat{\mathcal{H}}t}$ is constituted 
by two elementary time-evolution operators associated
with $\mathcal{\hat{H}}_A$ and $\mathcal{\hat{H}}_B$.  
Let us first introduce the exponential-{\sc swap} (e-{\sc swap}) gate
$\hat{\mathcal{U}}_{i,j}$~\cite{Loss1998,DiVincenzo2000,Brunner2011,Lloyd2014,Lau2016} 
\begin{equation}
  \hat{\mathcal{U}}_{i,j}(\theta)
  =\exp(-\imag \theta \hat{\mathcal{P}}_{i,j}/2),
  \label{eSWAP}
\end{equation}
where $\theta$ is a real-valued parameter. 
The e-{\sc swap} gate, 
which is equivalent to the {\sc swap}$^\alpha$ gate 
up to a two-qubit global phase factor~\cite{Fan2005,Balakrishnan2008,Gard2019,Liu2019},
is depicted schematically in Fig.~\ref{overview}(c) 
as a blue rectangular extended over two qubits.
The gate corresponding to Eq.~(\ref{eSWAP}) can be implemented with
three {\sc cnot} gates and few single-qubit rotations~\cite{Vidal2004,Chiesa2019,note_eSWAP}. 
The time-evolution operators of 
$\hat{\mathcal{H}}_{A}$ and $\hat{\mathcal{H}}_{B}$
are given respectively by
\begin{equation}
 \exp(-\imag \hat{\mathcal{H}}_A t)=\prod_{i=1}^{N/2} \hat{\mathcal{U}}_{2i,2i+1}(tJ) \label{UA} 
\end{equation}
and
\begin{equation}
  \exp(-\imag \hat{\mathcal{H}}_B t)=\prod_{i=1}^{N/2} \hat{\mathcal{U}}_{2i-1,2i}(tJ). \label{UB}  
\end{equation}
Since
$[\hat{\mathcal{U}}_{2i,2i+1},\hat{\mathcal{U}}_{2j,2j+1}]=0$ and 
$[\hat{\mathcal{U}}_{2i-1,2i},\hat{\mathcal{U}}_{2j-1,2j}]=0$ for $i \not=j$, 
the order of the product is not relevant in Eqs.~(\ref{UA}) and (\ref{UB}). 
Figure~\ref{overview}(c) illustrates a typical circuit structure  
that approximates the time-evolution operator $\hat{U}(\Delta_\tau)$, consisting of a product of 
$\exp(-\imag \hat{\mathcal{H}}_A \Delta_\tau s_i)$'s and 
$\exp(-\imag \hat{\mathcal{H}}_B \Delta_\tau s_i)$'s 
with real parameters $\{s_i\}$. 
As in Eq.~(\ref{ST2}) for the general case, 
the lowest-order
symmetric Suzuki-Trotter decomposition of $\hat{U}(\Delta_\tau)$
for the bipartitioned Hamiltonian
$\mathcal{\hat{H}}=\mathcal{\hat{H}}_A+\mathcal{\hat{H}}_B$ 
is given by 
\begin{equation}
  \hat{S}_2(\Delta_\tau)=
  \e^{-\imag \frac{\Delta_\tau}{2}\hat{\mathcal{H}}_{A}}
  \e^{-\imag \Delta_\tau    \hat{\mathcal{H}}_{B}}
  \e^{-\imag \frac{\Delta_\tau}{2}\hat{\mathcal{H}}_{A}}.
  \label{ST2AB}
\end{equation}

\subsection{Degree of approximation}\label{sec:error}

\begin{center}
  \begin{figure*}
    \includegraphics[width=1.95\columnwidth]{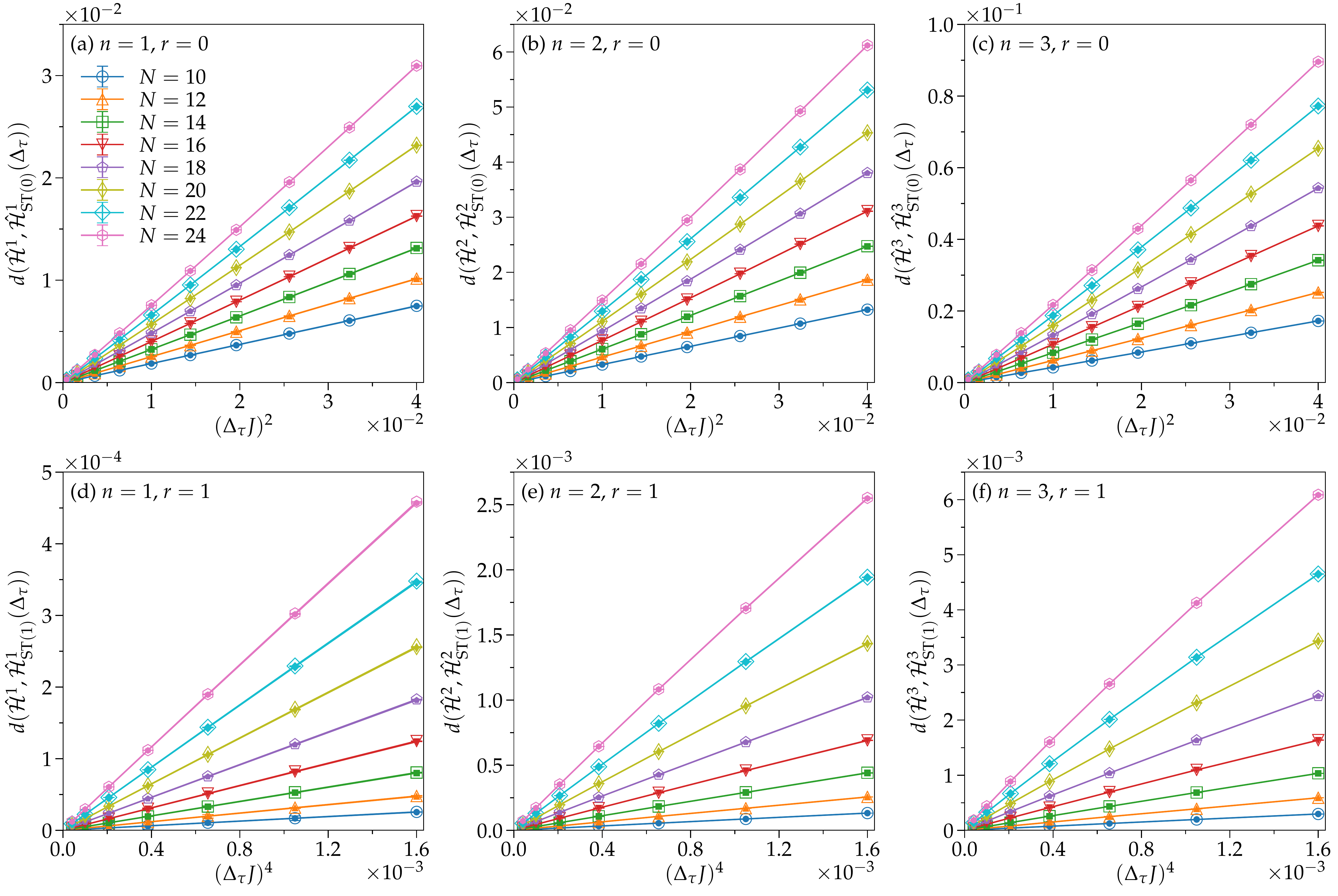}
    \caption{
      (a)--(c): 
      Distance $d\left( \hat{\mathcal{H}}^n, \hat{\mathcal{H}}^n_{{\rm ST}(r)}(\Delta_\tau) \right)$ between 
      the exact Hamiltonian power $\hat{\mathcal{H}}^n$ and
      the approximated Hamiltonian power 
      $\hat{\mathcal{H}}_{{\rm ST}(r)}^n(\Delta_\tau)$ given in Eq.~(\ref{HSTLC})
      without the Richardson extrapolation ($r=0$)
      as a function of $\Delta_\tau^2$ for
      (a) $n=1$,
      (b) $n=2$, and 
      (c) $n=3$.
      (d)--(f):
      Same as (a)--(c) but 
      with the first-order Richardson extrapolation ($r=1$) given in Eq.~(\ref{eq:Richardson}) 
      as a function of $\Delta_\tau^4$.  
      The Hamiltonian $\mathcal{\hat{H}}$ is for the spin-1/2 Heisenberg model on an $N$-qubit ring given in Eq.~(\ref{Ham_SWAP}). 
      The symmetric Suzuki-Trotter decompositions  
      $\hat{S}_{2}$ (empty symbols) and
      $\hat{S}_{4}^{(3)}$ (filled small symbols) are used.
      However, these results are on top of each other, as is expected. 
      The error bar indicates standard error of the mean. 
      The solid lines are guide for the eye.
      Note that each panel employs a different scale in the $y$ axis.
      \label{fig.norm}
    }
  \end{figure*}
\end{center}

We first examine quantitatively
how the Hamiltonian power $\hat{\mathcal{H}}^{n}$ is
approximated by $\hat{\mathcal{H}}_{{\rm ST}(r)}^{n}(\Delta_\tau)$.
For this purpose, we define a distance $d(\hat{A},\hat{B})$
between operators $\hat{A}$ and $\hat{B}$ as  
\begin{equation}
  d(\hat{A},\hat{B})
  =
  \sqrt{
  1- 
  \frac{
    \left|
    \left\langle 
    \hat{A}, \hat{B}
    \right\rangle_{\rm F}
    \right|
  }
       {
         \left|\left|
         \hat{A}
         \right|\right|_{\rm F}
         \left|\left|
         \hat{B}
         \right|\right|_{\rm F}
       }
  },
  \label{distance}
\end{equation}
where $\left\langle \hat{A}, \hat{B} \right\rangle_{\rm F}$ denotes
the Frobenius inner product between $\hat{A}$ and $\hat{B}$ defined by 
\begin{equation}
  \left\langle 
  \hat{A}, \hat{B}
  \right\rangle_{\rm F}
  =
  {\rm Tr}\left[
  \hat{A}^\dag
  \hat{B}
  \right]
  \label{Frobenius_inner}
\end{equation}
and $||\hat{A}||_{\rm F}$ denotes
the Frobenius norm of $\hat{A}$, i.e., 
\begin{equation}
  \left|\left|
  \hat{A}
  \right|\right|_{\rm F}
  =
  \sqrt{
  {\rm Tr}\left[
  \hat{A}^\dag
  \hat{A}
  \right]
  }.
  \label{Frobenius_norm}
\end{equation}
Note that
$\langle \hat{A},\hat{A}\rangle_{\rm F}=||\hat{A}||_{\rm F}^2$, 
$0\leqslant |\langle \hat{A},\hat{B}\rangle_{\rm F}| \leqslant ||\hat{A}||_{\rm F} ||\hat{B}||_{\rm F}$, 
$0\leqslant d(\hat{A},\hat{B}) \leqslant 1$, 
$d(\hat{A},\hat{B}) = d(a\hat{A},b\hat{B})$ with $a$ and $b$ being nonzero complex numbers,
and $d(\hat{A},\hat{B})=0$ if and only if $\hat{A}=\hat{B}$.
We compute the distance $d(\hat{A},\hat{B})$ for 
$\hat{A}=\hat{\mathcal{H}}^n$ and
$\hat{B}=\hat{\mathcal{H}}_{{\rm ST}(r)}^n(\Delta_\tau)$ 
given in Eq.~(\ref{HSTLC}) for $r=0$ 
and Eq.~(\ref{eq:Richardson}) for $r\geqslant 1$. 
The Hamiltonian $\mathcal{\hat{H}}$
is for the spin-1/2 Heisenberg model on an $N$-qubit ring  
given in Eq.~(\ref{Ham_SWAP}).

Evaluating the distance is costly as it demands 
matrix-matrix multiplications or diagonalizations. 
To avoid such costly operations, 
we employ a stochastic evaluation of the trace 
as~\cite{Drabold1993,Hams2000,IItaka2004,Weisse2006,Seki2020ftlm} 
\begin{equation}
  {\rm Tr}\left[
    \hat{\mathcal{X}}
  \right]
  =
  \lim_{R\to\infty}
  \frac{1}{R}
  \sum_{\zeta=1}^{R}\langle \phi_\zeta |\hat{\mathcal{X}}| \phi_\zeta \rangle, 
  \label{Trace}
\end{equation}
where 
$\hat{\mathcal{X}}\in\left\{\hat{A}^\dag \hat{A},\hat{B}^\dag \hat{B},\hat{A}^\dag \hat{B}\right\}$ 
and 
\begin{equation}
  |\phi_\zeta\rangle =
  \sum_{x} \e^{\imag \phi_{\zeta}(x)} |x\rangle 
\end{equation}
is a random-phase state 
with $\{|x\rangle\}$ being a complete orthonormal basis set
such that $\langle x|x'\rangle=\delta_{xx'}$ and
$\phi_\zeta(x)$ being a random variable
drawn uniformly from $[0,2\pi)$.
Note that 
$\langle \phi_\zeta|\phi_\zeta\rangle=2^N$, i.e., the dimension $N_{\rm D}$ of the Hilbert space.
We choose $\{|x\rangle\}$
as the orthonormal basis set that diagonalizes the local Pauli $Z$ operators.
The stochastic evaluation of the trace in Eq.~(\ref{Trace}) 
requires only sparse matrix-vector multiplications
and a single inner-product calculation for each $\zeta$, 
if $\hat{\mathcal{X}}$ is represented as a product of sparse matrices,
which is indeed the case here. 
Instead of taking the limit $R\to \infty$,
we fix $R=16$ for $N\geqslant 12$ and $R=256$ for $N=10$
and estimate error bars. 
Since
$\langle \phi_\zeta |\hat{A}^\dag \hat{A} |\phi_\zeta \rangle$,
$\langle \phi_\zeta |\hat{B}^\dag \hat{B} |\phi_\zeta \rangle$, and
$\langle \phi_\zeta |\hat{A}^\dag \hat{B} |\phi_\zeta \rangle$  
for 
$\hat{A}=\hat{\mathcal{H}}^n$ and
$\hat{B}=\hat{\mathcal{H}}^n_{{\rm ST}(r)}(\Delta_\tau)$ 
are highly correlated to each other,
error bars of $d(\hat{A},\hat{B})$
must be estimated using the corresponding
$3 \times 3$ covariance matrix.

Figure~\ref{fig.norm} shows the distance
as a function of $\Delta_\tau$ for $n=1,2,$ and $3$
with $N=10, 12, 14, 16, 18, 20, 22$ and $24$
using the symmetric Suzuki-Trotter decompositions 
$\hat{S}_{2}$ and $\hat{S}_{4}^{(3)}$.
Figures~\ref{fig.norm}(a)--\ref{fig.norm}(c) show the results
without the Richardson extrapolation ($r=0$). 
Since the leading systematic error in
$\hat{\mathcal{H}}^n_{{\rm ST}(r=0)}(\Delta_\tau)$ 
is  $O(\Delta_\tau^2)$, 
the distance scales almost linearly in $\Delta_\tau^2$ for each $N$.  
The distance simply increases with increasing $N$ and $n$. 
Figures~\ref{fig.norm}(d)--\ref{fig.norm}(f) show the results
with the first-order Richardson extrapolation ($r=1$). 
For each $n$, the distance with the Richardson extrapolation
is an order of magnitude smaller than that without the Richardson extrapolation. 
The leading systematic error in $\hat{\mathcal{H}}^n_{{\rm ST}(r=1)}(\Delta_\tau)$ 
is $O(\Delta_\tau^4)$, 
and the distance indeed scales almost linearly in $\Delta_\tau^4$.
As expected from Eq.~(\ref{QPWr}), essentially
no difference can be found between the results with 
$\hat{S}_{2}$ and 
$\hat{S}_{4}^{(3)}$, indicated respectively by empty and filled symbols in Fig.~\ref{fig.norm}.
These results clearly demonstrate that the systematic errors in
approximating the Hamiltonian power $\mathcal{\hat{H}}^n$ are 
well controlled.

\begin{center}
  \begin{figure}
    \includegraphics[width=1.0\columnwidth]{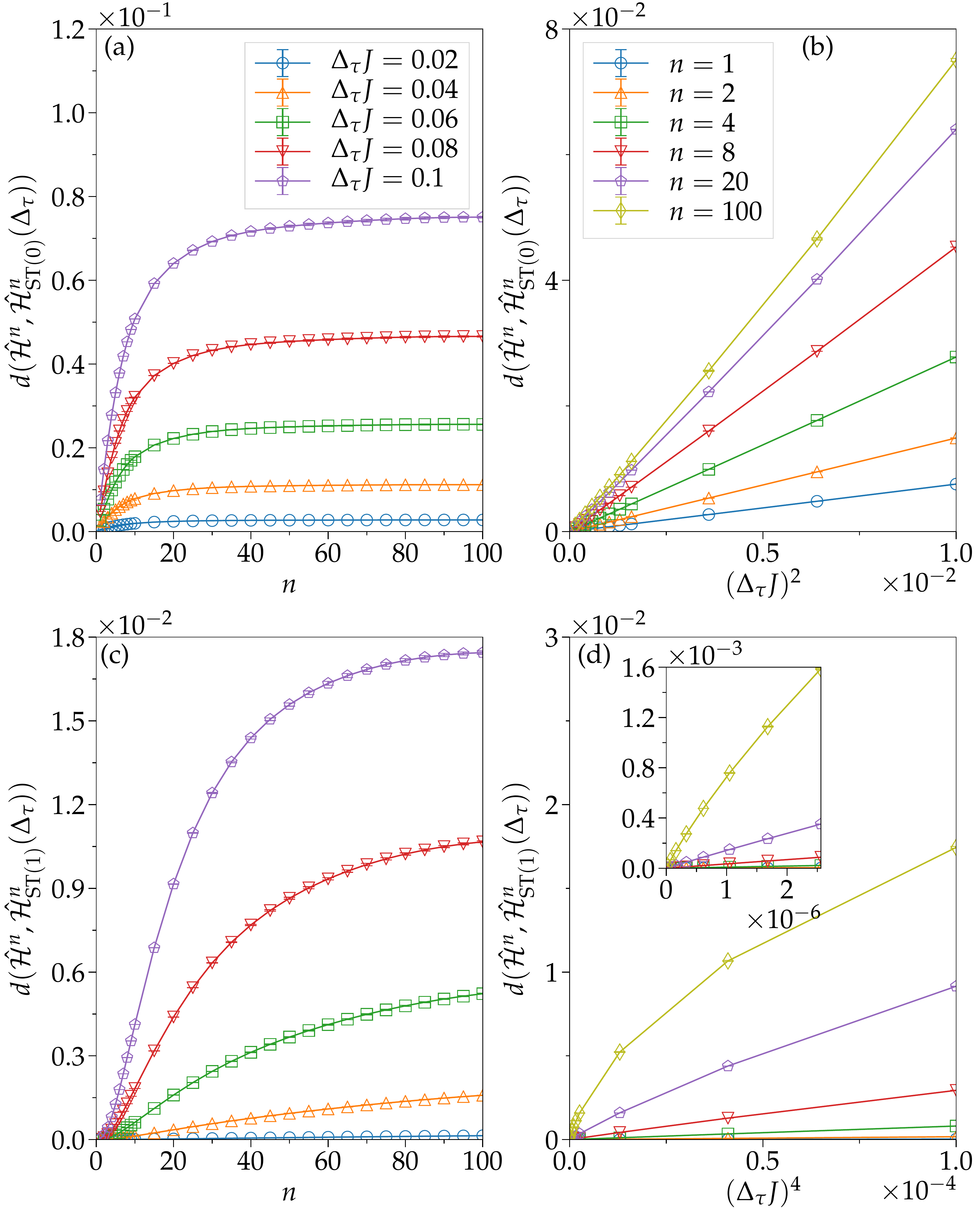}
    \caption{
      (a) Distance
      $d\left( \hat{\mathcal{H}}^n, \hat{\mathcal{H}}^n_{{\rm ST}(r)}(\Delta_\tau) \right)$ between 
      the exact Hamiltonian power $\hat{\mathcal{H}}^n$ and
      the approximated Hamiltonian power 
      $\hat{\mathcal{H}}_{{\rm ST}(r)}^n(\Delta_\tau)$ given in Eq.~(\ref{HSTLC}) 
      without the Richardson extrapolation ($r=0$)
      as a function of $n$ for different values of $\Delta_\tau$. 
      (b) Same as (a) but as a function of $\Delta_\tau^2$ for different values of the power $n$. 
      (c) Same as (a) but with the first-order Richardson extrapolation ($r=1$) given in Eq.~(\ref{eq:Richardson}). 
      (d) Same as (c) but as a function of $\Delta_\tau^4$ for different values of the power $n$.
      The inset in (d) shows the enlarged plot for
      $\Delta_\tau J \leqslant 0.04$.
      The Hamiltonian $\mathcal{\hat{H}}$ is for the spin-1/2 Heisenberg model on an 
      $N=24$ qubit ring 
      given in Eq.~(\ref{Ham_SWAP}). 
      The lowest-order symmetric Suzuki-Trotter decomposition 
      $\hat{S}_{2}$ is used. 
      The error bar indicates the standard error of the mean.
      The solid lines are guide for the eye.
      Note that each panel employs a different scale in the $y$ axis.
    } \label{fig.norm_ndep}
  \end{figure}
\end{center}

Figure~\ref{fig.norm_ndep}(a) 
shows the $n$ dependence of the distance
for $N=24$
with various values of $\Delta_\tau$ 
calculated using  
the lowest-order symmetric Suzuki-Trotter decomposition $\hat{S}_{2}$. 
The distance first increases with $n$ and tends to saturate at $n \sim 100$. 
It is remarkable to find in Fig.~\ref{fig.norm_ndep}(b) that,  
even with the large power exponents as large as $n=100$, 
the linear dependence of the distance on $\Delta_\tau^2$  
remains in a wide range of $\Delta_\tau$ ($\Delta_\tau J \alt 0.1$) 
and the distance is smoothly extrapolated to zero in the limit of $\Delta_\tau\to0$, 
clearly demonstrating the controlled accuracy of the
quantum power method.
Figures~\ref{fig.norm_ndep}(c) and \ref{fig.norm_ndep}(d) show the same results but obtained by 
using the first-order Richardson extrapolation ($r=1$), for which the systematic errors in approximating the 
Hamiltonian power $\mathcal{\hat{H}}^n$ are expected to be $O(\Delta_\tau^4)$. 
Indeed, our numerical simulations find the linear dependence of distance on $\Delta_\tau^4$ 
for at least $\Delta_\tau J \alt0.05$
when $n=100$ [see the inset in Fig.~\ref{fig.norm_ndep}(d)].
Notice also that the distance itself becomes smaller
by a factor of approximately $5$ even for large $n$  
when the first-order Richardson extrapolation is employed.

\subsection{Krylov-subspace diagonalization}\label{sec:egs}

We now perform numerical simulations of 
the Krylov-subspace diagonalization 
combined with the quantum power method  
to calculate the ground-state energy and fidelity of 
the spin-1/2 Heisenberg model described by the Hamiltonian $\hat{\mathcal{H}}$ in Eq.~(\ref{Ham_SWAP}) 
on a periodic chain of $N=16$ sites (i.e, qubits).

Considering
the Krylov-subspace diagonalization 
as an application of the quantum power method
on near-term quantum computers,
it is crucial to reduce the circuit depth.
As discussed in Sec.~\ref{sec:qpm} and Sec.~\ref{sec:ksm},
the depth of the circuit required for constructing
the block Krylov subspace
$\mathcal{K}_n\left(\hat{\mathcal{H}}_{{\rm ST}(r)}(\Delta_\tau),\{|q_k\rangle\}_{k=1}^{M_{\rm B}}\right)$
scales as
$O(n)$ with a prefactor $D_{2m}^{(p)}$.
Since $m$ and $p$ in the symmetric Suzuki-Trotter decomposition $\hat{S}_{2m}^{(p)}$
can be set to the minimum values 
$m=1$ and $p=3$, at least for the system sizes
examined in the previous section including $N=16$, 
the primary objective here is to reduce the power $n$.
For this purpose,
we first describe the selection of the reference states, aiming that
the block Krylov subspace
$\mathcal{K}_n\left(\hat{\mathcal{H}}_{{\rm ST}(r)}(\Delta_\tau),\{|q_k\rangle\}_{k=1}^{M_{\rm B}}\right)$
spanned by these reference states can approximate reasonably well
the target subspace,
which in the present case is 
the eigenspace of the ground state of $\mathcal{\hat{H}}$. 
Then, we show by numerical simulations 
how the selection of the reference states 
affects the convergence to the ground state with $n$.

\subsubsection{Selection of reference states}
Equation~(\ref{Psi_subspace}) suggests 
that the ground state $|\Psi_0\rangle$ 
can be well approximated 
if the reference states 
$\{|q_k\rangle\}_{k=1}^{M_{\rm B}}$
are chosen so that 
these states have substantial overlap
with the exact ground state.
Therefore, as the reference states,  
we introduce the following product states 
for the subspace diagonalization:  
\begin{alignat}{1}
  &|q_1\rangle = |\Phi_{A}\rangle = \otimes_{i=1}^{N/2} |s_{2i,2i+1}\rangle,  \label{eq:phi_a}\\
  &|q_2\rangle = |\Phi_{B}\rangle = \otimes_{i=1}^{N/2} |s_{2i-1,2i}\rangle,\\
  &|q_3\rangle = |X_{\rm AFM 1} \rangle = \otimes_{i=1}^{N/2} |+\rangle_{2i-1} |-\rangle_{2i},\\
  &|q_4\rangle = |X_{\rm AFM 2} \rangle = \otimes_{i=1}^{N/2} |+\rangle_{2i}   |-\rangle_{2i+1},\\
  &|q_5\rangle = |Y_{\rm AFM 1} \rangle = \otimes_{i=1}^{N/2} |R\rangle_{2i-1} |L\rangle_{2i},\\
  &|q_6\rangle = |Y_{\rm AFM 2} \rangle = \otimes_{i=1}^{N/2} |R\rangle_{2i}   |L\rangle_{2i+1},\\
  &|q_7\rangle = |Z_{\rm AFM 1} \rangle = \otimes_{i=1}^{N/2} |0\rangle_{2i-1} |1\rangle_{2i},\\
  &|q_8\rangle = |Z_{\rm AFM 2} \rangle = \otimes_{i=1}^{N/2} |0\rangle_{2i}   |1\rangle_{2i+1}. 
\end{alignat}
Here $|s_{i,j}\rangle=\frac{1}{\sqrt{2}}
(|0\rangle_i|1\rangle_j - |1\rangle_i|0\rangle_j)$ 
is the spin-singlet state and 
is an eigenstate of the {\sc swap} operator $\hat{\mathcal{P}}_{ij}$
with eigenvalue $-1$. It is also known as one of the Bell states.  
$|+\rangle_i=\frac{1}{\sqrt{2}}(|0\rangle_i + |1\rangle_i)$ and 
$|-\rangle_i=\frac{1}{\sqrt{2}}(|0\rangle_i - |1\rangle_i)$ are
the eigenstates of $\hat{X}_i$ with eigenvalues $\pm 1$, 
$|R\rangle_i=\frac{1}{\sqrt{2}}(|0\rangle_i + \imag |1\rangle_i)$ and 
$|L\rangle_i=\frac{1}{\sqrt{2}}(|0\rangle_i - \imag |1\rangle_i)$ are
the eigenstates of $\hat{Y}_i$ with eigenvalues $\pm 1$, and 
$|0\rangle_i$ and 
$|1\rangle_i$ are 
the eigenstates of $\hat{Z}_i$ with eigenvalues $\pm 1$.
$|\Phi_A\rangle$ and $|\Phi_B\rangle$ are the ground states of
$\hat{\mathcal{H}}_A$ and $\hat{\mathcal{H}}_B$, respectively, while 
others are the N\'{e}el states that are the ground states when
a mean-field theory is applied to the Hamiltonian. 
These product states
are expected to have a sizable overlap with
the exact ground state (also see Fig.~\ref{fig.fidelity}) and, moreover,  
they are easy to be prepared 
from $|0\rangle^{\otimes N}$ with appropriate combinations 
of Pauli, Hadamard, phase, and {\sc cnot} gates.

Another relevant candidate might be 
a variational state that has a substantial
overlap with the ground state. 
We thus introduce 
\begin{equation}
  |q_9\rangle=|\Psi_{\rm VQE}\rangle
  \label{eq:vqe}
\end{equation}
as another reference state,
where $|\Psi_{\rm VQE}\rangle$
is an approximate ground state 
prepared with a VQE scheme.
Specifically, 
we choose $|\Psi_{\rm VQE} \rangle$ as 
a resonating-valence-bond-type wave function without the symmetry projection operator, 
containing 64 optimized variational parameters for $N=16$ that do not reflect the spatial 
symmetry of the Hamiltonian, as reported in Ref.~\cite{Seki2020vqe}. 
While the exact ground-state energy is $E_0/NJ=-0.196393522$, 
our variational state $|\Psi_{\rm VQE} \rangle$ 
has the variational energy 
$\langle \Psi_{\rm VQE}|\hat{\mathcal{H}}|\Psi_{\rm VQE}\rangle/NJ=-0.1885$ (also see Fig.~\ref{fig.energy}) 
and the ground-state fidelity 
$|\langle \Psi_0|\Psi_{\rm VQE}\rangle|^2=0.771$ (also see Fig.~\ref{fig.fidelity}).

In our previous study~\cite{Seki2020vqe}, 
we have shown that
restoration of the spatial symmetry that is
broken by a circuit ansatz
greatly improves the ground-state-energy estimation
as well as the ground-state fidelity. 
Motivated by this finding, we introduce another set
of the reference states $\{|\bar{q}_k\rangle\}_{k=1}^N$ with
\begin{equation}
  |\bar{q}_k\rangle=\hat{\mathcal{T}}_{k-1} |\Psi_{\rm VQE}\rangle, 
\end{equation}
where
$\hat{\mathcal{T}}_{k}$ is a unitary operator representing
the one-dimensional $k$-lattice-space
translation with $\hat{\mathcal{T}}_0=\hat{I}$, and 
$ |\Psi_{\rm VQE}\rangle$ is the same state 
given in Eq.~(\ref{eq:vqe}).
With this set of the reference states, 
the translational symmetry that is broken in
the apparent circuit structure of $|\Psi_{\rm VQE}\rangle$
can be restored as a linear combination of the states
in the block Krylov subspace, 
without applying a projection operator to $|\Psi_{\rm VQE}\rangle$. 
For example, a simple sum of these $N$ reference states $\{|\bar{q}_k\rangle\}_{k=1}^N$, i.e., $\sum_{k=1}^N|\bar{q}_k\rangle$, is 
translationally symmetric with momentum zero.

The reference states
$|\Phi_A\rangle$, $|\Phi_B\rangle$, $|\Psi_{\rm VQE}\rangle$, and $\{|\overline{q}_k\rangle\}_{k=1}^{N}$ 
introduced above are all spin-singlet states, i.e, the total spin and the $Z$ component of the total spin 
being zero, while the $X$, $Y$, and $Z$ components of the total spin are zero for the reference states 
$|X_{\rm AFM 1(2)} \rangle$, $|Y_{\rm AFM 1(2)} \rangle$, and $|Z_{\rm AFM 1(2)} \rangle$, respectively.      
Because the Hamiltonian $\mathcal{\hat{H}}$ considered here is spin
${\rm SU}(2)$ symmetric and the quantum power method preserves the Hamiltonian symmetry as shown 
in Eq.~(\ref{eq:symmetry}), the Krylov subspace generated from these reference states remains 
in the same symmetry sector of the Hilbert space as the reference states. We select these reference states 
because it is known that the ground state of the spin-$1/2$ Heisenberg model
considered here is spin singlet~\cite{Marshall1955,Lieb1962}.

\subsubsection{Ground-state energy and fidelity}\label{sec:res_gs}

\begin{center}
  \begin{figure}
    \includegraphics[width=0.95\columnwidth]{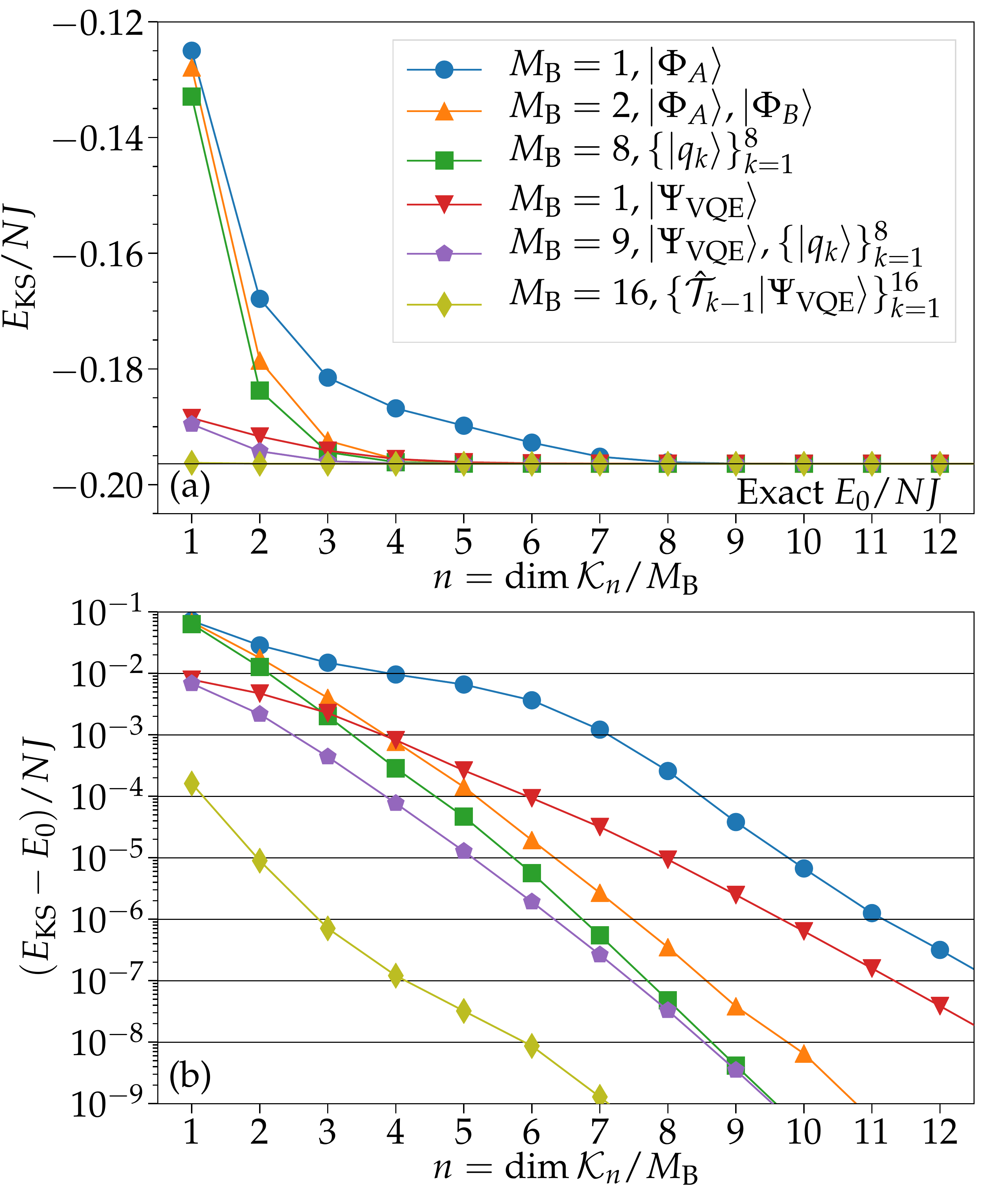}
    \caption{
      (a) Ground-state energy $E_{\rm KS}$ for 
      the spin-1/2 Heisenberg model on an $N=16$ qubit ring as a function of
      the dimension of the Krylov subspace $\mathcal{K}_n$ per block size $M_{\rm B}$,
      $n = \dim{\mathcal{K}}_n/M_{\rm B}$,
      with various sets of the reference states.
      The horizontal line indicates the exact
      ground-state energy $E_0$.
      The results are obtained with    
      $\Delta_\tau J=0.05$, $r=1$, $m=1$, and $p=3$.  
      (b) Same as (a) but a semilog plot of the energy difference $E_{\rm KS}-E_0$  
      as a function of $n$. 
      \label{fig.energy}
    }
  \end{figure}
\end{center}

\begin{center}
  \begin{figure}
    \includegraphics[width=0.95\columnwidth]{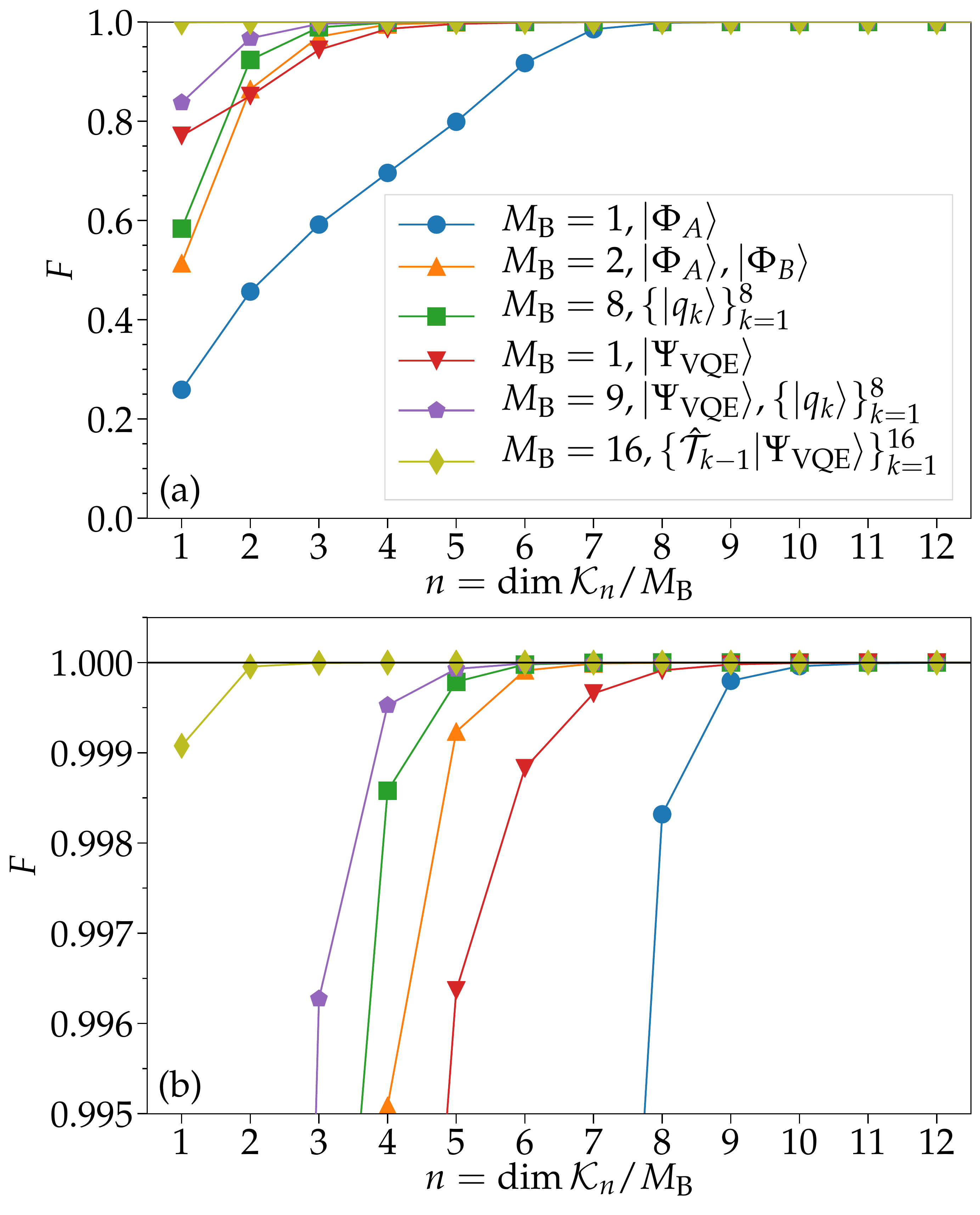}
    \caption{
      (a) Same as Fig.~\ref{fig.energy}(a) but
      for the ground-state fidelity $F=| \langle \Psi_0|\Psi_{\rm KS}\rangle |^2$. 
      Here, $|\Psi_0\rangle$ is the exact ground state and $|\Psi_{\rm KS}\rangle$ is the approximate 
      ground state with the corresponding energy $E_{\rm KS}$ 
      shown in Fig.~\ref{fig.energy}. Both states are assumed to be normalized. 
      (b) Enlarged plot of (a)
      \label{fig.fidelity}}
  \end{figure}
\end{center}

Figures~\ref{fig.energy} and \ref{fig.fidelity} 
show the estimated ground-state energy $E_{\rm KS}$ and 
the ground-state fidelity $F=|\langle\Psi_0|\Psi_{\rm KS}\rangle|^2$, 
obtained by solving Eq.~(\ref{standardeig}),
as a function of $n = \dim{\mathcal{K}}_n/M_{\rm B}$,
i.e., the dimension of the Krylov subspace $\mathcal{K}_n$ 
per block size $M_{\rm B}$. 
Note that $\hat{\mathcal{H}}_{{\rm ST}(r)}^{n-1}(\Delta_\tau)$ is
the maximum approximated Hamiltonian power multiplied to the reference states when 
the Krylov subspace $\mathcal{K}_{n}\left(\hat{\mathcal{H}}_{{\rm ST}(r)}(\Delta_\tau),\{|q_k\rangle\}_{k=1}^{M_{\rm B}}\right)$ 
is constructed in Eq.~(\ref{eq:ksub}). 
Here, the Krylov-subspace Hamiltonian matrix $[\tilde{\bs{H}}]_{ij}$ and
the overlap matrix $[\tilde{\bs{S}}]_{ij}$ are computed as
$\langle\tilde{u}_{i}| \hat{\mathcal{H}} |\tilde{u}_{j}\rangle$ and
$\langle\tilde{u}_{i}|\tilde{u}_{j}\rangle$ 
in Eqs.~(\ref{Hsubspace1}) and (\ref{Ssubspace1}), respectively. 
The first-order Richardson extrapolation ($r=1$) and  
the lowest-order symmetric Suzuki-Trotter decomposition
$\hat{S}_{2}$ are used for 
$\{\hat{\mathcal{H}}_{{\rm ST}(r)}^{l}(\Delta_\tau)\}_{l=1}^{n-1}$
with $\Delta_\tau J=0.05$,
in which the systematic errors are practically
negligible for our purpose, 
as discussed later in Fig.~\ref{fig.dtext}
(also see Figs.~\ref{fig.norm} and \ref{fig.norm_ndep}).

Let us first focus on the results for $n=1$, where
no Hamiltonian power is incorporated in the Krylov subspace. 
It is not surprising to find that the energy and the fidelity are substantially improved if 
the reference states include the VQE state $|\Psi_{\rm VQE}\rangle$.  
The improvement is even more significant 
if we incorporate the spatially translated VQE states $\{|\bar{q}_k\rangle\}_{k=1}^N$.
Note that, if $M_{\rm B}=1$, the energies plotted at $n=1$
are merely the expectation values of $\hat{\mathcal{H}}$ with
respect to the corresponding reference state, e.g., 
$\langle \Phi_{A}|\hat{\mathcal{H}}|\Phi_{\rm A}\rangle/NJ=-0.125$ and  
$\langle \Psi_{\rm VQE}|\hat{\mathcal{H}}|\Psi_{\rm VQE}\rangle/NJ=-0.1885$. 
The multireference scheme with $M_{\rm B} > 1$ further decreases the energy and improves the fidelity 
without applying the Hamiltonian power to the reference states.

With increasing $n$,    
the energy decreases monotonically
(Fig.~\ref{fig.energy})
and the fidelity keeps increasing towards one
(Fig.~\ref{fig.fidelity}), 
implying that the ground-state estimation can be improved
systematically over a chosen set of reference states  
without any parameter optimization,
at a cost of the linearly increasing circuit depth that scales at most as $2n-1$.
%
The nearly linear behavior of $E_{\rm KS}-E_0$
in the semilog plot shown in Fig.~\ref{fig.energy}(b) suggests 
the exponential convergence to the exact ground-state energy as a function of $n$,
as in the Lanczos method~\cite{Koch2019}. 
Notice also that the energy as well as the fidelity for $M_{\rm B}=16$ is consistently 
better than those for $M_{\rm B}\leqslant 9$ for every $n$.
Moreover, 
the slope in the semilog plot of $E_{\rm KS}-E_0$ 
and also the slope of the fidelity tend to be steeper 
for $M_{\rm B}>1$ than for $M_{\rm B}=1$
[see Fig.~\ref{fig.energy}(b) and Fig.~\ref{fig.fidelity}(a)],
implying that the convergence
towards the ground state is improved more efficiently in the multireference scheme 
with $M_{\rm B}>1$. 
Interestingly, even if $|\Psi_{\rm VQE}\rangle$ is not included in a set of 
reference states, the multireference schemes 
with $M_{\rm B}=2$ and $M_{\rm B}=8$ 
surpass the scheme including only $|\Psi_{\rm VQE}\rangle$ with $M_{\rm B}=1$ 
at $n=5$ and $3$, respectively, in terms of the ground-sate energy $E_{\rm KS}$. 
Therefore, the multireference scheme with $M_{\rm B}>1$ works 
effectively for reducing $n$ and 
hence the number of gates in a circuit, 
even if simple product states with no variational parameters are chosen for the reference states.
Table~\ref{table} summarizes
the minimum dimension $n$ of the Krylov subspace per block size
and the corresponding circuit depth required for converging the 
ground-state energy $E_{\rm KS}$ with an accuracy
$(E_{\rm KS}-E_0)/NJ \leqslant 10^{-4}$ for $N=16$. 
Note here that the commuting exponentials in
$[\hat{S}_{2}(\pm \Delta_\tau/2)]^{n-1}$
are contracted when the circuit depth is counted. 

\begin{table}
  \caption{
    Minimum dimension $n$ of the Krylov subspace 
    $\mathcal{K}_{n}\left(\hat{\mathcal{H}}_{{\rm ST}(r)}(\Delta_\tau),\{|q_k\rangle\}_{k=1}^{M_{\rm B}}\right)$
    per block size 
    $M_{\rm B}$ necessary to converge the 
    ground-state energy $E_{\rm KS}$ with an accuracy $(E_{\rm KS}-E_0)/NJ \leqslant 10^{-4}$ for 
    the spin-1/2 Heisenberg model on an $N=16$ qubit ring. 
    The third column indicates the maximum circuit depth to generate the corresponding Krylov subspace basis. Note that 
    8 additional layers are required to prepare the VQE state $|\Psi_{\rm VQE}\rangle$~\cite{Seki2020vqe} 
    and few additional gate operations are 
    necessary to generate each state of $\{ |q_k\rangle \}_{k=1}^8$, which are not counted in the maximum circuit depth listed in the third column.\label{table} }
  \begin{tabular}{lcc}
    \hline
    \hline
    Reference state(s)  & $n = \dim{\mathcal{K}}_n/M_{\rm B}$  & Circuit depth\\ \hline
    $M_{\rm B}=1$, $|\Phi_A\rangle$ & 9  & 17 \\
    $M_{\rm B}=2$, $|\Phi_A\rangle$, $|\Phi_B\rangle$ & 6 & 11\\
    $M_{\rm B}=8$, $\{ |q_k\rangle \}_{k=1}^8$ & 5 & 9\\
    $M_{\rm B}=1$, $|\Psi_{\rm VQE}\rangle$ & 7 & 13\\
    $M_{\rm B}=9$, $|\Psi_{\rm VQE}\rangle$, $\{ |q_k\rangle \}_{k=1}^8$ & 4 & 7\\
    $M_{\rm B}=16$, $\{\mathcal{\hat{T}}_{k-1} |\Psi_{\rm VQE}\rangle \}_{k=1}^{16}$ & 2 & 3\\
    \hline
    \hline
  \end{tabular}
\end{table}

\subsubsection{$\Delta_\tau$ dependence of the ground-state energy and extrapolation to $\Delta_\tau\to0$}\label{sec:extrapolation}

Finally, we describe a strategy for performing the Krylov-subspace
diagonalization combined with the quantum power method using a noisy near-term
quantum computer.
On a noisy quantum computer, 
a reasonably large $\Delta_\tau$ should be used in order to evaluate, 
without being buried in the noise,  
the approximated Hamiltonian power formulated on the basis of the central finite differentiation. 
For instance, the expectation value of the difference
$\hat{S}_{2}(\Delta_\tau/2)-\hat{S}_{2}(-\Delta_\tau/2)$
should be substantially more significant than the noise. 
However, as demonstrated above, the quantum power method can well control 
the systematic errors in approximating Hamiltonian power, 
and therefore 
one can accurately extrapolate the results to the limit of $\Delta_\tau \to 0$
even when a few results are available for relatively large values of $\Delta_\tau$ 
(also see, e.g., Refs.~\cite{Li2017PRX,Endo2018PRX}). 

\begin{center}
  \begin{figure}
    \includegraphics[width=0.95\columnwidth]{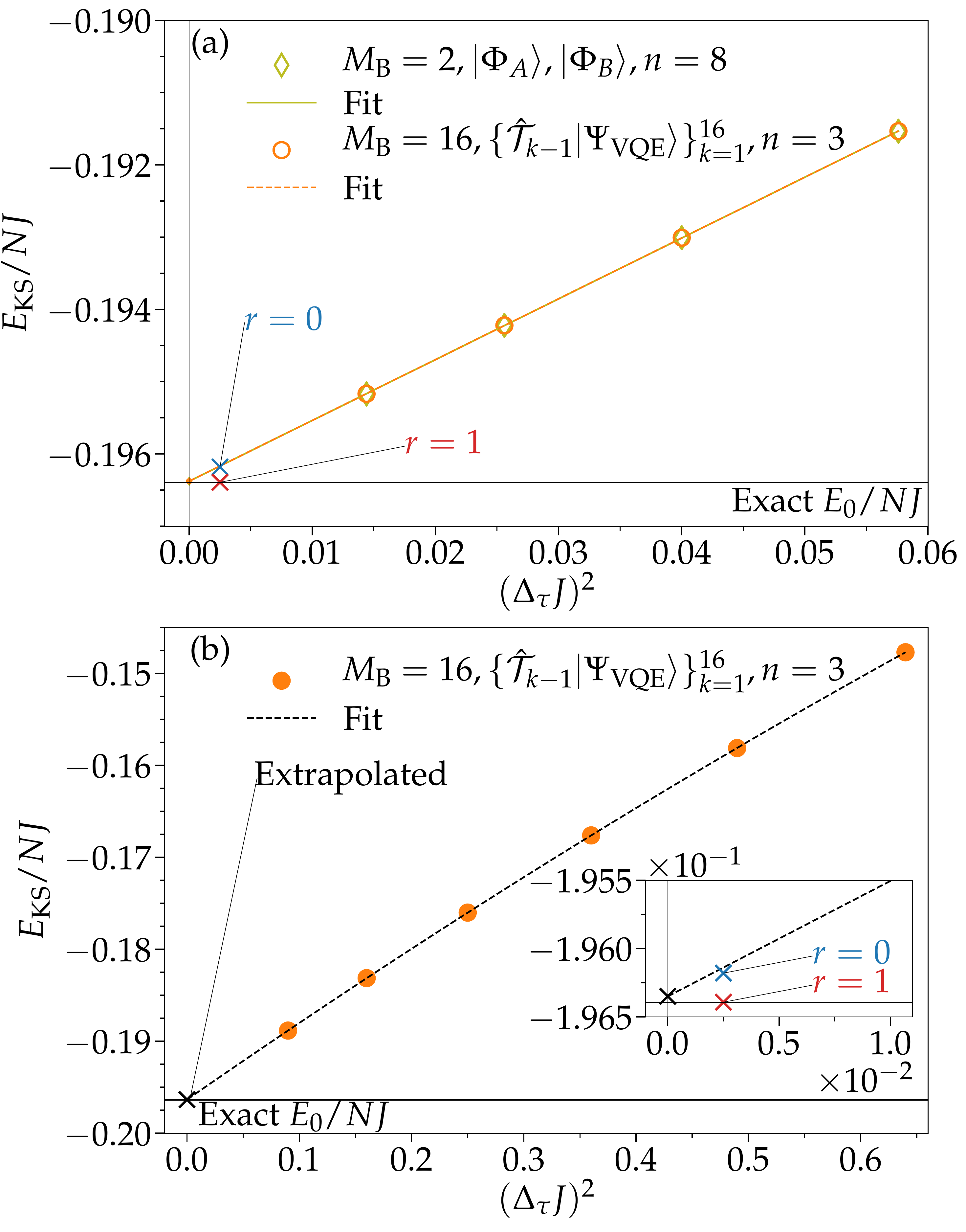}
    \caption{
      (a) The ground-state energy $E_{\rm KS}$ 
      evaluated for the spin-1/2 Heisenberg model on an $N=16$ qubit ring 
      by setting the Krylov-subspace parameters $(M_{\rm B},n)=(2,8)$ (diamonds) and 
      $(16,3)$ (circles)  
      at $\Delta_\tau J=0.12$, $0.16$, $0.2$, and $0.24$ without the Richardson extrapolation. 
      The lines are fits to the two sets of data obtained separately by assuming
      the form $E_{\rm KS}/NJ=a (\Delta_\tau J)^2 + b$
      with $a$ and $b$ being fitting parameters.
      (b) Same as (a) but for $(M_{\rm B},n)=(16,3)$ 
      at $\Delta_\tau J=0.3$, $0.4$, $0.5$, $0.6$, $0.7$, and $0.8$ 
      without the Richardson extrapolation. 
      The dashed line is a fit to the data obtained by assuming
      the form $E_{\rm KS}/NJ=a (\Delta_\tau J)^4 + b (\Delta_\tau J)^2 + c$
      with $a$, $b$,  and $c$ being fitting parameters, and the extrapolated value at 
      $\Delta_\tau=0$ is indicated by black cross. The inset is an enlarged plot. 
      For comparison, the results for $(M_{\rm B},n)=(16,3)$ at 
      $\Delta_\tau J=0.05$ with $r=0$ and $1$ 
      are also indicated by blue and red crosses, respectively, in (a) and the inset of (b). 
      Note that the result with $r=1$ corresponds to that shown in Fig.~\ref{fig.energy}.
      The horizontal line indicates 
      the exact ground-state energy $E_0$. 
    }
      \label{fig.dtext}
  \end{figure}
\end{center}

Figure~\ref{fig.dtext}(a) shows the numerical results of the ground-state energy 
$E_{\rm KS}$ evaluated by setting the Krylov-subspace parameters $(M_{\rm B},n)=(2,8)$ and $(16,3)$ 
without the Richardson extrapolation 
at $\Delta_\tau J=0.12, 0.16, 0.2$, and $0.24$, which are larger than $\Delta_\tau J=0.05$ used in 
Figs.~\ref{fig.energy} and \ref{fig.fidelity}.
For the results shown in Fig.~\ref{fig.dtext},
we adopt the approach based on Eqs.~(\ref{Hsubspace2}) and (\ref{Ssubspace2}) 
that requires a fewer number of state overlaps
than that based on Eqs.~(\ref{Hsubspace1}) and (\ref{Ssubspace1}).
The lines are fits to the data assuming the form 
$E_{\rm KS}/NJ=a (\Delta_\tau J)^2 + b$ with $a$ and $b$ being fitting parameters
determined by the least-squares method.
As shown in Fig.~\ref{fig.dtext}(a), the results are correctly extrapolated to
the exact energy within  2--3 standard deviations
of the fitting error (the error bars are not visible in the scale of the figure).
It is more striking to find in Fig.~\ref{fig.dtext}(b) 
that a similar extrapolation,
assuming the form $E_{\rm KS}/NJ=a (\Delta_\tau J)^4 + b (\Delta_\tau J)^2 + c$
with $a$, $b$,  and $c$ being fitting parameters,
is also satisfactory even when much larger values of $\Delta_\tau$, 
as large as $\Delta_\tau J=0.8$, are 
employed to evaluated the ground-state energy $E_{\rm KS}$.

These results corroborate that
the systematic errors due to a finite time interval $\Delta_\tau$ are well controlled 
also for the quantities evaluated in the Krylov-subspace diagonalization,  
thus allowing us to extrapolate the results
obtained for relatively large values of $\Delta_\tau$ to the limit of $\Delta_\tau\to0$.  
This could provide a promising error-mitigation strategy on a noisy near-term quantum computer. 
We should remark that, although the example shown here is 
the energy of the small system ($N=16$), 
the Suzuki-Trotter error is known to be well controlled even in larger systems   
not only for the energy but also for other quantities in general  
(for example, see Ref.~\cite{Seki2019}).

\section{Conclusion and discussion}\label{sec:conclusion}

We have proposed the quantum power method
that approximates the Hamiltonian power
$\hat{\mathcal{H}}^n$ with a linear combination
of the time-evolution operators. 
The key ingredients of the quantum power method
are the central-finite-difference scheme for the time derivatives and
the symmetric Suzuki-Trotter decomposition 
of the time-evolution operators, 
both of which are essential to 
retain the Hermiticity and the even parity in $\Delta_\tau$ of 
the approximated Hamiltonian power $\hat{\mathcal{H}}^n_{{\rm ST}}(\Delta_\tau)$, i.e., 
$\hat{\mathcal{H}}^n_{\rm ST}(\Delta_\tau)
=\left[\hat{\mathcal{H}}^n_{\rm ST}(\Delta_\tau)\right]^\dag
=\hat{\mathcal{H}}^n_{\rm ST}(-\Delta_\tau)$. 
The systematic errors due to the finite differentiation and the Suzuki-Trotter decomposition 
are well controlled in the quantum power method as    
$\mathcal{E}_{\rm FD}\sim O(\Delta_\tau^{2})$ and 
$\mathcal{E}_{\rm ST}\sim O(\Delta_\tau^{2m})$, respectively.
The number of gates
required for approximating the Hamiltonian power $\hat{\mathcal{H}}^n$ is $O(n \mathlcal{k} N)$, 
where $N$ is the number of qubits and
a $\mathlcal{k}$-local Hamiltonian $\hat{\mathcal{H}}$ in the qubit representation
composed of $O(N)$ terms is assumed, 
and thus it is at most $O(n N^2)$ for a fermion Hamiltonian
when the Jordan-Wigner transformation is used
(see Appendix~\ref{app:Hubbard}). 
This should be contrasted to the classical power method, in which the computational complexity
scales exponentially in $N$.

The $r$th-order Richardson extrapolation can be adopted to systematically improve the systematic errors as 
$\mathcal{E}_{\rm FD}\sim O(\Delta_\tau^{2+2r})$ and $\mathcal{E}_{\rm ST}\sim O(\Delta_\tau^{2m+2r})$ in 
the approximated Hamiltonian power $\hat{\mathcal{H}}^n_{{\rm ST}(r)}(\Delta_\tau)$, 
without increasing the number of gates required in each quantum circuit, 
although the number of terms in the linear combination, which can be treated classically, increases by the factor $r+1$. 
Thus, both with and without the Richardson extrapolation, the systematic errors $\mathcal{E}_{\rm FD}$ 
and $\mathcal{E}_{\rm ST}$ can be consistently treated with the lowest-order Suzuki-Trotter decomposition with $m=1$, 
independently of the power $n$, which reduces significantly the circuit depth. 
This is in sharp contrast to the algorithm that requires the 
higher-order Suzuki-Trotter decomposition with increasing the power $n$ (see Appendix~\ref{app:HST}). 
Therefore, the quantum power method proposed here is 
a potentially promising algorithm for near-term quantum devices.

By numerical simulations, we have tested the quantum power method 
and found that the Hamiltonian power $\mathcal{\hat{H}}^n$ for the spin-1/2 Heisenberg model 
can be well approximated by $\hat{\mathcal{H}}^n_{{\rm ST}(r)}(\Delta_\tau)$ with the controlled accuracy 
to be essentially exact 
for the power $n$ up to $100$ and $N$ as large as 
$24$ qubits, corresponding to the Hilbert space dimension 
$N_{\rm D}=2^N\approx10^7$.

As an application of the quantum power method, 
we have demonstrated, with noiseless numerical simulations, 
the multireference Krylov-subspace diagonalization combined with the quantum power method 
for the spin-$1/2$ Heisenberg model on an $N=16$ qubit ring to evaluate 
the ground-state energy and the ground-state fidelity. 
Considering the Hamiltonian power $\hat{\mathcal{H}}^n$ up to $n=11$, 
we have shown that  
the multireference Krylov-subspace diagonalization scheme with the block size $M_{\rm B}>1$ 
greatly accelerate 
the convergence to the ground state, even when simple parameter-free product states are employed for the reference states.
We have also found that the Krylov-subspace diagonalization scheme with $M_{\rm B}=1$, 
corresponding to a quantum version of the standard Lanczos method~\cite{Chatelin}, 
improves the ground-state energy of the VQE state $|\Psi_{\rm VQE}\rangle$
almost exponentially with increasing $n$. 
Thus, the Krylov-subspace diagonalization combined with the quantum power method, 
which satisfies the variational principle by definition, can 
provide a systematic way to further improve a VQE state that has already a reasonable overlap with an exact ground state. 
This is a quantum analog to the Lanczos iteration scheme in the variational Monte Carlo method on classical 
computers~\cite{Sorella2001}, 
but here one can treat higher powers of the Hamiltonian on quantum computers.

We have also demonstrated by numerical simulations the multireference Krylov-subspace diagonalization 
combined with the quantum power method for a Fermi-Hubbard model on a $4\times2$ lattice. 
In this case, the Hamiltonian can be divided into $N_\Gamma=4$ parts, instead of $N_\Gamma=2$ parts 
for the case of the one-dimensional spin-1/2 Heisenberg model. 
Moreover, the Hamiltonian mapped from the fermion representation 
to the qubit
representation on $N=16$ qubits contains terms with long-range Pauli strings. 
Even in this case, considering the Hamiltonian power $\hat{\mathcal{H}}^n$ up to $n=17$, 
we have found numerically that the ground-state energy 
converges almost exponentially with  $n$.

Although we have simulated only the ground-sate energy,
the expectation value of other observables that commute 
with the Hamiltonian $\mathcal{\hat{H}}$ can be evaluated similarly. 
When an observable $\hat{\mathcal{O}}$ does not commute with 
the Hamiltonian $\mathcal{\hat{H}}$, the expectation value 
with respect to the approximate ground state
$|\Psi_0\rangle \approx  |\tilde{\Psi}_{\rm KS}\rangle
\equiv \sum_{i=1}^{nM_{\rm B}} v_{i} |\tilde{u}_i\rangle$, 
with the coefficients $v_{i}$ already determined 
by solving Eq.~(\ref{geneig}) or Eq.~(\ref{standardeig}) in the block Krylov subspace
$\mathcal{K}_n\left(\hat{\mathcal{H}}_{{\rm ST}(r)}(\Delta_\tau),\{|q_k\rangle\}_{k=1}^{M_{\rm B}}\right) 
=  {\rm span}
\left(
\{
|\tilde{u}_i\rangle\}_{i=1}^{nM_{\rm B}}
\right)$,
can also be evaluated as
\begin{alignat}{3}
\langle \Psi_0| \hat{\mathcal{O}}|\Psi_0 \rangle 
&\approx
\sum_{i=1}^{nM_{\rm B}}
\sum_{j=1}^{nM_{\rm B}}v_i^* v_j
\langle\tilde{u}_i |\hat{\mathcal{O}} | \tilde{u}_j \rangle \notag \\
&=
\sum_{i=1}^{nM_{\rm B}}
\sum_{j=1}^{nM_{\rm B}}v_i^* v_j
\sum_{\nu=0}^{l-1}
\sum_{\nu'=0}^{l'-1}
c_{l-1,\nu}^*
c_{l'-1,\nu'}
\notag \\
&
\times
\langle q_k |
\left[\hat{S}_{2m}^{(p)}\left(-\frac{\Delta_\tau}{2}\right)\right]^{l-1-2\nu}
\hat{\mathcal{O}} 
\left[\hat{S}_{2m}^{(p)}\left(\frac{\Delta_\tau}{2}\right)\right]^{l'-1-2\nu'}
|q_{k'}\rangle, 
\end{alignat}
where
$|\tilde{u}_{i}\rangle = \hat{\mathcal{H}}^{l-1}_{{\rm ST}(r)}(\Delta_\tau)|q_k \rangle$,
as given in Eq.~(\ref{eq:ui}), and  
the explicit form of $\hat{\mathcal{H}}^{l-1}_{{\rm ST}(r)}(\Delta_\tau)$ 
with $r=0$ is used in the second line.  
Here, $i=k+(l-1)M_{\rm B}$ and $j=k'+(l'-1)M_{\rm B}$ for 
$1 \leqslant k,k' \leqslant M_{\rm B}$ and $1 \leqslant l,l' \leqslant n$.

Our numerical simulations clearly demonstrate a promising potential that 
the quantum power method  
combined with the multireference
Krylov-subspace diagonalization  
enables us to perform
systematic and optimization-free
calculations for quantum many-body systems, which 
is suitable for near-term quantum computers. 
Other applications of the quantum power method 
include various moment-based methods, which are briefly outlined in Appendix~\ref{app:moment} and 
Appendix~\ref{app:lanczos}. In these appendices, 
we show that the power method can evaluate $\langle\Psi|\hat{\mathcal{H}}^2|\Psi\rangle$ with exactly the same 
amount of resource that is required for $\langle\Psi|\hat{\mathcal{H}}|\Psi\rangle$, 
and therefore, for example, 
 the energy variance $\sigma^2=\langle\Psi|\hat{\mathcal{H}}^2|\Psi\rangle - \langle\Psi|\hat{\mathcal{H}}|\Psi\rangle^2$ 
can be easily obtained. Here, $|\Psi\rangle$ is a given quantum state. 
Using numerical simulations, we also demonstrate the CMX 
for the imaginary-time evolution. This formalism can be easily extended to other methods, e.g., 
the high-temperature series expansion~\cite{Oitmaa}. 

Finally, we remark that
the quantum power method 
proposed here can generally be applied to any sparse
Hermitian operator
$\hat{\mathcal{A}}$. In this case, the $n$th power of 
$\hat{\mathcal{A}}$ is given as 
\begin{equation}
  \hat{\mathcal{A}}^n
  =\imag^n \left.\frac{\dd^n \hat{V}(t)}{\dd t^n}\right|_{t=0}
\end{equation}
with the generating function $\hat{V}(t)=\e^{-\imag \hat{\mathcal{A}}t}$.
We can use the central finite-difference scheme for the time 
derivatives to represent $\hat{\mathcal{A}}^n$
as a linear combination of unitary operator $\hat{V}(t)$ at different time variables. 
The symmetric Suzuki-Trotter decomposition
is then used to decompose each unitary operator $\hat{V}(t)$.

\acknowledgements
Parts of numerical simulations are
on the HOKUSAI supercomputer at RIKEN 
(Project ID: G20015).
This work is supported by Grant-in-Aid for Research Activity start-up (No.~JP19K23433) and 
Grant-in-Aid for Scientific Research (B) (No.~JP18H01183) from MEXT, Japan.

{\it Note added:}
Recently, 
a related study by Bespalova and Kyriienko has been reported~\cite{bespalova2020hamiltonian}.

\appendix

\section{Comparison with other algorithms for the Krylov subspace diagonalization} \label{app:ksd}

In this appendix, we summarize the
distinctions between
the Krylov subspace diagonalization scheme 
described in Sec.~\ref{sec:ksd} and
other algorithms reported recently, i.e.,
the QLanczos method~\cite{Motta2019}, 
the MRSQK algorithm~\cite{Stair2020}, and
the QFD method~\cite{parrish2019quantum}.
However, before making a comparison,
we emphasize that the quantum power method 
proposed here allows us to evaluate
the Hamiltonian power $\hat{\mathcal{H}}^n$ directly
by approximating it with a linear 
combination of the time-evolution operators.
Consequently, the quantum power method finds many possible applications, 
some of which are described in this paper, and
the Krylov subspace diagonalization is one of the promising examples.  

\subsection{Brief review of the methods}

The QLanczos method is based on the quantum imaginary-time evolution (QITE) 
for generating states spanning a Krylov subspace. Namely,
the Krylov subspace to be approximated
in the QLanczos method is given as
\begin{equation}
  \mathcal{K}(\e^{-2\Delta_\tau \hat{\cal H}},|\psi\rangle)
  =\mathrm{span}
  \left(|\psi \rangle,
  \e^{-2 \Delta_\tau \hat{\mathcal{H}}} |\psi \rangle,
  \e^{-4 \Delta_\tau \hat{\mathcal{H}}} |\psi \rangle,
  \cdots \right), \label{KITE}
\end{equation}
where $|\psi\rangle$ is a given reference state and only even powers of
$\e^{- \Delta_\tau \hat{\mathcal{H}}}$ appear 
because it simplifies the evaluation of matrix elements of the Hamiltonian $\hat{\mathcal{H}}$ over the basis states in the 
Krylov subspace $\mathcal{K}$ and the overlap matrix. 
However, this is irrelevant for the discussion given here. 
Let us assume that a Hamiltonian $\hat{\mathcal{H}}$ 
is given as $\hat{\mathcal{H}}=\sum_{m} \hat{h}[m]$,
where $\hat{h}[m]$ represents the $m$th string of Pauli operators.
With the first-order Suzuki-Trotter decomposition, 
the imaginary-time evolution operator for an imaginary time $l\Delta_\tau$
can be written as 
$\e^{-l\Delta_\tau \hat{\mathcal{H}}}=
\left(
\prod_{m}
\e^{-\Delta_\tau \hat{h}[m]}\right)^l + O(l\Delta_\tau^2)$,
where $l$ is an integer.

The QITE approximates, for each Trotter step,
a normalized short imaginary-time evolved state 
$|\mathrm{ITE} \rangle \equiv
\e^{-\Delta_\tau \hat{h}[m]}|\psi\rangle/\sqrt{c}$ with
$c=\langle \psi|\e^{-2\Delta_\tau \hat{h}[m]} | \psi \rangle
\approx 1 - 2\Delta_\tau \langle \psi| \hat{h}[m] |\psi \rangle$  
by a unitary evolved state 
$|\mathrm{ITE}^\prime \rangle \equiv \e^{-i \Delta_\tau \hat{A}[m]}|\psi \rangle$,
where $\hat{A}[m]$ is a Hermitian operator of the form 
\begin{eqnarray}
  \hat{A}[m]&=&\sum_{i_{k_1}, i_{k_2}, \ldots, i_{k_D}} a_{i_{k_1} i_{k_2} \ldots i_{k_D}}[m]
    \hat{\sigma}_{i_{k_1}} \hat{\sigma}_{i_{k_2}} \ldots \hat{\sigma}_{i_{k_D}} \label{eq:am} \\
    &\equiv&
    \sum_{I} a_{I}[m] \hat{\sigma}_I. \label{eq:am2}
\end{eqnarray} 
Here, $\hat{\sigma}_{i_k} \in \{\hat{I}_k,\hat{X}_k,\hat{Y}_k,\hat{Z}_k\}$ in Eq.~(\ref{eq:am}), 
i.e., the identity and Pauli operators at the $k$th qubit, and  
$\hat{\sigma}_{I}$ in Eq.~(\ref{eq:am2}) represents a Pauli string of length $D$
($D$ is a parameter and is called domain size) 
with $I=\{ i_{k_1}, i_{k_2}, \ldots, i_{k_D} \}$ and $1\leqslant k_1 < k_2 < \cdots < k_D \leqslant N$.
The sum over $I$ runs at most up to
$4^D$ (see supplementary information of Ref.~\cite{Motta2019} for a precise counting).
The coefficients $a_{I}[m]$ should be determined by minimizing $||\ |\mathrm{ITE}\rangle-|\mathrm{ITE}^\prime\rangle \ ||$, 
which yields,  
to the first order of $\Delta_\tau$, 
the linear system 
$\bs{C a}=\bs{b}$ with  
$[\bs{C}]_{IJ}= \langle \psi|\hat{\sigma}_{I}^\dag  \hat{\sigma}_J |\psi\rangle $,
$[\bs{a}]_I=a_I[m]$, and 
$[\bs{b}]_{I}= -\frac{i}{\sqrt{c}}\langle \psi|\hat{\sigma}_{I}^\dag \hat{h}[m]|\psi \rangle $.  
Thus the solution vector $\bs{a}$ of the linear system gives the coefficients $a_I[m]$.
By repeating the above procedure for all the Trotter steps,
a state that approximates 
$\e^{-l\Delta_\tau \hat{\mathcal{H}}}|\psi\rangle/\sqrt{\langle \psi|\e^{-2 l \Delta_\tau \hat{\mathcal{H}}} | \psi \rangle}$ 
to the first order of $\Delta_\tau$ can be constructed. 
  
In the QLanczos method, one may wish to increase the time interval $\Delta_\tau$ to avoid the 
linear dependency of the bases generated in the Krylov subspace $\mathcal{K}$. However, 
there is a trade-off that the increase of $\Delta_\tau$ increases the systematic error in approximating 
the imaginary-time evolution operator with the Suzuki-Trotter decomposition and also requires to generally enlarge the 
qubit domain size $D$ in approximating the imaginary-time evolution with a unitary evolution. Instead, if one chooses 
a small value of $\Delta_\tau$, these difficulties may not occur. However, the states generated are most likely 
dependent on other states nearby in the imaginary time and thus many iterations of $l$ may be required.

In the QFD method and the MRSQK algorithm (with a single reference state),
the Krylov subspace to be approximated is given as 
\begin{equation}
  \mathcal{K}(\e^{-\imag \Delta_\tau \hat{\cal H}},|\psi\rangle)=\mathrm{span}
  \left(|\psi \rangle,
  \e^{-\imag \Delta_\tau \hat{\mathcal{H}}} |\psi \rangle,
  \e^{-\imag 2\Delta_\tau \hat{\mathcal{H}}} |\psi \rangle,
  \cdots \right),
  \label{KRTE}
\end{equation}
where the time-evolution operators $\e^{-\imag l \Delta_\tau \hat{\cal H}}$
are approximated by a Suzuki-Trotter decomposition
in practice.
The MRSQK algorithm allows for the use of not only a single state but also 
multiple states as the reference states.
The ground (or target) state is approximated as a linear combination of
the time-evolved states $\{\e^{-\imag l \Delta_\tau \hat{\mathcal{H}} }|\psi\rangle\}$ and the coefficients for the linear combination is
determined by the Rayleigh-Ritz variational principle, i.e., the subspace diagonalization.

For example, the time-evolved state $\e^{-\imag \Delta_\tau \hat{\mathcal{H}} }|\psi\rangle$ 
can be expanded as 
\begin{equation}
  \e^{-\imag \Delta_\tau \hat{\mathcal{H}} }|\psi\rangle = |\psi\rangle-\imag\Delta_\tau \hat{\mathcal{H}}|\psi\rangle + O(\Delta_\tau^2). 
\end{equation}
Therefore, if $\Delta_\tau$ is too small, a distance between the states 
$|\psi\rangle$ and $\e^{-\imag \Delta_\tau \hat{\mathcal{H}}}|\psi\rangle$ 
would be so small that a linear dependency problem may occur 
in the subspace diagonalization.
On the other hand,
if large $\Delta_\tau$ is chosen in order to reduce the
linear dependency problem, 
the corresponding Suzuki-Trotter error for approximating 
$\e^{-\imag \Delta_\tau \hat{\mathcal{H}}}$ becomes large. 
Hence, it is desirable to find an optimum value of $\Delta_\tau$, 
which is however unknown {\it a priori}.
This issue is similar to the case of the QLanczos method. 
The MRSQK algorithm can improve the linear-dependency problem by
virtue of the multireference states as compared to the single reference state.
However, in essence, the same issue may remain 
because the expansion process of the
Krylov subspace is
still based on a short real-time evolution.

In our method, the Krylov subspace is expanded by generating states 
$\hat{\mathcal{H}}_{\mathrm{ST}(r)}^n(\Delta_\tau)|\psi\rangle$ from a reference state $|\psi\rangle$. 
Namely, the Krylov subspace to be approximated
is given as 
\begin{equation}
    \mathcal{K}
  (\hat{\cal H},|\psi\rangle)
  =\mathrm{span}
  \left(|\psi \rangle,
  \hat{\mathcal{H}}^1 |\psi \rangle,
  \hat{\mathcal{H}}^2 |\psi \rangle, 
  \cdots \right),
  \label{KPOW}
\end{equation}
and 
the Hamiltonian power $\hat{\mathcal{H}}^n$ is approximated by 
a linear combination of the time-evolution operators, 
$\hat{\mathcal{H}}_{\mathrm{ST}(r)}^n(\Delta_\tau)$, 
with an $O(\Delta_\tau^{2+2r})$ error.
Therefore, in contrast to the MRSQK algorithm and the QFD method,  
the Krylov basis 
$\hat{\mathcal{H}}_{\mathrm{ST}(r)}^{n}(\Delta_\tau)|\psi \rangle$ 
itself is a linear combination of time-evolved states.
The coefficients $c_{n,k}$ for the linear combination 
in $\hat{\mathcal{H}}_{\mathrm{ST}(r)}^n(\Delta_\tau)|\psi \rangle$ are
already determined from the central finite-difference formula. 

More explicitly, 
the state $\hat{\mathcal{H}}_{\mathrm{ST}(r)}^n(\Delta_\tau)|\psi\rangle$ can be written as
\begin{equation}
  \hat{\mathcal{H}}_{\mathrm{ST}(r)}^n(\Delta_\tau)|\psi\rangle
  = \hat{\mathcal{H}}^n|\psi\rangle + \Delta_\tau^{2+2r} \hat{r}_n|\psi \rangle + O(\Delta_\tau^{4+2r}),
\end{equation}  
where
$\hat{r}_n$ is some Hermitian operator representing the leading error
(residual) term in $\hat{\mathcal{H}}_{\mathrm{ST}(r)}^n(\Delta_\tau)$. 
Thus, in our method, the linear-dependency
problem is expected to be less severe than in the other methods 
in the sense that the new basis in the Krylov subspace, e.g.,  
  $\hat{\mathcal{H}}_{\mathrm{ST}(r)}(\Delta_\tau)|\psi\rangle$,  
is in general linearly independent of $|\psi\rangle$, 
irrespectively of the value of $\Delta_\tau$
(unless $|\psi \rangle$ is an eigenstate, e.g., the ground state, of $\hat{\mathcal{H}}$,
which is the condition that indicates the convergence). 
Indeed, we have numerically found that the method is quite stable
against the values of $\Delta_\tau$ (see Fig.~\ref{fig.dtext} 
and also the next section).

\subsection{Condition number and residual ground-state energy}

While making a fair comparison of different methods
is not straightforward in any case, 
it would be instructive to examine by numerical simulations how the Krylov subspaces
${\cal K}_n(\e^{-\Delta_\tau \hat{\cal H}},|\psi\rangle)$,
${\cal K}_n(\e^{-\imag \Delta_\tau \hat{\cal H}},|\psi\rangle)$, and 
${\cal K}_n(\hat{\cal H}_{{\rm ST}(r)}(\Delta_\tau),|\psi\rangle)$
are different from each other, 
because it gains further insight into the different Krylov-subspace approaches
described above. 
This is precisely the purpose of this section.

To this end, 
we employ the imaginary-time version of the
second-order Suzuki-Trotter approximation (for the case of $N_{\Gamma}=2$) 
\begin{equation}
\e^{-l\Delta_\tau \hat{\cal H}} =
\left( \e^{-\frac{\Delta_\tau}{2} \hat{\cal H}_A}
\e^{-\Delta_\tau \hat{\cal H}_B}
\e^{-\frac{\Delta_\tau}{2} \hat{\cal H}_A} \right)^l +O(l\Delta_\tau^3)
\label{ITE}  
\end{equation}
to generate 
${\cal K}_n(\e^{-\Delta_\tau \hat{\cal H}},|\psi\rangle)$,
instead of the QITE reported originally in Ref.~\cite{Motta2019}.
With this treatment, we can remove arbitrariness in the QITE 
such as the choice of $\hat{A}[m]$ operators and the domain-size parameter $D$. 
We also employ the second-order Suzuki-Trotter decomposition
$\hat{S}_{2}(\Delta_\tau)$ to generate 
${\cal K}_n(\e^{-\imag \Delta_\tau \hat{\cal H}},|\psi\rangle)$ and 
${\cal K}_n(\hat{\cal H}_{{\rm ST}(r)}(\Delta_\tau),|\psi\rangle)$, 
i.e., with a similar decomposition for $\e^{-\imag l\Delta_\tau\hat{\cal H}}$ as in Eq.~(\ref{ITE}). 
Therefore, the Suzuki-Trotter error is at the same order
for all three cases.
Hereafter, we simply refer to 
${\cal K}_n(\e^{-\Delta_\tau \hat{\cal H}},|\psi\rangle)$
as imaginary-time evolution (ITE), 
${\cal K}_n(\e^{-\imag \Delta_\tau \hat{\cal H}},|\psi\rangle)$ 
as real-time evolution (RTE), and 
${\cal K}_n(\hat{\cal H}_{{\rm ST}(r)}(\Delta_\tau),|\psi\rangle)$ 
as the quantum power method (QPM).  
We use 
the spin-$1/2$ Heisenberg model on an $N=16$ qubit ring 
as the Hamiltonian $\hat{\cal H}$, 
and adopt the reference state 
$|\psi\rangle = |\Phi_A\rangle$ in Eq.~(\ref{eq:phi_a}) and 
$|\psi\rangle = |\Psi_{\rm VQE}\rangle$ in Eq.~(\ref{eq:vqe}). 
The overlap between these states and the exact ground state $|\Psi_0\rangle$ is 
$|\langle\Psi_0|\Phi_A\rangle|^2=0.259$  and
$|\langle\Psi_0|\Psi_{\rm VQE}\rangle|^2=0.771$ 
(see Fig.~\ref{fig.fidelity}). 
In the numerical simulations, we vary
$\Delta_\tau J = 0.01$, $0.1$, $0.3$, $0.5$, and $0.8$. 

The linear dependency of the basis states in a Krylov subspace can be examined by calculating the 
(Euclidean norm) condition number of the overlap matrix $\bs{S}$,
\begin{equation}
  {\rm cond}(\bs{S}) \equiv \frac{s_{\rm max}}{s_{\rm min}}, 
\end{equation}
where $s_{\rm max}$ ($s_{\rm min}$) is the maximum (minimum) singular value of
the overlap matrix $\bs{S}$ defined in each Krylov subspace [for example, see Eq.~(\ref{Ssubspace})].
To focus on the linear-dependency issue, 
we assume that the basis states are normalized, 
as it is always possible 
by replacing
$|u_i\rangle $ with $ |u_i\rangle/\sqrt{\langle u_i|u_i \rangle}$,
where $|u_i\rangle$ is an unnormalized basis state in ITE and QPM.
The Krylov-subspace diagonalization in ITE and QPM
for the normalized basis states
can be formulated simply by replacing the matrix elements 
$H_{ij}$ [for definition, see Eq.~(\ref{Hsubspace})] and $S_{ij}$ with 
$H_{ij}/\sqrt{S_{ii}S_{jj}}$ and 
$S_{ij}/\sqrt{S_{ii}S_{jj}}$,
or equivalently in  the matrix form as
\begin{equation}
  \bs{H} \mapsto
  \bs{\delta} \bs{H} \bs{\delta}
  \label{eqH}
\end{equation}
and
\begin{equation}
  \bs{S} \mapsto
  \bs{\delta} \bs{S} \bs{\delta},
  \label{eqS}
\end{equation}
where
\begin{equation}
    \bs{\delta}={\rm diag}\left(\frac{1}{\sqrt{S_{11}}}\ \frac{1}{\sqrt{S_{22}}}\ \cdots \right).   
\end{equation}
The reduction scheme of the condition number of a matrix
with a transformation by diagonal matrices as in 
Eqs.~(\ref{eqH}) and (\ref{eqS}) is a widely used procedure
known as equilibration of a matrix~\cite{Chatelin}.
Note that although the normalization of the basis states
alters
the condition number of the overlap matrix $\bs{S}$ in general,  
the resulting variational ground-state energy $E_{\rm KS}$ does not depend on
whether or not the basis states in the Krylov subspace are normalized,
as it can readily be confirmed from Eqs.~(\ref{geneig}), (\ref{eqH}), and (\ref{eqS}). 
In addition, the eigenvector $\bs{v}$ in Eq.~(\ref{geneig})
is simply given by replacing $\bs{v}$ with $\bs{\delta}^{-1} \bs{v}$ when 
the equilibration of the matrices in Eqs.~(\ref{eqH}) and (\ref{eqS}) is made.
If the basis states are orthonormalized,
the overlap matrix is an identity matrix,
and hence the condition number is one.
On the other hand,
if the basis states become linearly dependent,
$\bs{S}$ has zero singular value(s),
and hence the condition number diverges. 
We show the numerical results for ${\rm cond}(\bs{S})\leqslant 10^{13}$,
which is nearly the limit of double-precision floating-point number.

\begin{center}
  \begin{figure*}
    \includegraphics[width=1.\textwidth]{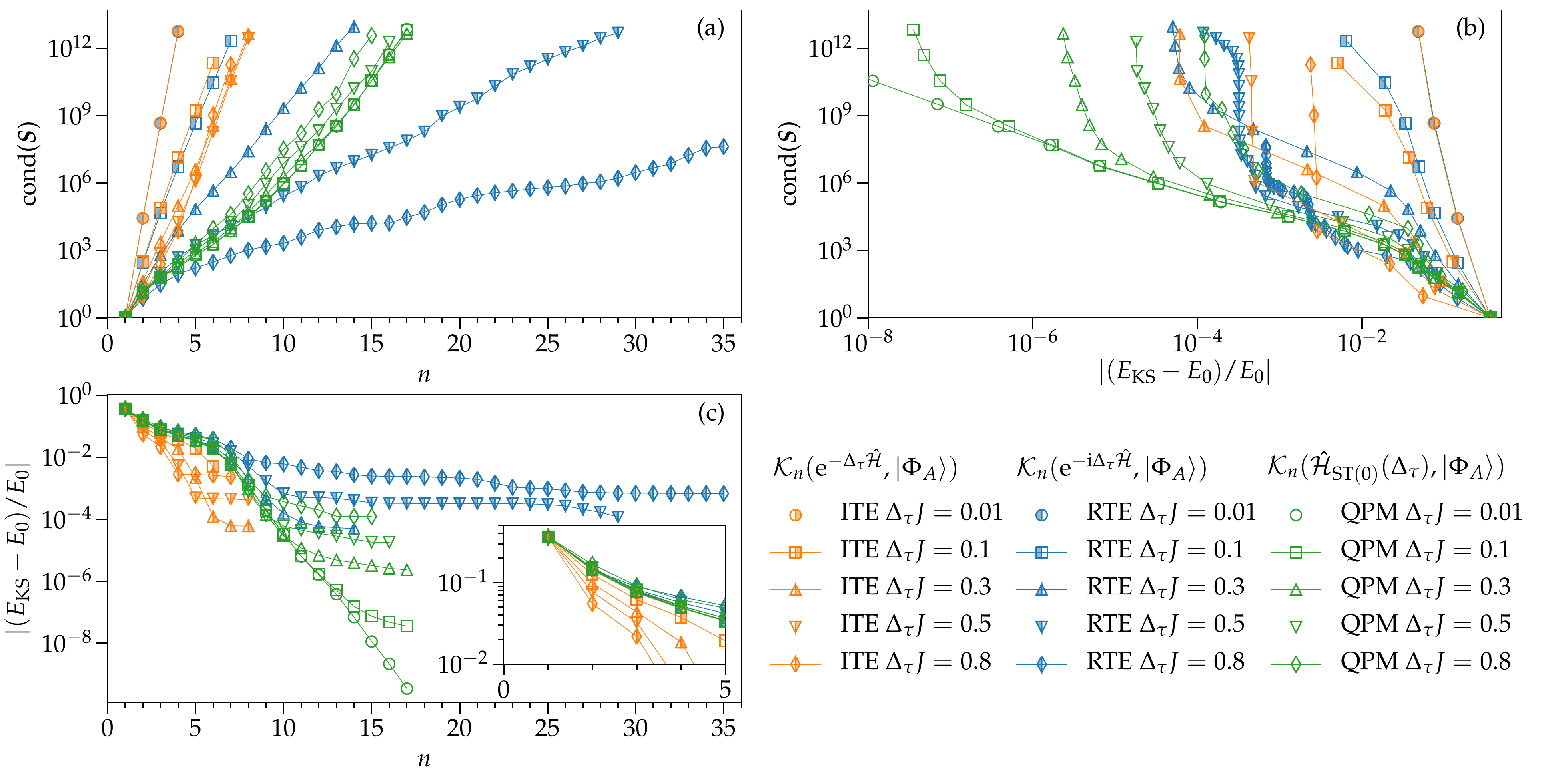}
    \caption{
      (a) Condition number ${\rm cond}(\bs{S})$ as a function of the subspace dimension $n$,
      (b) condition number ${\rm cond}(\bs{S})$ as a function of the error in energy
      $|(E_{\rm KS}-E_0)/E_0|$,  and
      (c) error in energy
      $|(E_{\rm KS}-E_0)/E_0|$
      as a function of the subspace dimension $n$ 
      for three different Krylov subspaces (ITE, RTE, and QPM) with several values of $\Delta_\tau$. 
      Here, the Hamiltonian $\hat{\cal H}$ is 
      the spin-$1/2$ Heisenberg model on an $N=16$ qubit ring with the exact ground-state energy $E_0$ and 
      $|\psi\rangle=|\Phi_A\rangle$ is used as the reference state.
      For QPM, the Richardson extrapolation is not used ($r=0$). 
      Notice that the results for ITE and RTE with $\Delta_\tau J=0.01$ are identical in this scale.
      The inset in (c) is the enlarged plot for small $n$.
      \label{condition_r0}
    }
  \end{figure*}
\end{center}

\begin{center}
  \begin{figure*}
    \includegraphics[width=1.\textwidth]{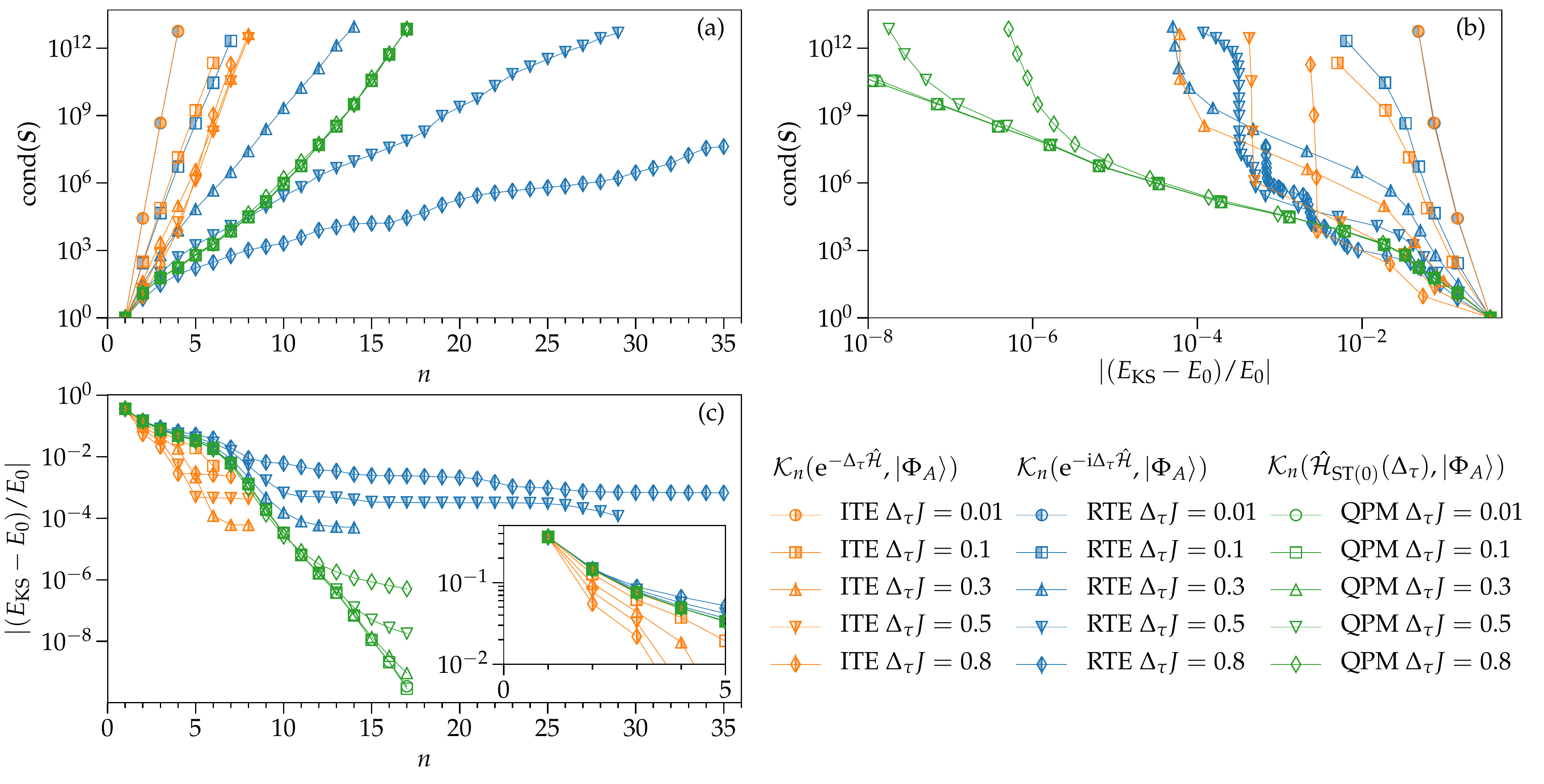}
    \caption{
      Same as Fig.~\ref{condition_r0} but with the first-order Richardson extrapolation ($r=1$) for QPM.  
      \label{condition_r1}
    }
  \end{figure*}
\end{center}

Figure~\ref{condition_r0} shows the numerical results that summarize relations
between 
the condition number ${\rm cond}(\bs{S})$,  
the error in energy
$|(E_{\rm KS}-E_0)/E_0|$, and
the dimension of the Krylov subspace $n$ 
for several $\Delta_\tau$ values 
with the reference state $|\psi\rangle = |\Phi_A\rangle$.
As shown in Fig.~\ref{condition_r0}(a), 
when $\Delta_\tau$ is small,
the condition number grows rapidly in $n$ for ITE and RTE. 
With increasing $\Delta_\tau$, the condition number decreases
significantly for RTE, while the decrease is less pronounced for ITE.
On the other hand, the condition number for
QPM is much less sensitive to $\Delta_\tau$ than that for the other Krylov subspaces, 
even without the Richardson extrapolation ($r=0$) 
[also see Fig.~\ref{condition_r1}(a) for the results with the first-order Richardson extrapolation].  
Remarkably, RTE shows the smaller condition number than QPM
for $\Delta_\tau J = 0.5$ and $0.8$ as a function of $n$.
However,
as we shall discuss below, 
the smaller condition number does not necessarily guarantee 
the better approximation to the ground state.

Figure~\ref{condition_r0}(b)
shows the condition number as a function of
$|(E_{\rm KS}-E_0)/E_0|$. 
It is found that, in all the Krylov subspaces,
the condition number becomes larger as the 
error in energy becomes smaller. 
Such a behavior can be understood as follows. 
If a basis state $|u_i \rangle$ in the subspace
is close to the exact ground state,
$|u_i \rangle \sim |\Psi_0\rangle$,
a good variational energy can be obtained, while
the overlap matrix tends to be more ill conditioned because the generated states
$|u_{i+1}\rangle$ can be almost parallel to $|u_i \rangle$.
We should also note that the condition number
as a function of
$|(E_{\rm KS}-E_0)/E_0|$
in Fig.~\ref{condition_r0}(b) 
behaves rather similarly between ITE and RTE,
especially when $\Delta_\tau$ is small, and indeed it is almost 
identical when $\Delta_\tau J=0.01$.
As discussed above, this is expected because these two Krylov subspaces 
should be equivalent in a region of small $\Delta_\tau$.
Instead, the results for QPM are
very different from those for ITE and RTE, 
even when $\Delta_\tau$ is small.
This is simply because the Krylov subspace generated in QPM is different from those 
in ITE and RTE, including in the limit of $\Delta_\tau\to0$.

Figure~\ref{condition_r0}(c)
shows
$|(E_{\rm KS}-E_0)/E_0|$
as a function of $n$.
While ITE gives the better energy than the others when $n$ is small,
it is difficult to reach larger $n$ because of the
large condition number. 
On the other hand, RTE can achieve the largest dimension of
$n=35$ (or more) when $\Delta_\tau J=0.8$ owing to 
the smaller condition number. 
However, the significant improvement of
$|(E_{\rm KS}-E_0)/E_0|$
is not observed, 
despite the fact that the dimension of the subspace is substantially increased.  
We also find that both ITE and RTE give nearly the same error in energy at the maximum
dimension $n$ when the same $\Delta_\tau$ value is used 
[also see Fig.~\ref{condition_r0}(b)].  
The results for QPM are located somewhere between those for ITE and RTE, and, depending on
$\Delta_\tau$ values, QPM can achieve the best accuracy
in the ground-state energy among the three Krylov subspaces considered here. 
We should also note that 
the $\Delta_\tau$ dependence of the results is somewhat scattered for ITE and RTE, 
but it is more systematic for QPM.

\begin{center}
  \begin{figure*}
    \includegraphics[width=1.\textwidth]{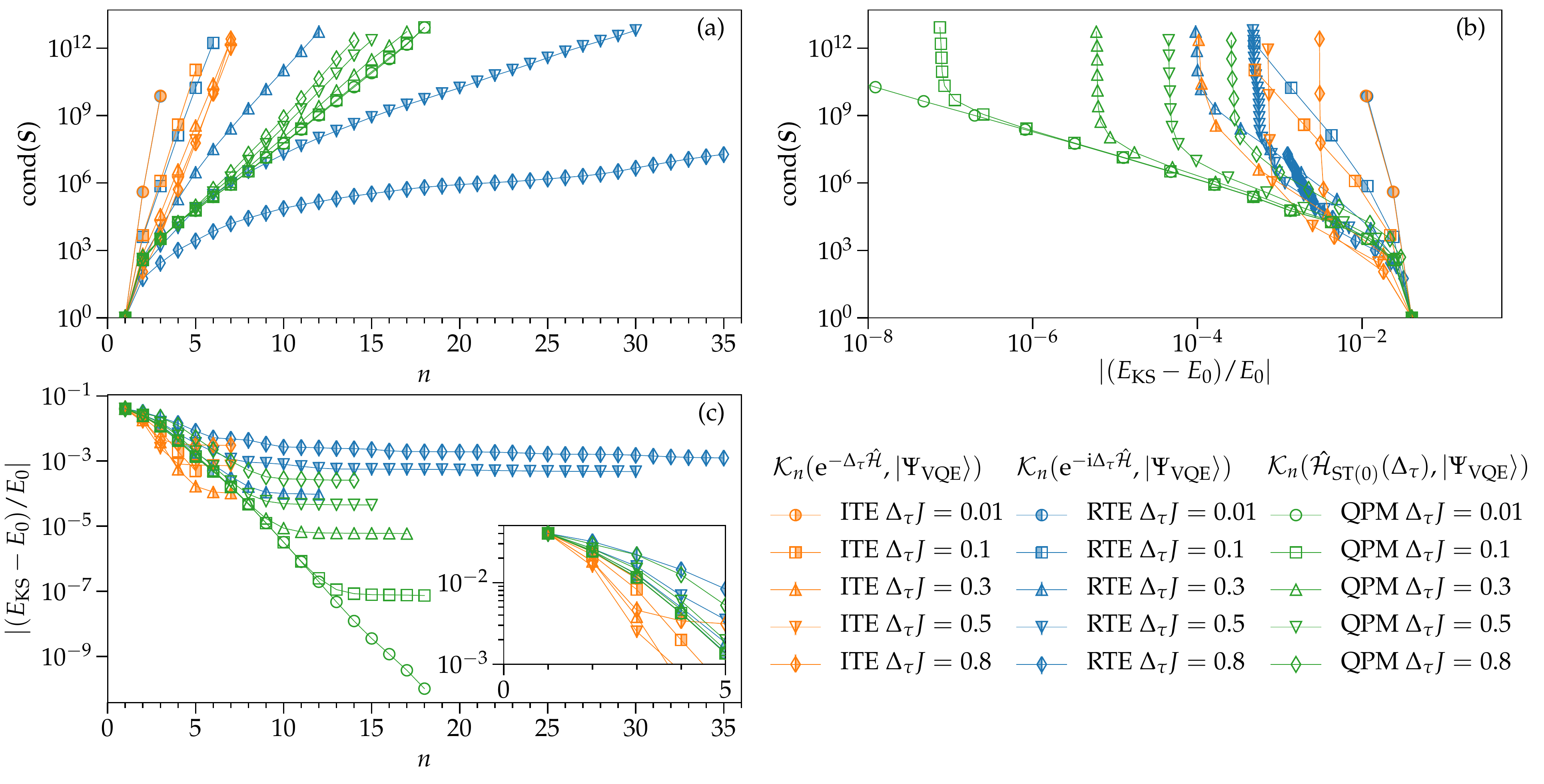}
    \caption{
      Same as Fig.~\ref{condition_r0} but for the reference state $|\psi\rangle=|\Psi_{\rm VQE}\rangle$.  
      Notice that the results for ITE and RTE with $\Delta_\tau J=0.01$ are identical in this scale.
      \label{condition_vqe_r0}
    }
  \end{figure*}
\end{center}

These features do not depend significantly on the choice of the reference state. 
Figure~\ref{condition_vqe_r0} summarizes the same results but for the reference state 
$|\psi\rangle = |\Psi_{\rm VQE}\rangle$. 
Since the VQE state $|\Psi_{\rm VQE}\rangle$ has a larger overlap with the exact ground state, 
a better convergence is generally expected in a Krylov-subspace diagonalization. 
Indeed, the error in energy
$|(E_{\rm KS}-E_0)/E_0|$
at $n=1$ is already one order of magnitude smaller 
than that for $|\psi\rangle = |\Psi_A\rangle$. 
Apart from this, the results are qualitatively the same as those in Fig.~\ref{condition_r0}

\begin{center}
  \begin{figure*}
    \includegraphics[width=1.\textwidth]{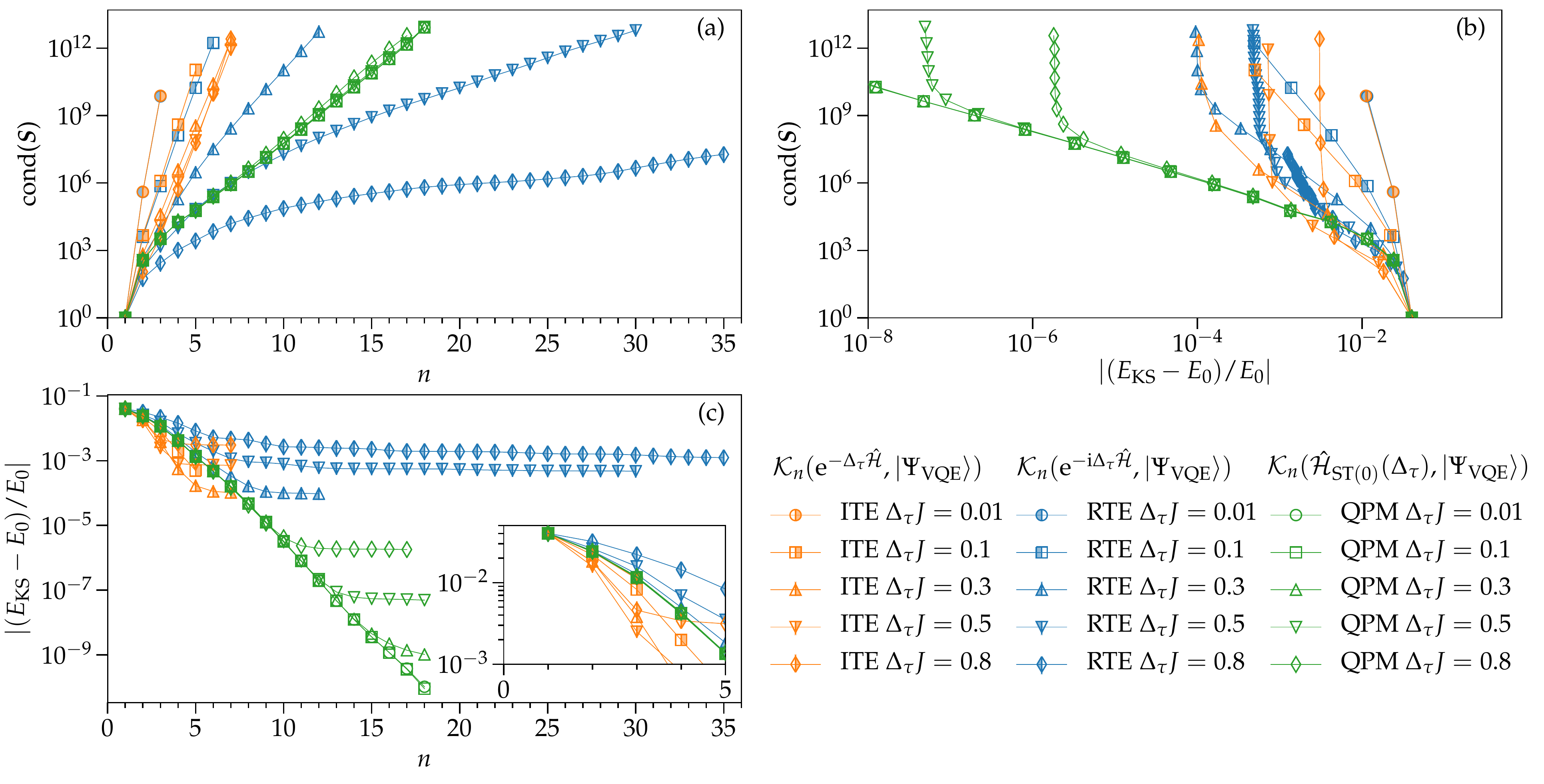}
    \caption{
      Same as Fig.~\ref{condition_vqe_r0} but with the first-order Richardson extrapolation ($r=1$) for QPM.  
      \label{condition_vqe_r1}
    }
  \end{figure*}
\end{center}

As demonstrated in some details in Sec.~\ref{sec:result},
the rather systematic dependence on $\Delta_\tau$ for QPM 
allows us to perform the 
Richardson extrapolation with an increase of measurements, e.g., 
$(r+1)^2=4$ times more for the first-order Richardson extrapolation. 
Figures~\ref{condition_r1} and ~\ref{condition_vqe_r1}
show the same results as in
Figs.~\ref{condition_r0} and ~\ref{condition_vqe_r0}
but with the first-order Richardson extrapolation for QPM.
With the Richardson extrapolation, 
the QPM results almost collapse on a single curve
(aside from the error in energy for large $n$ and large $\Delta_\tau$), 
implying that the $\Delta_\tau$ dependence becomes almost negligible.
Although ITE and RTE may perform better than QPM
in terms of the ground-state energy 
if an optimal $\Delta_\tau$ value can be found,
such an optimal value is not known {\it a priori}.
In addition, considering that the exact solution is generally unknown, 
the systematic dependence on $\Delta_\tau$ that 
guarantees the convergence to the exact is indispensable.
We also recall that performing an imaginary-time
evolution on a quantum computer is rather involved
when the QITE is used.
Thus, the application of the quantum power method
for the Krylov-subspace diagonalization would be
a promising alternative for the Krylov-subspace approaches
with quantum computers.

\subsection{Summary}

In short, as compared to the QLanczos method, 
our method is basically free from optimization of several parameters, 
such as the domain size $D$, the form of Hermitian operator $\hat{A}[m]$, and
the coefficients $\bs{a}[m]$, 
which requires measurements of Pauli strings for
$\bs{C}$ and $\bs{b}$ and solving the linear system $\bs{Ca}=\bs{b}$
of dimension approximately $4^D$. 
The only but crucial parameter in our method is $\Delta_\tau$, which 
introduces the time-discretization error, but the systematic error is well controlled, 
as demonstrated numerically above and throughout this paper. 
In addition, our method can separate the choice of $\Delta_\tau$ 
from the issue on the linear dependency of the basis states in the Krylov subspace, 
which is thus different from the other methods.
In general, smaller $\Delta_\tau$ approximates the ground state better. 
However, the well-controlled behavior of $\Delta_\tau$ allows us to
use large values of $\Delta_\tau$ and then 
extrapolate the results to $\Delta_\tau \to 0$. 
This is also advantageous when a simulation
on a noisy quantum computer is considered (see Sec.~\ref{sec:extrapolation}). 

\section{Fermi-Hubbard model}~\label{app:Hubbard}

\begin{center}
  \begin{figure}
    \includegraphics[width=.9\columnwidth]{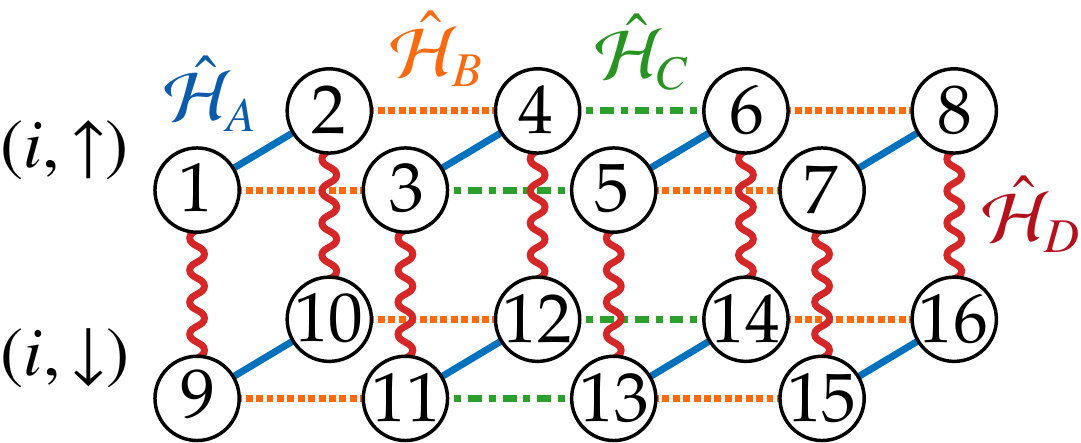}
    \caption{
      The qubit indexing used for the Fermi-Hubbard model on a square lattice 
      with a ladderlike $4\times 2$ cluster under open boundary conditions.
      A circle with a number denotes a qubit and the lines between
      qubits indicate the terms of the subdivided Hamiltonians
      $\hat{\mathcal{H}}_A$, $\hat{\mathcal{H}}_B$, $\hat{\mathcal{H}}_C$, and $\hat{\mathcal{H}}_D$ 
      in the fermion representation.
      Qubits 1--8 are assigned to fermions 
      at site $i$ ($1 \leqslant i \leqslant 8$) with spin $\uparrow$ 
     (upper layer), and
      qubits 9--16 are assigned to fermions 
      at site $i$ ($1 \leqslant i \leqslant 8$) with spin $\downarrow$ 
      (lower layer).
      The subdivided Hamiltonians
      $\hat{\mathcal{H}}_A$,
      $\hat{\mathcal{H}}_B$, and 
      $\hat{\mathcal{H}}_C$ correspond to the hopping ($J$) terms, while
      $\hat{\mathcal{H}}_D$ corresponds to the interaction ($U_{\rm H}$) terms,
      indicated respectively by
      the solid lines (blue),
      the dashed lines (orange),
      the dash-dotted lines (green), and
      the wavy lines (red).
    } \label{fig.Hubbard}
  \end{figure}
\end{center}

In this appendix, we demonstrate 
the Krylov-subspace diagonalization combined
with the quantum power method for 
a spin-1/2 Fermi-Hubbard model on a square lattice with a ladderlike $4\times 2$ cluster 
(i.e., $N=16$ qubits) under open boundary conditions (Fig.~\ref{fig.Hubbard}). 
The Hamiltonian of the Fermi-Hubbard model is given by 
\begin{equation}
  \hat{\mathcal{H}}
  = - J\sum_{\langle i,j \rangle,\,\sigma}
  \left(\hat{c}_{i\sigma}^\dag \hat{c}_{j\sigma} + {\rm H.c.} \right)
  + U_{\rm H}\sum_{i=1}^{N/2}
  \left(\hat{n}_{i \uparrow} -\frac{1}{2}\right)
  \left(\hat{n}_{i \downarrow} -\frac{1}{2}\right),
  \label{Ham_Hubbard}
\end{equation}
where
$\hat{c}_{i\sigma}^\dag$ 
($\hat{c}_{i\sigma}$) is a creation (annihilation) 
operator of a fermion at site $i$ with spin $\sigma\,(=\uparrow,\downarrow)$  
and satisfies the canonical anticommutation relations 
$\{\hat{c}_{i\sigma}, \hat{c}_{j\sigma^\prime}\}=
\{\hat{c}_{i\sigma}^\dag, \hat{c}_{j\sigma^\prime}^\dag\}=0$ and 
$\{\hat{c}_{i\sigma}, \hat{c}_{j\sigma^\prime}^\dag\}
= \delta_{ij} \delta_{\sigma\sigma^\prime}$. 
$\hat{n}_{i\sigma}=\hat{c}_{i\sigma}^\dag \hat{c}_{i\sigma}$ is the density operator and
$\langle i,j \rangle$ runs over all pairs of nearest-neighbor sites $i$ and $j$ on a square lattice. 
We assume that $J>0$, $U_{\rm H}>0$, and the fermion density is one, i.e., at half filling.

Before applying a fermion-to-qubit mapping, 
we first subdivide the Hamiltonian $\hat{\mathcal{H}}$ into noncommuting parts.  
For the $4 \times 2$ cluster, the Hamiltonian $\hat{\mathcal{H}}$ 
can be subdivided into $N_\Gamma=4$ parts as
\begin{equation}
  \hat{\mathcal{H}}=
  \hat{\mathcal{H}}_A
  +\hat{\mathcal{H}}_B
  +\hat{\mathcal{H}}_C
  +\hat{\mathcal{H}}_D, 
\end{equation}
where
$\hat{\mathcal{H}}_A$ denotes the hopping terms along the rung direction, 
$\hat{\mathcal{H}}_B$ the hopping terms on the odd bonds along the leg direction,
$\hat{\mathcal{H}}_C$ the hopping terms on the even bonds along the leg direction, and
$\hat{\mathcal{H}}_D$ the on-site interaction terms, as schematically shown in Fig.~\ref{fig.Hubbard}. 
Note that all terms in each subdivided Hamiltonian $\hat{\mathcal{H}}_\Gamma$ commute with each other, 
although $[\hat{\mathcal{H}}_\Gamma,\hat{\mathcal{H}}_{\Gamma'}]\ne0$ when $\Gamma\ne\Gamma'$.  
From Eq.~(\ref{ST2}), the lowest-order 
symmetric Suzuki-Trotter decomposition $\hat{S}_{2}(\Delta_\tau)$ 
of the time-evolution operator $\hat{U}(\Delta_\tau)=\e^{-\imag \Delta_\tau \hat{\mathcal{H}} }$ can be given as
\begin{equation}
  \hat{S}_{2}(\Delta_\tau)=
  \e^{-\imag\frac{\Delta_\tau}{2}\hat{\mathcal{H}}_A}
  \e^{-\imag\frac{\Delta_\tau}{2}\hat{\mathcal{H}}_B}
  \e^{-\imag\frac{\Delta_\tau}{2}\hat{\mathcal{H}}_C}
  \e^{-\imag \Delta_\tau\hat{\mathcal{H}}_D}
  \e^{-\imag\frac{\Delta_\tau}{2}\hat{\mathcal{H}}_C}
  \e^{-\imag\frac{\Delta_\tau}{2}\hat{\mathcal{H}}_B}
  \e^{-\imag\frac{\Delta_\tau}{2}\hat{\mathcal{H}}_A}. 
  \label{S2Hub}
\end{equation}

We now assign the fermion indexes to the qubit indexes as 
$(i,\uparrow) \mapsto  i_\uparrow \equiv i$ and  
$(i,\downarrow) \mapsto i_\downarrow \equiv i+N/2$ for $1 \leqslant i \leqslant N/2$~\cite{Reiner2016}. 
The full indexing for the $4\times2$ cluster is given in Fig.~\ref{fig.Hubbard}. 
Let us now apply the Jordan-Wigner transformation to represent  
the fermion creation and annihilation operators by Pauli operators as~\cite{Jordan1928,Rodriguez1959} 
\begin{equation}
  \hat{c}_{i\sigma}^\dag = \hat{\sigma}_{i_\sigma}^{-} \prod_{k < {i_\sigma}} \hat{Z}_k
\end{equation}
and
\begin{equation}
  \hat{c}_{i\sigma} = \prod_{k < {i_\sigma}} \hat{Z}_k \ \hat{\sigma}_{i_\sigma}^{+} 
  = \hat{\sigma}_{i_\sigma}^{+} \prod_{k < {i_\sigma}} \hat{Z}_k,
\end{equation}
where $\hat{\sigma}_i^{\pm} \equiv \frac{1}{2}(\hat{X}_{i} \pm \imag \hat{Y}_{i})$. 
The fermion density operator is given by
$\hat{n}_{i\sigma}=\hat{\sigma}_{i_\sigma}^- \hat{\sigma}_{i_\sigma}^+
=\frac{1}{2}(1-\hat{Z}_{i_\sigma})$,
implying that the single-particle state $(i,\sigma)$ is
occupied (unoccupied)
if the $i_\sigma$th qubit state is
$|1\rangle_{i_\sigma}$ ($|0\rangle_{i_\sigma}$).
Now the Fermi-Hubbard Hamiltonian in the qubit representation reads 
\begin{equation}
  \hat{\mathcal{H}}=-\frac{J}{2} \sum_\sigma \sum_{\langle i_\sigma, j_\sigma \rangle}
  \left(
  \hat{X}_{i_\sigma} \hat{X}_{j_\sigma} +
  \hat{Y}_{i_\sigma} \hat{Y}_{j_\sigma} 
  \right)
  \hat{Z}_{{\rm JW},i_\sigma j_\sigma} 
  + \frac{U_{\rm H}}{4} \sum_{i=1}^{N/2} \hat{Z}_i \hat{Z}_{i+N/2}.
  \label{JWHub}
\end{equation}
Here, 
$\hat{Z}_{{\rm JW},ij}=\prod_{i \lessgtr k \lessgtr j} \hat{Z}_k$
is the Jordan-Wigner string for $i \lessgtr j$,
which supplies the fermion sign $\pm1$, depending
on the population parity of the fermion occupation
between the $i$th and $j$th qubits. 
Note that the subdivided Hamiltonians $\hat{\mathcal{H}}_\Gamma$ with $\Gamma=A,B,C,D$ 
are similarly transformed in the qubit representation (see Fig.~\ref{fig.Hubbard}) and apparently 
all terms in each $\hat{\mathcal{H}}_\Gamma$ in the qubit representation still commute with each other.

\begin{center}
  \begin{figure*}
    \includegraphics[width=1.95\columnwidth]{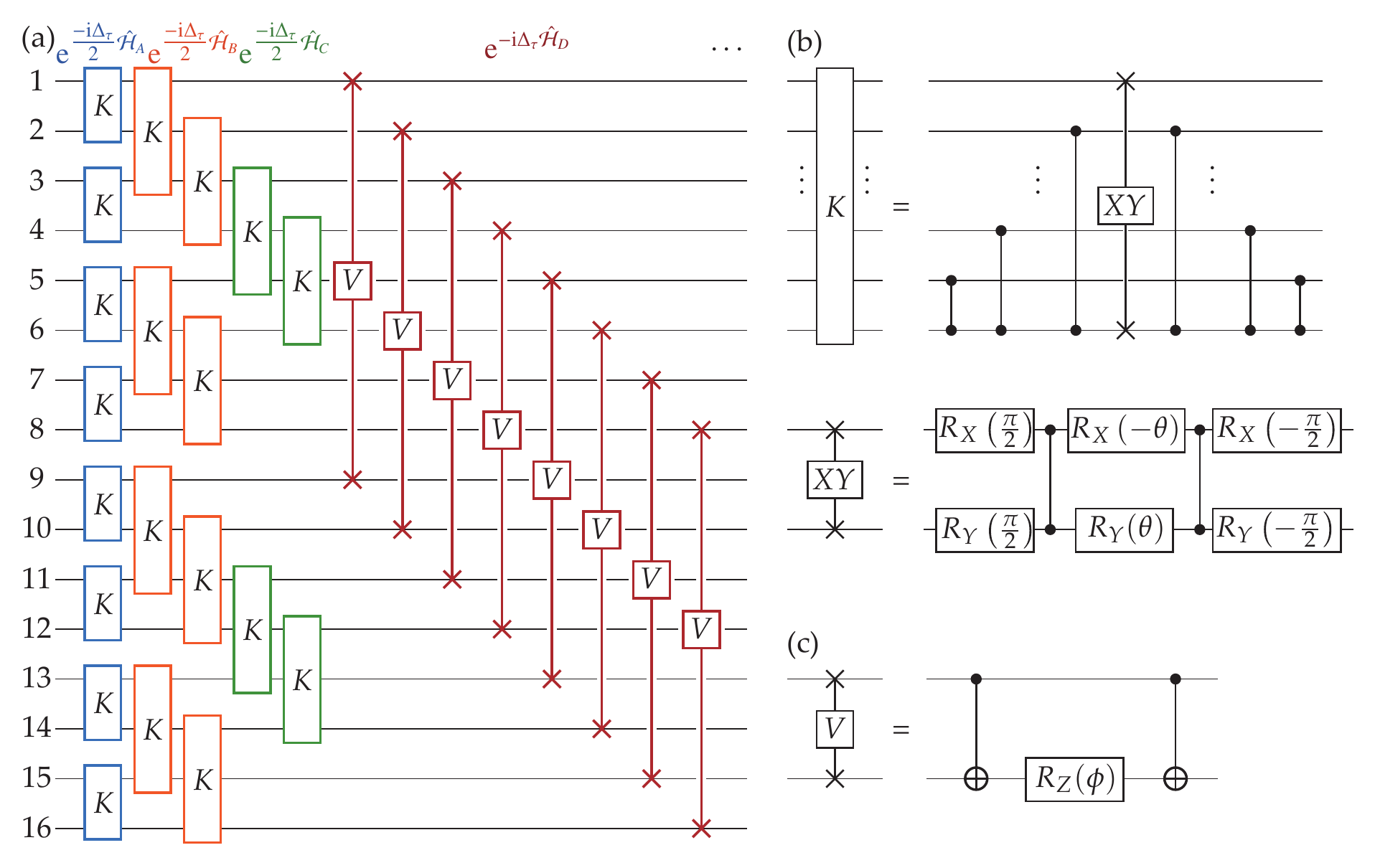}
    \caption{
      (a) Circuit structure of the lowest-order 
      Suzuki-Trotter-decomposed time-evolution operator $\hat{S}_{2}(\Delta_\tau)$  
      in Eq.~(\ref{S2Hub}) 
      for the Fermi-Hubbard model considered here in Eq.~(\ref{JWHub}). 
      The gate $K$ 
      denotes the exponentiated hopping term defined in (b), while 
      the gate $V$ denotes the exponentiated interaction term defined in (c). 
      Here, only the part corresponding to the first four exponentials is shown. 
      (b) Decomposition of the $K$ gate that represents 
      $\exp[-\imag\theta(\hat{X}_i\hat{X}_j+\hat{Y}_i\hat{Y}_j)\hat{Z}_{{\rm JW},ij}/2]$ 
      operating at qubits $i$ and $j$ as well as all qubits between these two qubits. 
      The gate $XY$ represents 
      $\exp[-\imag\theta(\hat{X}_i\hat{X}_j+\hat{Y}_i\hat{Y}_j)/2]$ 
      operating at qubits $i$ and $j$. 
      The gate $R_{X(Y)}(\theta)$ is given by 
      $R_{X(Y)}(\theta)=\exp[-\imag\theta\hat{X}_i(\hat{Y}_i)/2]$ 
      operating at qubit $i$.
      (c) Decomposition of the $V$ gate that represents 
      $\exp[-\imag\phi\hat{Z}_i\hat{Z}_j/2]$ 
      operating at qubits $i$ and $j$. 
      The gate $R_Z(\phi)$ is given by 
      $R_{Z}(\phi)=\exp[-\imag\phi\hat{Z}_i/2]$ 
      operating at qubit $i$. 
      Here, $\theta=-\Delta_\tau J/2$ and $\phi=\Delta_\tau U_{\rm H}/2$ for the Fermi-Hubbard model given in Eq.~(\ref{JWHub}). 
      }
  \label{fig.Hubbardcircuit}
  \end{figure*}
\end{center}

  Figure~\ref{fig.Hubbardcircuit} illustrates a
  circuit structure of the lowest-order Suzuki-Trotter-decomposed
  time-evolution operator $\hat{S}_{2}(\Delta_\tau)$ in Eq.~(\ref{S2Hub}) 
  for the Fermi-Hubbard model given in Eq.~(\ref{JWHub}) obtained by the Jordan-Wigner transformation. 
  To implement $\hat{S}_{2}(\Delta_\tau)$ in a circuit, 
  we first exponentiate the hopping term, i.e., 
  $\exp[-\imag \theta (\hat{X}_i\hat{X}_j+\hat{Y}_i\hat{Y}_j)\hat{Z}_{{\rm JW},ij}/2]$,  
  by using a circuit representation of 
  $\exp[-\imag \theta (\hat{X}_i\hat{X}_j+\hat{Y}_i\hat{Y}_j)/2]$ (denoted as the $XY$ gate in Fig.~\ref{fig.Hubbardcircuit}), 
  which is composed of 2 {\sc cz} gates and 6 single-qubit rotations~\cite{fossfeig2020holographic}, 
  sandwiched by $2(|i-j|-1)$ {\sc cz} gates~\cite{Reiner2019} 
  (also see Refs.~\cite{Vidal2004,Wecker2015,Dallaire-Demers2016gates,Reiner2018,dallairedemers2020application,arute2020observation}
  for other possible circuit
  realizations of the same operator). 
  Here, the rotation angle is given by $\theta = -\Delta_\tau J/2$. 
  This gate is denoted as the $K$ gate in Fig.~\ref{fig.Hubbardcircuit}. 
  The interaction term is transformed
  into the Ising interaction and hence it can be exponentiated
  with a single-qubit rotation $\exp[-\imag \phi \hat{Z}_{i+N/2}/2]$
  sandwiched by 2 {\sc c}$_i${\sc not}$_{i+N/2}$ gates~\cite{NielsenChuang} 
  (denoted as the $V$ gate in Fig.~\ref{fig.Hubbardcircuit}). 
  Here, the rotation angle is given by $\phi = \Delta_\tau U_{\rm H} /2$.

  As the reference states for the Krylov-subspace diagonalization, we consider the following 
  product states:
  \begin{alignat}{1}
    &|\Phi_A\rangle = \otimes_{i=1}^{N/2} |t_{2i-1,2i} \rangle, \\
    &|Z_{\rm AFM 1}\rangle = \otimes_{i=1}^{N/4} |0\rangle_{2i-1} |1_{2i}\rangle \otimes_{i=1}^{N/4}|1\rangle_{2i-1+N/2} |0\rangle_{2i+N/2},\\
    &|Z_{\rm AFM 2}\rangle = \otimes_{i=1}^{N/4} |1\rangle_{2i-1} |0_{2i}\rangle \otimes_{i=1}^{N/4}|0\rangle_{2i-1+N/2} |1\rangle_{2i+N/2},
  \end{alignat}
  where 
  $|t_{i,j} \rangle = \frac{1}{\sqrt{2}}(|0\rangle_{i} |1\rangle_{j} + |1 \rangle_{i} |0 \rangle_{j})$
  is one of the Bell states and can be interpreted
  as a ``spin-triplet state'' in the qubit representation or 
  as a ``bonding state'' in the fermion representation. 
  $|\Phi_A\rangle$ is the ground state of the subdivided Hamiltonian $\hat{\mathcal{H}}_A$, 
  because $\hat{\mathcal{H}}_A$ after the Jordan-Wigner
  transformation is merely a direct sum of two-site $XY$ models 
  with the ferromagnetic exchange interaction $-J<0$, 
  and hence the ground state is given by the direct product of $|t_{i,j}\rangle$. 
  $|Z_{\rm AFM 1}\rangle$ and $|Z_{\rm AFM 2}\rangle$ are the N\'{e}el states 
  (both in the qubit and fermion representations) 
  with the staggered moments pointing alternatively along the spin-$Z$ axis.  

  All these three states are within the subspace of the
  half filling and zero magnetization
  because they have
  $N/4$\ $|0\rangle_{i_\sigma}$'s for $1 \leqslant i_\sigma \leqslant N/2$ and
  $N/4$\ $|0\rangle_{i_\sigma}$'s for $N/2+1 \leqslant i_\sigma \leqslant N$. 
  Moreover, these states  
  can be easily generated from $|0\rangle^{\otimes N}$ with
  appropriate combinations of Pauli $X$, Hadamard, and {\sc cnot} gates. 
  The particle number and magnetization are conserved 
  even after applying the Hamiltonian power to these states. 
  In addition to these three states, we adopt 
  the ground state of $\hat{\mathcal{H}}$ at $U_{\rm H}=0$, 
  $|\Psi_{U_{\rm H}=0}\rangle$, 
  as a reference state.
  Since $|\Psi_{U_{\rm H}=0}\rangle$ is a Slater determinant,
  i.e., a particular case of fermionic Gaussian states,
  it can in principle be prepared on a quantum circuit
  with at most $O(N^2)$ gates~\cite{Wecker2015,Kivlichan2018,Jiang2018,Shirakawa2020}.  
  $|\Psi_{U_{\rm H}=0}\rangle$ is also within the subspace of the half filling and zero magnetization.

\begin{center}
    \begin{figure}
      \includegraphics[width=0.95\columnwidth]{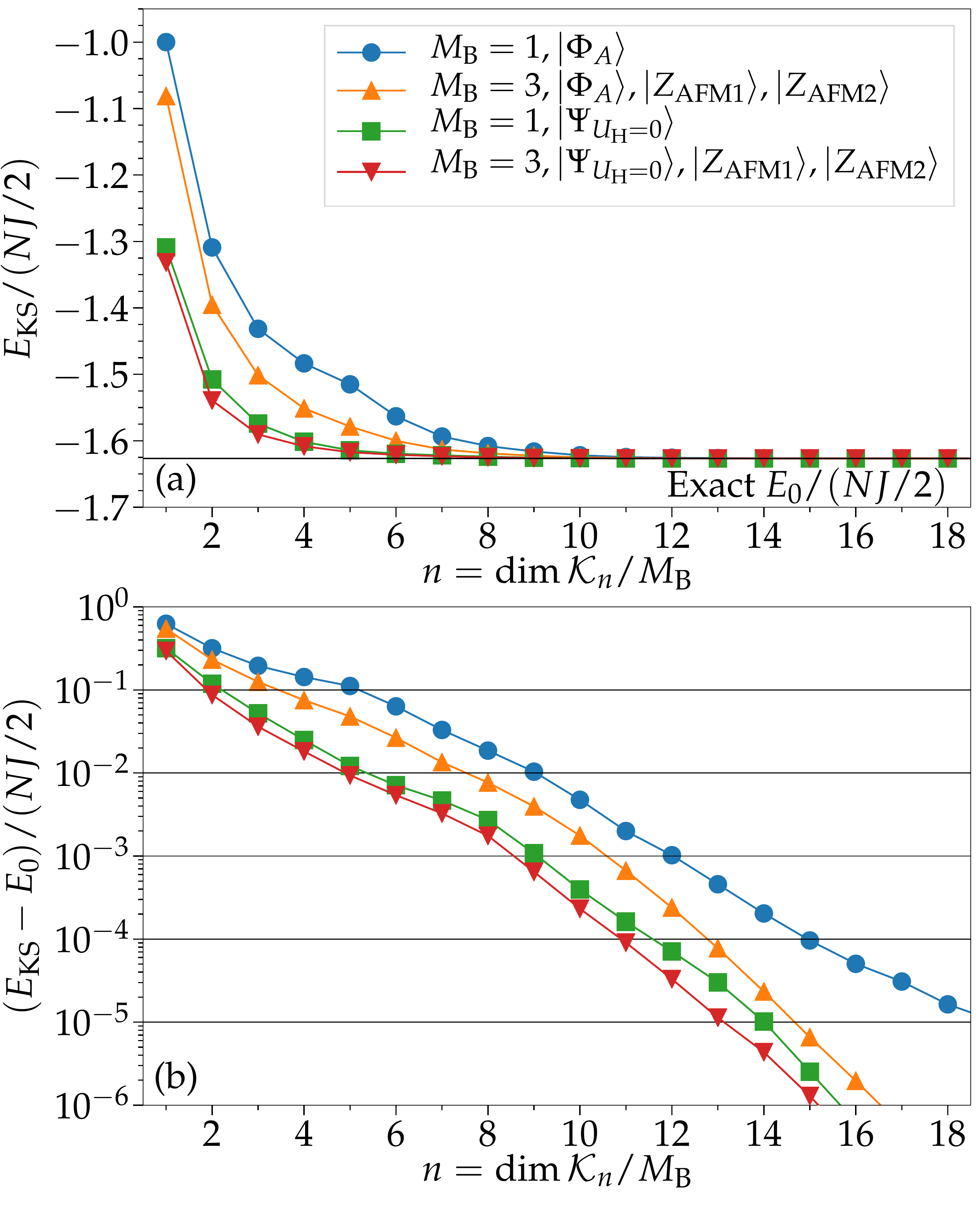}
      \caption{
        Same as Fig.~\ref{fig.energy} 
        but for the Fermi-Hubbard model with $U_{\rm H}/J=4$ at half filling. 
      }
      \label{fig.Hubenergy}
    \end{figure}
  \end{center}

\begin{center}
    \begin{figure}
      \includegraphics[width=0.95\columnwidth]{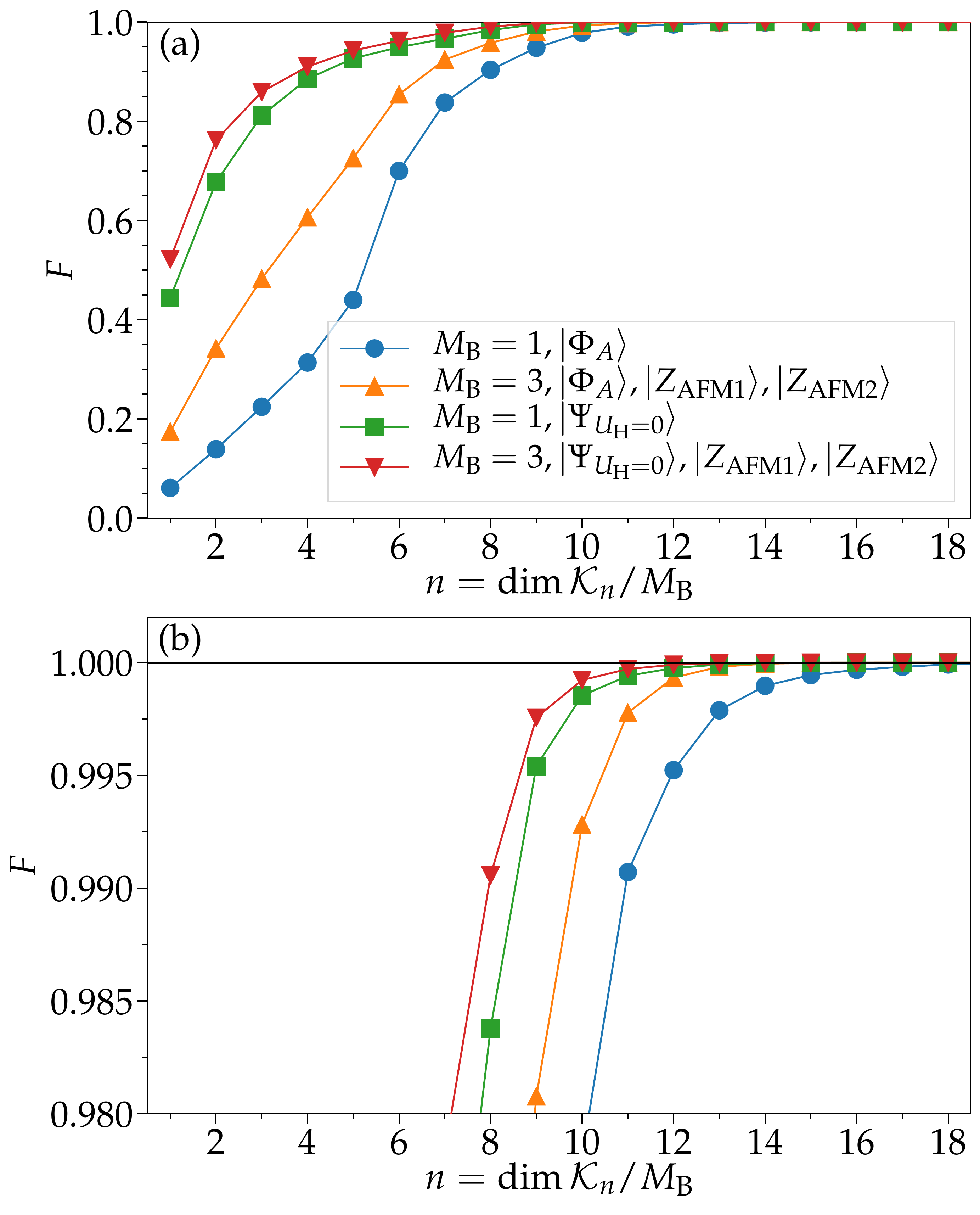}
      \caption{
        Same as Fig.~\ref{fig.fidelity}   
        but for the Fermi-Hubbard model with $U_{\rm H}/J=4$ at half filling.
      }
      \label{fig.Hubfidelity}
    \end{figure}
  \end{center}

  Figures~\ref{fig.Hubenergy} and \ref{fig.Hubfidelity} 
  show the numerical results of the estimated ground-state energy $E_{\rm KS}$ and 
  the ground-state fidelity $F=|\langle\Psi_0|\Psi_{\rm KS}\rangle|^2$, respectively, 
  for the Fermi-Hubbard model with $U_{\rm H}/J=4$, obtained 
  by the same procedures as in the case of the spin-1/2 Heisenberg model discussed in Sec.~\ref{sec:egs}. 
  Here, $|\Psi_0\rangle$ is the exact ground state and we set the time interval $\Delta_\tau J=0.05$ with 
  $r=1$, $m=1$, $p=3$, and $N_\Gamma=4$ for 
  approximating the Hamiltonian power in the numerical simulations.   
  The exact ground-state energy per site is
  $E_0/(NJ/2)=-1.626562894$. Note that $N$ is the number of qubits and the number of lattice sites 
  of the Fermi-Hubbard model is given by $N/2$. 

  As shown in Figs.~\ref{fig.Hubenergy} and \ref{fig.Hubfidelity},
  the convergence to the ground state 
  is improved with increasing the block size $M_{\rm B}$, 
  which is similar to the case of the spin-1/2 Heisenberg model found in Figs.~\ref{fig.energy} and \ref{fig.fidelity}. 
  We also observe in Fig.~\ref{fig.Hubenergy}(b) the exponential convergence of the
  energy with respect to $n$ for any set of reference states. 
  We can also notice in Fig.~\ref{fig.Hubfidelity}(a) that the noninteracting ground state $|\Psi_{U_{\rm H}=0}\rangle$  
  has a significantly larger overlap with $|\Psi_0\rangle$
  than $|\Phi_{A}\rangle$, 
  and indeed the results obtained with $|\Psi_{U_{\rm H}=0}\rangle$  
  shows the faster convergence than those with $|\Phi_A\rangle$. 
  A relatively slower convergence found here for the Fermi-Hubbard model as compared to 
  the case of the spin-1/2 Heisenberg model, 
  when only the simple product states
  other than $|\Psi_{U_{\rm H}=0}\rangle$ are used, 
  might be due to 
  the smaller state overlaps between
  the reference states and the exact ground state. 
  These results clearly demonstrate that 
  the quantum power method can also be effective for
  fermion systems where a transformed Hamiltonian in the qubit representation 
  involves more complex terms.

  Finally, it should be noted that although 
  the Bravyi-Kitaev transformation is
  known to scale asymptotically better than
  the Jordan-Wigner transformation in gate count,  
  the difference in efficiency between the two transformations
  is not obvious for small systems such as that considered here with $N=16$ 
  (see, for example, Refs.~\cite{Tranter2018,cade2019strategies}). 
  In addition, since the
  Jordan-Wigner transformation is based on the
  occupation basis of fermions, 
  density-density interactions such as the 
  Hubbard interaction can be written simply as
  an Ising Hamiltonian [see Eq.~(\ref{JWHub})],
  while Pauli strings might be involved 
  in the interaction term when 
  the Bravyi-Kitaev transformation is employed 
  (see, for example, Refs.~\cite{Seeley2012,Havlicek2017}). 
  A comparative study of fermion-to-qubit mappings 
  for the Hubbard model is, however, beyond the scope of this paper,  
  and we have adopted the Jordan-Wigner transformation
  in favor of the simple form of the interaction term.

\section{Higher-order symmetric Suzuki-Trotter decompositions $\hat{S}_{2m}^{(p)}(\Delta_\tau)$}\label{app:ST}

In this appendix,
we provide a {\sc Python} program that generates
coefficients required for the
higher-order symmetric Suzuki-Trotter decompositions $\hat{S}_{2m}^{(p)}(\Delta_\tau)$ 
introduced in Sec.~\ref{sec:rec_ST}, and examine numerically the systematic errors due to the Suzuki-Trotter decompositions 
$\hat{S}_{2m}^{(p)}(\Delta_\tau)$ with different parameters $m$ and $p$. 
Note that $m$ is an integer with $m\geqslant 1$ and $p$ is an odd integer with $p\geqslant 3$.

\subsection{Coefficients for higher-order Suzuki-Trotter decompositions}\label{app:hoST}
Listing~1 shows a {\sc Python} program
that generates the coefficients $\{s_i\}_{i=1}^{D^{(p)}_{2m}}$
for a given set of parameters $m$ and $p$ in the symmetric Suzuki-Trotter decompositions $\hat{S}_{2m}^{(p)}(\Delta_\tau)$: 
\begin{alignat}{1}
  \hat{S}_{2m}^{(p)}(\Delta_\tau) &=
  \e^{x s_1\hat{\mathcal{H}}_A}
  \e^{x s_2\hat{\mathcal{H}}_B}
  \e^{x s_3\hat{\mathcal{H}}_C} \notag \\
  &\times
  \cdots 
  \times
  \e^{x s_{D^{(p)}_{2m}-2}\hat{\mathcal{H}}_C}
  \e^{x s_{D^{(p)}_{2m}-1}\hat{\mathcal{H}}_B}
  \e^{x s_{D^{(p)}_{2m}}\hat{\mathcal{H}}_A},
  \label{si}
\end{alignat}
where $x=-\imag \Delta_\tau$ and
$D^{(p)}_{2m}=2(N_\Gamma-1)p^{m-1}+1$ as given in Eq.~(\ref{depth}).
The program includes  
an example for $m=2$, $p=5$, and $N_\Gamma=2$. 
In this case, 
the symmetric Suzuki-Trotter decomposition has a form 
\begin{alignat}{1}
  \hat{S}_{4}^{(5)}(\Delta_\tau) &=
  \e^{x s_1\hat{\mathcal{H}}_A}
  \e^{x s_2\hat{\mathcal{H}}_B}
  \e^{x s_3\hat{\mathcal{H}}_A}
  \e^{x s_4\hat{\mathcal{H}}_B}\notag \\
  &\times
  \e^{x s_5\hat{\mathcal{H}}_A}
  \e^{x s_6\hat{\mathcal{H}}_B}
  \e^{x s_7\hat{\mathcal{H}}_A}\notag \\
  &\times
  \e^{x s_8\hat{\mathcal{H}}_B}
  \e^{x s_9\hat{\mathcal{H}}_A}
  \e^{x s_{10}\hat{\mathcal{H}}_B}
  \e^{x s_{11}\hat{\mathcal{H}}_A} 
\end{alignat}
and the output of the program gives the $11$ 
coefficients 
\begin{alignat}{1}
s_1 &=0.20724538589718786, \notag \\
s_2 &=0.4144907717943757, \notag \\
s_3 &=0.4144907717943757, \notag \\
s_4 &=0.4144907717943757,\notag \\
s_5 &=-0.12173615769156357, \notag \\
s_6 &=-0.6579630871775028, \notag \\
s_7 &=-0.12173615769156357, \notag \\
s_8 &=0.4144907717943757, \notag \\
s_9 &=0.4144907717943757, \notag \\
s_{10} &=0.4144907717943757, \notag \\
s_{11} &=0.20724538589718786. \notag
\end{alignat}
By modifying lines 21-23 in the program,
one can obtain $\{s_{i}\}_{i=1}^{D^{(p)}_{2m}}$
for other values of $m$, $p$, and $N_\Gamma$. 

Notice that the coefficients $\{s_i\}_{i=1}^{D^{(p)}_{2m}}$ are symmetric, i.e., 
\begin{equation}
  s_i=s_{D^{(p)}_{2m}-i+1}
\end{equation}
and satisfy the following sum rule:  
\begin{equation}
  \sum_{i=1}^{D^{(p)}_{2m}}s_i=N_\Gamma 
\end{equation}
for any $m$ and $p$.
Although it is sufficient to find the coefficients $\{s_{i}\}_{i=1}^{D^{(p)}_{2m}}$ 
for our purpose, 
the program can also output a cumulative
sum $T_i$ of the coefficient $s_i$ defined as
\begin{equation}
  T_{i}=\sum_{k=1}^{i}s_k.  
\end{equation}
By plotting $T_{i}$ as a function of $i$ (or $i/D_{2m}^{(p)}$)
for several values of $m$ with a fixed $p$, one can find 
a fractal feature appearing in the higher-order 
Suzuki-Trotter decompositions~\cite{Suzuki1990,Hatano2005}.

\begin{minipage}[!t]{1.0\columnwidth}
  \begin{lstlisting}[language=Python, caption=A {\sc Python} program for generating the coefficients $\{s_i\}_{i=1}^{D^{(p)}_{2m}}$ 
  in the symmetric Suzuki-Trotter decomposition $\hat{S}_{2m}^{(p)}$.]
import numpy
    
def Suzuki_Trotter(m,p,NGamma):
    s=numpy.array([0.5 for _ in range(2*NGamma-1)])
    s[NGamma-1]=1.0
    pl=int((p-1)/2)
    for mm in range(m-1):
        k0=1.0/(p-1-(p-1)**(1.0/(2*mm+3)))
        k1=1.0-(p-1)*k0
        sl=s[:-1]*k0
        sl[0]=sl[0]*2.0
        sl=numpy.concatenate([sl for _ in range(pl)])
        sl[0]=sl[0]/2.0
        s=s*k1
        s[0]=s[0]+sl[0]
        s[-1]=s[0]
        s=numpy.concatenate([sl,s,sl[::-1]])
    return s

# example                                                                                                      
m=2
p=5
NGamma=2
s=Suzuki_Trotter(m,p,NGamma)
T=numpy.cumsum(s)
print('order =',2*m)
print('depth =',len(s))
print('sum(s) =',numpy.sum(s))
for i in range(len(s)):
    print(i+1,s[i],T[i])
\end{lstlisting}
\end{minipage}

\subsection{Numerical examination of a Suzuki-Trotter error}\label{ST_numerical}

Here we numerically examine the systematic errors due to the Suzuki-Trotter decompositions 
$\hat{S}_{2m}^{(p)}(\Delta_\tau)$ with different parameters $m$ and $p$.  
The Trotter formula~\cite{Trotter1959,Suzuki1976PTP,Suzuki1976}
combined with $\hat{S}_{2m}^{(p)}(\Delta_\tau)$ yields 
\begin{equation}
  \hat{U}(t) = \left[\hat{S}_{2m}^{(p)}(\Delta_\tau)\right]^{M} + O(t \Delta_\tau^{2m}),
  \label{Uapprox}
\end{equation}
where $M$ is an integer such that $t=M\Delta_\tau$.

\begin{center}
  \begin{figure*}
    \includegraphics[width=1.95\columnwidth]{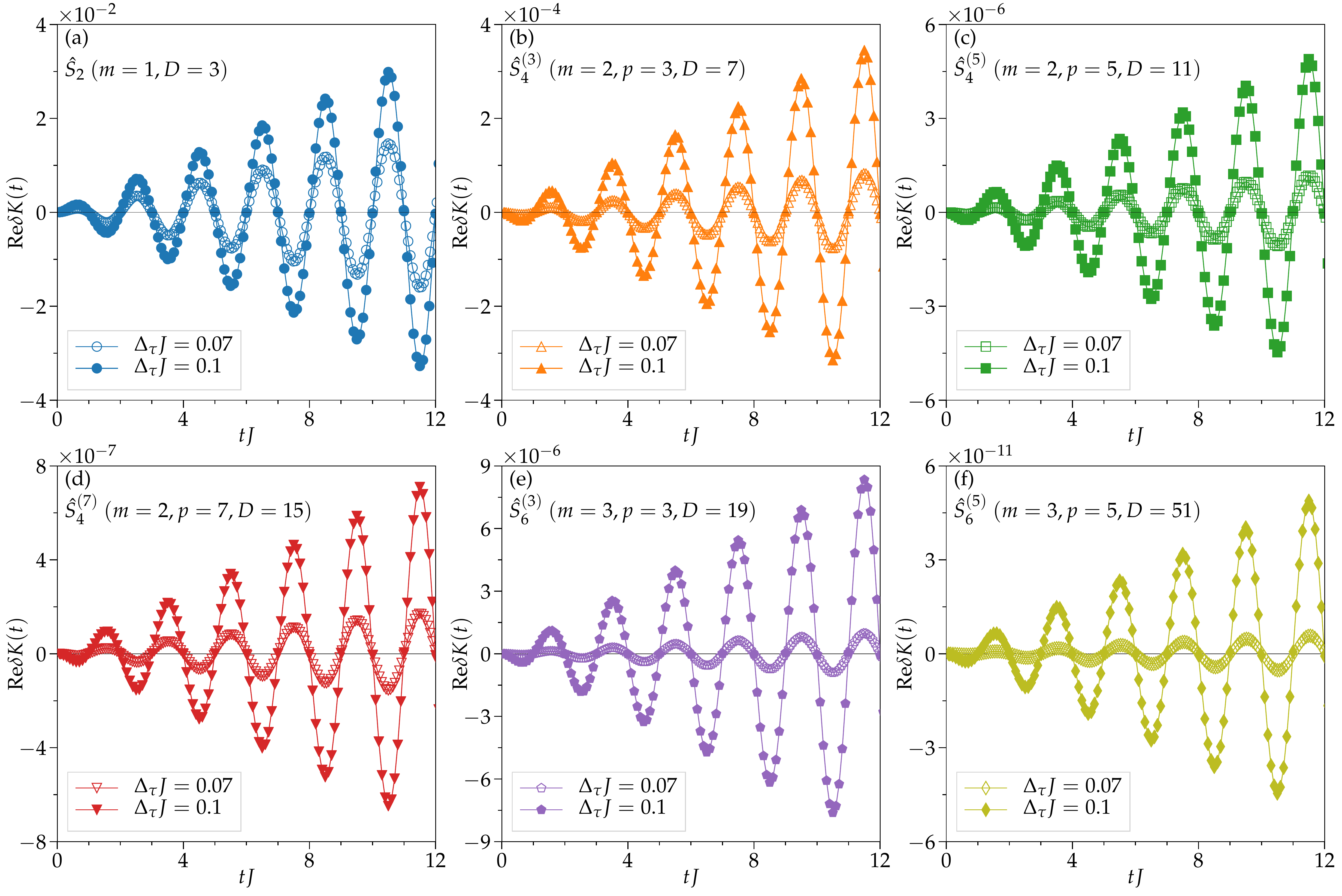}
    \caption{
      Deviation of the real part of the propagator 
      from the exact value, ${\rm Re} \delta K(t)$, for
      $\Delta_\tau J=0.07$ (empty symbols) and
      $\Delta_\tau J=0.1$ (filled symbols) 
      with different approximation schemes of the symmetric Suzuki-Trotter decomposition 
      $\hat{S}_{2m}^{(p)}(\Delta_\tau)$ for the time-evolution operator. 
      (a) $m=1$,        $D=3$,  
      (b) $m=2$, $p=3$, $D=7$,  
      (c) $m=2$, $p=5$, $D=11$,  
      (d) $m=2$, $p=7$, $D=15$,  
      (e) $m=3$, $p=3$, $D=19$,  and 
      (f) $m=3$, $p=5$, $D=51$,  
      where $D=D^{(p)}_{2m}$ is the depth of a single
      $\hat{S}_{2m}^{(p)}(\Delta_\tau)$ given in Eq.~(\ref{depth}). 
      Note that $\hat{S}_2(\Delta_\tau)$ corresponds to
      $\hat{S}_{2m}^{(p)}(\Delta_\tau)$ with 
      $m=1$ and $p=3$. 
      The results are for the spin-1/2 Heisenberg model on an $N=16$ qubit ring described 
      by the Hamiltonian $\mathcal{\hat{H}}$ in Eq.~(\ref{Ham_SWAP}) and $N_\Gamma=2$.
      The solid lines are guide for the eye.
    } \label{fig.time_evolve_nodivision}
  \end{figure*}
\end{center}

Figure~\ref{fig.time_evolve_nodivision} shows 
the real part of the difference between the
exact propagator 
\begin{equation}
  K(t)= \langle \Psi_0 |\hat{U}(t) |\Psi_0 \rangle
\end{equation}
and the approximated propagator
\begin{equation}
  \tilde{K}(t)= \langle \Psi_0 |\left[\hat{S}_{2m}^{(p)}(\Delta_\tau)\right]^{M} |\Psi_0 \rangle, 
\end{equation}
i.e., 
\begin{equation}
  {\rm Re} \delta K(t) =
  {\rm Re} \tilde{K}(t) - {\rm Re} K(t),
  \label{deltaK}
\end{equation}
with $\Delta_\tau J=$ 0.07 and 0.1 for the spin-1/2 Heisenberg model on an $N=16$ qubit ring described by the Hamiltonian 
$\mathcal{\hat{H}}$ in Eq.~(\ref{Ham_SWAP}), 
in which the Hamiltonian is subdivided into $N_\Gamma=2$ parts.  
Here, $|\Psi_0\rangle$ is the exact ground state.  
The exact propagator is simply given by 
$K(t)=\e^{-\imag E_0 t}$, 
where $E_0$ is the exact ground-state energy. 
As expected, when $p$ is fixed, 
the error decreases by orders of magnitude with increasing $m$.
It is also found that, when $m$ is fixed, 
the error decreases by orders of magnitude with increasing $p$.
Although we only show ${\rm Re}\delta K(t)$, the imaginary part of the difference, 
${\rm Im}\delta K(t)$, behaves similarly.

We should emphasize here that while the deviation of
the approximated propagator $\tilde{K}(t)$ from the exact one 
$K(t)$ becomes larger in the long time limit ($tJ \gg 1$),
the quantum power method proposed here is formulated on the basis of 
the time-evolution operators $\hat{U}(t)$ at time $t$ close to zero,
for which the deviation is small. 
Therefore, this is another advantage of the quantum power 
method in controlling the Suzuki-Trotter error over other
quantum algorithms that require the long-time dynamics 
approximately described by the Suzuki-Trotter-decomposed
time-evolution operators.

\begin{center}
  \begin{figure*}
    \includegraphics[width=1.95\columnwidth]{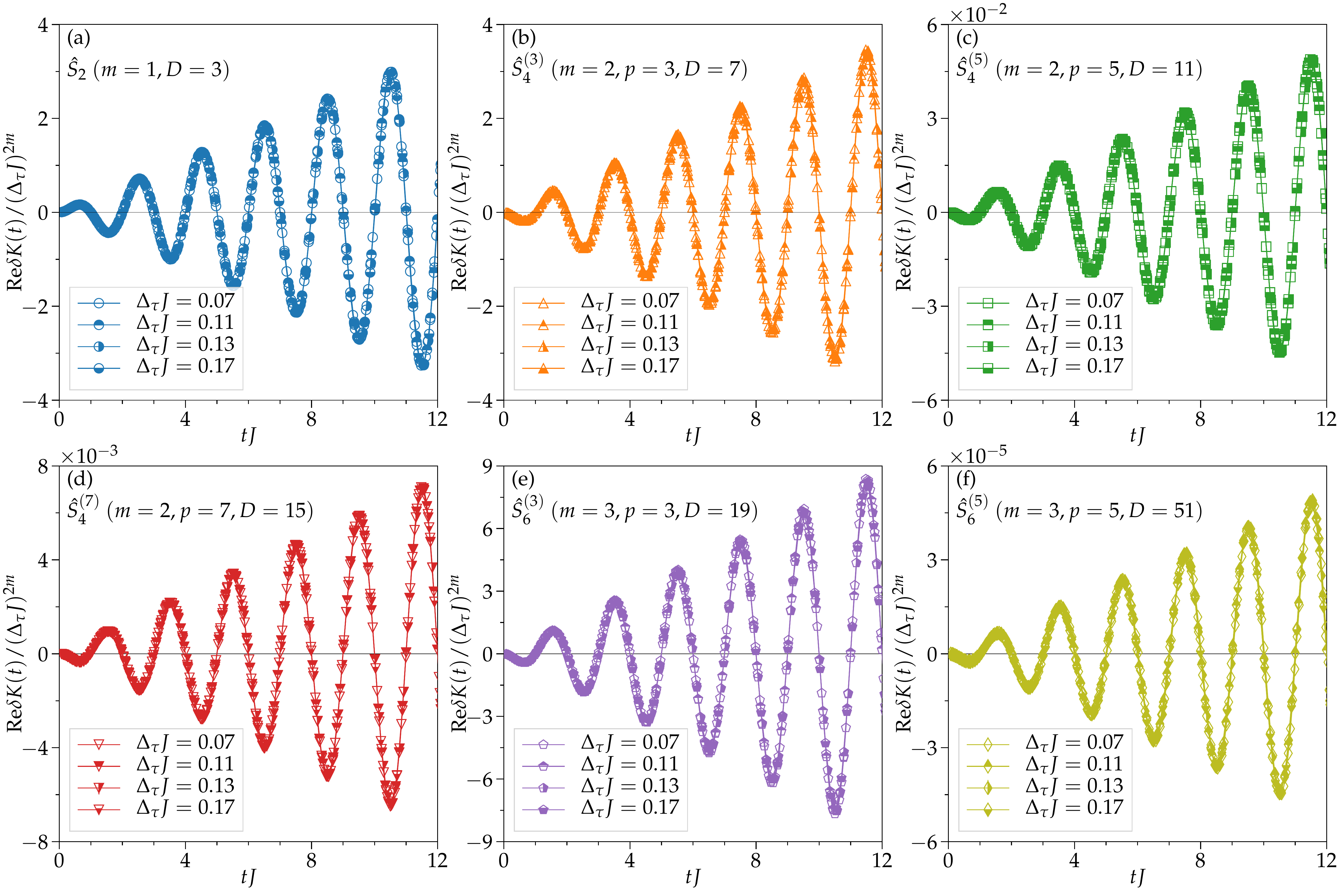}
    \caption{
      Same as Fig.~\ref{fig.time_evolve_nodivision}
      but the deviation ${\rm Re} \delta K(t)$ divided by $(\Delta_\tau J)^{2m}$      
      for several values of $\Delta_\tau$ as indicated in the figures.  
      \label{fig.time_evolve}
    }
  \end{figure*}
\end{center}

Figure~\ref{fig.time_evolve} shows 
${\rm Re}\delta K(t)$ divided by $(\Delta_\tau J)^{2m}$
for $\Delta_\tau J=$ 0.07, 0.11, 0.13, and 0.17. 
As expected from Eq.~(\ref{Uapprox}),
the values of 
$\delta K(t)/(\Delta_\tau J)^{2m}$ 
for different $\Delta_\tau$ 
are almost on the same curve. 
It is also found that the error decreases
with increasing $p$ for a fixed $m$, 
independently of $\Delta_\tau$.  
This suggests that the increase of $p$
reduces the coefficient of the
leading-error term by orders of magnitude. 
However, as shown in Eq.~(\ref{depth}),
$p$ is the base of the exponential which determines the circuit depth. 
Thus, as far as noisy near-term quantum computers are concerned, 
$p=3$ might be a more suitable value than $p\geqslant 5$. 

We also examine the systematic errors due to 
the Suzuki-Trotter decompositions $\hat{S}_{2m}^{(p)}(\Delta_\tau)$ 
in approximating the time evolution operator for 
the Fermi-Hubbard model on a square lattice with a ladderlike $4\times 2$ cluster 
under open boundary conditions, in which  
the Hamiltonian is subdivided into 
$N_\Gamma=4$ parts, as described in appendix~\ref{app:Hubbard}.
Figure~\ref{fig.time_evolve_Hub} shows the numerical results of 
${\rm Re}\delta K(t)$ divided by $(\Delta_\tau J)^{2m}$ for different values of $\Delta_\tau$ 
for the Fermi-Hubbard model
with $U_{\rm H}/J=4$ at half filling.
As in the case of the spin-1/2 Heisenberg model shown in Fig.~\ref{fig.time_evolve}, 
these values 
for different values of $\Delta_\tau$ 
are almost on the same curve.
In addition,  
the systematic errors decrease with increasing $p$ for a fixed $m$, 
independently of $\Delta_\tau$,  
suggesting that the increase of $p$
reduces the coefficient of the
leading-error term by orders of magnitude
also for $N_\Gamma=4$.

\begin{center}
  \begin{figure*}
    \includegraphics[width=1.95\columnwidth]{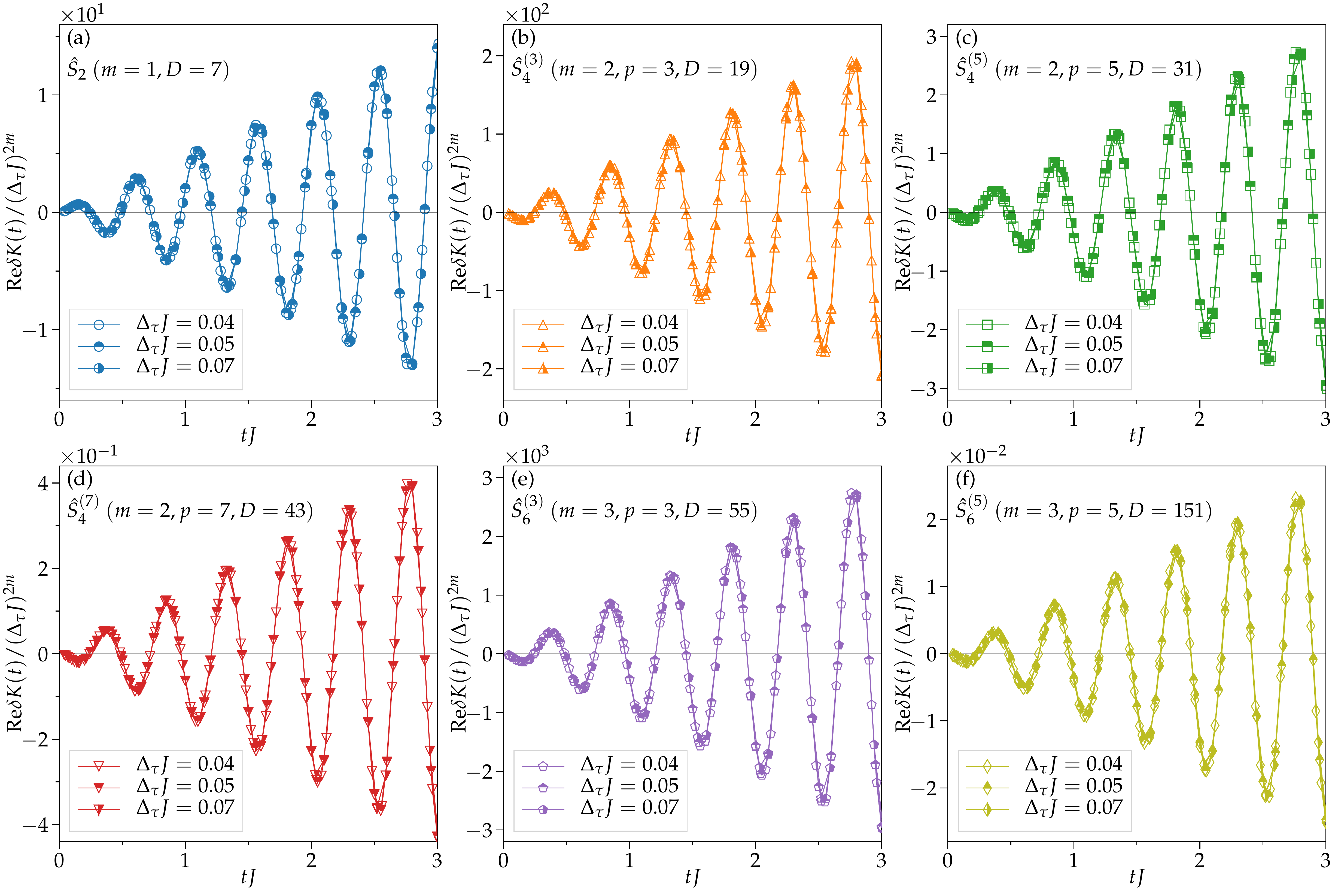}
    \caption{
      Same as Fig.~\ref{fig.time_evolve}      
      but for the Fermi-Hubbard model on a square lattice with a $4\times 2$ cluster under open boundary conditions 
      described by the Hamiltonian $\hat{\mathcal{H}}$ in Eq.~(\ref{JWHub}). 
      The other parameters of the model are $U_{\rm H}/J=4$ and $N=16$ at half filling. 
      $N_\Gamma=4$ and $\Delta_\tau J=$ 0.04, 0.05, and 0.07 are used.    
      \label{fig.time_evolve_Hub}
    }
  \end{figure*}
\end{center}

Finally, we note that 
several exponential-product formulas,
not limited to those found by Suzuki, 
up to the depth $\leqslant 11$ with an error analysis 
can be found in Ref.~\cite{Omelyan2003}. 
Other error analysis of the Suzuki-Trotter 
decomposition devoted for quantum computing 
can be found in Refs.~\cite{Papageorgiou2012,Heyl2019,childs2019theory}.

\section{Another formalism for approximating the Hamiltonian power}\label{app:HST}

As discussed in Sec.~\ref{ST}, we can formulate at least two different algorithms for evaluating 
the Hamiltonian power $\mathcal{\hat{H}}^n$, depending on in which stage the time-evolution operators 
in the approximated Hamiltonian power $\hat{\mathcal{H}}^n(\Delta_\tau)$ are replaced with 
the symmetric Suzuki-Trotter decomposition, either in Eq.~(\ref{naiveHFD}) or in Eq.~(\ref{powerHFD}). 
In the quantum power method described in Sec.~\ref{sec:qpm}, the time-evolution operators in Eq.~(\ref{powerHFD}) are 
approximated by the symmetric Suzuki-Trotter decomposition. 
In this appendix, we describe the other formalism by approximating the time-evolution operators in Eq.~(\ref{naiveHFD}) and 
show that the resulting algorithm scales differently from the one formulated in Sec.~\ref{sec:qpm}.

By incorporating the symmetric Suzuki-Trotter decomposition $\hat{S}_{2m}^{(p)}$ 
into the approximated Hamiltonian power $\hat{\mathcal{H}}^n(\Delta_\tau)$ in Eq.~(\ref{naiveHFD}),  
the Hamiltonian power $\mathcal{\hat{H}}^n$ is now approximated as 
\begin{alignat}{1}
  \hat{\mathcal{H}}^n
  &=
  \hat{\mathcal{H}}^n_{\underline{\rm ST}}(\Delta_\tau)
  + O(\Delta_\tau^{2})
  + \mathcal{E}_{\rm ST},
  \label{QPW2}
\end{alignat}
where 
\begin{equation}
  \hat{\mathcal{H}}^n_{\underline{\rm ST}}(\Delta_\tau)
  =
  \sum_{k=0}^n
  c_{n,k}
  \hat{S}_{2m}^{(p)}\left(\left(\frac{n}{2}-k\right)\Delta_\tau\right),  \label{naiveHST}
\end{equation}
$O(\Delta_\tau^{2})$ represents the systematic error $\mathcal{E}_{\rm FD}$ 
due to the finite-difference scheme for the time derivatives, 
and 
$\mathcal{E}_{\rm ST}$ denotes the systematic error due to the Suzuki-Trotter decomposition 
of the time-evolution operators.
The order of $\mathcal{E}_{\rm ST}$ is discussed below.
We should emphasize here that  
$\hat{\mathcal{H}}^n_{\underline{\rm ST}}(\Delta_\tau) \ne \hat{\mathcal{H}}^n_{\rm ST}(\Delta_\tau)$ 
for $n\geqslant 2$, 
where $\hat{\mathcal{H}}^n_{\rm ST}(\Delta_\tau)$ is defined in Eq.~(\ref{HSTLC}), because  
\begin{equation}
  \hat{S}_{2m}^{(p)}(\Delta_\tau)\hat{S}_{2m}^{(p)}(\Delta_\tau^\prime) \ne \hat{S}_{2m}^{(p)}(\Delta_\tau+\Delta_\tau^\prime)
  \label{eq:SS}
\end{equation}
for $\Delta_\tau\ne-\Delta_\tau^\prime$,  
although the exact time-evolution operators satisfy the multiplication law  
$\hat{U}(\Delta_\tau)\hat{U}(\Delta_\tau^\prime) = \hat{U}(\Delta_\tau+\Delta_\tau^\prime)$.
Note also that 
$\hat{\mathcal{H}}^1_{\underline{\rm ST}}(\Delta_\tau)=
\hat{\mathcal{H}}^1_{\rm ST}(\Delta_\tau)$.

We can readily confirm that 
$\hat{\mathcal{H}}^n_{\underline{\rm ST}}(\Delta_\tau)$
is Hermitian and an even function of $\Delta_\tau$, i.e., 
\begin{equation}
  \hat{\mathcal{H}}^n_{\underline{\rm ST}}(\Delta_\tau)
  =\left[\hat{\mathcal{H}}^n_{\underline{\rm ST}}(\Delta_\tau)\right]^\dag
  =\hat{\mathcal{H}}^n_{\underline{\rm ST}}(-\Delta_\tau),   
\end{equation}
as in the case of $\hat{\mathcal{H}}^n_{\rm ST}(\Delta_\tau)$ given in Eq.~(\ref{eq:Hst})
and hence the systematic error $\mathcal{E}_{\rm ST}$ 
(as well as the systematic error $\mathcal{E}_{\rm FD}$, see Sec.~\ref{sec:hp})
in odd powers of $\Delta_\tau$ is absent in Eq.~(\ref{QPW2}).
We can also show that  
$\hat{\mathcal{H}}^n_{\underline{\rm ST}}(\Delta_\tau)$ 
does {\it not} satisfy the law of exponents, i.e.,  
\begin{equation}
  \hat{\mathcal{H}}^n_{\underline{\rm ST}}(\Delta_\tau) \not =
  \left[\hat{\mathcal{H}}^1_{\underline{\rm ST}}(\Delta_\tau)\right]^n
  \label{notpower}
\end{equation}
for $n\geqslant 2$, simply because of Eq.~(\ref{eq:SS}), 
but only satisfies it approximately within the systematic errors. 
This is in sharp contrast to the case of $\hat{\mathcal{H}}^n_{\rm ST}(\Delta_\tau)$, which 
satisfies exactly the law of exponents in Eq.~(\ref{HSTpower}).

At first glance, one would tend to conclude that $\hat{\mathcal{H}}^n_{\underline{\rm ST}}(\Delta_\tau)$ in Eq.~(\ref{naiveHST}) is 
more suitable to approximate the Hamiltonian power $\mathcal{\hat{H}}^n$  
than $\hat{\mathcal{H}}^n_{\rm ST}(\Delta_\tau)$ in Eq.~(\ref{HSTLC}),
because each term in 
$\hat{\mathcal{H}}^n_{\underline{\rm ST}}(\Delta_\tau)$ contains
a single $\hat{S}_{2m}^{(p)}$, not a product 
of multiple $\hat{S}_{2m}^{(p)}$'s
as in $\hat{\mathcal{H}}^n_{\rm ST}(\Delta_\tau)$,
thus expecting fewer gates in the circuit. 
However, the disadvantage of
$\hat{\mathcal{H}}^n_{\underline{\rm ST}}(\Delta_\tau)$ in Eq.~(\ref{naiveHST})
is that the higher-order Suzuki-Trotter decompositions are required
for approximating the Hamiltonian power $\mathcal{\hat{H}}^n$ with larger $n$.

This can be understood by recalling
that $\hat{S}_{2m}^{(p)}(t)$ has a form of 
Eq.~(\ref{residual}): 
\begin{alignat}{1}
  \hat{S}_{2m}^{(p)}(t) 
  =\exp\left[-\imag t \hat{\mathcal{H}}
    + (-\imag t)^{2m+1} \hat{R}_{2m+1}
    + \cdots\right].
\end{alignat}
Accordingly, the higher-order derivative of $\hat{S}_{2m}^{(p)}(t)$ 
at $t=0$ is given by
\begin{equation}
  \imag^n \left.\frac{\dd^n \hat{S}_{2m}^{(p)}(t)}{\dd t^n}\right|_{t=0} 
  =\hat{\mathcal{H}}^n
  \label{powersmalln}
\end{equation}
for $n\leqslant 2m$, but 
\begin{equation}
  \imag^n \left.\frac{\dd^n \hat{S}_{2m}^{(p)}(t)}{\dd t^n}\right|_{t=0} 
  \not=\hat{\mathcal{H}}^n
  \label{nopower}
\end{equation}
for $n>2m$.
For example, if $n=2m+1$, the derivative reads
\begin{equation}
  \imag^{2m+1} \left.\frac{\dd^{2m+1} \hat{S}_{2m}^{(p)}(t)}{\dd t^{2m+1}}\right|_{t=0} 
  =\hat{\mathcal{H}}^{2m+1} + (2m+1)!\hat{R}_{2m+1}.
  \label{nopower2mp1}
\end{equation} 
It is now important to notice that 
the right-hand side of 
Eq.~(\ref{naiveHST}) corresponds to
the central finite-difference approximation of 
$ \imag^n \left.\frac{\dd^n \hat{S}_{2m}^{(p)}(t)}{\dd t^n}\right|_{t=0} $, i.e., 
\begin{alignat}{1}
\imag^n \left.\frac{\dd^n \hat{S}_{2m}^{(p)}(t)}{\dd t^n}\right|_{t=0} = 
\sum_{k=0}^n
  c_{n,k}
  \hat{S}_{2m}^{(p)}\left(\left(\frac{n}{2}-k\right)\Delta_\tau\right) + O(\Delta_\tau^2). 
  \label{eq:Stn}
\end{alignat}
In other words, the approximated Hamiltonian power $\hat{\mathcal{H}}_{\underline{\rm ST}}^n(\Delta_\tau)$ 
in Eq.~(\ref{naiveHST})  is given  
by the higher-order derivative of  $\hat{S}_{2m}^{(p)}(t)$ at $t=0$ as 
\begin{equation}
\hat{\mathcal{H}}_{\underline{\rm ST}}^n(\Delta_\tau) = \imag^n \left.\frac{\dd^n \hat{S}_{2m}^{(p)}(t)}{\dd t^n}\right|_{t=0}  
+ O(\Delta_\tau^2). 
\end{equation}

Now, it is obvious that the formalism in Eq.~(\ref{naiveHST}) breaks down if
$n>2m$ because in this case, according to Eq.~(\ref{nopower}),  
$\lim_{\Delta_\tau \to 0}
\hat{\mathcal{H}}_{\underline{\rm ST}}^n(\Delta_\tau) \ne  \hat{\mathcal{H}}^n$, which contradicts to Eq.~(\ref{QPW2}). 
Therefore, 
\begin{equation}
  2m \geqslant n
  \label{eq:setm}
\end{equation}
is required for approximating the Hamiltonian power $\mathcal{\hat{H}}^n$
by $\hat{\mathcal{H}}_{\underline{\rm ST}}^n(\Delta_\tau)$ under a controlled accuracy 
with the systematic error 
\begin{equation}
\mathcal{E}_{\rm ST}\sim O(\Delta_\tau^{2}).
\label{eq:est}
\end{equation}
This is the most important difference from the algorithm described in Sec.~\ref{sec:qpm},
where the lowest-order 
Suzuki-Trotter decomposition with $m=1$ is adequate for any power $n$.

There are two remarks in order. First, the approximated Hamiltonian power $\hat{\mathcal{H}}_{\rm ST}^n(\Delta_\tau)$ in 
Eq.~(\ref{HST}) can be considered as 
\begin{alignat}{1}
\hat{\mathcal{H}}_{\rm ST}^n(\Delta_\tau) = 
    \left[
 \imag \left.\frac{\dd \hat{S}_{2m}^{(p)}(t)}{\dd t}\right|_{t=0}  
+ O(\Delta_\tau^2)      \right]^n. 
\end{alignat}
Therefore, the lowest-order symmetric Suzuki-Trotter decomposition with $m=1$ is adequate to satisfy 
Eq.~(\ref{eq:setm}) and indeed, as discussed in Sec.~\ref{sec:qpm}, it approximates the Hamiltonian power 
$\hat{\mathcal{H}}^n$ with the controlled accuracy. 
Second, although we have emphasized that the violation of the multiplication law 
$\hat{S}_{2m}^{(p)}\left(\frac{\Delta_\tau}{2}\right) \hat{S}_{2m}^{(p)}\left(\frac{\Delta_\tau}{2}\right) \ne \hat{S}_{2m}^{(p)}\left(\Delta_\tau\right)$ is the essential point that distinguishes the two algorithms described here and in Sec.~\ref{sec:qpm}, 
this equation is satisfied within the systematic error. i.e., 
\begin{equation}
\hat{S}_{2m}^{(p)}\left(\frac{\Delta_\tau}{2}\right) \hat{S}_{2m}^{(p)}\left(\frac{\Delta_\tau}{2}\right) 
=  \hat{S}_{2m}^{(p)}\left(\Delta_\tau\right) + O(\Delta_\tau^{2m+1}).
\end{equation}
Accordingly, the two algorithms described here and in Sec.~\ref{sec:qpm} should be the same within the systematic error. 
In fact, the approximated Hamiltonian powers $\hat{\mathcal{H}}_{\rm ST}^n(\Delta_\tau)$ and 
$\hat{\mathcal{H}}_{\underline{\rm ST}}^n(\Delta_\tau) $ in Eqs.~(\ref{HSTLC}) and (\ref{naiveHST}), respectively, 
are equivalent within the systematic error
because 
\begin{alignat}{1}
  \hat{\mathcal{H}}_{\rm ST}^n(\Delta_\tau)
  &=
  \sum_{k=0}^n
  c_{n,k}
  \left[\hat{S}_{2m}^{(p)}
    \left(\frac{\Delta_\tau}{2}\right)\right]^{n-2k}  \\ 
  &= 
  \sum_{k=0}^n
  c_{n,k}
  \left[ \hat{S}_{2m}^{(p)}\left(\left(\frac{n}{2}-k\right)\Delta_\tau\right) + O(\Delta_\tau^{2m+1}) \right]\\
&=
  \hat{\mathcal{H}}^n_{\underline{\rm ST}}(\Delta_\tau) + O(\Delta_\tau^{2m+1-n}),  
\end{alignat}
provided that $2m+1>n$, which is consistent with Eq.~(\ref{eq:setm}).

\begin{center}
  \begin{figure}
    \includegraphics[width=0.9\columnwidth] {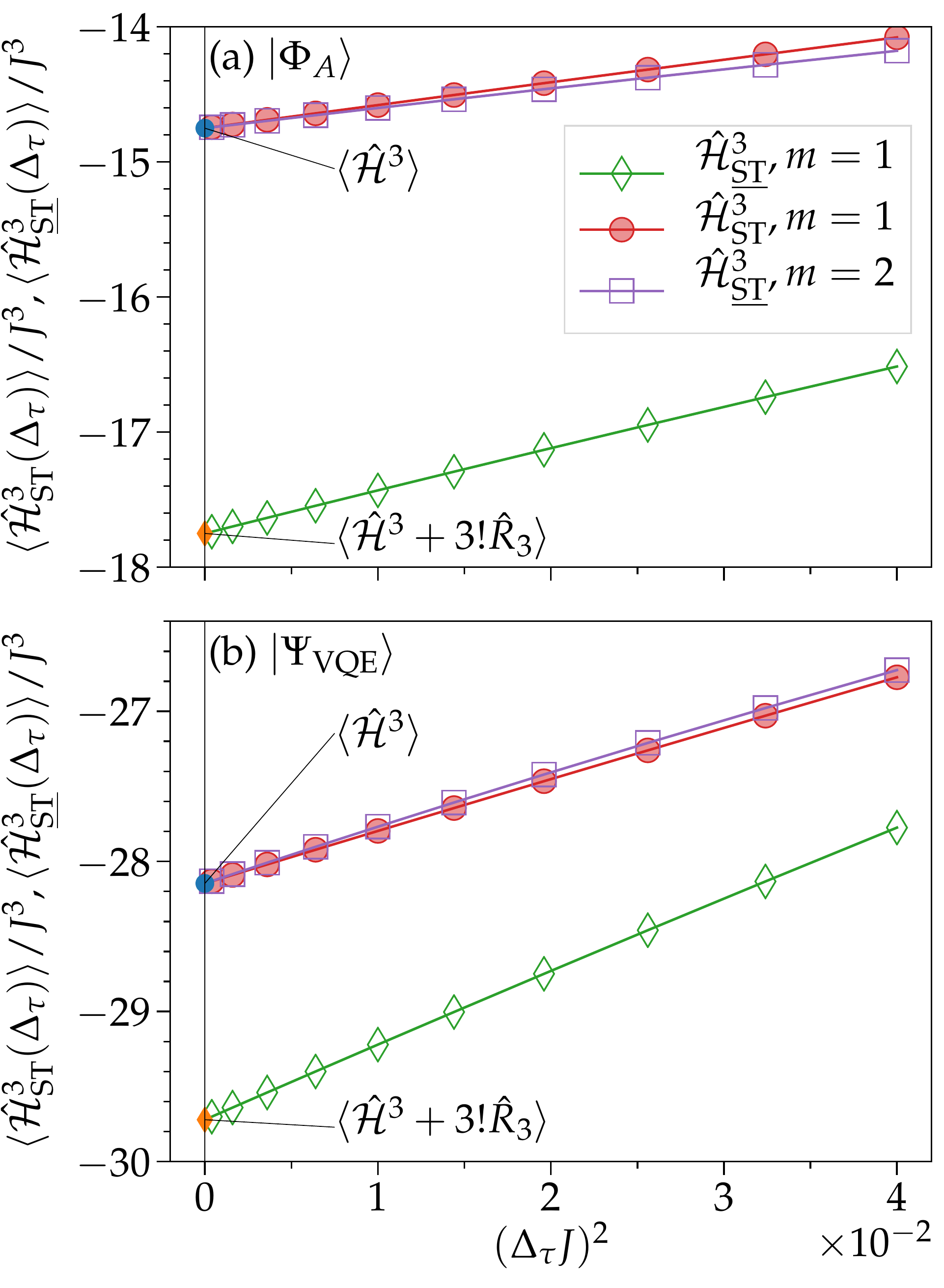}
    \caption{
      $\langle \hat{\mathcal{H}}_{\underline{\rm ST}}^3(\Delta_\tau) \rangle$ and 
      $\langle \hat{\mathcal{H}}_{{\rm ST}}^3(\Delta_\tau)\rangle$
      as a function of $\Delta_\tau^2$ evaluated 
      for the spin-$1/2$ Heisenberg model on an $N=16$ qubit ring, 
      using different orders of the symmetric Suzuki-Trotter decomposition $\hat{S}_{2m}^{(p)}$
      with $m=1,2$ and $p=3$. 
      For a quantum state $|\Psi\rangle$, we choose
      (a) the singlet-pair product state $|\Phi_A\rangle$ in Eq.~(\ref{eq:phi_a}) and
      (b) the VQE state $|\Psi_{\rm VQE}\rangle$ in Eq.~(\ref{eq:vqe}).
      The exact values  
      are indicated at $\Delta_\tau=0$ with the filled symbols (blue and orange).   
    }\label{fig:conv_mu3}
  \end{figure}
\end{center}

As an example, we show in Fig.~\ref{fig:conv_mu3}
the expectation values 
$\langle \hat{\mathcal{H}}_{\underline{\rm ST}}^n(\Delta_\tau) \rangle$ and 
$\langle \hat{\mathcal{H}}_{{\rm ST}}^n(\Delta_\tau)\rangle$ 
with respect to the quantum states 
$|\Phi_A\rangle$ and $|\Psi_{\rm VQE}\rangle$
of the spin-$1/2$ Heisenberg model on an $N=16$ qubit ring 
for the power $n=3$.
Here, a simplified notation of the expectation value 
\begin{equation}
  \langle  \cdots \rangle
  \equiv
  \langle \Psi | \cdots | \Psi \rangle
  \label{expec}
\end{equation}
is introduced with $|\Psi\rangle \in \{|\Phi_A\rangle, |\Psi_{\rm VQE}\rangle\}$ 
given in Eqs.~(\ref{eq:phi_a}) and (\ref{eq:vqe}).
According to 
Eqs.~(\ref{powersmalln})--(\ref{nopower2mp1}),
$\langle \hat{\mathcal{H}}_{\underline{\rm ST}}^3(\Delta_\tau)\rangle$
in the limit of $\Delta_\tau \to 0$ should converge as  
\begin{equation}
  \lim_{\Delta_\tau \to 0}
  \left\langle\hat{\mathcal{H}}_{\underline{\rm ST}}^3(\Delta_\tau) \right\rangle
  =
  \left\langle \hat{\mathcal{H}}^3 \right\rangle
  \label{H3}
\end{equation}
for $m\geqslant 2$, but 
\begin{equation}
  \lim_{\Delta_\tau \to 0}
  \left\langle\hat{\mathcal{H}}_{\underline{\rm ST}}^3(\Delta_\tau) \right\rangle
  =
  \left\langle \hat{\mathcal{H}}^3+3!\hat{R}_3 \right\rangle
  \label{R3}
\end{equation}
for $m=1$. Here, the explicit form of the residual term $\hat{R}_3$ in Eq.~(\ref{R3}) 
for $N_\Gamma=2$
can be derived by using the Baker-Campbell-Hausdorff formula
for $\hat{S}_{2}$ as~\cite{Yoshida1990,Omelyan2003,Becca_Sorella_book}
\begin{equation}
  \hat{R}_3= -\frac{1}{24}
  \left[\hat{\mathcal{H}}_A,
    \left[\hat{\mathcal{H}}_A,\hat{\mathcal{H}}_B\right]\right]
  +\frac{1}{12}
    \left[\hat{\mathcal{H}}_B,
    \left[\hat{\mathcal{H}}_B,\hat{\mathcal{H}}_A\right]\right].
\end{equation}
The numerical results in Fig.~\ref{fig:conv_mu3} confirm 
Eqs.~(\ref{H3}) and~(\ref{R3}), as well as
the expected behavior
$\lim_{\Delta_\tau \to 0} \langle \hat{\mathcal{H}}_{\rm ST}^3(\Delta_\tau)\rangle
=\langle \hat{\mathcal{H}}^3 \rangle$ for $m=1$. 
Note also that the linear convergence of these quantities to the exact values as a function of $\Delta_\tau^2$ shown in Fig.~\ref{fig:conv_mu3} corroborates the systematic errors expected for 
$\langle \hat{\mathcal{H}}_{\underline{\rm ST}}^n(\Delta_\tau) \rangle$ in Eqs.~(\ref{QPW2}) and (\ref{eq:est}) 
and $\langle \hat{\mathcal{H}}_{{\rm ST}}^n(\Delta_\tau)\rangle$ in Eq.~(\ref{QPW}).

Let us now discuss the gate count for approximating $\hat{\mathcal{H}}^n$ 
with $\hat{\mathcal{H}}_{\underline{\rm ST}}^n(\Delta_\tau)$. 
As described above, Eq.~(\ref{eq:setm}) sets the order of the Suzuki-Trotter decomposition such that
$2m\geqslant n$, i.e., the smallest order $m$ of the Suzuki-Trotter 
decomposition to evaluate 
$\hat{\mathcal{H}}^n$ being $m=\lceil n/2\rceil$,
where $\lceil \cdot\rceil$ is the ceiling function
that returns the minimum integer larger 
than or equal to the argument.
Therefore,
assuming that
a $\mathlcal{k}$-local Hamiltonian $\hat{\mathcal{H}}$ composed of $O(N)$ terms, 
the number of gates required for approximating $\hat{\mathcal{H}}^n$ with $\hat{\mathcal{H}}_{\underline{\rm ST}}^n(\Delta)$ is
$O(p^{n/2} \mathlcal{k} N)$
because
the circuit depth $D_{2m}^{(p)}$ for the single Suzuki-Trotter-decomposed time-evolution operator $\hat{S}_{2m}^{(p)}$ is given by Eq.~(\ref{Dm}),
and thus it increases exponentially in the power $n$.
In contrast, as described in Sec.~\ref{sec:qp_gc}, 
the number of gates required for approximating $\hat{\mathcal{H}}^n$
with $\hat{\mathcal{H}}_{{\rm ST}}^{n}(\Delta_\tau)$ is 
$O(n \mathlcal{k} N)$ with a prefactor
$D_{2}^{(p)}=2N_\Gamma-1\sim O(1)$,
i.e., increasing polynomially in $N$ and $n$.

This indicates that the algorithm based on 
$\hat{\mathcal{H}}_{\underline{\rm ST}}^n(\Delta_\tau)$ 
suffers from the exponential increase of the number of gates for large $n$.  
However, 
the algorithm based on $\hat{\mathcal{H}}_{\underline{\rm ST}}^n(\Delta_\tau)$
can be more favorable than that based on 
$\hat{\mathcal{H}}_{{\rm ST}}^n(\Delta_\tau)$ 
when the power $n$ is small. 
To be more specific, let us consider the case of $p=3$
and $N_\Gamma=2$.
Then, the circuit depth 
for $\hat{\mathcal{H}}_{\underline{\rm ST}}^n(\Delta_\tau)$
is given by 
$D_{2\lceil n/2 \rceil}^{(3)}=3, 3, 7,  7, 19, 19, 55, 55, 163, \cdots$,
while the largest circuit depth for
$\hat{\mathcal{H}}_{{\rm ST}}^n(\Delta_\tau)$
involving $[\hat{S}_{2}^{(p)}(\pm \Delta/2)]^n$ is 
$n(D_{2}^{(3)}-1)+1=2n+1    =3, 5, 7, 9, 11, 13, 15, 17, 19, \cdots$,
for the Hamiltonian power 
$n=1, 2, 3,  4,  5,  6,  7,  8, 9, \cdots$.
Here, for the latter,
the depth is counted
by assuming that the commuting exponentials
in $[\hat{S}_{2}^{(p)}(\pm \Delta/2)]^n$ are contracted
(also see Table.~\ref{table}).
Therefore, in this case with $p=3$
and $N_\Gamma=2$, 
the algorithm based on $\hat{\mathcal{H}}_{\underline{\rm ST}}^n(\Delta_\tau)$
is more preferable than that based on 
$\hat{\mathcal{H}}_{{\rm ST}}^n(\Delta_\tau)$
as long as the power $n\leqslant4$. 
In fact, one can readily show that this is generally the case, 
irrespectively of the value of $N_\Gamma$, when $p=3$~\cite{gate4}. 
As shown in appendix~\ref{app:1st2nd}, the algorithm based on $\hat{\mathcal{H}}_{\underline{\rm ST}}^n(\Delta_\tau)$ 
is indeed particularly useful when the lowest-order moments are evaluated.

To apply 
the quantum power method formulated in this appendix 
to the Krylov-subspace diagonalization scheme,
it is crucial to reduce the maximum power $n$ appearing in the formalism.
Defining 
\begin{equation}
  |\underline{\tilde{u}}_{i}\rangle
  = \hat{\mathcal{H}}^{l-1}_{\underline{\rm ST}}(\Delta_\tau)|q_k \rangle  
  \label{eq:ui2}
\end{equation}
for the basis set generated in the block Krylov subspace
$\mathcal{K}_n\left(\hat{\mathcal{H}}_{\underline{\rm ST}}(\Delta_\tau),\{|q_k\rangle\}_{k=1}^{M_{\rm B}}\right)$, 
the matrix elements $\bs{H}$ and $\bs{S}$ in Eqs.~(\ref{Hsubspace}) and (\ref{Ssubspace}) are now approximated 
by replacing $|u_{i}\rangle$ with $|\underline{\tilde{u}}_{i}\rangle$ as 
\begin{alignat}{1}
  \underline{\tilde{H}}_{ij}
  &=\langle \underline{\tilde{u}}_i|\hat{\mathcal{H}}|\underline{\tilde{u}}_j\rangle 
  =
  \langle q_k |
  \hat{\mathcal{H}}^{l-1}_{\underline{\rm ST}}(\Delta_\tau)
  \hat{\mathcal{H}}
  \hat{\mathcal{H}}^{l'-1}_{\underline{\rm ST}}(\Delta_\tau)
  |q_{k'}\rangle 
   \label{Hsubspace4}
\end{alignat}
and 
\begin{equation}
  \underline{\tilde{S}}_{ij}
  = \langle \underline{\tilde{u}}_i|\underline{\tilde{u}}_j \rangle
  =
  \langle q_k |
  \hat{\mathcal{H}}^{l-1}_{\underline{\rm ST}}(\Delta_\tau)
  \hat{\mathcal{H}}^{l'-1}_{\underline{\rm ST}}(\Delta_\tau)
  |q_{k'}\rangle, 
  \label{Ssubspace4}
\end{equation}
where $i=k+(l-1)M_{\rm B}$ and $j=k'+(l'-1)M_{\rm B}$ for
$1 \leqslant k,k' \leqslant M_{\rm B}$ and
$1 \leqslant l,l' \leqslant n$.
Similarly to Eqs.~(\ref{Hsubspace1}) and (\ref{Ssubspace1}),
the power exponents are distributed 
to the left and the right basis states.

To be more specific, 
$\underline{\tilde{H}}_{ij}$ and $\underline{\tilde{S}}_{ij}$ 
in terms of $\hat{S}_{2m}^{(p)}$ 
are given as
\begin{alignat}{1}
  \underline{\tilde{H}}_{ij}
  &=
  \sum_{\nu=0}^{l-1}
  \sum_{\nu'=0}^{l'-1}
  c_{l-1,\nu}^*
  c_{l'-1,\nu'}
  \notag \\
  &
  \times
  \langle q_k |
  \hat{S}_{2m}^{(p)}\left(-t^{(l-1)}_{\nu}\right)
  \hat{\mathcal{H}}
  \hat{S}_{2m}^{(p)}\left(t^{(l'-1)}_{\nu'}\right)
  |q_{k'}\rangle
  \label{Hsubspace5}
\end{alignat}
and 
\begin{alignat}{1}
  \underline{\tilde{S}}_{ij}
  &=
  \sum_{\nu=0}^{l-1}
  \sum_{\nu'=0}^{l'-1}
  c_{l-1,\nu}^*
  c_{l'-1,\nu'}
  \notag \\
  &
  \times
  \langle q_k |
  \hat{S}_{2m}^{(p)}\left(-t^{(l-1)}_{\nu}\right)
  \hat{S}_{2m}^{(p)}\left(t^{(l'-1)}_{\nu'}\right)
  | q_{k'} \rangle,
  \label{Ssubspace5}
\end{alignat}
where $t^{(l-1)}_{\nu}=\left(\frac{l-1}{2}-\nu\right)\Delta_\tau$ and Eq.~(\ref{unitarityS}) is used.
The number of terms in 
Eqs.~(\ref{Hsubspace5}) and (\ref{Ssubspace5}) is
$O(Nll')$ and $O(ll')$, respectively. 
Here, we assume that $\hat{\mathcal{H}}$ consists of $O(N)$
local terms. 
In total, 
$O(n^2 M_{\rm B}^2 N)$ and $O(n^2 M_{\rm B}^2)$ 
state overlaps are required to be evaluated 
for constructing all matrix elements of the $n M_{\rm B} \times n M_{\rm B}$ matrices
$\underline{\tilde{\bs{H}}}$ and $\underline{\tilde{\bs{S}}}$, 
respectively.

Finally, we note that the systematic errors
$\mathcal{E}_{\rm FD}$ and $\mathcal{E}_{\rm ST}$ in
Eqs.~(\ref{QPW2}) and (\ref{eq:est}) can be 
improved systematically, without increasing the gate count of each circuit,
by adopting the Richardson extrapolation as 
\begin{alignat}{1}
  \hat{\mathcal{H}}^n
  &=
  \hat{\mathcal{H}}^n_{\underline{\rm ST}(r)}(\Delta_\tau)
  + O(\Delta_\tau^{2+2r})
  \label{QPW2_re}
\end{alignat}
where 
$\hat{\mathcal{H}}^n_{\underline{\rm ST}(r)}(\Delta_\tau)$ 
is the $r$th-order Richardson extrapolation
of the approximated Hamiltonian power, i.e., 
\begin{equation}
  \hat{\mathcal{H}}^n_{\underline{\rm ST}(r)}(\Delta_\tau) =
  \frac{h^{2r} \hat{\mathcal{H}}^n_{\underline{\rm ST}(r-1)}(\Delta_\tau/h)-
    \hat{\mathcal{H}}^n_{\underline{\rm ST}(r-1)}(\Delta_\tau)}{h^{2r}-1},
  \label{eq:Richardson2}
\end{equation}
with 
$\hat{\mathcal{H}}^n_{\underline{\rm ST}(0)}(\Delta_\tau)
\equiv\hat{\mathcal{H}}^n_{\underline{\rm ST}}(\Delta_\tau)$.
Since $\hat{\mathcal{H}}^n_{\underline{\rm ST}(0)}(\Delta_\tau)$ is
a linear combination of $n+1$ unitaries 
$\left\{ \hat{S}_{2m}^{(p)}(\frac{n-2k}{2}\Delta_\tau) \right\}_{k=0}^n$, 
$\hat{\mathcal{H}}^n_{\underline{\rm ST}(r)}(\Delta_\tau)$ is 
a linear combination of $(r+1)(n+1)$ unitaries 
$\left\{\left\{ \hat{S}_{2m}^{(p)}(\frac{n-2k}{2h^l}\Delta_\tau) \right\}_{k=0}^n\right\}_{l=0}^r$.

\section{Moment methods}~\label{app:moment}

In this appendix, we outline moment methods as other applications of the quantum power method 
to evaluate the moments and cumulants of the Hamiltonian. By using numerical simulations, we demonstrate 
the CMX
for a short-time imaginary-time evolution 
and estimate the ground-state energy 
of the spin-1/2 Heisenberg model. These numerical results are 
compared with those obtained by the multireference Krylov-subspace diagonalization combined with the 
quantum power method discussed in Sec.~\ref{sec:egs}. 
We also show that the quantum power method can particularly simply evaluate 
the lowest order moments.

\subsection{Moment and cumulant}
The Feynman propagator with respect to a state $|\Psi \rangle$ 
can be written as 
\begin{equation}
  K(t)
  = \langle \hat{U}(t) \rangle 
  = \sum_{n=0}^{\infty} \frac{(-\imag t)^n}{n!} \mu_n,  
\end{equation}
where $\hat{U}(t)$ is the time-evolution operator given in Eq.~(\ref{unitary}) and   
\begin{equation}
  \mu_n = \langle \hat{\mathcal{H}}^n \rangle
  \label{moment_n}
\end{equation} 
is the $n$th Hamiltonian moment.  
We also define
the generating function $\Phi(t)$
of the cumulants $\{\kappa_n\}$ as 
\begin{equation}
  \Phi(t)
  \equiv \ln K(t)
  =\ln\langle  \e^{-\imag \hat{\mathcal{H}} t}  \rangle
  \equiv \sum_{n=0}^{\infty} \frac{(-\imag t)^n}{n!} \kappa_n. 
\end{equation}
Thus, the $n$th moment $\mu_n$ and cumulant $\kappa_n$ are given 
by the $n$th time derivative of generating functions $K(t)$ and $\Phi(t)$, respectively, 
as 
\begin{equation}
  \mu_n=\imag^n \left.\frac{\dd^n K(t)}{\dd t^n}\right|_{t=0} \label{moment}
\end{equation}
and
\begin{equation}
  \kappa_n=\imag^n \left.\frac{\dd^n \Phi(t)}{\dd t^n}\right|_{t=0}.\label{cumulant}
\end{equation}
We note that, recently,  
a method making use of 
the expectation value of the time-evolution operator   
has been proposed for evaluating eigenvalues of the Hamiltonian~\cite{rol2019quantum}.

It should be noticed~\cite{Horn1984} that the $n$th moment $\mu_n$ can be expressed as
\begin{alignat}{1}
  \mu_{n}
  =\kappa_{n}+\sum_{k=1}^{n-1}\binom{n-1}{k-1} \kappa_{k} \mu_{n-k} 
  \label{moments_from_cumulants}
\end{alignat}
and, equivalently, the $n$th cumulant $\kappa_n$ can be expressed as 
\begin{alignat}{1}
  \kappa_{n}
  =\mu_{n}-\sum_{k=1}^{n-1}\binom{n-1}{k-1} \kappa_{k} \mu_{n-k}.
  \label{cumulants_from_moments}
\end{alignat}
Therefore, from the moments  $\{\mu_k\}_{k\leqslant n}$, one can obtain 
the cumulants $\{\kappa_{k}\}_{k\leqslant n}$ and vice versa.  
A remarkable difference between these two quantities is that
the magnitude of the moment
grows exponentially in $n$ as 
$\mu_n \sim O(N^n)$,
while the magnitude of the cumulant remains as 
$\kappa_n \sim O(N)$~\cite{Kubo1962}.

\subsection{Finite-difference approximation}
By approximating 
the derivative in Eq.~(\ref{moment})
with the central-finite-difference 
method, we obtain that 
\begin{equation}
  \mu_n
  =\mu_n(\Delta_\tau)
  + O(\Delta_\tau^2),  
   \label{muDelta}
\end{equation}
where 
\begin{alignat}{1}
  \mu_n(\Delta_\tau)
  &=
  \sum_{i=0}^n c_{n,i}
  K\left( \left(\frac{n}{2}-i\right)\Delta_\tau \right). 
  \label{centralFD}
\end{alignat}
Using 
$K(-\Delta_\tau)=K(\Delta_\tau)^*$,  
$\mu_n(\Delta_\tau)$ for $n$ odd and $n$ even
can be expressed, respectively, as 
\begin{equation}
  \mu_{2m+1}(\Delta_\tau)
  =
  2\imag
  \sum_{i=0}^{m}
  c_{2m+1,i}
  {\rm Im}
  K\left( \left(m+\frac{1}{2}-i\right)\Delta_\tau \right)
  \label{mu_odd}
\end{equation}
and
\begin{equation}
  \mu_{2m}(\Delta_\tau)
  =
  c_{2m,m}
  + 2
  \sum_{i=0}^{m-1}
  c_{2m,i}
  {\rm Re}
  K\left( \left(m-i\right)\Delta_\tau \right),
  \label{mu_even}
\end{equation}
where $K(0)=1$ and $c_{2m,m} =  \frac{1}{\Delta_\tau^{2m}}\binom{2m}{m}$ are used in Eq.~(\ref{mu_even}). 
Thus,
for obtaining $\mu_{n}(\Delta_\tau)$,
it suffices to evaluate $K(t)$
at equally spaced $\lceil n/2 \rceil$ different points, 
where $\lceil \cdot \rceil$ denotes the ceiling function 
defined previously. 
Moreover, if $\{K(l\Delta_\tau/2)\}_{l=1}^{n-1}$ used
for evaluating the moments $\{\mu_{l}(\Delta_\tau)\}_{l=1}^{n-1}$ are
all stored, only $K(n\Delta_\tau/2)$ has to be evaluated
for $\mu_n(\Delta_\tau)$.
Therefore, for obtaining all $n$ moments $\{\mu_l(\Delta_\tau)\}_{l=1}^{n}$, 
it is sufficient to evaluate the propagator at $n$ different points, i.e., 
$\{K(l\Delta_\tau/2)\}_{l=1}^{n}$, only once. 

We can apply the same argument for the cumulants. 
Using the central-finite-difference method, the $n$th cumulant is evaluated as 
\begin{equation}
 \kappa_{n}=\kappa_n(\Delta_\tau) + O(\Delta_\tau^2), \label{kappaDelta}
\end{equation}
where 
\begin{equation}
  \kappa_n(\Delta_\tau) =
  \sum_{i=0}^n c_{n,i}
  \Phi\left( \left(\frac{n}{2}-i\right)\Delta_\tau \right). 
  \label{centralFD2}
\end{equation}
Because $\Phi(-\Delta_\tau)=\Phi(\Delta_\tau)^*$, 
$\kappa_n(\Delta_\tau)$ for $n$ odd and $n$ even
can be expressed, respectively, as

\begin{equation}
  \kappa_{2m+1}(\Delta_\tau)
  =
  2\imag
  \sum_{i=0}^{m}
  c_{2m+1,i}
  {\rm Im}
  \Phi \left( \left(m+\frac{1}{2}-i\right)\Delta_\tau \right)
  \label{cum_odd}
\end{equation}
and
\begin{equation}
  \kappa_{2m}(\Delta_\tau)
  =
  2
  \sum_{i=0}^{m-1}
  c_{2m,i}
  {\rm Re}
  \Phi \left( \left(m-i\right)\Delta_\tau \right),
  \label{cum_even}
\end{equation}
where $\Phi(0)=\ln{K(0)}=0$ is used in Eq.~(\ref{cum_even}). 
Note that, if we write the propagator as
$K(t)=a(t)\e^{\imag \varphi(t)}$
with $a(t)$ and $\varphi(t)$ real, then
${\rm Re}\Phi(t)=\ln a(t)$ and
${\rm Im}\Phi(t)=\varphi(t)$, implying that 
the cumulants with odd order are related to the phase of $K(t)$, while
the cumulants with even order are related to the amplitude of $K(t)$. 
Recently, an efficient method for estimating
the overlap amplitude of two pure states has been proposed~\cite{Fanizza2020}.
Such a method might be utilized for evaluating the cumulants with even order.

\subsection{Quantum power method for moment and cumulant}\label{sec:qpm_mc}

As shown explicitly in the previous section, the $n$th moment $\mu_n$ can be approximated as 
a linear combination of the Feynman propagator $K(t)$, i.e., the expectation value of
the time-evolution operator 
$\hat{U}(t)$, evaluated at different time variables
$t_i^{(n)}=\left(\frac{n}{2}-i \right)\Delta_\tau$ for $i=0,1,\dots,n$. 
Similarly, the $n$th cumulant $\kappa_n$ can be approximated as 
a linear combination of $\Phi(t)$, i.e., logarithm of
the Feynman propagator $K(t)$, evaluated at different time variables 
$t_i^{(n)}$. Therefore, an important quantity here is again
the time-evolution operator $\hat{U}(t)$.

To implement on quantum computers,
the time-evolution operator is further decomposed approximately 
by using the symmetric Suzuki-Trotter decomposition as in Eq.~(\ref{Suzuki}). 
However, at this point, it is crucially important to recall the argument given in Sec.~\ref{sec:tworoutes} and appendix~\ref{app:HST}. 
Although the time evolution operator $\hat{U}(t)$ evaluated at time $t_i^{(n)}=\left(\frac{n}{2}-i \right)\Delta_\tau$ satisfies that 
\begin{equation}
\hat{U}\left(\left(\frac{n}{2}-i \right)\Delta_\tau \right) 
= \left[ \hat{U}\left( \frac{\Delta_\tau}{2} \right) \right]^{n-2i},
\end{equation} 
and thus the approximated $n$th moment $\mu_n(\Delta_\tau)$ in Eq.~(\ref{centralFD}) is equivalent to  
\begin{alignat}{1}
  \mu_n(\Delta_\tau)
  &=
  \sum_{i=0}^n c_{n,i}
  \left\langle \left[ \hat{U}\left( \frac{\Delta_\tau}{2} \right) \right]^{n-2i} \right\rangle,
\end{alignat}
these are no longer generally correct when the time-evolution operators are approximated by the Suzuki-Trotter decomposition, i.e., 
\begin{equation}
\hat{S}_{2m}^{(p)}\left(\left(\frac{n}{2}-i \right)\Delta_\tau \right) 
\ne \left[ \hat{S}_{2m}^{(p)}\left( \frac{\Delta_\tau}{2} \right) \right]^{n-2i}.
\end{equation} 
Therefore, the Feynman propagator $K(t_i^{(n)})$ in Eq.~(\ref{centralFD}) can be approximated either as 
\begin{equation}
K(t_i^{(n)}) 
= \left\langle \hat{S}_{2m}^{(p)}\left(\left(\frac{n}{2}-i \right)\Delta_\tau \right) \right\rangle + 
O(\Delta_\tau^{2m+1})
\label{eq:K2}
\end{equation}
or 
\begin{equation}
K(t_i^{(n)}) 
= \left\langle  \left[ \hat{S}_{2m}^{(p)}\left( \frac{\Delta_\tau}{2} \right) \right]^{n-2i} \right\rangle + 
O(\Delta_\tau^{2m+1}).
\label{eq:K1}
\end{equation}

If the Feynman propagator $K(t_i^{(n)})$ is approximated as in Eq.~(\ref{eq:K1}),
the $n$th moment $\mu_n$ is given by  
\begin{equation}
\mu_n = \sum_{i=0}^n c_{n,i}
  \left\langle  \left[ \hat{S}_{2m}^{(p)}\left( \frac{\Delta_\tau}{2} \right) \right]^{n-2i} \right\rangle 
  + O(\Delta_\tau^2) + O(\Delta_\tau^{2m})
\end{equation}
and thus the lowest-order symmetric Suzuki-Trotter decomposition $\hat{S}_{2m}^{(p)}$ with $m=1$ 
can be adopted (see Sec.~\ref{sec:qpm}). 
This approach is suitable for the calculations of higher-order moments and cumulants. 
On the other hand, if the Feynman propagator $K(t_i^{(n)})$ is approximated as in Eq.~(\ref{eq:K2}), the higher-order 
symmetric Suzuki-Trotter decomposition $\hat{S}_{2m}^{(p)}$ is required. As discussed in appendix~\ref{app:HST}, 
in order to evaluate the $n$th moment $\mu_n$ 
with the controlled accuracy, 
the order of the symmetric Suzuki-Trotter decomposition $\hat{S}_{2m}^{(p)}$ must be $2m \geqslant n$ 
[see Eq.~(\ref{eq:setm})]. In this case, the systematic error is $O(\Delta_\tau^2)$, i.e., 
\begin{equation}
\mu_n = \sum_{i=0}^n c_{n,i}
  \left\langle \hat{S}_{2m}^{(p)}\left(\left(\frac{n}{2}-i \right)\Delta_\tau \right) \right\rangle 
  + O(\Delta_\tau^2). 
\end{equation}
Therefore, this approach is not suitable for large $n$ 
but is more preferable than the other approach when $n\leqslant4$, 
assuming $p=3$ in the symmetric Suzuki-Trotter decomposition. 
Since the cumulant $\kappa_n$ can be expressed
in terms of the moments $\{\mu_k\}_{k \leqslant n}$ as in Eq.~(\ref{cumulants_from_moments}), 
the same argument is applied for the cumulant.

\subsection{First and second moments}\label{app:1st2nd}

The first and second moments are the most fundamental quantities for many practical purposes because 
$\mu_1=\langle \hat{\mathcal{H}}\rangle$ is the average of the energy and 
$\mu_2 = \langle \hat{\mathcal{H}}^2\rangle$ is related to the variance of the energy. 
The first moment $\langle \hat{\mathcal{H}}\rangle$ is directly evaluated by measuring each term 
of the Hamiltonian $ \hat{\mathcal{H}}$ on quantum computers. Perhaps,
$\langle \hat{\mathcal{H}}^2\rangle$ 
could also be evaluated in the same way, although terms to be measured are increased by a factor of $O(N)$, assuming that 
a Hamiltonian $\hat{\mathcal{H}}$ is local. 
The quantum power method can provide an alternative approach to evaluate these quantities with the 
same amount of resource.

\begin{center}
  \begin{figure}
    \includegraphics[width=0.9\columnwidth] {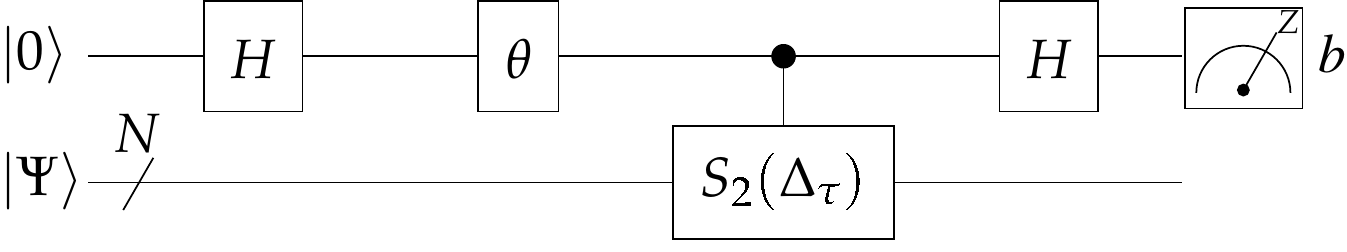}
    \caption{
      Quantum circuit to evaluate
      ${\rm Re} \langle\Psi| \hat{S}_2(\Delta_\tau ) |\Psi\rangle$ or 
      ${\rm Im} \langle\Psi| \hat{S}_2(\Delta_\tau ) |\Psi\rangle$.
      $\theta$ in the circuit denotes the phase gate such that
      $\hat{\theta}|0\rangle=|0\rangle$ and 
      $\hat{\theta}|1\rangle=\e^{\imag \theta}|1\rangle$.
      Since 
      $P_0-P_1={\rm Re} [\e^{\imag \theta}\langle\Psi| \hat{S}_2(\Delta_\tau ) |\Psi\rangle]$,    
      one can evaluate 
      ${\rm Re} \langle\Psi| \hat{S}_2(\Delta_\tau ) |\Psi\rangle$ if $\theta=0$ and 
      ${\rm Im} \langle\Psi| \hat{S}_2(\Delta_\tau ) |\Psi\rangle$ if $\theta=-\pi/2$
      from the difference of the probabilities $P_0$ and $P_1$.
      Here, $P_b$ is the probability for finding a bit $b\,(=0,1)$ by measuring
      the ancilla qubit.    
    }\label{fig:ancilla}
  \end{figure}
\end{center}

From Eqs.~(\ref{mu_odd}) and (\ref{mu_even}),
we can approximate the first and second moments $\mu_1$ and $\mu_2$ as 
\begin{equation}
\mu_1(\Delta_\tau)=-\frac{2}{\Delta_\tau} {\rm Im}  \left\langle \hat{U}\left( \frac{\Delta_\tau}{2} \right) \right\rangle
\label{eq:mu1}
\end{equation}
and 
\begin{equation}
  \mu_2(\Delta_\tau)=\frac{2}{\Delta_\tau^2} \left[ 1 - {\rm Re}
    \left\langle \hat{U}\left(\Delta_\tau \right) \right\rangle \right],
\label{eq:mu2}
\end{equation}
respectively. This is already remarkable because the second moment $\mu_2$ is also estimated simply by 
the expectation value of a single time-evolution operator. 
To evaluate these quantities on quantum computers, the time-evolution operator $\hat{U}(\Delta_\tau)$ is approximated by 
the lowest-order symmetric Suzuki-Trotter decomposition $\hat{S}_2(\Delta_\tau)$ (see appendix~\ref{sec:qpm_mc}). 
Therefore, in the quantum power method, the first and second moments $\mu_1$ and $\mu_2$ are estimated simply by evaluating 
${\rm Im}  \left\langle \hat{S}_2\left( \frac{\Delta_\tau}{2} \right) \right\rangle$ and 
${\rm Re} \left\langle \hat{S}_2\left(\Delta_\tau \right) \right\rangle$, i.e., 
\begin{equation}
\mu_1(\Delta_\tau)\approx -\frac{2}{\Delta_\tau} {\rm Im}  \left\langle \hat{S}_2\left( \frac{\Delta_\tau}{2} \right) \right\rangle
\label{eq:ap_m1}
\end{equation}
and 
\begin{equation}
  \mu_2(\Delta_\tau)\approx \frac{2}{\Delta_\tau^2} \left[ 1 - {\rm Re}
    \left\langle \hat{S}_2\left(\Delta_\tau \right) \right\rangle \right],
\label{eq:ap_m2}
\end{equation}
respectively.   
Although we have to introduce an ancilla qubit 
(see Fig.~\ref{fig:ancilla}),  
$\mu_1 = \langle \hat{\mathcal{H}}\rangle$ and
$\mu_2 = \langle \hat{\mathcal{H}}^2 \rangle$ 
can thus be estimated with exactly the same amount of resource. 
If noise in quantum devices is not destructively serious, this approach based on the quantum power method might be more 
suitable than the direct approach measuring all terms in $\hat{\mathcal{H}}$ and $\hat{\mathcal{H}}^2$.

\begin{center}
  \begin{figure}
    \includegraphics[width=0.9\columnwidth] {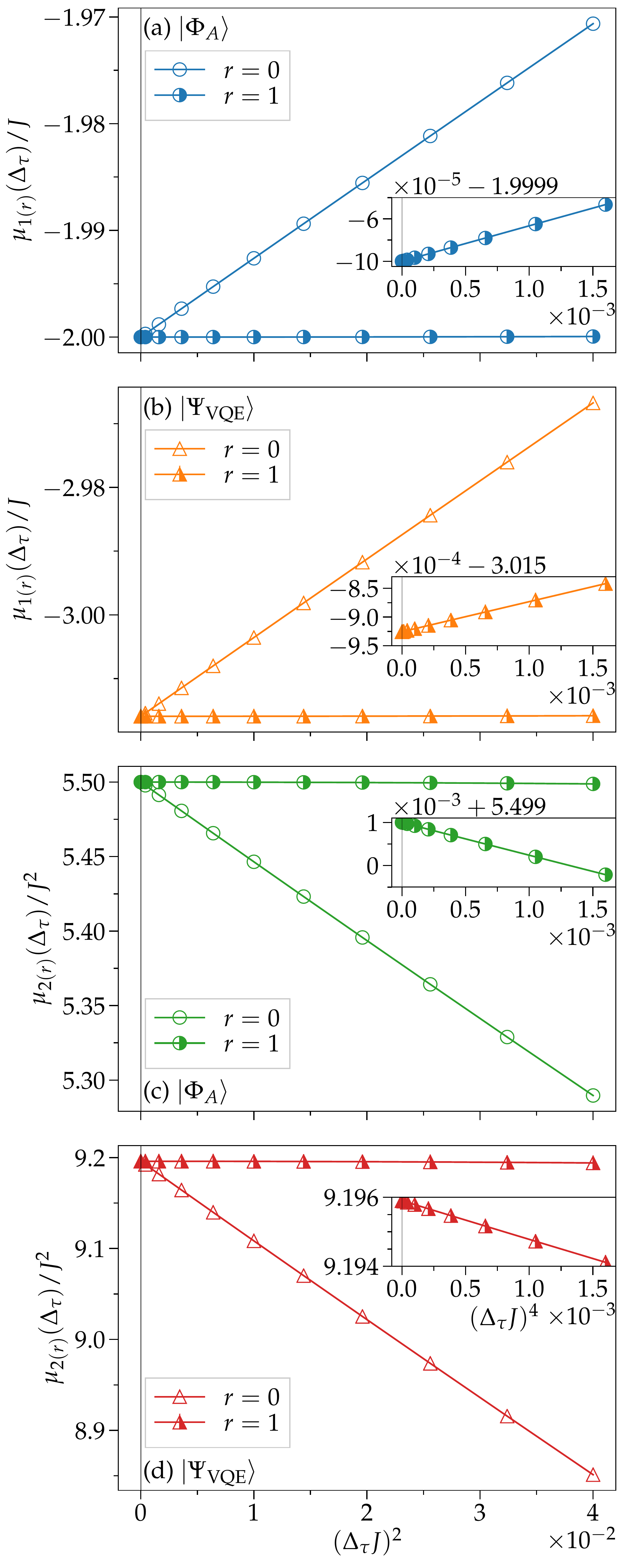}
    \caption{
      (a,b) The first moment $\mu_1$ and
      (c,d) the second moment $\mu_2$
      as a function of $\Delta_\tau^2$ evaluated from Eqs.~(\ref{eq:ap_m1}) and 
      (\ref{eq:ap_m2}), respectively,
      by numerical simulations for the spin-$1/2$ Heisenberg model on an $N=16$ qubit ring. 
      For a quantum state $|\Psi\rangle$, we choose
      (a,c) the singlet-pair product state $|\Phi_A\rangle$ in Eq.~(\ref{eq:phi_a}) and
      (b,d) the VQE state $|\Psi_{\rm VQE}\rangle$ in Eq.~(\ref{eq:vqe}).
      The results obtained by the first-order Richardson extrapolation
      ($r=1$) are also plotted. The insets show the same results for
      the first-order Richardson extrapolation but 
      plotted against $\Delta_\tau^4$.
      The exact values  
      are indicated at $\Delta_\tau=0$ with the filled symbols.     
    }\label{fig:conv_moment}
  \end{figure}
\end{center}

Figure~\ref{fig:conv_moment} shows
the numerical results of $\mu_1$ and $\mu_2$ evaluated from Eqs.~(\ref{eq:ap_m1}) and 
(\ref{eq:ap_m2}) for the spin-$1/2$ Heisenberg model defined in Eq.~(\ref{Ham_SWAP}) with two different quantum states. 
We also show the results obtained by 
employing the first-order Richardson extrapolation, i.e., 
\begin{equation}
  \mu_{n(1)}(\Delta_\tau)=\frac{h^2\mu_n(\Delta_\tau/h)-\mu_n(\Delta_\tau)}{h^2-1}   
\end{equation}
for $n=1$ and $2$,
which expects that the systematic error scales as $O(\Delta_\tau^4)$,
instead of $O(\Delta_\tau^2)$ without the Richardson extrapolation. 
Our numerical simulations clearly demonstrate that the systematic errors are well controlled 
and the results converge smoothly to the exact values in the limit of $\Delta_\tau\to 0$. 
The quantum power method for the first and second moments could be useful to, e.g.,
the energy variance minimization for optimizing a parametrized quantum circuit~\cite{zhang2020variational}.

Here, we only consider the first and second moments,
but the higher-order moments can be similarly evaluated. 
For example, the third and fourth moments are given as 
\begin{equation}
  \mu_3(\Delta_\tau)=\frac{2}{\Delta_\tau^3} \left[ {\rm Im}  \left\langle \hat{U}\left( \frac{3\Delta_\tau}{2} \right) \right\rangle
    - 3 {\rm Im}  \left\langle
    \hat{U}\left( \frac{\Delta_\tau}{2} \right)
    \right\rangle \right ]
\end{equation}
and
\begin{equation}
  \mu_4(\Delta_\tau)=\frac{2}{\Delta_\tau^4} \left[ {\rm Re} \left\langle
    \hat{U}\left(2\Delta_\tau \right) \right\rangle 
    - 4{\rm Re} \left\langle \hat{U}\left(\Delta_\tau \right) \right\rangle +3 \right],
\end{equation}
respectively.
To implement these on quantum computers, $\hat{U}\left(\Delta_\tau \right)$ is now approximated by using 
the higher-order Suzuki-Trotter decomposition $\hat{S}_4^{(p)}(\Delta_\tau)$ with $m=2$ (also see Fig.~\ref{fig:conv_mu3}), 
which is still affordable.

\subsection{Imaginary-time evolution}

For an application of the cumulants, 
we now consider the imaginary-time evolution of a quantum state $|\Psi\rangle$, i.e., 
\begin{equation}
  |\Psi(\tau)\rangle = 
  \frac{\e^{-\tau \hat{\mathcal{H}}/2} |\Psi\rangle}
  {\sqrt{\langle \Psi|\e^{-\tau \hat{\mathcal{H}}}|\Psi\rangle}}   
  \label{eq:psi_tau}
\end{equation}
for $\tau$ real.
We introduce a simplified notation for the 
imaginary-time-dependent expectation value as 
$\langle \cdots \rangle_\tau \equiv
\langle \Psi(\tau)|\cdots|\Psi(\tau)\rangle$. 
Then, the energy expectation value with respect to
$|\Psi(\tau)\rangle$ is given as 
\begin{alignat}{1}
  E(\tau)
  &=
  \langle 
  \hat{\mathcal{H}}
  \rangle_\tau \notag \\
&=
\frac
  {\langle \Psi|\e^{-\tau \hat{\mathcal{H}}/2}\hat{\mathcal{H}}\e^{-\tau \hat{\mathcal{H}}/2} |\Psi\rangle}
  {\langle \Psi|\e^{-\tau \hat{\mathcal{H}}}|\Psi\rangle} \notag \\
  &=
  \frac{\langle \hat{\mathcal{H}}\e^{-\tau \hat{\mathcal{H}}}\rangle}{\langle\e^{-\tau \hat{\mathcal{H}}} \rangle}.
\end{alignat}
Observing that 
$E(\tau) = -\frac{\dd}{\dd \tau}
\ln \langle \e^{-\tau \hat{\mathcal{H}}} \rangle
  =
  -\frac{\dd }{\dd \tau}\sum_{n=0}^{\infty}\frac{(-\tau)^n}{n!}\kappa_n$, 
the CMX of the energy is given as~\cite{Horn1984} 
\begin{equation}
  E(\tau) 
  =\sum_{n=0}^{\infty}
  \frac{(-\tau)^n}{n!}\kappa_{n+1}.
\end{equation}

\begin{center}
  \begin{figure*}
    \includegraphics[width=1.95\columnwidth]{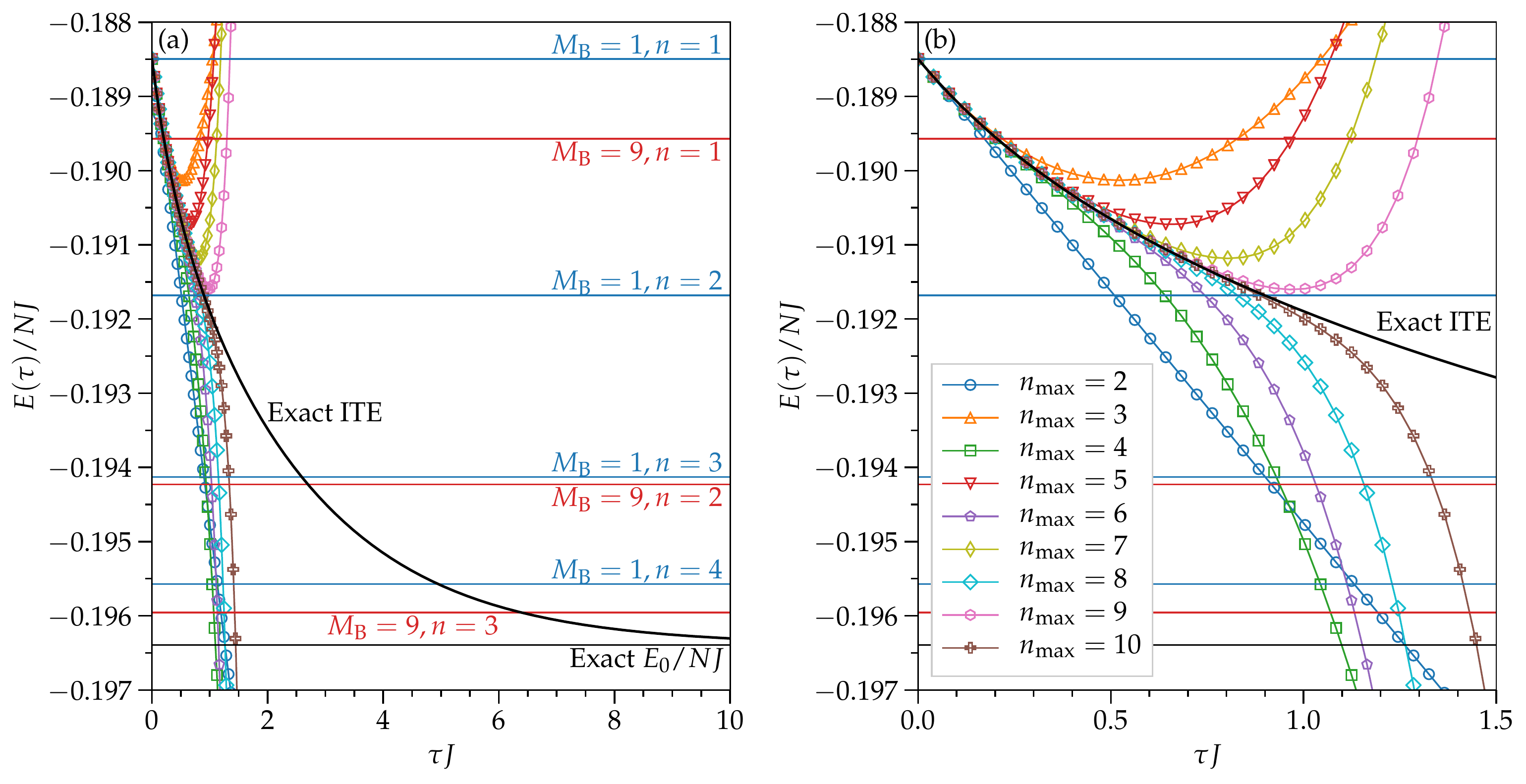}
    \caption{
      (a) 
      The energy expectation value $E(\tau)$
      with respect to the exact imaginary-time evolution
      of $|\Psi_{\rm VQE}\rangle$ 
      (black solid line) and 
      the CMX energy $E_{n_{\rm max}}(\tau)$
      with various truncation order $n_{\rm max}$ (symbols) 
      as a function of the imaginary time $\tau$ 
      for the spin-1/2 Heisenberg model on an $N=16$ qubit ring given in Eq.~(\ref{Ham_SWAP}). 
      For comparison, 
      the energies estimated with 
      the Krylov-subspace diagonalization 
      involving
      $|\Psi_{\rm VQE}\rangle$ as a reference state 
      with $M_{\rm B}=1$ for
      $1 \leqslant n\leqslant 4$ (blue horizontal lines)
      and 
      with $M_{\rm B}=9$ for
      $1 \leqslant n\leqslant 3$ (red horizontal lines) 
      are shown. These are the same results in Fig.~\ref{fig.energy}. 
      Here, $n = \dim{\mathcal{K}}_n/M_{\rm B}$, i.e., 
      the dimension of the Krylov subspace $\mathcal{K}_n$ per block size $M_{\rm B}$ 
      (for details, see Sec.~\ref{sec:egs}). 
      (b) Same as (a) but
      an enlarged plot for $0\leqslant \tau J \leqslant 1.5$. 
      \label{fig.cmx}
    }
  \end{figure*}
\end{center}

Figure~\ref{fig.cmx} shows the exact $E(\tau)$
and the CMX of the energy truncated at the $n_{\rm max}$th cumulant 
\begin{equation}
  E_{n_{\rm max}}(\tau)
  =
  \sum_{n=0}^{n_{\rm max}-1}
  \frac{(-\tau)^n}{n!}\kappa_{n+1} 
  \label{CMX}
\end{equation}
for the spin-1/2 Heisenberg model defined in Eq.~(\ref{Ham_SWAP}), 
where the VQE state $|\Psi_{\rm VQE}\rangle$ in Eq.~(\ref{eq:vqe}) is selected for the quantum state $|\Psi\rangle$ 
in Eq.~(\ref{eq:psi_tau}). 
The energy $E(\tau)$ with the exact ITE decreases monotonically
in $\tau$, because the first derivative of $E(\tau)$ is minus of 
the energy fluctuation~\cite{Horn1984} 
\begin{equation}
  \frac{\dd E(\tau)}{\dd \tau} =
  -
  \left(
    \langle \hat{\mathcal{H}}^2 \rangle_\tau 
    -
    \langle \hat{\mathcal{H}}\rangle_\tau^2
  \right)
  \leqslant 0, 
\end{equation}
where the equality satisfies if and only if $|\Psi_{\rm VQE}(\tau)\rangle$
is an exact eigenstate (e.g., the ground state)
of $\hat{\mathcal{H}}$. 
On the other hand, 
due to the truncation of the series at finite order, 
$E_{n_{\rm max}}(\tau)$ at large $\tau$ diverges 
to $-\infty$ for even $n_{\rm max} \geqslant 2$ or
to $+\infty$ for odd  $n_{\rm max} \geqslant 3$. 
Note that 
$E_{2}(\tau)=\kappa_1-\kappa_2 \tau$  
is the tangent line of $E(\tau)$ at $\tau=0$.
We also find that the convergence of $E_{n_{\rm max}}(\tau)$ 
to the exact ground-state energy $E_0$ 
with respect to the power exponents $n_{\rm max}$ in the cumulants required  
is rather slower, as compared with the Krylov-subspace
diagonalization with either
$M_{\rm B}=1$ or $M_{\rm B}=9$ discussed in Sec.~\ref{sec:egs}. 
This is not quite surprising because the form of $E(\tau)$
in Eq.~(\ref{CMX}) is an expansion around $\tau=0$, 
which is analogous to the high-temperature expansion. 

Finally, it should be noted that recently    
several schemes different from the quantum power method 
to evaluate $\langle \hat{\mathcal{H}}^n \rangle$
are proposed and demonstrated
with various CMX methods for quantum-chemistry Hamiltonians~\cite{kowalski2020quantum} and
with the Lanczos method for Heisenberg Hamiltonians~\cite{vallury2020quantum}.

\section{Lanczos method}\label{app:lanczos}

In this appendix, we briefly outline the Lanczos method
with an emphasis on its
aspect as a moment method~\cite{Witte1994,Haxton2005}, 
i.e., a potential application of the quantum power method.

\subsection{Lanczos tridiagonal matrix and Hamiltonian moment}
The Lanczos method generates a sequence of
orthonormalized states $\{|q_i\rangle\}$,
satisfying $\langle q_i|q_j\rangle=\delta_{ij}$, 
from an initial (reference) state $|q_1\rangle = |\Psi\rangle$
recursively as 
\begin{alignat}{1}
  \hat{\mathcal{H}} |q_i\rangle
  = 
  \beta_{i-1}|q_{i-1}\rangle
  + \alpha_i |q_i \rangle
  + \beta_{i}|q_{i+1}\rangle,
  \label{Lanczos}
\end{alignat}
with
$\alpha_i=\langle q_i |\hat{\mathcal{H}}|q_i \rangle$, 
$\beta_i=\langle q_{i} |\hat{\mathcal{H}}|q_{i+1} \rangle$,  
$\beta_{0}\equiv 0$, and $|q_{0}\rangle\equiv 0$.
After obtaining $\{|q_{i}\rangle\}_{i=1}^{n}$, 
the Hamiltonian $\hat{\mathcal{H}}$ can be
represented as a tridiagonal matrix 
$[\bs{T}_n]_{ij}=\langle q_i|\hat{\mathcal{H}}|q_j\rangle $ as 
\begin{equation}
  \bs{T}_n=
  \begin{bmatrix}
    \alpha_1 & \beta_1  &          &        &  &  \\
    \beta_1  & \alpha_2 & \beta_2  &        &  &  \\
             & \beta_2  & \alpha_3 & \ddots &  &  \\
             &          & \ddots   & \ddots &  &   \\
             &          &          &        &             & \beta_{n-1} \\
             &          &          &        & \beta_{n-1} & \alpha_n 
  \end{bmatrix}.
\end{equation}

The matrix elements $\{\alpha_{i}\}$ and $\{\beta_{i}\}$ 
can also be constructed recursively using the Hamiltonian moments~\cite{Witte1994,Haxton2005}.
Following Ref.~\cite{Witte1994},
$\{\alpha_i\}$ and $\{\beta_i\}$ are given
in terms of $\{\mu_n\}$ as 
\begin{equation}
  \alpha_{i}=
  \left(
  \frac{\mathcal{L}_{i-1}}{\mathcal{L}_{i-2}}
  \right)
  \left(
  \frac{\mathcal{M}_{i-2}}{\mathcal{M}_{i-3}}
  \right)^{-1}
  +
  \left(
  \frac{\mathcal{M}_{i-1}}{\mathcal{M}_{i-2}}
  \right)
  \left(
  \frac{\mathcal{L}_{i-1}}{\mathcal{L}_{i-2}}
  \right)^{-1} \label{lanA}
\end{equation}
and 
\begin{equation}
  \beta_{i}^2=
  \left(
  \frac{\mathcal{L}_{i}}{\mathcal{L}_{i-1}}
  \right)
  \left(
  \frac{\mathcal{L}_{i-1}}{\mathcal{L}_{i-2}}
  \right)^{-1}, \label{lanB}
\end{equation}
where
$\mathcal{L}_n \equiv \det{\bs{\mathcal{L}}_n}$ and 
$\mathcal{M}_n \equiv \det{\bs{\mathcal{M}}_n}$ are 
determinants of $(n+1) \times (n+1)$ Hankel matrices
defined respectively as
$[\bs{\mathcal{L}}_{n}]_{ij} = \mu_{i+j-2}$ and 
$[\bs{\mathcal{M}}_{n}]_{ij} = \mu_{i+j-1}$,
or more explicitly 
\begin{equation}
  \bs{\mathcal{L}}_{n}
  =
  \begin{bmatrix}
    \mu_0 & \mu_1 & \mu_2 & \cdots & \mu_{n-1} & \mu_{n} \\
    \mu_1 & \mu_2 & \mu_3 & \cdots & \mu_{n}   & \mu_{n+1} \\
    \mu_2 & \mu_3 & \mu_4 & \cdots & \mu_{n+1} & \mu_{n+2} \\
    \vdots & \vdots & \vdots & \vdots & \vdots & \vdots\\
    \mu_{n-1} & \mu_{n}   & \mu_{n+1} & \cdots & \mu_{2n-2} & \mu_{2n-1} \\
    \mu_{n}   & \mu_{n+1} & \mu_{n+2} & \cdots & \mu_{2n-1} & \mu_{2n} \\
  \end{bmatrix}
  \label{eq:hankel_l}
\end{equation}
for $n \geqslant 0$ and
\begin{equation}
  \bs{\mathcal{M}}_{n}
  =
  \begin{bmatrix}
    \mu_1 & \mu_2 & \mu_3 & \cdots & \mu_{n}   & \mu_{n+1} \\
    \mu_2 & \mu_3 & \mu_4 & \cdots & \mu_{n+1} & \mu_{n+2} \\
    \mu_3 & \mu_4 & \mu_5 & \cdots & \mu_{n+2} & \mu_{n+3} \\
    \vdots & \vdots & \vdots & \vdots & \vdots & \vdots\\
    \mu_{n}   & \mu_{n+1} & \mu_{n+2} & \cdots & \mu_{2n-1} & \mu_{2n} \\
    \mu_{n+1} & \mu_{n+2} & \mu_{n+3} & \cdots & \mu_{2n}   & \mu_{2n+1} \\
  \end{bmatrix}
  \label{eq:hankel_m}
\end{equation}
for $n \geqslant 0$.
Equations~(\ref{lanA}) and (\ref{lanB})
hold for $i\geqslant 1$ provided that
${\cal M}_n$ and ${\cal L}_n$ with negative indices are defined as 
${\cal M}_{-1}=1$, 
${\cal M}_{-2}=0$, and
${\cal L}_{-1}=1$.
The Hankel matrices
$\bs{\mathcal{L}}_{n-1}$  and
$\bs{\mathcal{M}}_{n-1}$ are identical respectively to 
$\bs{S}$ in Eq.~(\ref{Ssubspace}) and
$\bs{H}$ in Eq.~(\ref{Hsubspace}) if $M_{\rm B}=1$. 
It is noticed in Eqs.~(\ref{lanA}) and (\ref{lanB}) that
the Lanczos matrix elements $\alpha_i$ and $\beta_i$ are
expressed in terms of the ratios of the Hankel determinants 
whose matrix dimensions differ only by 1.
The particular structure of the Hankel matrices
$\bs{\mathcal{L}}_n$ and  $\bs{\mathcal{M}}_n$ 
allows us to evaluate the ratios of the determinants
appearing in Eqs.~(\ref{lanA}) and (\ref{lanB}) 
recursively, as described in appendix~\ref{app:lanczos2}.

It is instructive to give the explicit forms of 
the first few matrix elements of $\bs{T}_n$.
The first three matrix elements required for constructing
the $2\times 2$ matrix $\bs{T}_2$ are given by 
\begin{alignat}{1}
  \alpha_{1}&= \langle \hat{\mathcal{H}} \rangle, \\
  \beta_{1}&=\sqrt{\langle \hat{\mathcal{H}}^2 \rangle - \langle \hat{\mathcal{H}}\rangle^2  }, \\
  \alpha_{2}& =
  \frac{
  \langle \hat{\mathcal{H}}^3\rangle
  -2\langle \hat{\mathcal{H}}^2\rangle \langle \hat{\mathcal{H}}\rangle
  + \langle \hat{\mathcal{H}}\rangle^3}
       {\langle \hat{\mathcal{H}}^2 \rangle - \langle \hat{\mathcal{H}}\rangle^2  },  
\end{alignat}
where $\langle\cdots\rangle=\langle q_1|\cdots|q_1\rangle$. 
Therefore, $\alpha_1$ and $\beta_1^2$ are  
the energy expectation value and
the energy variance with respect to the initial state $|q_1\rangle$, respectively.

\subsection{Ratio of Hankel determinants}\label{app:lanczos2}
We now describe a way to 
calculate recursively the ratio of the determinants appearing 
in Eqs.~(\ref{lanA}) and (\ref{lanB}). 
Let us first review the determinant and
the matrix-inversion formulas for general matrices. 
Let
$\bs{A}_n$ be an $n\times n$ matrix,
$\bs{b}$ be an $n\times 1$ matrix,
$\bs{c}$ be an $n\times 1$ matrix, and
$d$ be a $1\times 1$ matrix (i.e., a scalar),
and let us consider an $(n+1)\times(n+1)$ matrix $\bs{A}_{n+1}$
of the form 
\begin{equation}
  \bs{A}_{n+1}=
  \begin{bmatrix}
    \bs{A}_{n} &\bs{b} \\
    \bs{c}^T & d
  \end{bmatrix}. 
\end{equation}
If we define 
\begin{equation}
  r=d-\bs{c}^T \bs{A}_{n}^{-1} \bs{b}, 
\end{equation}
the determinant of $\bs{A}_{n+1}$ is given by 
\begin{equation}
  \det \bs{A}_{n+1}
  =
  \det
  \begin{bmatrix}
    \bs{A}_n &\bs{b} \\
    \bs{c}^T & d
  \end{bmatrix}
  =r \det{\bs{A}}_n,
  \label{detA}
\end{equation}
and the inverse $\bs{A}_{n+1}^{-1}$ is given by 
\begin{alignat}{1}
  \bs{A}_{n+1}^{-1}
  &=
  \begin{bmatrix}
    \bs{A}_n  &\bs{b} \\
    \bs{c}^T & d
  \end{bmatrix}^{-1}\notag \\
  &=
  \begin{bmatrix}
    \bs{A}^{-1}_n +
    (\bs{A}^{-1}_n\bs{b})
    (\bs{c}^T\bs{A}_n^{-1})/r
    &
    -\bs{A}^{-1}_n\bs{b}/r \\
    -\bs{c}^T\bs{A}^{-1}_n/r & 1/r
  \end{bmatrix}.
  \label{invA}
\end{alignat}

Now we apply the above formulas to 
recursively evaluate the ratios
of the determinants of
$\bs{\mathcal{L}}_n$ and $\bs{\mathcal{L}}_{n-1}$. 
Due to its particular structure, $\bs{\mathcal{L}}_n$
can be expressed in terms of $\bs{\mathcal{L}}_{n-1}$ as 
\begin{equation}
  \bs{\mathcal{L}}_n 
  =
  \begin{bmatrix}
    \bs{\mathcal{L}}_{n-1} & \bs{m}_n \\
    \bs{m}_n^T & \mu_{2n}
  \end{bmatrix}
\end{equation}
with the following $n$-dimensional vector:
\begin{equation}
  \bs{m}_n^{T}=(\mu_n,\mu_{n+1},\dots,\mu_{2n-1}). 
\end{equation}
From the formula in Eq.~(\ref{detA}), 
the ratio of the determinants is given by 
\begin{equation}
  \frac{\mathcal{L}_n}{\mathcal{L}_{n-1}}= 
  \frac{\det\bs{\mathcal{L}}_n}{\det\bs{\mathcal{L}}_{n-1}}=  r_n 
  \label{eq:ratio_det_L}
\end{equation}
with 
\begin{equation}
  r_n = \mu_{2n} - \bs{m}_n^T \bs{\mathcal{L}}_{n-1}^{-1} \bs{m}_n,
  \label{detratio}
\end{equation}
which involves the inverse $\bs{\mathcal{L}}_{n-1}^{-1}$ whose
dimension is less than that of $\bs{\mathcal{L}}_n^{-1}$ by 1.

The inverse matrix $\bs{\mathcal{L}}_n^{-1}$ can be calculated
using Eq.~(\ref{invA}). 
Starting with
\begin{equation}
  \bs{\mathcal{L}}_0^{-1}=\mu_0^{-1},
\end{equation}
$\bs{\mathcal{L}}_n^{-1}$ for $n\geqslant 1$
can be constructed from
$\bs{\mathcal{L}}_{n-1}^{-1}$ and $\bs{m}_n$
recursively as  
\begin{alignat}{1}
  \bs{\mathcal{L}}_n^{-1}
  &=
  \begin{bmatrix}
    \bs{\mathcal{L}}_{n-1} & \bs{m}_n \\
    \bs{m}_n^T        & \mu_{2n}
  \end{bmatrix}^{-1} \notag \\
  &=
  \begin{bmatrix}
    \bs{\mathcal{L}}_{n-1}^{-1} +
    (\bs{\mathcal{L}}_{n-1}^{-1}\bs{m}_n)
    (\bs{\mathcal{L}}_{n-1}^{-1}\bs{m}_n)^{T}/r_n 
    &
    -\bs{\mathcal{L}}_{n-1}^{-1}\bs{m}_n/r_n \\
    -(\bs{\mathcal{L}}_{n-1}^{-1}\bs{m}_n)^{T}/r_n 
    & 1/r_n
  \end{bmatrix},
  \label{Linv}
\end{alignat}
where $(\bs{\mathcal{L}}_n^{-1})^{T}=\bs{\mathcal{L}}_n^{-1}$ is used. 
Thus, starting with the known $\bs{\mathcal{L}}_{0}^{-1}$
and using Eqs.~(\ref{detratio}) and (\ref{Linv}), 
one can obtain $\{r_n\}$ recursively as
$
\bs{\mathcal{L}}_0^{-1} \to r_{1} \to
\bs{\mathcal{L}}_1^{-1} \to r_{2} \to
\bs{\mathcal{L}}_2^{-1} \to r_{3} \to \cdots 
$.

It should be noted that 
Eq.~(\ref{detratio}) involves a matrix-vector multiplication and, 
in addition, Eq.~(\ref{Linv}) involves a rank-1 update. 
Therefore, the complexity for computing the
ratio of determinants in Eq.~(\ref{eq:ratio_det_L}) is $O(n^2)$. 
This is more efficient when $n$ is large because 
the direct calculation of a determinant from scratch, e.g., by using the LU decomposition,  
requires $O(n^3)$ operations. 
Noticing that 
$[\bs{\mathcal{L}}_{n}]_{ij}=\mu_{i+j-2}$ while 
$[\bs{\mathcal{M}}_{n}]_{ij}=\mu_{i+j-1}$, 
the similar recursive formula for $\bs{\mathcal{M}}_{n}$
can be readily derived
simply by replacing the indexes for the moments 
in the above as $\{\mu_{i}\}_{i=0}^{2n}\to\{\mu_{i+1}\}_{i=0}^{2n}$.

\bibliography{biball}

\begin{thebibliography}{133}%
\makeatletter
\providecommand \@ifxundefined [1]{%
 \@ifx{#1\undefined}
}%
\providecommand \@ifnum [1]{%
 \ifnum #1\expandafter \@firstoftwo
 \else \expandafter \@secondoftwo
 \fi
}%
\providecommand \@ifx [1]{%
 \ifx #1\expandafter \@firstoftwo
 \else \expandafter \@secondoftwo
 \fi
}%
\providecommand \natexlab [1]{#1}%
\providecommand \enquote  [1]{``#1''}%
\providecommand \bibnamefont  [1]{#1}%
\providecommand \bibfnamefont [1]{#1}%
\providecommand \citenamefont [1]{#1}%
\providecommand \href@noop [0]{\@secondoftwo}%
\providecommand \href [0]{\begingroup \@sanitize@url \@href}%
\providecommand \@href[1]{\@@startlink{#1}\@@href}%
\providecommand \@@href[1]{\endgroup#1\@@endlink}%
\providecommand \@sanitize@url [0]{\catcode `\\12\catcode `\$12\catcode
  `\&12\catcode `\#12\catcode `\^12\catcode `\_12\catcode `\%12\relax}%
\providecommand \@@startlink[1]{}%
\providecommand \@@endlink[0]{}%
\providecommand \url  [0]{\begingroup\@sanitize@url \@url }%
\providecommand \@url [1]{\endgroup\@href {#1}{\urlprefix }}%
\providecommand \urlprefix  [0]{URL }%
\providecommand \Eprint [0]{\href }%
\providecommand \doibase [0]{https://doi.org/}%
\providecommand \selectlanguage [0]{\@gobble}%
\providecommand \bibinfo  [0]{\@secondoftwo}%
\providecommand \bibfield  [0]{\@secondoftwo}%
\providecommand \translation [1]{[#1]}%
\providecommand \BibitemOpen [0]{}%
\providecommand \bibitemStop [0]{}%
\providecommand \bibitemNoStop [0]{.\EOS\space}%
\providecommand \EOS [0]{\spacefactor3000\relax}%
\providecommand \BibitemShut  [1]{\csname bibitem#1\endcsname}%
\let\auto@bib@innerbib\@empty
\bibitem [{\citenamefont {LeBlanc}\ \emph {et~al.}(2015)\citenamefont
  {LeBlanc}, \citenamefont {Antipov}, \citenamefont {Becca}, \citenamefont
  {Bulik}, \citenamefont {Chan}, \citenamefont {Chung}, \citenamefont {Deng},
  \citenamefont {Ferrero}, \citenamefont {Henderson}, \citenamefont
  {Jim\'enez-Hoyos}, \citenamefont {Kozik}, \citenamefont {Liu}, \citenamefont
  {Millis}, \citenamefont {Prokof'ev}, \citenamefont {Qin}, \citenamefont
  {Scuseria}, \citenamefont {Shi}, \citenamefont {Svistunov}, \citenamefont
  {Tocchio}, \citenamefont {Tupitsyn}, \citenamefont {White}, \citenamefont
  {Zhang}, \citenamefont {Zheng}, \citenamefont {Zhu},\ and\ \citenamefont
  {Gull}}]{LeBlanc2015}%
  \BibitemOpen
  \bibfield  {author} {\bibinfo {author} {\bibfnamefont {J.~P.~F.}\
  \bibnamefont {LeBlanc}}, \bibinfo {author} {\bibfnamefont {A.~E.}\
  \bibnamefont {Antipov}}, \bibinfo {author} {\bibfnamefont {F.}~\bibnamefont
  {Becca}}, \bibinfo {author} {\bibfnamefont {I.~W.}\ \bibnamefont {Bulik}},
  \bibinfo {author} {\bibfnamefont {G.~K.-L.}\ \bibnamefont {Chan}}, \bibinfo
  {author} {\bibfnamefont {C.-M.}\ \bibnamefont {Chung}}, \bibinfo {author}
  {\bibfnamefont {Y.}~\bibnamefont {Deng}}, \bibinfo {author} {\bibfnamefont
  {M.}~\bibnamefont {Ferrero}}, \bibinfo {author} {\bibfnamefont {T.~M.}\
  \bibnamefont {Henderson}}, \bibinfo {author} {\bibfnamefont {C.~A.}\
  \bibnamefont {Jim\'enez-Hoyos}}, \bibinfo {author} {\bibfnamefont
  {E.}~\bibnamefont {Kozik}}, \bibinfo {author} {\bibfnamefont {X.-W.}\
  \bibnamefont {Liu}}, \bibinfo {author} {\bibfnamefont {A.~J.}\ \bibnamefont
  {Millis}}, \bibinfo {author} {\bibfnamefont {N.~V.}\ \bibnamefont
  {Prokof'ev}}, \bibinfo {author} {\bibfnamefont {M.}~\bibnamefont {Qin}},
  \bibinfo {author} {\bibfnamefont {G.~E.}\ \bibnamefont {Scuseria}}, \bibinfo
  {author} {\bibfnamefont {H.}~\bibnamefont {Shi}}, \bibinfo {author}
  {\bibfnamefont {B.~V.}\ \bibnamefont {Svistunov}}, \bibinfo {author}
  {\bibfnamefont {L.~F.}\ \bibnamefont {Tocchio}}, \bibinfo {author}
  {\bibfnamefont {I.~S.}\ \bibnamefont {Tupitsyn}}, \bibinfo {author}
  {\bibfnamefont {S.~R.}\ \bibnamefont {White}}, \bibinfo {author}
  {\bibfnamefont {S.}~\bibnamefont {Zhang}}, \bibinfo {author} {\bibfnamefont
  {B.-X.}\ \bibnamefont {Zheng}}, \bibinfo {author} {\bibfnamefont
  {Z.}~\bibnamefont {Zhu}},\ and\ \bibinfo {author} {\bibfnamefont
  {E.}~\bibnamefont {Gull}} (\bibinfo {collaboration} {Simons Collaboration on
  the Many-Electron Problem}),\ }\bibfield  {title} {\bibinfo {title}
  {{Solutions of the Two-Dimensional Hubbard Model: Benchmarks and Results from
  a Wide Range of Numerical Algorithms}},\ }\href
  {https://doi.org/10.1103/PhysRevX.5.041041} {\bibfield  {journal} {\bibinfo
  {journal} {Phys. Rev. X}\ }\textbf {\bibinfo {volume} {5}},\ \bibinfo {pages}
  {041041} (\bibinfo {year} {2015})}\BibitemShut {NoStop}%
\bibitem [{\citenamefont {Motta}\ \emph {et~al.}(2017)\citenamefont {Motta},
  \citenamefont {Ceperley}, \citenamefont {Chan}, \citenamefont {Gomez},
  \citenamefont {Gull}, \citenamefont {Guo}, \citenamefont {Jim\'enez-Hoyos},
  \citenamefont {Lan}, \citenamefont {Li}, \citenamefont {Ma}, \citenamefont
  {Millis}, \citenamefont {Prokof'ev}, \citenamefont {Ray}, \citenamefont
  {Scuseria}, \citenamefont {Sorella}, \citenamefont {Stoudenmire},
  \citenamefont {Sun}, \citenamefont {Tupitsyn}, \citenamefont {White},
  \citenamefont {Zgid},\ and\ \citenamefont {Zhang}}]{Motta2017}%
  \BibitemOpen
  \bibfield  {author} {\bibinfo {author} {\bibfnamefont {M.}~\bibnamefont
  {Motta}}, \bibinfo {author} {\bibfnamefont {D.~M.}\ \bibnamefont {Ceperley}},
  \bibinfo {author} {\bibfnamefont {G.~K.-L.}\ \bibnamefont {Chan}}, \bibinfo
  {author} {\bibfnamefont {J.~A.}\ \bibnamefont {Gomez}}, \bibinfo {author}
  {\bibfnamefont {E.}~\bibnamefont {Gull}}, \bibinfo {author} {\bibfnamefont
  {S.}~\bibnamefont {Guo}}, \bibinfo {author} {\bibfnamefont {C.~A.}\
  \bibnamefont {Jim\'enez-Hoyos}}, \bibinfo {author} {\bibfnamefont {T.~N.}\
  \bibnamefont {Lan}}, \bibinfo {author} {\bibfnamefont {J.}~\bibnamefont
  {Li}}, \bibinfo {author} {\bibfnamefont {F.}~\bibnamefont {Ma}}, \bibinfo
  {author} {\bibfnamefont {A.~J.}\ \bibnamefont {Millis}}, \bibinfo {author}
  {\bibfnamefont {N.~V.}\ \bibnamefont {Prokof'ev}}, \bibinfo {author}
  {\bibfnamefont {U.}~\bibnamefont {Ray}}, \bibinfo {author} {\bibfnamefont
  {G.~E.}\ \bibnamefont {Scuseria}}, \bibinfo {author} {\bibfnamefont
  {S.}~\bibnamefont {Sorella}}, \bibinfo {author} {\bibfnamefont {E.~M.}\
  \bibnamefont {Stoudenmire}}, \bibinfo {author} {\bibfnamefont
  {Q.}~\bibnamefont {Sun}}, \bibinfo {author} {\bibfnamefont {I.~S.}\
  \bibnamefont {Tupitsyn}}, \bibinfo {author} {\bibfnamefont {S.~R.}\
  \bibnamefont {White}}, \bibinfo {author} {\bibfnamefont {D.}~\bibnamefont
  {Zgid}},\ and\ \bibinfo {author} {\bibfnamefont {S.}~\bibnamefont {Zhang}}
  (\bibinfo {collaboration} {Simons Collaboration on the Many-Electron
  Problem}),\ }\bibfield  {title} {\bibinfo {title} {{Towards the Solution of
  the Many-Electron Problem in Real Materials: Equation of State of the
  Hydrogen Chain with State-of-the-Art Many-Body Methods}},\ }\href
  {https://doi.org/10.1103/PhysRevX.7.031059} {\bibfield  {journal} {\bibinfo
  {journal} {Phys. Rev. X}\ }\textbf {\bibinfo {volume} {7}},\ \bibinfo {pages}
  {031059} (\bibinfo {year} {2017})}\BibitemShut {NoStop}%
\bibitem [{\citenamefont {Motta}\ \emph {et~al.}(2020)\citenamefont {Motta},
  \citenamefont {Genovese}, \citenamefont {Ma}, \citenamefont {Cui},
  \citenamefont {Sawaya}, \citenamefont {Chan}, \citenamefont {Chepiga},
  \citenamefont {Helms}, \citenamefont {Jim\'enez-Hoyos}, \citenamefont
  {Millis}, \citenamefont {Ray}, \citenamefont {Ronca}, \citenamefont {Shi},
  \citenamefont {Sorella}, \citenamefont {Stoudenmire}, \citenamefont {White},\
  and\ \citenamefont {Zhang}}]{motta2019groundstate}%
  \BibitemOpen
  \bibfield  {author} {\bibinfo {author} {\bibfnamefont {M.}~\bibnamefont
  {Motta}}, \bibinfo {author} {\bibfnamefont {C.}~\bibnamefont {Genovese}},
  \bibinfo {author} {\bibfnamefont {F.}~\bibnamefont {Ma}}, \bibinfo {author}
  {\bibfnamefont {Z.-H.}\ \bibnamefont {Cui}}, \bibinfo {author} {\bibfnamefont
  {R.}~\bibnamefont {Sawaya}}, \bibinfo {author} {\bibfnamefont {G.~K.-L.}\
  \bibnamefont {Chan}}, \bibinfo {author} {\bibfnamefont {N.}~\bibnamefont
  {Chepiga}}, \bibinfo {author} {\bibfnamefont {P.}~\bibnamefont {Helms}},
  \bibinfo {author} {\bibfnamefont {C.}~\bibnamefont {Jim\'enez-Hoyos}},
  \bibinfo {author} {\bibfnamefont {A.~J.}\ \bibnamefont {Millis}}, \bibinfo
  {author} {\bibfnamefont {U.}~\bibnamefont {Ray}}, \bibinfo {author}
  {\bibfnamefont {E.}~\bibnamefont {Ronca}}, \bibinfo {author} {\bibfnamefont
  {H.}~\bibnamefont {Shi}}, \bibinfo {author} {\bibfnamefont {S.}~\bibnamefont
  {Sorella}}, \bibinfo {author} {\bibfnamefont {E.~M.}\ \bibnamefont
  {Stoudenmire}}, \bibinfo {author} {\bibfnamefont {S.~R.}\ \bibnamefont
  {White}},\ and\ \bibinfo {author} {\bibfnamefont {S.}~\bibnamefont {Zhang}}
  (\bibinfo {collaboration} {Simons Collaboration on the Many-Electron
  Problem}),\ }\bibfield  {title} {\bibinfo {title} {{Ground-State Properties
  of the Hydrogen Chain: Dimerization, Insulator-to-Metal Transition, and
  Magnetic Phases}},\ }\href {https://doi.org/10.1103/PhysRevX.10.031058}
  {\bibfield  {journal} {\bibinfo  {journal} {Phys. Rev. X}\ }\textbf {\bibinfo
  {volume} {10}},\ \bibinfo {pages} {031058} (\bibinfo {year}
  {2020})}\BibitemShut {NoStop}%
\bibitem [{\citenamefont {Eriksen}\ \emph {et~al.}(2020)\citenamefont
  {Eriksen}, \citenamefont {Anderson}, \citenamefont {Deustua}, \citenamefont
  {Ghanem}, \citenamefont {Hait}, \citenamefont {Hoffmann}, \citenamefont
  {Lee}, \citenamefont {Levine}, \citenamefont {Magoulas}, \citenamefont
  {Shen}, \citenamefont {Tubman}, \citenamefont {Whaley}, \citenamefont {Xu},
  \citenamefont {Yao}, \citenamefont {Zhang}, \citenamefont {Alavi},
  \citenamefont {Chan}, \citenamefont {Head-Gordon}, \citenamefont {Liu},
  \citenamefont {Piecuch}, \citenamefont {Sharma}, \citenamefont {Ten-no},
  \citenamefont {Umrigar},\ and\ \citenamefont {Gauss}}]{Eriksen2020}%
  \BibitemOpen
  \bibfield  {author} {\bibinfo {author} {\bibfnamefont {J.~J.}\ \bibnamefont
  {Eriksen}}, \bibinfo {author} {\bibfnamefont {T.~A.}\ \bibnamefont
  {Anderson}}, \bibinfo {author} {\bibfnamefont {J.~E.}\ \bibnamefont
  {Deustua}}, \bibinfo {author} {\bibfnamefont {K.}~\bibnamefont {Ghanem}},
  \bibinfo {author} {\bibfnamefont {D.}~\bibnamefont {Hait}}, \bibinfo {author}
  {\bibfnamefont {M.~R.}\ \bibnamefont {Hoffmann}}, \bibinfo {author}
  {\bibfnamefont {S.}~\bibnamefont {Lee}}, \bibinfo {author} {\bibfnamefont
  {D.~S.}\ \bibnamefont {Levine}}, \bibinfo {author} {\bibfnamefont
  {I.}~\bibnamefont {Magoulas}}, \bibinfo {author} {\bibfnamefont
  {J.}~\bibnamefont {Shen}}, \bibinfo {author} {\bibfnamefont {N.~M.}\
  \bibnamefont {Tubman}}, \bibinfo {author} {\bibfnamefont {K.~B.}\
  \bibnamefont {Whaley}}, \bibinfo {author} {\bibfnamefont {E.}~\bibnamefont
  {Xu}}, \bibinfo {author} {\bibfnamefont {Y.}~\bibnamefont {Yao}}, \bibinfo
  {author} {\bibfnamefont {N.}~\bibnamefont {Zhang}}, \bibinfo {author}
  {\bibfnamefont {A.}~\bibnamefont {Alavi}}, \bibinfo {author} {\bibfnamefont
  {G.~K.-L.}\ \bibnamefont {Chan}}, \bibinfo {author} {\bibfnamefont
  {M.}~\bibnamefont {Head-Gordon}}, \bibinfo {author} {\bibfnamefont
  {W.}~\bibnamefont {Liu}}, \bibinfo {author} {\bibfnamefont {P.}~\bibnamefont
  {Piecuch}}, \bibinfo {author} {\bibfnamefont {S.}~\bibnamefont {Sharma}},
  \bibinfo {author} {\bibfnamefont {S.~L.}\ \bibnamefont {Ten-no}}, \bibinfo
  {author} {\bibfnamefont {C.~J.}\ \bibnamefont {Umrigar}},\ and\ \bibinfo
  {author} {\bibfnamefont {J.}~\bibnamefont {Gauss}},\ }\bibfield  {title}
  {\bibinfo {title} {{The Ground State Electronic Energy of Benzene}},\ }\href
  {https://doi.org/10.1021/acs.jpclett.0c02621} {\bibfield  {journal} {\bibinfo
   {journal} {The Journal of Physical Chemistry Letters}\ }\textbf {\bibinfo
  {volume} {11}},\ \bibinfo {pages} {8922} (\bibinfo {year}
  {2020})}\BibitemShut {NoStop}%
\bibitem [{\citenamefont {Liesen}\ and\ \citenamefont
  {Strako\v{s}}(2013)}]{Liesen}%
  \BibitemOpen
  \bibfield  {author} {\bibinfo {author} {\bibfnamefont {J.}~\bibnamefont
  {Liesen}}\ and\ \bibinfo {author} {\bibfnamefont {Z.}~\bibnamefont
  {Strako\v{s}}},\ }\href@noop {} {\emph {\bibinfo {title} {{Krylov Subspace
  Methods}}}}\ (\bibinfo  {publisher} {Oxford University Press},\ \bibinfo
  {address} {Oxford},\ \bibinfo {year} {2013})\ Chap.~\bibinfo {chapter}
  {3}\BibitemShut {NoStop}%
\bibitem [{\citenamefont {Dagotto}(1994)}]{Dagotto1994}%
  \BibitemOpen
  \bibfield  {author} {\bibinfo {author} {\bibfnamefont {E.}~\bibnamefont
  {Dagotto}},\ }\bibfield  {title} {\bibinfo {title} {{Correlated electrons in
  high-temperature superconductors}},\ }\href
  {https://doi.org/10.1103/RevModPhys.66.763} {\bibfield  {journal} {\bibinfo
  {journal} {Rev. Mod. Phys.}\ }\textbf {\bibinfo {volume} {66}},\ \bibinfo
  {pages} {763} (\bibinfo {year} {1994})}\BibitemShut {NoStop}%
\bibitem [{\citenamefont {Jakli\v{c}}\ and\ \citenamefont
  {Prelo\v{s}ek}(1994)}]{Jaklic1994}%
  \BibitemOpen
  \bibfield  {author} {\bibinfo {author} {\bibfnamefont {J.}~\bibnamefont
  {Jakli\v{c}}}\ and\ \bibinfo {author} {\bibfnamefont {P.}~\bibnamefont
  {Prelo\v{s}ek}},\ }\bibfield  {title} {\bibinfo {title} {{Lanczos method for
  the calculation of finite-temperature quantities in correlated systems}},\
  }\href {https://doi.org/10.1103/PhysRevB.49.5065} {\bibfield  {journal}
  {\bibinfo  {journal} {Phys. Rev. B}\ }\textbf {\bibinfo {volume} {49}},\
  \bibinfo {pages} {5065} (\bibinfo {year} {1994})}\BibitemShut {NoStop}%
\bibitem [{\citenamefont {Jakli\v{c}}\ and\ \citenamefont
  {Prelov\v{s}ek}(2000)}]{Jaklic2000}%
  \BibitemOpen
  \bibfield  {author} {\bibinfo {author} {\bibfnamefont {J.}~\bibnamefont
  {Jakli\v{c}}}\ and\ \bibinfo {author} {\bibfnamefont {P.}~\bibnamefont
  {Prelov\v{s}ek}},\ }\bibfield  {title} {\bibinfo {title} {{Finite-temperature
  properties of doped antiferromagnets}},\ }\href
  {http://dx.doi.org/10.1080/000187300243381} {\bibfield  {journal} {\bibinfo
  {journal} {Adv. Phys.}\ }\textbf {\bibinfo {volume} {49}},\ \bibinfo {pages}
  {1} (\bibinfo {year} {2000})}\BibitemShut {NoStop}%
\bibitem [{\citenamefont {Wei{\ss}e}\ and\ \citenamefont
  {Fehske}(2008{\natexlab{a}})}]{Weisse}%
  \BibitemOpen
  \bibfield  {author} {\bibinfo {author} {\bibfnamefont {A.}~\bibnamefont
  {Wei{\ss}e}}\ and\ \bibinfo {author} {\bibfnamefont {H.}~\bibnamefont
  {Fehske}},\ }\bibinfo {title} {{Exact Diagonalization Techniques}},\ in\
  \href {https://doi.org/10.1007/978-3-540-74686-7_18} {\emph {\bibinfo
  {booktitle} {{Computational Many-Particle Physics}}}},\ Vol.\ \bibinfo
  {volume} {739},\ \bibinfo {editor} {edited by\ \bibinfo {editor}
  {\bibfnamefont {H.}~\bibnamefont {Fehske}}, \bibinfo {editor} {\bibfnamefont
  {R.}~\bibnamefont {Schneider}},\ and\ \bibinfo {editor} {\bibfnamefont
  {A.}~\bibnamefont {Wei{\ss}e}}}\ (\bibinfo  {publisher} {Springer Berlin
  Heidelberg},\ \bibinfo {address} {Berlin, Heidelberg},\ \bibinfo {year}
  {2008})\ pp.\ \bibinfo {pages} {529--544}\BibitemShut {NoStop}%
\bibitem [{\citenamefont {Prelov{\v{s}}ek}\ and\ \citenamefont
  {Bon{\v{c}}a}(2013)}]{Prelovsek}%
  \BibitemOpen
  \bibfield  {author} {\bibinfo {author} {\bibfnamefont {P.}~\bibnamefont
  {Prelov{\v{s}}ek}}\ and\ \bibinfo {author} {\bibfnamefont {J.}~\bibnamefont
  {Bon{\v{c}}a}},\ }\bibinfo {title} {{Ground State and Finite Temperature
  Lanczos Methods}},\ in\ \href {https://doi.org/10.1007/978-3-642-35106-8_1}
  {\emph {\bibinfo {booktitle} {Strongly Correlated Systems: Numerical
  Methods}}},\ \bibinfo {editor} {edited by\ \bibinfo {editor} {\bibfnamefont
  {A.}~\bibnamefont {Avella}}\ and\ \bibinfo {editor} {\bibfnamefont
  {F.}~\bibnamefont {Mancini}}}\ (\bibinfo  {publisher} {Springer Berlin
  Heidelberg},\ \bibinfo {address} {Berlin, Heidelberg},\ \bibinfo {year}
  {2013})\ pp.\ \bibinfo {pages} {1--30}\BibitemShut {NoStop}%
\bibitem [{\citenamefont {Koch}(2019)}]{Koch2019}%
  \BibitemOpen
  \bibfield  {author} {\bibinfo {author} {\bibfnamefont {E.}~\bibnamefont
  {Koch}},\ }\bibinfo {title} {{Exact Diagonalization and Lanczos Method}},\
  in\ \href {https://juser.fz-juelich.de/record/864818} {\emph {\bibinfo
  {booktitle} {{M}any-{B}ody {M}ethods for {R}eal {M}aterials}}},\ \bibinfo
  {series} {Schriften des Forschungszentrums J\"ulich Modeling and Simulation},
  Vol.~\bibinfo {volume} {9},\ \bibinfo {editor} {edited by\ \bibinfo {editor}
  {\bibfnamefont {E.}~\bibnamefont {Pavarini}}\ and\ \bibinfo {editor}
  {\bibfnamefont {S.}~\bibnamefont {Zhang}}}\ (\bibinfo  {publisher}
  {Forschungszentrum J\"ulich GmbH Zentralbibliothek, Verlag},\ \bibinfo
  {address} {J\"ulich},\ \bibinfo {year} {2019})\ Chap.~\bibinfo {chapter}
  {7}\BibitemShut {NoStop}%
\bibitem [{\citenamefont {Wei{\ss}e}\ and\ \citenamefont
  {Fehske}(2008{\natexlab{b}})}]{Weisse_book}%
  \BibitemOpen
  \bibfield  {author} {\bibinfo {author} {\bibfnamefont {A.}~\bibnamefont
  {Wei{\ss}e}}\ and\ \bibinfo {author} {\bibfnamefont {H.}~\bibnamefont
  {Fehske}},\ }\bibinfo {title} {{Chebyshev Expansion Techniques}},\ in\ \href
  {https://doi.org/10.1007/978-3-540-74686-7_19} {\emph {\bibinfo {booktitle}
  {{Computational Many-Particle Physics}}}},\ \bibinfo {editor} {edited by\
  \bibinfo {editor} {\bibfnamefont {H.}~\bibnamefont {Fehske}}, \bibinfo
  {editor} {\bibfnamefont {R.}~\bibnamefont {Schneider}},\ and\ \bibinfo
  {editor} {\bibfnamefont {A.}~\bibnamefont {Wei{\ss}e}}}\ (\bibinfo
  {publisher} {Springer Berlin Heidelberg},\ \bibinfo {address} {Berlin,
  Heidelberg},\ \bibinfo {year} {2008})\ pp.\ \bibinfo {pages}
  {545--577}\BibitemShut {NoStop}%
\bibitem [{\citenamefont {Tal-Ezer}\ and\ \citenamefont
  {Kosloff}(1984)}]{TalEzer1984}%
  \BibitemOpen
  \bibfield  {author} {\bibinfo {author} {\bibfnamefont {H.}~\bibnamefont
  {Tal-Ezer}}\ and\ \bibinfo {author} {\bibfnamefont {R.}~\bibnamefont
  {Kosloff}},\ }\bibfield  {title} {\bibinfo {title} {{An accurate and
  efficient scheme for propagating the time dependent Schr\"odinger
  equation}},\ }\href {https://doi.org/10.1063/1.448136} {\bibfield  {journal}
  {\bibinfo  {journal} {The Journal of Chemical Physics}\ }\textbf {\bibinfo
  {volume} {81}},\ \bibinfo {pages} {3967} (\bibinfo {year}
  {1984})}\BibitemShut {NoStop}%
\bibitem [{\citenamefont {Park}\ and\ \citenamefont {Light}(1986)}]{Park1986}%
  \BibitemOpen
  \bibfield  {author} {\bibinfo {author} {\bibfnamefont {T.~J.}\ \bibnamefont
  {Park}}\ and\ \bibinfo {author} {\bibfnamefont {J.~C.}\ \bibnamefont
  {Light}},\ }\bibfield  {title} {\bibinfo {title} {{Unitary quantum time
  evolution by iterative Lanczos reduction}},\ }\href
  {https://doi.org/10.1063/1.451548} {\bibfield  {journal} {\bibinfo  {journal}
  {The Journal of Chemical Physics}\ }\textbf {\bibinfo {volume} {85}},\
  \bibinfo {pages} {5870} (\bibinfo {year} {1986})}\BibitemShut {NoStop}%
\bibitem [{\citenamefont {Vijay}\ and\ \citenamefont
  {Metiu}(2002)}]{Vijay2002}%
  \BibitemOpen
  \bibfield  {author} {\bibinfo {author} {\bibfnamefont {A.}~\bibnamefont
  {Vijay}}\ and\ \bibinfo {author} {\bibfnamefont {H.}~\bibnamefont {Metiu}},\
  }\bibfield  {title} {\bibinfo {title} {{A polynomial expansion of the quantum
  propagator, the Green’s function, and the spectral density operator}},\
  }\href {https://doi.org/10.1063/1.1425824} {\bibfield  {journal} {\bibinfo
  {journal} {The Journal of Chemical Physics}\ }\textbf {\bibinfo {volume}
  {116}},\ \bibinfo {pages} {60} (\bibinfo {year} {2002})}\BibitemShut
  {NoStop}%
\bibitem [{\citenamefont {Iitaka}\ and\ \citenamefont
  {Ebisuzaki}(2003)}]{Iitaka2003}%
  \BibitemOpen
  \bibfield  {author} {\bibinfo {author} {\bibfnamefont {T.}~\bibnamefont
  {Iitaka}}\ and\ \bibinfo {author} {\bibfnamefont {T.}~\bibnamefont
  {Ebisuzaki}},\ }\bibfield  {title} {\bibinfo {title} {{Algorithm for Linear
  Response Functions at Finite Temperatures: Application to ESR Spectrum of
  $s=\frac{1}{2}$ Antiferromagnet Cu Benzoate}},\ }\href
  {https://doi.org/10.1103/PhysRevLett.90.047203} {\bibfield  {journal}
  {\bibinfo  {journal} {Phys. Rev. Lett.}\ }\textbf {\bibinfo {volume} {90}},\
  \bibinfo {pages} {047203} (\bibinfo {year} {2003})}\BibitemShut {NoStop}%
\bibitem [{\citenamefont {Mohankumar}\ and\ \citenamefont
  {Auerbach}(2006)}]{Mohankumar2006}%
  \BibitemOpen
  \bibfield  {author} {\bibinfo {author} {\bibfnamefont {N.}~\bibnamefont
  {Mohankumar}}\ and\ \bibinfo {author} {\bibfnamefont {S.~M.}\ \bibnamefont
  {Auerbach}},\ }\bibfield  {title} {\bibinfo {title} {{On time-step bounds in
  unitary quantum evolution using the Lanczos method}},\ }\href
  {https://doi.org/https://doi.org/10.1016/j.cpc.2006.07.005} {\bibfield
  {journal} {\bibinfo  {journal} {Computer Physics Communications}\ }\textbf
  {\bibinfo {volume} {175}},\ \bibinfo {pages} {473 } (\bibinfo {year}
  {2006})}\BibitemShut {NoStop}%
\bibitem [{\citenamefont {Sorella}(2001)}]{Sorella2001}%
  \BibitemOpen
  \bibfield  {author} {\bibinfo {author} {\bibfnamefont {S.}~\bibnamefont
  {Sorella}},\ }\bibfield  {title} {\bibinfo {title} {{Generalized Lanczos
  algorithm for variational quantum Monte Carlo}},\ }\href
  {https://doi.org/10.1103/PhysRevB.64.024512} {\bibfield  {journal} {\bibinfo
  {journal} {Phys. Rev. B}\ }\textbf {\bibinfo {volume} {64}},\ \bibinfo
  {pages} {024512} (\bibinfo {year} {2001})}\BibitemShut {NoStop}%
\bibitem [{\citenamefont {Feynman}(1982)}]{Feynman1982}%
  \BibitemOpen
  \bibfield  {author} {\bibinfo {author} {\bibfnamefont {R.~P.}\ \bibnamefont
  {Feynman}},\ }\bibfield  {title} {\bibinfo {title} {Simulating physics with
  computers},\ }\href {https://doi.org/10.1007/BF02650179} {\bibfield
  {journal} {\bibinfo  {journal} {International Journal of Theoretical
  Physics}\ }\textbf {\bibinfo {volume} {21}},\ \bibinfo {pages} {467}
  (\bibinfo {year} {1982})}\BibitemShut {NoStop}%
\bibitem [{\citenamefont {Aspuru-Guzik}\ \emph {et~al.}(2005)\citenamefont
  {Aspuru-Guzik}, \citenamefont {Dutoi}, \citenamefont {Love},\ and\
  \citenamefont {Head-Gordon}}]{Aspuru-Guzik2005}%
  \BibitemOpen
  \bibfield  {author} {\bibinfo {author} {\bibfnamefont {A.}~\bibnamefont
  {Aspuru-Guzik}}, \bibinfo {author} {\bibfnamefont {A.~D.}\ \bibnamefont
  {Dutoi}}, \bibinfo {author} {\bibfnamefont {P.~J.}\ \bibnamefont {Love}},\
  and\ \bibinfo {author} {\bibfnamefont {M.}~\bibnamefont {Head-Gordon}},\
  }\bibfield  {title} {\bibinfo {title} {{Simulated Quantum Computation of
  Molecular Energies}},\ }\href {https://doi.org/10.1126/science.1113479}
  {\bibfield  {journal} {\bibinfo  {journal} {Science}\ }\textbf {\bibinfo
  {volume} {309}},\ \bibinfo {pages} {1704} (\bibinfo {year}
  {2005})}\BibitemShut {NoStop}%
\bibitem [{\citenamefont {Wecker}\ \emph
  {et~al.}(2015{\natexlab{a}})\citenamefont {Wecker}, \citenamefont {Hastings},
  \citenamefont {Wiebe}, \citenamefont {Clark}, \citenamefont {Nayak},\ and\
  \citenamefont {Troyer}}]{Wecker2015}%
  \BibitemOpen
  \bibfield  {author} {\bibinfo {author} {\bibfnamefont {D.}~\bibnamefont
  {Wecker}}, \bibinfo {author} {\bibfnamefont {M.~B.}\ \bibnamefont
  {Hastings}}, \bibinfo {author} {\bibfnamefont {N.}~\bibnamefont {Wiebe}},
  \bibinfo {author} {\bibfnamefont {B.~K.}\ \bibnamefont {Clark}}, \bibinfo
  {author} {\bibfnamefont {C.}~\bibnamefont {Nayak}},\ and\ \bibinfo {author}
  {\bibfnamefont {M.}~\bibnamefont {Troyer}},\ }\bibfield  {title} {\bibinfo
  {title} {{Solving strongly correlated electron models on a quantum
  computer}},\ }\href {https://doi.org/10.1103/PhysRevA.92.062318} {\bibfield
  {journal} {\bibinfo  {journal} {Phys. Rev. A}\ }\textbf {\bibinfo {volume}
  {92}},\ \bibinfo {pages} {062318} (\bibinfo {year}
  {2015}{\natexlab{a}})}\BibitemShut {NoStop}%
\bibitem [{\citenamefont {McArdle}\ \emph {et~al.}(2020)\citenamefont
  {McArdle}, \citenamefont {Endo}, \citenamefont {Aspuru-Guzik}, \citenamefont
  {Benjamin},\ and\ \citenamefont {Yuan}}]{McArdle2020}%
  \BibitemOpen
  \bibfield  {author} {\bibinfo {author} {\bibfnamefont {S.}~\bibnamefont
  {McArdle}}, \bibinfo {author} {\bibfnamefont {S.}~\bibnamefont {Endo}},
  \bibinfo {author} {\bibfnamefont {A.}~\bibnamefont {Aspuru-Guzik}}, \bibinfo
  {author} {\bibfnamefont {S.~C.}\ \bibnamefont {Benjamin}},\ and\ \bibinfo
  {author} {\bibfnamefont {X.}~\bibnamefont {Yuan}},\ }\bibfield  {title}
  {\bibinfo {title} {{Quantum computational chemistry}},\ }\href
  {https://doi.org/10.1103/RevModPhys.92.015003} {\bibfield  {journal}
  {\bibinfo  {journal} {Rev. Mod. Phys.}\ }\textbf {\bibinfo {volume} {92}},\
  \bibinfo {pages} {015003} (\bibinfo {year} {2020})}\BibitemShut {NoStop}%
\bibitem [{\citenamefont {Bauer}\ \emph {et~al.}(2020)\citenamefont {Bauer},
  \citenamefont {Bravyi}, \citenamefont {Motta},\ and\ \citenamefont
  {Kin-Lic~Chan}}]{bauer2020quantum}%
  \BibitemOpen
  \bibfield  {author} {\bibinfo {author} {\bibfnamefont {B.}~\bibnamefont
  {Bauer}}, \bibinfo {author} {\bibfnamefont {S.}~\bibnamefont {Bravyi}},
  \bibinfo {author} {\bibfnamefont {M.}~\bibnamefont {Motta}},\ and\ \bibinfo
  {author} {\bibfnamefont {G.}~\bibnamefont {Kin-Lic~Chan}},\ }\bibfield
  {title} {\bibinfo {title} {{Quantum Algorithms for Quantum Chemistry and
  Quantum Materials Science}},\ }\bibfield  {journal} {\bibinfo  {journal}
  {Chemical Reviews}\ }\href {https://doi.org/10.1021/acs.chemrev.9b00829}
  {10.1021/acs.chemrev.9b00829} (\bibinfo {year} {2020})\BibitemShut {NoStop}%
\bibitem [{\citenamefont {Nakamura}\ \emph {et~al.}(1999)\citenamefont
  {Nakamura}, \citenamefont {Pashkin},\ and\ \citenamefont
  {Tsai}}]{Nakamura1999}%
  \BibitemOpen
  \bibfield  {author} {\bibinfo {author} {\bibfnamefont {Y.}~\bibnamefont
  {Nakamura}}, \bibinfo {author} {\bibfnamefont {Y.~A.}\ \bibnamefont
  {Pashkin}},\ and\ \bibinfo {author} {\bibfnamefont {J.~S.}\ \bibnamefont
  {Tsai}},\ }\bibfield  {title} {\bibinfo {title} {{Coherent control of
  macroscopic quantum states in a single-Cooper-pair box}},\ }\href
  {https://doi.org/10.1038/19718} {\bibfield  {journal} {\bibinfo  {journal}
  {Nature}\ }\textbf {\bibinfo {volume} {398}},\ \bibinfo {pages} {786}
  (\bibinfo {year} {1999})}\BibitemShut {NoStop}%
\bibitem [{\citenamefont {Kok}\ \emph {et~al.}(2007)\citenamefont {Kok},
  \citenamefont {Munro}, \citenamefont {Nemoto}, \citenamefont {Ralph},
  \citenamefont {Dowling},\ and\ \citenamefont {Milburn}}]{optical_RMP2007}%
  \BibitemOpen
  \bibfield  {author} {\bibinfo {author} {\bibfnamefont {P.}~\bibnamefont
  {Kok}}, \bibinfo {author} {\bibfnamefont {W.~J.}\ \bibnamefont {Munro}},
  \bibinfo {author} {\bibfnamefont {K.}~\bibnamefont {Nemoto}}, \bibinfo
  {author} {\bibfnamefont {T.~C.}\ \bibnamefont {Ralph}}, \bibinfo {author}
  {\bibfnamefont {J.~P.}\ \bibnamefont {Dowling}},\ and\ \bibinfo {author}
  {\bibfnamefont {G.~J.}\ \bibnamefont {Milburn}},\ }\bibfield  {title}
  {\bibinfo {title} {Linear optical quantum computing with photonic qubits},\
  }\href {https://doi.org/10.1103/RevModPhys.79.135} {\bibfield  {journal}
  {\bibinfo  {journal} {Rev. Mod. Phys.}\ }\textbf {\bibinfo {volume} {79}},\
  \bibinfo {pages} {135} (\bibinfo {year} {2007})}\BibitemShut {NoStop}%
\bibitem [{\citenamefont {Ladd}\ \emph {et~al.}(2010)\citenamefont {Ladd},
  \citenamefont {Jelezko}, \citenamefont {Laflamme}, \citenamefont {Nakamura},
  \citenamefont {Monroe},\ and\ \citenamefont {O'Brien}}]{Ladd2010}%
  \BibitemOpen
  \bibfield  {author} {\bibinfo {author} {\bibfnamefont {T.~D.}\ \bibnamefont
  {Ladd}}, \bibinfo {author} {\bibfnamefont {F.}~\bibnamefont {Jelezko}},
  \bibinfo {author} {\bibfnamefont {R.}~\bibnamefont {Laflamme}}, \bibinfo
  {author} {\bibfnamefont {Y.}~\bibnamefont {Nakamura}}, \bibinfo {author}
  {\bibfnamefont {C.}~\bibnamefont {Monroe}},\ and\ \bibinfo {author}
  {\bibfnamefont {J.~L.}\ \bibnamefont {O'Brien}},\ }\bibfield  {title}
  {\bibinfo {title} {Quantum computers},\ }\href
  {https://doi.org/10.1038/nature08812} {\bibfield  {journal} {\bibinfo
  {journal} {Nature}\ }\textbf {\bibinfo {volume} {464}},\ \bibinfo {pages}
  {45} (\bibinfo {year} {2010})}\BibitemShut {NoStop}%
\bibitem [{\citenamefont {Xiang}\ \emph {et~al.}(2013)\citenamefont {Xiang},
  \citenamefont {Ashhab}, \citenamefont {You},\ and\ \citenamefont
  {Nori}}]{RevModPhys.85.623}%
  \BibitemOpen
  \bibfield  {author} {\bibinfo {author} {\bibfnamefont {Z.-L.}\ \bibnamefont
  {Xiang}}, \bibinfo {author} {\bibfnamefont {S.}~\bibnamefont {Ashhab}},
  \bibinfo {author} {\bibfnamefont {J.~Q.}\ \bibnamefont {You}},\ and\ \bibinfo
  {author} {\bibfnamefont {F.}~\bibnamefont {Nori}},\ }\bibfield  {title}
  {\bibinfo {title} {Hybrid quantum circuits: Superconducting circuits
  interacting with other quantum systems},\ }\href
  {https://doi.org/10.1103/RevModPhys.85.623} {\bibfield  {journal} {\bibinfo
  {journal} {Rev. Mod. Phys.}\ }\textbf {\bibinfo {volume} {85}},\ \bibinfo
  {pages} {623} (\bibinfo {year} {2013})}\BibitemShut {NoStop}%
\bibitem [{\citenamefont {Barends}\ \emph {et~al.}(2014)\citenamefont
  {Barends}, \citenamefont {Kelly}, \citenamefont {Megrant}, \citenamefont
  {Veitia}, \citenamefont {Sank}, \citenamefont {Jeffrey}, \citenamefont
  {White}, \citenamefont {Mutus}, \citenamefont {Fowler}, \citenamefont
  {Campbell}, \citenamefont {Chen}, \citenamefont {Chen}, \citenamefont
  {Chiaro}, \citenamefont {Dunsworth}, \citenamefont {Neill}, \citenamefont
  {O'Malley}, \citenamefont {Roushan}, \citenamefont {Vainsencher},
  \citenamefont {Wenner}, \citenamefont {Korotkov}, \citenamefont {Cleland},\
  and\ \citenamefont {Martinis}}]{Barends2014}%
  \BibitemOpen
  \bibfield  {author} {\bibinfo {author} {\bibfnamefont {R.}~\bibnamefont
  {Barends}}, \bibinfo {author} {\bibfnamefont {J.}~\bibnamefont {Kelly}},
  \bibinfo {author} {\bibfnamefont {A.}~\bibnamefont {Megrant}}, \bibinfo
  {author} {\bibfnamefont {A.}~\bibnamefont {Veitia}}, \bibinfo {author}
  {\bibfnamefont {D.}~\bibnamefont {Sank}}, \bibinfo {author} {\bibfnamefont
  {E.}~\bibnamefont {Jeffrey}}, \bibinfo {author} {\bibfnamefont {T.~C.}\
  \bibnamefont {White}}, \bibinfo {author} {\bibfnamefont {J.}~\bibnamefont
  {Mutus}}, \bibinfo {author} {\bibfnamefont {A.~G.}\ \bibnamefont {Fowler}},
  \bibinfo {author} {\bibfnamefont {B.}~\bibnamefont {Campbell}}, \bibinfo
  {author} {\bibfnamefont {Y.}~\bibnamefont {Chen}}, \bibinfo {author}
  {\bibfnamefont {Z.}~\bibnamefont {Chen}}, \bibinfo {author} {\bibfnamefont
  {B.}~\bibnamefont {Chiaro}}, \bibinfo {author} {\bibfnamefont
  {A.}~\bibnamefont {Dunsworth}}, \bibinfo {author} {\bibfnamefont
  {C.}~\bibnamefont {Neill}}, \bibinfo {author} {\bibfnamefont
  {P.}~\bibnamefont {O'Malley}}, \bibinfo {author} {\bibfnamefont
  {P.}~\bibnamefont {Roushan}}, \bibinfo {author} {\bibfnamefont
  {A.}~\bibnamefont {Vainsencher}}, \bibinfo {author} {\bibfnamefont
  {J.}~\bibnamefont {Wenner}}, \bibinfo {author} {\bibfnamefont {A.~N.}\
  \bibnamefont {Korotkov}}, \bibinfo {author} {\bibfnamefont {A.~N.}\
  \bibnamefont {Cleland}},\ and\ \bibinfo {author} {\bibfnamefont {J.~M.}\
  \bibnamefont {Martinis}},\ }\bibfield  {title} {\bibinfo {title}
  {Superconducting quantum circuits at the surface code threshold for fault
  tolerance},\ }\href {https://doi.org/10.1038/nature13171} {\bibfield
  {journal} {\bibinfo  {journal} {Nature}\ }\textbf {\bibinfo {volume} {508}},\
  \bibinfo {pages} {500} (\bibinfo {year} {2014})}\BibitemShut {NoStop}%
\bibitem [{\citenamefont {Chow}\ \emph {et~al.}(2014)\citenamefont {Chow},
  \citenamefont {Gambetta}, \citenamefont {Magesan}, \citenamefont {Abraham},
  \citenamefont {Cross}, \citenamefont {Johnson}, \citenamefont {Masluk},
  \citenamefont {Ryan}, \citenamefont {Smolin}, \citenamefont {Srinivasan},\
  and\ \citenamefont {Steffen}}]{Chow2014}%
  \BibitemOpen
  \bibfield  {author} {\bibinfo {author} {\bibfnamefont {J.~M.}\ \bibnamefont
  {Chow}}, \bibinfo {author} {\bibfnamefont {J.~M.}\ \bibnamefont {Gambetta}},
  \bibinfo {author} {\bibfnamefont {E.}~\bibnamefont {Magesan}}, \bibinfo
  {author} {\bibfnamefont {D.~W.}\ \bibnamefont {Abraham}}, \bibinfo {author}
  {\bibfnamefont {A.~W.}\ \bibnamefont {Cross}}, \bibinfo {author}
  {\bibfnamefont {B.~R.}\ \bibnamefont {Johnson}}, \bibinfo {author}
  {\bibfnamefont {N.~A.}\ \bibnamefont {Masluk}}, \bibinfo {author}
  {\bibfnamefont {C.~A.}\ \bibnamefont {Ryan}}, \bibinfo {author}
  {\bibfnamefont {J.~A.}\ \bibnamefont {Smolin}}, \bibinfo {author}
  {\bibfnamefont {S.~J.}\ \bibnamefont {Srinivasan}},\ and\ \bibinfo {author}
  {\bibfnamefont {M.}~\bibnamefont {Steffen}},\ }\bibfield  {title} {\bibinfo
  {title} {Implementing a strand of a scalable fault-tolerant quantum computing
  fabric},\ }\href {https://doi.org/10.1038/ncomms5015} {\bibfield  {journal}
  {\bibinfo  {journal} {Nature Communications}\ }\textbf {\bibinfo {volume}
  {5}},\ \bibinfo {pages} {4015} (\bibinfo {year} {2014})}\BibitemShut
  {NoStop}%
\bibitem [{\citenamefont {Kelly}\ \emph {et~al.}(2015)\citenamefont {Kelly},
  \citenamefont {Barends}, \citenamefont {Fowler}, \citenamefont {Megrant},
  \citenamefont {Jeffrey}, \citenamefont {White}, \citenamefont {Sank},
  \citenamefont {Mutus}, \citenamefont {Campbell}, \citenamefont {Chen},
  \citenamefont {Chen}, \citenamefont {Chiaro}, \citenamefont {Dunsworth},
  \citenamefont {Hoi}, \citenamefont {Neill}, \citenamefont {O'Malley},
  \citenamefont {Quintana}, \citenamefont {Roushan}, \citenamefont
  {Vainsencher}, \citenamefont {Wenner}, \citenamefont {Cleland},\ and\
  \citenamefont {Martinis}}]{Kelly2015}%
  \BibitemOpen
  \bibfield  {author} {\bibinfo {author} {\bibfnamefont {J.}~\bibnamefont
  {Kelly}}, \bibinfo {author} {\bibfnamefont {R.}~\bibnamefont {Barends}},
  \bibinfo {author} {\bibfnamefont {A.~G.}\ \bibnamefont {Fowler}}, \bibinfo
  {author} {\bibfnamefont {A.}~\bibnamefont {Megrant}}, \bibinfo {author}
  {\bibfnamefont {E.}~\bibnamefont {Jeffrey}}, \bibinfo {author} {\bibfnamefont
  {T.~C.}\ \bibnamefont {White}}, \bibinfo {author} {\bibfnamefont
  {D.}~\bibnamefont {Sank}}, \bibinfo {author} {\bibfnamefont {J.~Y.}\
  \bibnamefont {Mutus}}, \bibinfo {author} {\bibfnamefont {B.}~\bibnamefont
  {Campbell}}, \bibinfo {author} {\bibfnamefont {Y.}~\bibnamefont {Chen}},
  \bibinfo {author} {\bibfnamefont {Z.}~\bibnamefont {Chen}}, \bibinfo {author}
  {\bibfnamefont {B.}~\bibnamefont {Chiaro}}, \bibinfo {author} {\bibfnamefont
  {A.}~\bibnamefont {Dunsworth}}, \bibinfo {author} {\bibfnamefont {I.-C.}\
  \bibnamefont {Hoi}}, \bibinfo {author} {\bibfnamefont {C.}~\bibnamefont
  {Neill}}, \bibinfo {author} {\bibfnamefont {P.~J.~J.}\ \bibnamefont
  {O'Malley}}, \bibinfo {author} {\bibfnamefont {C.}~\bibnamefont {Quintana}},
  \bibinfo {author} {\bibfnamefont {P.}~\bibnamefont {Roushan}}, \bibinfo
  {author} {\bibfnamefont {A.}~\bibnamefont {Vainsencher}}, \bibinfo {author}
  {\bibfnamefont {J.}~\bibnamefont {Wenner}}, \bibinfo {author} {\bibfnamefont
  {A.~N.}\ \bibnamefont {Cleland}},\ and\ \bibinfo {author} {\bibfnamefont
  {J.~M.}\ \bibnamefont {Martinis}},\ }\bibfield  {title} {\bibinfo {title}
  {State preservation by repetitive error detection in a superconducting
  quantum circuit},\ }\href {https://doi.org/10.1038/nature14270} {\bibfield
  {journal} {\bibinfo  {journal} {Nature}\ }\textbf {\bibinfo {volume} {519}},\
  \bibinfo {pages} {66} (\bibinfo {year} {2015})}\BibitemShut {NoStop}%
\bibitem [{\citenamefont {Rist{\`e}}\ \emph {et~al.}(2015)\citenamefont
  {Rist{\`e}}, \citenamefont {Poletto}, \citenamefont {Huang}, \citenamefont
  {Bruno}, \citenamefont {Vesterinen}, \citenamefont {Saira},\ and\
  \citenamefont {DiCarlo}}]{Riste2015}%
  \BibitemOpen
  \bibfield  {author} {\bibinfo {author} {\bibfnamefont {D.}~\bibnamefont
  {Rist{\`e}}}, \bibinfo {author} {\bibfnamefont {S.}~\bibnamefont {Poletto}},
  \bibinfo {author} {\bibfnamefont {M.-Z.}\ \bibnamefont {Huang}}, \bibinfo
  {author} {\bibfnamefont {A.}~\bibnamefont {Bruno}}, \bibinfo {author}
  {\bibfnamefont {V.}~\bibnamefont {Vesterinen}}, \bibinfo {author}
  {\bibfnamefont {O.-P.}\ \bibnamefont {Saira}},\ and\ \bibinfo {author}
  {\bibfnamefont {L.}~\bibnamefont {DiCarlo}},\ }\bibfield  {title} {\bibinfo
  {title} {Detecting bit-flip errors in a logical qubit using stabilizer
  measurements},\ }\href {https://doi.org/10.1038/ncomms7983} {\bibfield
  {journal} {\bibinfo  {journal} {Nature Communications}\ }\textbf {\bibinfo
  {volume} {6}},\ \bibinfo {pages} {6983} (\bibinfo {year} {2015})}\BibitemShut
  {NoStop}%
\bibitem [{\citenamefont {Arute}\ \emph {et~al.}(2019)\citenamefont {Arute},
  \citenamefont {Arya}, \citenamefont {Babbush}, \citenamefont {Bacon},
  \citenamefont {Bardin}, \citenamefont {Barends}, \citenamefont {Biswas},
  \citenamefont {Boixo}, \citenamefont {Brandao}, \citenamefont {Buell},
  \citenamefont {Burkett}, \citenamefont {Chen}, \citenamefont {Chen},
  \citenamefont {Chiaro}, \citenamefont {Collins}, \citenamefont {Courtney},
  \citenamefont {Dunsworth}, \citenamefont {Farhi}, \citenamefont {Foxen},
  \citenamefont {Fowler}, \citenamefont {Gidney}, \citenamefont {Giustina},
  \citenamefont {Graff}, \citenamefont {Guerin}, \citenamefont {Habegger},
  \citenamefont {Harrigan}, \citenamefont {Hartmann}, \citenamefont {Ho},
  \citenamefont {Hoffmann}, \citenamefont {Huang}, \citenamefont {Humble},
  \citenamefont {Isakov}, \citenamefont {Jeffrey}, \citenamefont {Jiang},
  \citenamefont {Kafri}, \citenamefont {Kechedzhi}, \citenamefont {Kelly},
  \citenamefont {Klimov}, \citenamefont {Knysh}, \citenamefont {Korotkov},
  \citenamefont {Kostritsa}, \citenamefont {Landhuis}, \citenamefont
  {Lindmark}, \citenamefont {Lucero}, \citenamefont {Lyakh}, \citenamefont
  {Mandr{\`a}}, \citenamefont {McClean}, \citenamefont {McEwen}, \citenamefont
  {Megrant}, \citenamefont {Mi}, \citenamefont {Michielsen}, \citenamefont
  {Mohseni}, \citenamefont {Mutus}, \citenamefont {Naaman}, \citenamefont
  {Neeley}, \citenamefont {Neill}, \citenamefont {Niu}, \citenamefont {Ostby},
  \citenamefont {Petukhov}, \citenamefont {Platt}, \citenamefont {Quintana},
  \citenamefont {Rieffel}, \citenamefont {Roushan}, \citenamefont {Rubin},
  \citenamefont {Sank}, \citenamefont {Satzinger}, \citenamefont {Smelyanskiy},
  \citenamefont {Sung}, \citenamefont {Trevithick}, \citenamefont
  {Vainsencher}, \citenamefont {Villalonga}, \citenamefont {White},
  \citenamefont {Yao}, \citenamefont {Yeh}, \citenamefont {Zalcman},
  \citenamefont {Neven},\ and\ \citenamefont {Martinis}}]{Arute2019}%
  \BibitemOpen
  \bibfield  {author} {\bibinfo {author} {\bibfnamefont {F.}~\bibnamefont
  {Arute}}, \bibinfo {author} {\bibfnamefont {K.}~\bibnamefont {Arya}},
  \bibinfo {author} {\bibfnamefont {R.}~\bibnamefont {Babbush}}, \bibinfo
  {author} {\bibfnamefont {D.}~\bibnamefont {Bacon}}, \bibinfo {author}
  {\bibfnamefont {J.~C.}\ \bibnamefont {Bardin}}, \bibinfo {author}
  {\bibfnamefont {R.}~\bibnamefont {Barends}}, \bibinfo {author} {\bibfnamefont
  {R.}~\bibnamefont {Biswas}}, \bibinfo {author} {\bibfnamefont
  {S.}~\bibnamefont {Boixo}}, \bibinfo {author} {\bibfnamefont {F.~G. S.~L.}\
  \bibnamefont {Brandao}}, \bibinfo {author} {\bibfnamefont {D.~A.}\
  \bibnamefont {Buell}}, \bibinfo {author} {\bibfnamefont {B.}~\bibnamefont
  {Burkett}}, \bibinfo {author} {\bibfnamefont {Y.}~\bibnamefont {Chen}},
  \bibinfo {author} {\bibfnamefont {Z.}~\bibnamefont {Chen}}, \bibinfo {author}
  {\bibfnamefont {B.}~\bibnamefont {Chiaro}}, \bibinfo {author} {\bibfnamefont
  {R.}~\bibnamefont {Collins}}, \bibinfo {author} {\bibfnamefont
  {W.}~\bibnamefont {Courtney}}, \bibinfo {author} {\bibfnamefont
  {A.}~\bibnamefont {Dunsworth}}, \bibinfo {author} {\bibfnamefont
  {E.}~\bibnamefont {Farhi}}, \bibinfo {author} {\bibfnamefont
  {B.}~\bibnamefont {Foxen}}, \bibinfo {author} {\bibfnamefont
  {A.}~\bibnamefont {Fowler}}, \bibinfo {author} {\bibfnamefont
  {C.}~\bibnamefont {Gidney}}, \bibinfo {author} {\bibfnamefont
  {M.}~\bibnamefont {Giustina}}, \bibinfo {author} {\bibfnamefont
  {R.}~\bibnamefont {Graff}}, \bibinfo {author} {\bibfnamefont
  {K.}~\bibnamefont {Guerin}}, \bibinfo {author} {\bibfnamefont
  {S.}~\bibnamefont {Habegger}}, \bibinfo {author} {\bibfnamefont {M.~P.}\
  \bibnamefont {Harrigan}}, \bibinfo {author} {\bibfnamefont {M.~J.}\
  \bibnamefont {Hartmann}}, \bibinfo {author} {\bibfnamefont {A.}~\bibnamefont
  {Ho}}, \bibinfo {author} {\bibfnamefont {M.}~\bibnamefont {Hoffmann}},
  \bibinfo {author} {\bibfnamefont {T.}~\bibnamefont {Huang}}, \bibinfo
  {author} {\bibfnamefont {T.~S.}\ \bibnamefont {Humble}}, \bibinfo {author}
  {\bibfnamefont {S.~V.}\ \bibnamefont {Isakov}}, \bibinfo {author}
  {\bibfnamefont {E.}~\bibnamefont {Jeffrey}}, \bibinfo {author} {\bibfnamefont
  {Z.}~\bibnamefont {Jiang}}, \bibinfo {author} {\bibfnamefont
  {D.}~\bibnamefont {Kafri}}, \bibinfo {author} {\bibfnamefont
  {K.}~\bibnamefont {Kechedzhi}}, \bibinfo {author} {\bibfnamefont
  {J.}~\bibnamefont {Kelly}}, \bibinfo {author} {\bibfnamefont {P.~V.}\
  \bibnamefont {Klimov}}, \bibinfo {author} {\bibfnamefont {S.}~\bibnamefont
  {Knysh}}, \bibinfo {author} {\bibfnamefont {A.}~\bibnamefont {Korotkov}},
  \bibinfo {author} {\bibfnamefont {F.}~\bibnamefont {Kostritsa}}, \bibinfo
  {author} {\bibfnamefont {D.}~\bibnamefont {Landhuis}}, \bibinfo {author}
  {\bibfnamefont {M.}~\bibnamefont {Lindmark}}, \bibinfo {author}
  {\bibfnamefont {E.}~\bibnamefont {Lucero}}, \bibinfo {author} {\bibfnamefont
  {D.}~\bibnamefont {Lyakh}}, \bibinfo {author} {\bibfnamefont
  {S.}~\bibnamefont {Mandr{\`a}}}, \bibinfo {author} {\bibfnamefont {J.~R.}\
  \bibnamefont {McClean}}, \bibinfo {author} {\bibfnamefont {M.}~\bibnamefont
  {McEwen}}, \bibinfo {author} {\bibfnamefont {A.}~\bibnamefont {Megrant}},
  \bibinfo {author} {\bibfnamefont {X.}~\bibnamefont {Mi}}, \bibinfo {author}
  {\bibfnamefont {K.}~\bibnamefont {Michielsen}}, \bibinfo {author}
  {\bibfnamefont {M.}~\bibnamefont {Mohseni}}, \bibinfo {author} {\bibfnamefont
  {J.}~\bibnamefont {Mutus}}, \bibinfo {author} {\bibfnamefont
  {O.}~\bibnamefont {Naaman}}, \bibinfo {author} {\bibfnamefont
  {M.}~\bibnamefont {Neeley}}, \bibinfo {author} {\bibfnamefont
  {C.}~\bibnamefont {Neill}}, \bibinfo {author} {\bibfnamefont {M.~Y.}\
  \bibnamefont {Niu}}, \bibinfo {author} {\bibfnamefont {E.}~\bibnamefont
  {Ostby}}, \bibinfo {author} {\bibfnamefont {A.}~\bibnamefont {Petukhov}},
  \bibinfo {author} {\bibfnamefont {J.~C.}\ \bibnamefont {Platt}}, \bibinfo
  {author} {\bibfnamefont {C.}~\bibnamefont {Quintana}}, \bibinfo {author}
  {\bibfnamefont {E.~G.}\ \bibnamefont {Rieffel}}, \bibinfo {author}
  {\bibfnamefont {P.}~\bibnamefont {Roushan}}, \bibinfo {author} {\bibfnamefont
  {N.~C.}\ \bibnamefont {Rubin}}, \bibinfo {author} {\bibfnamefont
  {D.}~\bibnamefont {Sank}}, \bibinfo {author} {\bibfnamefont {K.~J.}\
  \bibnamefont {Satzinger}}, \bibinfo {author} {\bibfnamefont {V.}~\bibnamefont
  {Smelyanskiy}}, \bibinfo {author} {\bibfnamefont {K.~J.}\ \bibnamefont
  {Sung}}, \bibinfo {author} {\bibfnamefont {M.~D.}\ \bibnamefont
  {Trevithick}}, \bibinfo {author} {\bibfnamefont {A.}~\bibnamefont
  {Vainsencher}}, \bibinfo {author} {\bibfnamefont {B.}~\bibnamefont
  {Villalonga}}, \bibinfo {author} {\bibfnamefont {T.}~\bibnamefont {White}},
  \bibinfo {author} {\bibfnamefont {Z.~J.}\ \bibnamefont {Yao}}, \bibinfo
  {author} {\bibfnamefont {P.}~\bibnamefont {Yeh}}, \bibinfo {author}
  {\bibfnamefont {A.}~\bibnamefont {Zalcman}}, \bibinfo {author} {\bibfnamefont
  {H.}~\bibnamefont {Neven}},\ and\ \bibinfo {author} {\bibfnamefont {J.~M.}\
  \bibnamefont {Martinis}},\ }\bibfield  {title} {\bibinfo {title} {Quantum
  supremacy using a programmable superconducting processor},\ }\href
  {https://doi.org/10.1038/s41586-019-1666-5} {\bibfield  {journal} {\bibinfo
  {journal} {Nature}\ }\textbf {\bibinfo {volume} {574}},\ \bibinfo {pages}
  {505} (\bibinfo {year} {2019})}\BibitemShut {NoStop}%
\bibitem [{\citenamefont {Asavanant}\ \emph {et~al.}(2019)\citenamefont
  {Asavanant}, \citenamefont {Shiozawa}, \citenamefont {Yokoyama},
  \citenamefont {Charoensombutamon}, \citenamefont {Emura}, \citenamefont
  {Alexander}, \citenamefont {Takeda}, \citenamefont {Yoshikawa}, \citenamefont
  {Menicucci}, \citenamefont {Yonezawa},\ and\ \citenamefont
  {Furusawa}}]{Asavanant2019}%
  \BibitemOpen
  \bibfield  {author} {\bibinfo {author} {\bibfnamefont {W.}~\bibnamefont
  {Asavanant}}, \bibinfo {author} {\bibfnamefont {Y.}~\bibnamefont {Shiozawa}},
  \bibinfo {author} {\bibfnamefont {S.}~\bibnamefont {Yokoyama}}, \bibinfo
  {author} {\bibfnamefont {B.}~\bibnamefont {Charoensombutamon}}, \bibinfo
  {author} {\bibfnamefont {H.}~\bibnamefont {Emura}}, \bibinfo {author}
  {\bibfnamefont {R.~N.}\ \bibnamefont {Alexander}}, \bibinfo {author}
  {\bibfnamefont {S.}~\bibnamefont {Takeda}}, \bibinfo {author} {\bibfnamefont
  {J.-i.}\ \bibnamefont {Yoshikawa}}, \bibinfo {author} {\bibfnamefont {N.~C.}\
  \bibnamefont {Menicucci}}, \bibinfo {author} {\bibfnamefont {H.}~\bibnamefont
  {Yonezawa}},\ and\ \bibinfo {author} {\bibfnamefont {A.}~\bibnamefont
  {Furusawa}},\ }\bibfield  {title} {\bibinfo {title} {Generation of
  time-domain-multiplexed two-dimensional cluster state},\ }\href
  {https://doi.org/10.1126/science.aay2645} {\bibfield  {journal} {\bibinfo
  {journal} {Science}\ }\textbf {\bibinfo {volume} {366}},\ \bibinfo {pages}
  {373} (\bibinfo {year} {2019})}\BibitemShut {NoStop}%
\bibitem [{\citenamefont {Preskill}(2018)}]{Preskill2018}%
  \BibitemOpen
  \bibfield  {author} {\bibinfo {author} {\bibfnamefont {J.}~\bibnamefont
  {Preskill}},\ }\bibfield  {title} {\bibinfo {title} {Quantum {C}omputing in
  the {NISQ} era and beyond},\ }\href
  {https://doi.org/10.22331/q-2018-08-06-79} {\bibfield  {journal} {\bibinfo
  {journal} {{Quantum}}\ }\textbf {\bibinfo {volume} {2}},\ \bibinfo {pages}
  {79} (\bibinfo {year} {2018})}\BibitemShut {NoStop}%
\bibitem [{\citenamefont {Peruzzo}\ \emph {et~al.}(2014)\citenamefont
  {Peruzzo}, \citenamefont {McClean}, \citenamefont {Shadbolt}, \citenamefont
  {Yung}, \citenamefont {Zhou}, \citenamefont {Love}, \citenamefont
  {Aspuru-Guzik},\ and\ \citenamefont {O'Brien}}]{Peruzzo2014}%
  \BibitemOpen
  \bibfield  {author} {\bibinfo {author} {\bibfnamefont {A.}~\bibnamefont
  {Peruzzo}}, \bibinfo {author} {\bibfnamefont {J.}~\bibnamefont {McClean}},
  \bibinfo {author} {\bibfnamefont {P.}~\bibnamefont {Shadbolt}}, \bibinfo
  {author} {\bibfnamefont {M.-H.}\ \bibnamefont {Yung}}, \bibinfo {author}
  {\bibfnamefont {X.-Q.}\ \bibnamefont {Zhou}}, \bibinfo {author}
  {\bibfnamefont {P.~J.}\ \bibnamefont {Love}}, \bibinfo {author}
  {\bibfnamefont {A.}~\bibnamefont {Aspuru-Guzik}},\ and\ \bibinfo {author}
  {\bibfnamefont {J.~L.}\ \bibnamefont {O'Brien}},\ }\bibfield  {title}
  {\bibinfo {title} {A variational eigenvalue solver on a photonic quantum
  processor},\ }\href {https://doi.org/10.1038/ncomms5213} {\bibfield
  {journal} {\bibinfo  {journal} {Nature Communications}\ }\textbf {\bibinfo
  {volume} {5}},\ \bibinfo {pages} {4213} (\bibinfo {year} {2014})}\BibitemShut
  {NoStop}%
\bibitem [{\citenamefont {Wecker}\ \emph
  {et~al.}(2015{\natexlab{b}})\citenamefont {Wecker}, \citenamefont
  {Hastings},\ and\ \citenamefont {Troyer}}]{Wecker2015vqe}%
  \BibitemOpen
  \bibfield  {author} {\bibinfo {author} {\bibfnamefont {D.}~\bibnamefont
  {Wecker}}, \bibinfo {author} {\bibfnamefont {M.~B.}\ \bibnamefont
  {Hastings}},\ and\ \bibinfo {author} {\bibfnamefont {M.}~\bibnamefont
  {Troyer}},\ }\bibfield  {title} {\bibinfo {title} {{Progress towards
  practical quantum variational algorithms}},\ }\href
  {https://doi.org/10.1103/PhysRevA.92.042303} {\bibfield  {journal} {\bibinfo
  {journal} {Phys. Rev. A}\ }\textbf {\bibinfo {volume} {92}},\ \bibinfo
  {pages} {042303} (\bibinfo {year} {2015}{\natexlab{b}})}\BibitemShut
  {NoStop}%
\bibitem [{\citenamefont {O'Malley}\ \emph {et~al.}(2016)\citenamefont
  {O'Malley}, \citenamefont {Babbush}, \citenamefont {Kivlichan}, \citenamefont
  {Romero}, \citenamefont {McClean}, \citenamefont {Barends}, \citenamefont
  {Kelly}, \citenamefont {Roushan}, \citenamefont {Tranter}, \citenamefont
  {Ding}, \citenamefont {Campbell}, \citenamefont {Chen}, \citenamefont {Chen},
  \citenamefont {Chiaro}, \citenamefont {Dunsworth}, \citenamefont {Fowler},
  \citenamefont {Jeffrey}, \citenamefont {Lucero}, \citenamefont {Megrant},
  \citenamefont {Mutus}, \citenamefont {Neeley}, \citenamefont {Neill},
  \citenamefont {Quintana}, \citenamefont {Sank}, \citenamefont {Vainsencher},
  \citenamefont {Wenner}, \citenamefont {White}, \citenamefont {Coveney},
  \citenamefont {Love}, \citenamefont {Neven}, \citenamefont {Aspuru-Guzik},\
  and\ \citenamefont {Martinis}}]{O'Malley2016}%
  \BibitemOpen
  \bibfield  {author} {\bibinfo {author} {\bibfnamefont {P.~J.~J.}\
  \bibnamefont {O'Malley}}, \bibinfo {author} {\bibfnamefont {R.}~\bibnamefont
  {Babbush}}, \bibinfo {author} {\bibfnamefont {I.~D.}\ \bibnamefont
  {Kivlichan}}, \bibinfo {author} {\bibfnamefont {J.}~\bibnamefont {Romero}},
  \bibinfo {author} {\bibfnamefont {J.~R.}\ \bibnamefont {McClean}}, \bibinfo
  {author} {\bibfnamefont {R.}~\bibnamefont {Barends}}, \bibinfo {author}
  {\bibfnamefont {J.}~\bibnamefont {Kelly}}, \bibinfo {author} {\bibfnamefont
  {P.}~\bibnamefont {Roushan}}, \bibinfo {author} {\bibfnamefont
  {A.}~\bibnamefont {Tranter}}, \bibinfo {author} {\bibfnamefont
  {N.}~\bibnamefont {Ding}}, \bibinfo {author} {\bibfnamefont {B.}~\bibnamefont
  {Campbell}}, \bibinfo {author} {\bibfnamefont {Y.}~\bibnamefont {Chen}},
  \bibinfo {author} {\bibfnamefont {Z.}~\bibnamefont {Chen}}, \bibinfo {author}
  {\bibfnamefont {B.}~\bibnamefont {Chiaro}}, \bibinfo {author} {\bibfnamefont
  {A.}~\bibnamefont {Dunsworth}}, \bibinfo {author} {\bibfnamefont {A.~G.}\
  \bibnamefont {Fowler}}, \bibinfo {author} {\bibfnamefont {E.}~\bibnamefont
  {Jeffrey}}, \bibinfo {author} {\bibfnamefont {E.}~\bibnamefont {Lucero}},
  \bibinfo {author} {\bibfnamefont {A.}~\bibnamefont {Megrant}}, \bibinfo
  {author} {\bibfnamefont {J.~Y.}\ \bibnamefont {Mutus}}, \bibinfo {author}
  {\bibfnamefont {M.}~\bibnamefont {Neeley}}, \bibinfo {author} {\bibfnamefont
  {C.}~\bibnamefont {Neill}}, \bibinfo {author} {\bibfnamefont
  {C.}~\bibnamefont {Quintana}}, \bibinfo {author} {\bibfnamefont
  {D.}~\bibnamefont {Sank}}, \bibinfo {author} {\bibfnamefont {A.}~\bibnamefont
  {Vainsencher}}, \bibinfo {author} {\bibfnamefont {J.}~\bibnamefont {Wenner}},
  \bibinfo {author} {\bibfnamefont {T.~C.}\ \bibnamefont {White}}, \bibinfo
  {author} {\bibfnamefont {P.~V.}\ \bibnamefont {Coveney}}, \bibinfo {author}
  {\bibfnamefont {P.~J.}\ \bibnamefont {Love}}, \bibinfo {author}
  {\bibfnamefont {H.}~\bibnamefont {Neven}}, \bibinfo {author} {\bibfnamefont
  {A.}~\bibnamefont {Aspuru-Guzik}},\ and\ \bibinfo {author} {\bibfnamefont
  {J.~M.}\ \bibnamefont {Martinis}},\ }\bibfield  {title} {\bibinfo {title}
  {{Scalable Quantum Simulation of Molecular Energies}},\ }\href
  {https://doi.org/10.1103/PhysRevX.6.031007} {\bibfield  {journal} {\bibinfo
  {journal} {Phys. Rev. X}\ }\textbf {\bibinfo {volume} {6}},\ \bibinfo {pages}
  {031007} (\bibinfo {year} {2016})}\BibitemShut {NoStop}%
\bibitem [{\citenamefont {McClean}\ \emph {et~al.}(2016)\citenamefont
  {McClean}, \citenamefont {Romero}, \citenamefont {Babbush},\ and\
  \citenamefont {Aspuru-Guzik}}]{McClean2016}%
  \BibitemOpen
  \bibfield  {author} {\bibinfo {author} {\bibfnamefont {J.~R.}\ \bibnamefont
  {McClean}}, \bibinfo {author} {\bibfnamefont {J.}~\bibnamefont {Romero}},
  \bibinfo {author} {\bibfnamefont {R.}~\bibnamefont {Babbush}},\ and\ \bibinfo
  {author} {\bibfnamefont {A.}~\bibnamefont {Aspuru-Guzik}},\ }\bibfield
  {title} {\bibinfo {title} {The theory of variational hybrid quantum-classical
  algorithms},\ }\href {https://doi.org/10.1088/1367-2630/18/2/023023}
  {\bibfield  {journal} {\bibinfo  {journal} {New Journal of Physics}\ }\textbf
  {\bibinfo {volume} {18}},\ \bibinfo {pages} {023023} (\bibinfo {year}
  {2016})}\BibitemShut {NoStop}%
\bibitem [{\citenamefont {Kandala}\ \emph {et~al.}(2017)\citenamefont
  {Kandala}, \citenamefont {Mezzacapo}, \citenamefont {Temme}, \citenamefont
  {Takita}, \citenamefont {Brink}, \citenamefont {Chow},\ and\ \citenamefont
  {Gambetta}}]{Kandala2017}%
  \BibitemOpen
  \bibfield  {author} {\bibinfo {author} {\bibfnamefont {A.}~\bibnamefont
  {Kandala}}, \bibinfo {author} {\bibfnamefont {A.}~\bibnamefont {Mezzacapo}},
  \bibinfo {author} {\bibfnamefont {K.}~\bibnamefont {Temme}}, \bibinfo
  {author} {\bibfnamefont {M.}~\bibnamefont {Takita}}, \bibinfo {author}
  {\bibfnamefont {M.}~\bibnamefont {Brink}}, \bibinfo {author} {\bibfnamefont
  {J.~M.}\ \bibnamefont {Chow}},\ and\ \bibinfo {author} {\bibfnamefont
  {J.~M.}\ \bibnamefont {Gambetta}},\ }\bibfield  {title} {\bibinfo {title}
  {Hardware-efficient variational quantum eigensolver for small molecules and
  quantum magnets},\ }\href {https://doi.org/10.1038/nature23879} {\bibfield
  {journal} {\bibinfo  {journal} {Nature}\ }\textbf {\bibinfo {volume} {549}},\
  \bibinfo {pages} {242} (\bibinfo {year} {2017})}\BibitemShut {NoStop}%
\bibitem [{\citenamefont {Li}\ \emph {et~al.}(2017)\citenamefont {Li},
  \citenamefont {Yang}, \citenamefont {Peng},\ and\ \citenamefont
  {Sun}}]{Li2017}%
  \BibitemOpen
  \bibfield  {author} {\bibinfo {author} {\bibfnamefont {J.}~\bibnamefont
  {Li}}, \bibinfo {author} {\bibfnamefont {X.}~\bibnamefont {Yang}}, \bibinfo
  {author} {\bibfnamefont {X.}~\bibnamefont {Peng}},\ and\ \bibinfo {author}
  {\bibfnamefont {C.-P.}\ \bibnamefont {Sun}},\ }\bibfield  {title} {\bibinfo
  {title} {{Hybrid Quantum-Classical Approach to Quantum Optimal Control}},\
  }\href {https://doi.org/10.1103/PhysRevLett.118.150503} {\bibfield  {journal}
  {\bibinfo  {journal} {Phys. Rev. Lett.}\ }\textbf {\bibinfo {volume} {118}},\
  \bibinfo {pages} {150503} (\bibinfo {year} {2017})}\BibitemShut {NoStop}%
\bibitem [{\citenamefont {Mazzola}\ \emph {et~al.}(2019)\citenamefont
  {Mazzola}, \citenamefont {Ollitrault}, \citenamefont {Barkoutsos},\ and\
  \citenamefont {Tavernelli}}]{Mazzola2019}%
  \BibitemOpen
  \bibfield  {author} {\bibinfo {author} {\bibfnamefont {G.}~\bibnamefont
  {Mazzola}}, \bibinfo {author} {\bibfnamefont {P.~J.}\ \bibnamefont
  {Ollitrault}}, \bibinfo {author} {\bibfnamefont {P.~K.}\ \bibnamefont
  {Barkoutsos}},\ and\ \bibinfo {author} {\bibfnamefont {I.}~\bibnamefont
  {Tavernelli}},\ }\bibfield  {title} {\bibinfo {title} {Nonunitary operations
  for ground-state calculations in near-term quantum computers},\ }\href
  {https://doi.org/10.1103/PhysRevLett.123.130501} {\bibfield  {journal}
  {\bibinfo  {journal} {Phys. Rev. Lett.}\ }\textbf {\bibinfo {volume} {123}},\
  \bibinfo {pages} {130501} (\bibinfo {year} {2019})}\BibitemShut {NoStop}%
\bibitem [{\citenamefont {Arute}\ \emph
  {et~al.}(2020{\natexlab{a}})\citenamefont {Arute}, \citenamefont {Arya},
  \citenamefont {Babbush}, \citenamefont {Bacon}, \citenamefont {Bardin},
  \citenamefont {Barends}, \citenamefont {Boixo}, \citenamefont {Broughton},
  \citenamefont {Buckley}, \citenamefont {Buell}, \citenamefont {Burkett},
  \citenamefont {Bushnell}, \citenamefont {Chen}, \citenamefont {Chen},
  \citenamefont {Chiaro}, \citenamefont {Collins}, \citenamefont {Courtney},
  \citenamefont {Demura}, \citenamefont {Dunsworth}, \citenamefont {Farhi},
  \citenamefont {Fowler}, \citenamefont {Foxen}, \citenamefont {Gidney},
  \citenamefont {Giustina}, \citenamefont {Graff}, \citenamefont {Habegger},
  \citenamefont {Harrigan}, \citenamefont {Ho}, \citenamefont {Hong},
  \citenamefont {Huang}, \citenamefont {Huggins}, \citenamefont {Ioffe},
  \citenamefont {Isakov}, \citenamefont {Jeffrey}, \citenamefont {Jiang},
  \citenamefont {Jones}, \citenamefont {Kafri}, \citenamefont {Kechedzhi},
  \citenamefont {Kelly}, \citenamefont {Kim}, \citenamefont {Klimov},
  \citenamefont {Korotkov}, \citenamefont {Kostritsa}, \citenamefont
  {Landhuis}, \citenamefont {Laptev}, \citenamefont {Lindmark}, \citenamefont
  {Lucero}, \citenamefont {Martin}, \citenamefont {Martinis}, \citenamefont
  {McClean}, \citenamefont {McEwen}, \citenamefont {Megrant}, \citenamefont
  {Mi}, \citenamefont {Mohseni}, \citenamefont {Mruczkiewicz}, \citenamefont
  {Mutus}, \citenamefont {Naaman}, \citenamefont {Neeley}, \citenamefont
  {Neill}, \citenamefont {Neven}, \citenamefont {Niu}, \citenamefont
  {O{\textquoteright}Brien}, \citenamefont {Ostby}, \citenamefont {Petukhov},
  \citenamefont {Putterman}, \citenamefont {Quintana}, \citenamefont {Roushan},
  \citenamefont {Rubin}, \citenamefont {Sank}, \citenamefont {Satzinger},
  \citenamefont {Smelyanskiy}, \citenamefont {Strain}, \citenamefont {Sung},
  \citenamefont {Szalay}, \citenamefont {Takeshita}, \citenamefont
  {Vainsencher}, \citenamefont {White}, \citenamefont {Wiebe}, \citenamefont
  {Yao}, \citenamefont {Yeh},\ and\ \citenamefont
  {Zalcman}}]{arute2020hartreefock}%
  \BibitemOpen
  \bibfield  {author} {\bibinfo {author} {\bibfnamefont {F.}~\bibnamefont
  {Arute}}, \bibinfo {author} {\bibfnamefont {K.}~\bibnamefont {Arya}},
  \bibinfo {author} {\bibfnamefont {R.}~\bibnamefont {Babbush}}, \bibinfo
  {author} {\bibfnamefont {D.}~\bibnamefont {Bacon}}, \bibinfo {author}
  {\bibfnamefont {J.~C.}\ \bibnamefont {Bardin}}, \bibinfo {author}
  {\bibfnamefont {R.}~\bibnamefont {Barends}}, \bibinfo {author} {\bibfnamefont
  {S.}~\bibnamefont {Boixo}}, \bibinfo {author} {\bibfnamefont
  {M.}~\bibnamefont {Broughton}}, \bibinfo {author} {\bibfnamefont {B.~B.}\
  \bibnamefont {Buckley}}, \bibinfo {author} {\bibfnamefont {D.~A.}\
  \bibnamefont {Buell}}, \bibinfo {author} {\bibfnamefont {B.}~\bibnamefont
  {Burkett}}, \bibinfo {author} {\bibfnamefont {N.}~\bibnamefont {Bushnell}},
  \bibinfo {author} {\bibfnamefont {Y.}~\bibnamefont {Chen}}, \bibinfo {author}
  {\bibfnamefont {Z.}~\bibnamefont {Chen}}, \bibinfo {author} {\bibfnamefont
  {B.}~\bibnamefont {Chiaro}}, \bibinfo {author} {\bibfnamefont
  {R.}~\bibnamefont {Collins}}, \bibinfo {author} {\bibfnamefont
  {W.}~\bibnamefont {Courtney}}, \bibinfo {author} {\bibfnamefont
  {S.}~\bibnamefont {Demura}}, \bibinfo {author} {\bibfnamefont
  {A.}~\bibnamefont {Dunsworth}}, \bibinfo {author} {\bibfnamefont
  {E.}~\bibnamefont {Farhi}}, \bibinfo {author} {\bibfnamefont
  {A.}~\bibnamefont {Fowler}}, \bibinfo {author} {\bibfnamefont
  {B.}~\bibnamefont {Foxen}}, \bibinfo {author} {\bibfnamefont
  {C.}~\bibnamefont {Gidney}}, \bibinfo {author} {\bibfnamefont
  {M.}~\bibnamefont {Giustina}}, \bibinfo {author} {\bibfnamefont
  {R.}~\bibnamefont {Graff}}, \bibinfo {author} {\bibfnamefont
  {S.}~\bibnamefont {Habegger}}, \bibinfo {author} {\bibfnamefont {M.~P.}\
  \bibnamefont {Harrigan}}, \bibinfo {author} {\bibfnamefont {A.}~\bibnamefont
  {Ho}}, \bibinfo {author} {\bibfnamefont {S.}~\bibnamefont {Hong}}, \bibinfo
  {author} {\bibfnamefont {T.}~\bibnamefont {Huang}}, \bibinfo {author}
  {\bibfnamefont {W.~J.}\ \bibnamefont {Huggins}}, \bibinfo {author}
  {\bibfnamefont {L.}~\bibnamefont {Ioffe}}, \bibinfo {author} {\bibfnamefont
  {S.~V.}\ \bibnamefont {Isakov}}, \bibinfo {author} {\bibfnamefont
  {E.}~\bibnamefont {Jeffrey}}, \bibinfo {author} {\bibfnamefont
  {Z.}~\bibnamefont {Jiang}}, \bibinfo {author} {\bibfnamefont
  {C.}~\bibnamefont {Jones}}, \bibinfo {author} {\bibfnamefont
  {D.}~\bibnamefont {Kafri}}, \bibinfo {author} {\bibfnamefont
  {K.}~\bibnamefont {Kechedzhi}}, \bibinfo {author} {\bibfnamefont
  {J.}~\bibnamefont {Kelly}}, \bibinfo {author} {\bibfnamefont
  {S.}~\bibnamefont {Kim}}, \bibinfo {author} {\bibfnamefont {P.~V.}\
  \bibnamefont {Klimov}}, \bibinfo {author} {\bibfnamefont {A.}~\bibnamefont
  {Korotkov}}, \bibinfo {author} {\bibfnamefont {F.}~\bibnamefont {Kostritsa}},
  \bibinfo {author} {\bibfnamefont {D.}~\bibnamefont {Landhuis}}, \bibinfo
  {author} {\bibfnamefont {P.}~\bibnamefont {Laptev}}, \bibinfo {author}
  {\bibfnamefont {M.}~\bibnamefont {Lindmark}}, \bibinfo {author}
  {\bibfnamefont {E.}~\bibnamefont {Lucero}}, \bibinfo {author} {\bibfnamefont
  {O.}~\bibnamefont {Martin}}, \bibinfo {author} {\bibfnamefont {J.~M.}\
  \bibnamefont {Martinis}}, \bibinfo {author} {\bibfnamefont {J.~R.}\
  \bibnamefont {McClean}}, \bibinfo {author} {\bibfnamefont {M.}~\bibnamefont
  {McEwen}}, \bibinfo {author} {\bibfnamefont {A.}~\bibnamefont {Megrant}},
  \bibinfo {author} {\bibfnamefont {X.}~\bibnamefont {Mi}}, \bibinfo {author}
  {\bibfnamefont {M.}~\bibnamefont {Mohseni}}, \bibinfo {author} {\bibfnamefont
  {W.}~\bibnamefont {Mruczkiewicz}}, \bibinfo {author} {\bibfnamefont
  {J.}~\bibnamefont {Mutus}}, \bibinfo {author} {\bibfnamefont
  {O.}~\bibnamefont {Naaman}}, \bibinfo {author} {\bibfnamefont
  {M.}~\bibnamefont {Neeley}}, \bibinfo {author} {\bibfnamefont
  {C.}~\bibnamefont {Neill}}, \bibinfo {author} {\bibfnamefont
  {H.}~\bibnamefont {Neven}}, \bibinfo {author} {\bibfnamefont {M.~Y.}\
  \bibnamefont {Niu}}, \bibinfo {author} {\bibfnamefont {T.~E.}\ \bibnamefont
  {O{\textquoteright}Brien}}, \bibinfo {author} {\bibfnamefont
  {E.}~\bibnamefont {Ostby}}, \bibinfo {author} {\bibfnamefont
  {A.}~\bibnamefont {Petukhov}}, \bibinfo {author} {\bibfnamefont
  {H.}~\bibnamefont {Putterman}}, \bibinfo {author} {\bibfnamefont
  {C.}~\bibnamefont {Quintana}}, \bibinfo {author} {\bibfnamefont
  {P.}~\bibnamefont {Roushan}}, \bibinfo {author} {\bibfnamefont {N.~C.}\
  \bibnamefont {Rubin}}, \bibinfo {author} {\bibfnamefont {D.}~\bibnamefont
  {Sank}}, \bibinfo {author} {\bibfnamefont {K.~J.}\ \bibnamefont {Satzinger}},
  \bibinfo {author} {\bibfnamefont {V.}~\bibnamefont {Smelyanskiy}}, \bibinfo
  {author} {\bibfnamefont {D.}~\bibnamefont {Strain}}, \bibinfo {author}
  {\bibfnamefont {K.~J.}\ \bibnamefont {Sung}}, \bibinfo {author}
  {\bibfnamefont {M.}~\bibnamefont {Szalay}}, \bibinfo {author} {\bibfnamefont
  {T.~Y.}\ \bibnamefont {Takeshita}}, \bibinfo {author} {\bibfnamefont
  {A.}~\bibnamefont {Vainsencher}}, \bibinfo {author} {\bibfnamefont
  {T.}~\bibnamefont {White}}, \bibinfo {author} {\bibfnamefont
  {N.}~\bibnamefont {Wiebe}}, \bibinfo {author} {\bibfnamefont {Z.~J.}\
  \bibnamefont {Yao}}, \bibinfo {author} {\bibfnamefont {P.}~\bibnamefont
  {Yeh}},\ and\ \bibinfo {author} {\bibfnamefont {A.}~\bibnamefont {Zalcman}},\
  }\bibfield  {title} {\bibinfo {title} {{Hartree-Fock on a superconducting
  qubit quantum computer}},\ }\href {https://doi.org/10.1126/science.abb9811}
  {\bibfield  {journal} {\bibinfo  {journal} {Science}\ }\textbf {\bibinfo
  {volume} {369}},\ \bibinfo {pages} {1084} (\bibinfo {year}
  {2020}{\natexlab{a}})}\BibitemShut {NoStop}%
\bibitem [{\citenamefont {Liu}\ \emph {et~al.}(2019)\citenamefont {Liu},
  \citenamefont {Zhang}, \citenamefont {Wan},\ and\ \citenamefont
  {Wang}}]{Liu2019}%
  \BibitemOpen
  \bibfield  {author} {\bibinfo {author} {\bibfnamefont {J.-G.}\ \bibnamefont
  {Liu}}, \bibinfo {author} {\bibfnamefont {Y.-H.}\ \bibnamefont {Zhang}},
  \bibinfo {author} {\bibfnamefont {Y.}~\bibnamefont {Wan}},\ and\ \bibinfo
  {author} {\bibfnamefont {L.}~\bibnamefont {Wang}},\ }\bibfield  {title}
  {\bibinfo {title} {Variational quantum eigensolver with fewer qubits},\
  }\href {https://doi.org/10.1103/PhysRevResearch.1.023025} {\bibfield
  {journal} {\bibinfo  {journal} {Phys. Rev. Research}\ }\textbf {\bibinfo
  {volume} {1}},\ \bibinfo {pages} {023025} (\bibinfo {year}
  {2019})}\BibitemShut {NoStop}%
\bibitem [{\citenamefont {Foss-Feig}\ \emph {et~al.}(2020)\citenamefont
  {Foss-Feig}, \citenamefont {Hayes}, \citenamefont {Dreiling}, \citenamefont
  {Figgatt}, \citenamefont {Gaebler}, \citenamefont {Moses}, \citenamefont
  {Pino},\ and\ \citenamefont {Potter}}]{fossfeig2020holographic}%
  \BibitemOpen
  \bibfield  {author} {\bibinfo {author} {\bibfnamefont {M.}~\bibnamefont
  {Foss-Feig}}, \bibinfo {author} {\bibfnamefont {D.}~\bibnamefont {Hayes}},
  \bibinfo {author} {\bibfnamefont {J.~M.}\ \bibnamefont {Dreiling}}, \bibinfo
  {author} {\bibfnamefont {C.}~\bibnamefont {Figgatt}}, \bibinfo {author}
  {\bibfnamefont {J.~P.}\ \bibnamefont {Gaebler}}, \bibinfo {author}
  {\bibfnamefont {S.~A.}\ \bibnamefont {Moses}}, \bibinfo {author}
  {\bibfnamefont {J.~M.}\ \bibnamefont {Pino}},\ and\ \bibinfo {author}
  {\bibfnamefont {A.~C.}\ \bibnamefont {Potter}},\ }\href@noop {} {\bibinfo
  {title} {{Holographic quantum algorithms for simulating correlated spin
  systems}}} (\bibinfo {year} {2020}),\ \Eprint
  {https://arxiv.org/abs/2005.03023} {arXiv:2005.03023 [quant-ph]} \BibitemShut
  {NoStop}%
\bibitem [{\citenamefont {Motta}\ \emph {et~al.}(2019)\citenamefont {Motta},
  \citenamefont {Sun}, \citenamefont {Tan}, \citenamefont {O'Rourke},
  \citenamefont {Ye}, \citenamefont {Minnich}, \citenamefont {Brand{\~a}o},\
  and\ \citenamefont {Chan}}]{Motta2019}%
  \BibitemOpen
  \bibfield  {author} {\bibinfo {author} {\bibfnamefont {M.}~\bibnamefont
  {Motta}}, \bibinfo {author} {\bibfnamefont {C.}~\bibnamefont {Sun}}, \bibinfo
  {author} {\bibfnamefont {A.~T.~K.}\ \bibnamefont {Tan}}, \bibinfo {author}
  {\bibfnamefont {M.~J.}\ \bibnamefont {O'Rourke}}, \bibinfo {author}
  {\bibfnamefont {E.}~\bibnamefont {Ye}}, \bibinfo {author} {\bibfnamefont
  {A.~J.}\ \bibnamefont {Minnich}}, \bibinfo {author} {\bibfnamefont {F.~G.
  S.~L.}\ \bibnamefont {Brand{\~a}o}},\ and\ \bibinfo {author} {\bibfnamefont
  {G.~K.-L.}\ \bibnamefont {Chan}},\ }\bibfield  {title} {\bibinfo {title}
  {Determining eigenstates and thermal states on a quantum computer using
  quantum imaginary time evolution},\ }\bibfield  {journal} {\bibinfo
  {journal} {Nature Physics}\ }\href
  {https://doi.org/10.1038/s41567-019-0704-4} {10.1038/s41567-019-0704-4}
  (\bibinfo {year} {2019})\BibitemShut {NoStop}%
\bibitem [{\citenamefont {Yeter-Aydeniz}\ \emph
  {et~al.}(2020{\natexlab{a}})\citenamefont {Yeter-Aydeniz}, \citenamefont
  {Pooser},\ and\ \citenamefont {Siopsis}}]{Yeter-Aydeniz2020}%
  \BibitemOpen
  \bibfield  {author} {\bibinfo {author} {\bibfnamefont {K.}~\bibnamefont
  {Yeter-Aydeniz}}, \bibinfo {author} {\bibfnamefont {R.~C.}\ \bibnamefont
  {Pooser}},\ and\ \bibinfo {author} {\bibfnamefont {G.}~\bibnamefont
  {Siopsis}},\ }\bibfield  {title} {\bibinfo {title} {{Practical quantum
  computation of chemical and nuclear energy levels using quantum imaginary
  time evolution and Lanczos algorithms}},\ }\href
  {https://doi.org/10.1038/s41534-020-00290-1} {\bibfield  {journal} {\bibinfo
  {journal} {npj Quantum Information}\ }\textbf {\bibinfo {volume} {6}},\
  \bibinfo {pages} {63} (\bibinfo {year} {2020}{\natexlab{a}})}\BibitemShut
  {NoStop}%
\bibitem [{\citenamefont {Nishi}\ \emph {et~al.}(2020)\citenamefont {Nishi},
  \citenamefont {Kosugi},\ and\ \citenamefont {Matsushita}}]{Nishi2020}%
  \BibitemOpen
  \bibfield  {author} {\bibinfo {author} {\bibfnamefont {H.}~\bibnamefont
  {Nishi}}, \bibinfo {author} {\bibfnamefont {T.}~\bibnamefont {Kosugi}},\ and\
  \bibinfo {author} {\bibfnamefont {Y.}~\bibnamefont {Matsushita}},\
  }\href@noop {} {\bibinfo {title} {{Implementation of quantum imaginary-time
  evolution method on NISQ devices: Nonlocal approximation}}} (\bibinfo {year}
  {2020}),\ \Eprint {https://arxiv.org/abs/2005.12715} {arXiv:2005.12715
  [quant-ph]} \BibitemShut {NoStop}%
\bibitem [{\citenamefont {Gomes}\ \emph {et~al.}(2020)\citenamefont {Gomes},
  \citenamefont {Zhang}, \citenamefont {Berthusen}, \citenamefont {Wang},
  \citenamefont {Ho}, \citenamefont {Orth},\ and\ \citenamefont
  {Yao}}]{Gomes2020}%
  \BibitemOpen
  \bibfield  {author} {\bibinfo {author} {\bibfnamefont {N.}~\bibnamefont
  {Gomes}}, \bibinfo {author} {\bibfnamefont {F.}~\bibnamefont {Zhang}},
  \bibinfo {author} {\bibfnamefont {N.~F.}\ \bibnamefont {Berthusen}}, \bibinfo
  {author} {\bibfnamefont {C.-Z.}\ \bibnamefont {Wang}}, \bibinfo {author}
  {\bibfnamefont {K.-M.}\ \bibnamefont {Ho}}, \bibinfo {author} {\bibfnamefont
  {P.~P.}\ \bibnamefont {Orth}},\ and\ \bibinfo {author} {\bibfnamefont
  {Y.}~\bibnamefont {Yao}},\ }\bibfield  {title} {\bibinfo {title} {Efficient
  step-merged quantum imaginary time evolution algorithm for quantum
  chemistry},\ }\href {https://doi.org/10.1021/acs.jctc.0c00666} {\bibfield
  {journal} {\bibinfo  {journal} {Journal of Chemical Theory and Computation}\
  }\textbf {\bibinfo {volume} {16}},\ \bibinfo {pages} {6256} (\bibinfo {year}
  {2020})}\BibitemShut {NoStop}%
\bibitem [{\citenamefont {Yeter-Aydeniz}\ \emph
  {et~al.}(2020{\natexlab{b}})\citenamefont {Yeter-Aydeniz}, \citenamefont
  {Siopsis},\ and\ \citenamefont {Pooser}}]{yeteraydeniz2020scattering}%
  \BibitemOpen
  \bibfield  {author} {\bibinfo {author} {\bibfnamefont {K.}~\bibnamefont
  {Yeter-Aydeniz}}, \bibinfo {author} {\bibfnamefont {G.}~\bibnamefont
  {Siopsis}},\ and\ \bibinfo {author} {\bibfnamefont {R.~C.}\ \bibnamefont
  {Pooser}},\ }\href@noop {} {\bibinfo {title} {{Scattering in the Ising Model
  Using Quantum Lanczos Algorithm}}} (\bibinfo {year} {2020}{\natexlab{b}}),\
  \Eprint {https://arxiv.org/abs/2008.08763} {arXiv:2008.08763 [quant-ph]}
  \BibitemShut {NoStop}%
\bibitem [{\citenamefont {Stair}\ \emph {et~al.}(2020)\citenamefont {Stair},
  \citenamefont {Huang},\ and\ \citenamefont {Evangelista}}]{Stair2020}%
  \BibitemOpen
  \bibfield  {author} {\bibinfo {author} {\bibfnamefont {N.~H.}\ \bibnamefont
  {Stair}}, \bibinfo {author} {\bibfnamefont {R.}~\bibnamefont {Huang}},\ and\
  \bibinfo {author} {\bibfnamefont {F.~A.}\ \bibnamefont {Evangelista}},\
  }\bibfield  {title} {\bibinfo {title} {{A Multireference Quantum Krylov
  Algorithm for Strongly Correlated Electrons}},\ }\href
  {https://doi.org/10.1021/acs.jctc.9b01125} {\bibfield  {journal} {\bibinfo
  {journal} {Journal of Chemical Theory and Computation}\ }\textbf {\bibinfo
  {volume} {16}},\ \bibinfo {pages} {2236} (\bibinfo {year}
  {2020})}\BibitemShut {NoStop}%
\bibitem [{\citenamefont {Parrish}\ and\ \citenamefont
  {McMahon}(2019)}]{parrish2019quantum}%
  \BibitemOpen
  \bibfield  {author} {\bibinfo {author} {\bibfnamefont {R.~M.}\ \bibnamefont
  {Parrish}}\ and\ \bibinfo {author} {\bibfnamefont {P.~L.}\ \bibnamefont
  {McMahon}},\ }\href@noop {} {\bibinfo {title} {{Quantum Filter
  Diagonalization: Quantum Eigendecomposition without Full Quantum Phase
  Estimation}}} (\bibinfo {year} {2019}),\ \Eprint
  {https://arxiv.org/abs/1909.08925} {arXiv:1909.08925 [quant-ph]} \BibitemShut
  {NoStop}%
\bibitem [{\citenamefont {Kyriienko}(2020)}]{Kyriienko_2020}%
  \BibitemOpen
  \bibfield  {author} {\bibinfo {author} {\bibfnamefont {O.}~\bibnamefont
  {Kyriienko}},\ }\bibfield  {title} {\bibinfo {title} {Quantum inverse
  iteration algorithm for programmable quantum simulators},\ }\href
  {https://doi.org/10.1038/s41534-019-0239-7} {\bibfield  {journal} {\bibinfo
  {journal} {npj Quantum Information}\ }\textbf {\bibinfo {volume} {6}},\
  \bibinfo {pages} {7} (\bibinfo {year} {2020})}\BibitemShut {NoStop}%
\bibitem [{\citenamefont {Childs}\ \emph {et~al.}(2017)\citenamefont {Childs},
  \citenamefont {Kothari},\ and\ \citenamefont {Somma}}]{Childs2017}%
  \BibitemOpen
  \bibfield  {author} {\bibinfo {author} {\bibfnamefont {A.~M.}\ \bibnamefont
  {Childs}}, \bibinfo {author} {\bibfnamefont {R.}~\bibnamefont {Kothari}},\
  and\ \bibinfo {author} {\bibfnamefont {R.~D.}\ \bibnamefont {Somma}},\
  }\bibfield  {title} {\bibinfo {title} {{Quantum Algorithm for Systems of
  Linear Equations with Exponentially Improved Dependence on Precision}},\
  }\href {https://doi.org/10.1137/16m1087072} {\bibfield  {journal} {\bibinfo
  {journal} {SIAM Journal on Computing}\ }\textbf {\bibinfo {volume} {46}},\
  \bibinfo {pages} {1920} (\bibinfo {year} {2017})}\BibitemShut {NoStop}%
\bibitem [{\citenamefont {Liu}\ \emph {et~al.}(2020)\citenamefont {Liu},
  \citenamefont {Liu}, \citenamefont {Wang},\ and\ \citenamefont
  {Fan}}]{Liu2020}%
  \BibitemOpen
  \bibfield  {author} {\bibinfo {author} {\bibfnamefont {T.}~\bibnamefont
  {Liu}}, \bibinfo {author} {\bibfnamefont {J.-G.}\ \bibnamefont {Liu}},
  \bibinfo {author} {\bibfnamefont {L.}~\bibnamefont {Wang}},\ and\ \bibinfo
  {author} {\bibfnamefont {H.}~\bibnamefont {Fan}},\ }\href@noop {} {\bibinfo
  {title} {{Probabilistic Nonunitary Gate in Imaginary Time Evolution}}}
  (\bibinfo {year} {2020}),\ \Eprint {https://arxiv.org/abs/2006.09726}
  {arXiv:2006.09726 [quant-ph]} \BibitemShut {NoStop}%
\bibitem [{\citenamefont {McClean}\ \emph {et~al.}(2017)\citenamefont
  {McClean}, \citenamefont {Kimchi-Schwartz}, \citenamefont {Carter},\ and\
  \citenamefont {de~Jong}}]{McClean2017}%
  \BibitemOpen
  \bibfield  {author} {\bibinfo {author} {\bibfnamefont {J.~R.}\ \bibnamefont
  {McClean}}, \bibinfo {author} {\bibfnamefont {M.~E.}\ \bibnamefont
  {Kimchi-Schwartz}}, \bibinfo {author} {\bibfnamefont {J.}~\bibnamefont
  {Carter}},\ and\ \bibinfo {author} {\bibfnamefont {W.~A.}\ \bibnamefont
  {de~Jong}},\ }\bibfield  {title} {\bibinfo {title} {{Hybrid quantum-classical
  hierarchy for mitigation of decoherence and determination of excited
  states}},\ }\href {https://doi.org/10.1103/PhysRevA.95.042308} {\bibfield
  {journal} {\bibinfo  {journal} {Phys. Rev. A}\ }\textbf {\bibinfo {volume}
  {95}},\ \bibinfo {pages} {042308} (\bibinfo {year} {2017})}\BibitemShut
  {NoStop}%
\bibitem [{\citenamefont {Colless}\ \emph {et~al.}(2018)\citenamefont
  {Colless}, \citenamefont {Ramasesh}, \citenamefont {Dahlen}, \citenamefont
  {Blok}, \citenamefont {Kimchi-Schwartz}, \citenamefont {McClean},
  \citenamefont {Carter}, \citenamefont {de~Jong},\ and\ \citenamefont
  {Siddiqi}}]{Colless2018}%
  \BibitemOpen
  \bibfield  {author} {\bibinfo {author} {\bibfnamefont {J.~I.}\ \bibnamefont
  {Colless}}, \bibinfo {author} {\bibfnamefont {V.~V.}\ \bibnamefont
  {Ramasesh}}, \bibinfo {author} {\bibfnamefont {D.}~\bibnamefont {Dahlen}},
  \bibinfo {author} {\bibfnamefont {M.~S.}\ \bibnamefont {Blok}}, \bibinfo
  {author} {\bibfnamefont {M.~E.}\ \bibnamefont {Kimchi-Schwartz}}, \bibinfo
  {author} {\bibfnamefont {J.~R.}\ \bibnamefont {McClean}}, \bibinfo {author}
  {\bibfnamefont {J.}~\bibnamefont {Carter}}, \bibinfo {author} {\bibfnamefont
  {W.~A.}\ \bibnamefont {de~Jong}},\ and\ \bibinfo {author} {\bibfnamefont
  {I.}~\bibnamefont {Siddiqi}},\ }\bibfield  {title} {\bibinfo {title}
  {{Computation of Molecular Spectra on a Quantum Processor with an
  Error-Resilient Algorithm}},\ }\href
  {https://doi.org/10.1103/PhysRevX.8.011021} {\bibfield  {journal} {\bibinfo
  {journal} {Phys. Rev. X}\ }\textbf {\bibinfo {volume} {8}},\ \bibinfo {pages}
  {011021} (\bibinfo {year} {2018})}\BibitemShut {NoStop}%
\bibitem [{\citenamefont {Parrish}\ \emph {et~al.}(2019)\citenamefont
  {Parrish}, \citenamefont {Hohenstein}, \citenamefont {McMahon},\ and\
  \citenamefont {Mart\'{\i}nez}}]{Parrish2019}%
  \BibitemOpen
  \bibfield  {author} {\bibinfo {author} {\bibfnamefont {R.~M.}\ \bibnamefont
  {Parrish}}, \bibinfo {author} {\bibfnamefont {E.~G.}\ \bibnamefont
  {Hohenstein}}, \bibinfo {author} {\bibfnamefont {P.~L.}\ \bibnamefont
  {McMahon}},\ and\ \bibinfo {author} {\bibfnamefont {T.~J.}\ \bibnamefont
  {Mart\'{\i}nez}},\ }\bibfield  {title} {\bibinfo {title} {{Quantum
  Computation of Electronic Transitions Using a Variational Quantum
  Eigensolver}},\ }\href {https://doi.org/10.1103/PhysRevLett.122.230401}
  {\bibfield  {journal} {\bibinfo  {journal} {Phys. Rev. Lett.}\ }\textbf
  {\bibinfo {volume} {122}},\ \bibinfo {pages} {230401} (\bibinfo {year}
  {2019})}\BibitemShut {NoStop}%
\bibitem [{\citenamefont {Nakanishi}\ \emph {et~al.}(2019)\citenamefont
  {Nakanishi}, \citenamefont {Mitarai},\ and\ \citenamefont
  {Fujii}}]{nakanishi2018subspacesearch}%
  \BibitemOpen
  \bibfield  {author} {\bibinfo {author} {\bibfnamefont {K.~M.}\ \bibnamefont
  {Nakanishi}}, \bibinfo {author} {\bibfnamefont {K.}~\bibnamefont {Mitarai}},\
  and\ \bibinfo {author} {\bibfnamefont {K.}~\bibnamefont {Fujii}},\ }\bibfield
   {title} {\bibinfo {title} {Subspace-search variational quantum eigensolver
  for excited states},\ }\href
  {https://doi.org/10.1103/PhysRevResearch.1.033062} {\bibfield  {journal}
  {\bibinfo  {journal} {Phys. Rev. Research}\ }\textbf {\bibinfo {volume}
  {1}},\ \bibinfo {pages} {033062} (\bibinfo {year} {2019})}\BibitemShut
  {NoStop}%
\bibitem [{\citenamefont {Heya}\ \emph {et~al.}(2019)\citenamefont {Heya},
  \citenamefont {Nakanishi}, \citenamefont {Mitarai},\ and\ \citenamefont
  {Fujii}}]{heya2019subspace}%
  \BibitemOpen
  \bibfield  {author} {\bibinfo {author} {\bibfnamefont {K.}~\bibnamefont
  {Heya}}, \bibinfo {author} {\bibfnamefont {K.~M.}\ \bibnamefont {Nakanishi}},
  \bibinfo {author} {\bibfnamefont {K.}~\bibnamefont {Mitarai}},\ and\ \bibinfo
  {author} {\bibfnamefont {K.}~\bibnamefont {Fujii}},\ }\href@noop {} {\bibinfo
  {title} {{Subspace Variational Quantum Simulator}}} (\bibinfo {year}
  {2019}),\ \Eprint {https://arxiv.org/abs/1904.08566} {arXiv:1904.08566
  [quant-ph]} \BibitemShut {NoStop}%
\bibitem [{\citenamefont {Huggins}\ \emph {et~al.}(2020)\citenamefont
  {Huggins}, \citenamefont {Lee}, \citenamefont {Baek}, \citenamefont
  {O'Gorman},\ and\ \citenamefont {Whaley}}]{huggins2019nonorthogonal}%
  \BibitemOpen
  \bibfield  {author} {\bibinfo {author} {\bibfnamefont {W.~J.}\ \bibnamefont
  {Huggins}}, \bibinfo {author} {\bibfnamefont {J.}~\bibnamefont {Lee}},
  \bibinfo {author} {\bibfnamefont {U.}~\bibnamefont {Baek}}, \bibinfo {author}
  {\bibfnamefont {B.}~\bibnamefont {O'Gorman}},\ and\ \bibinfo {author}
  {\bibfnamefont {K.~B.}\ \bibnamefont {Whaley}},\ }\bibfield  {title}
  {\bibinfo {title} {{A non-orthogonal variational quantum eigensolver}},\
  }\href {https://doi.org/10.1088/1367-2630/ab867b} {\bibfield  {journal}
  {\bibinfo  {journal} {New Journal of Physics}\ }\textbf {\bibinfo {volume}
  {22}},\ \bibinfo {pages} {073009} (\bibinfo {year} {2020})}\BibitemShut
  {NoStop}%
\bibitem [{\citenamefont {Nielsen}\ and\ \citenamefont
  {Chuang}(2000)}]{NielsenChuang}%
  \BibitemOpen
  \bibfield  {author} {\bibinfo {author} {\bibfnamefont {M.~A.}\ \bibnamefont
  {Nielsen}}\ and\ \bibinfo {author} {\bibfnamefont {I.~L.}\ \bibnamefont
  {Chuang}},\ }\href@noop {} {\emph {\bibinfo {title} {Quantum Computation and
  Quantum Information}}}\ (\bibinfo  {publisher} {Cambridge University Press},\
  \bibinfo {address} {New York},\ \bibinfo {year} {2000})\BibitemShut {NoStop}%
\bibitem [{\citenamefont {Jordan}\ and\ \citenamefont
  {Wigner}(1928)}]{Jordan1928}%
  \BibitemOpen
  \bibfield  {author} {\bibinfo {author} {\bibfnamefont {P.}~\bibnamefont
  {Jordan}}\ and\ \bibinfo {author} {\bibfnamefont {E.}~\bibnamefont
  {Wigner}},\ }\bibfield  {title} {\bibinfo {title} {{{\"U}ber das Paulische
  {\"A}quivalenzverbot}},\ }\href {https://doi.org/10.1007/BF01331938}
  {\bibfield  {journal} {\bibinfo  {journal} {Zeitschrift f{\"u}r Physik}\
  }\textbf {\bibinfo {volume} {47}},\ \bibinfo {pages} {631} (\bibinfo {year}
  {1928})}\BibitemShut {NoStop}%
\bibitem [{\citenamefont {Bravyi}\ and\ \citenamefont
  {Kitaev}(2002)}]{Bravyi2002}%
  \BibitemOpen
  \bibfield  {author} {\bibinfo {author} {\bibfnamefont {S.~B.}\ \bibnamefont
  {Bravyi}}\ and\ \bibinfo {author} {\bibfnamefont {A.~Y.}\ \bibnamefont
  {Kitaev}},\ }\bibfield  {title} {\bibinfo {title} {Fermionic quantum
  computation},\ }\href
  {https://doi.org/https://doi.org/10.1006/aphy.2002.6254} {\bibfield
  {journal} {\bibinfo  {journal} {Annals of Physics}\ }\textbf {\bibinfo
  {volume} {298}},\ \bibinfo {pages} {210 } (\bibinfo {year}
  {2002})}\BibitemShut {NoStop}%
\bibitem [{\citenamefont {Seeley}\ \emph {et~al.}(2012)\citenamefont {Seeley},
  \citenamefont {Richard},\ and\ \citenamefont {Love}}]{Seeley2012}%
  \BibitemOpen
  \bibfield  {author} {\bibinfo {author} {\bibfnamefont {J.~T.}\ \bibnamefont
  {Seeley}}, \bibinfo {author} {\bibfnamefont {M.~J.}\ \bibnamefont
  {Richard}},\ and\ \bibinfo {author} {\bibfnamefont {P.~J.}\ \bibnamefont
  {Love}},\ }\bibfield  {title} {\bibinfo {title} {{The Bravyi-Kitaev
  transformation for quantum computation of electronic structure}},\ }\href
  {https://doi.org/10.1063/1.4768229} {\bibfield  {journal} {\bibinfo
  {journal} {The Journal of Chemical Physics}\ }\textbf {\bibinfo {volume}
  {137}},\ \bibinfo {pages} {224109} (\bibinfo {year} {2012})}\BibitemShut
  {NoStop}%
\bibitem [{\citenamefont {Tranter}\ \emph {et~al.}(2015)\citenamefont
  {Tranter}, \citenamefont {Sofia}, \citenamefont {Seeley}, \citenamefont
  {Kaicher}, \citenamefont {McClean}, \citenamefont {Babbush}, \citenamefont
  {Coveney}, \citenamefont {Mintert}, \citenamefont {Wilhelm},\ and\
  \citenamefont {Love}}]{Tranter2013}%
  \BibitemOpen
  \bibfield  {author} {\bibinfo {author} {\bibfnamefont {A.}~\bibnamefont
  {Tranter}}, \bibinfo {author} {\bibfnamefont {S.}~\bibnamefont {Sofia}},
  \bibinfo {author} {\bibfnamefont {J.}~\bibnamefont {Seeley}}, \bibinfo
  {author} {\bibfnamefont {M.}~\bibnamefont {Kaicher}}, \bibinfo {author}
  {\bibfnamefont {J.}~\bibnamefont {McClean}}, \bibinfo {author} {\bibfnamefont
  {R.}~\bibnamefont {Babbush}}, \bibinfo {author} {\bibfnamefont {P.~V.}\
  \bibnamefont {Coveney}}, \bibinfo {author} {\bibfnamefont {F.}~\bibnamefont
  {Mintert}}, \bibinfo {author} {\bibfnamefont {F.}~\bibnamefont {Wilhelm}},\
  and\ \bibinfo {author} {\bibfnamefont {P.~J.}\ \bibnamefont {Love}},\
  }\bibfield  {title} {\bibinfo {title} {{The Bravyi-Kitaev transformation:
  Properties and applications}},\ }\href {https://doi.org/10.1002/qua.24969}
  {\bibfield  {journal} {\bibinfo  {journal} {International Journal of Quantum
  Chemistry}\ }\textbf {\bibinfo {volume} {115}},\ \bibinfo {pages} {1431}
  (\bibinfo {year} {2015})}\BibitemShut {NoStop}%
\bibitem [{\citenamefont {Havl\'{\i}\ifmmode\check{c}\else\v{c}\fi{}ek}\ \emph
  {et~al.}(2017)\citenamefont {Havl\'{\i}\ifmmode\check{c}\else\v{c}\fi{}ek},
  \citenamefont {Troyer},\ and\ \citenamefont {Whitfield}}]{Havlicek2017}%
  \BibitemOpen
  \bibfield  {author} {\bibinfo {author} {\bibfnamefont {V.}~\bibnamefont
  {Havl\'{\i}\ifmmode\check{c}\else\v{c}\fi{}ek}}, \bibinfo {author}
  {\bibfnamefont {M.}~\bibnamefont {Troyer}},\ and\ \bibinfo {author}
  {\bibfnamefont {J.~D.}\ \bibnamefont {Whitfield}},\ }\bibfield  {title}
  {\bibinfo {title} {{Operator locality in the quantum simulation of fermionic
  models}},\ }\href {https://doi.org/10.1103/PhysRevA.95.032332} {\bibfield
  {journal} {\bibinfo  {journal} {Phys. Rev. A}\ }\textbf {\bibinfo {volume}
  {95}},\ \bibinfo {pages} {032332} (\bibinfo {year} {2017})}\BibitemShut
  {NoStop}%
\bibitem [{\citenamefont {Childs}\ and\ \citenamefont
  {Weibe}(2012)}]{Childs2012}%
  \BibitemOpen
  \bibfield  {author} {\bibinfo {author} {\bibfnamefont {A.~M.}\ \bibnamefont
  {Childs}}\ and\ \bibinfo {author} {\bibfnamefont {N.}~\bibnamefont {Weibe}},\
  }\bibfield  {title} {\bibinfo {title} {{Hamiltonian simulation using linear
  combinations of unitary operations}},\ }\href
  {http://dx.doi.org/10.26421/QIC12.11-12} {\bibfield  {journal} {\bibinfo
  {journal} {Quantum Information and Computation}\ }\textbf {\bibinfo {volume}
  {12}},\ \bibinfo {pages} {901} (\bibinfo {year} {2012})}\BibitemShut
  {NoStop}%
\bibitem [{\citenamefont {Kosugi}\ and\ \citenamefont
  {Matsushita}(2020{\natexlab{a}})}]{kosugi2019construction}%
  \BibitemOpen
  \bibfield  {author} {\bibinfo {author} {\bibfnamefont {T.}~\bibnamefont
  {Kosugi}}\ and\ \bibinfo {author} {\bibfnamefont {Y.}~\bibnamefont
  {Matsushita}},\ }\bibfield  {title} {\bibinfo {title} {Construction of
  green's functions on a quantum computer: Quasiparticle spectra of
  molecules},\ }\href {https://doi.org/10.1103/PhysRevA.101.012330} {\bibfield
  {journal} {\bibinfo  {journal} {Phys. Rev. A}\ }\textbf {\bibinfo {volume}
  {101}},\ \bibinfo {pages} {012330} (\bibinfo {year}
  {2020}{\natexlab{a}})}\BibitemShut {NoStop}%
\bibitem [{\citenamefont {Kosugi}\ and\ \citenamefont
  {Matsushita}(2020{\natexlab{b}})}]{kosugi2019charge}%
  \BibitemOpen
  \bibfield  {author} {\bibinfo {author} {\bibfnamefont {T.}~\bibnamefont
  {Kosugi}}\ and\ \bibinfo {author} {\bibfnamefont {Y.-i.}\ \bibnamefont
  {Matsushita}},\ }\bibfield  {title} {\bibinfo {title} {Linear-response
  functions of molecules on a quantum computer: Charge and spin responses and
  optical absorption},\ }\href
  {https://doi.org/10.1103/PhysRevResearch.2.033043} {\bibfield  {journal}
  {\bibinfo  {journal} {Phys. Rev. Research}\ }\textbf {\bibinfo {volume}
  {2}},\ \bibinfo {pages} {033043} (\bibinfo {year}
  {2020}{\natexlab{b}})}\BibitemShut {NoStop}%
\bibitem [{odd()}]{odd_power}%
  \BibitemOpen
  \href@noop {} {}\bibinfo {note} {The expectation value of
  $\hat{\mathcal{H}}_{\rm ST}^{n}(\Delta_\tau)$ with an odd power $n$ might be
  evaluated with an Hadamard-test like circuit by introducing an additional
  ancilla qubit}\BibitemShut {NoStop}%
\bibitem [{\citenamefont {Suzuki}(1990)}]{Suzuki1990}%
  \BibitemOpen
  \bibfield  {author} {\bibinfo {author} {\bibfnamefont {M.}~\bibnamefont
  {Suzuki}},\ }\bibfield  {title} {\bibinfo {title} {{Fractal decomposition of
  exponential operators with applications to many-body theories and Monte Carlo
  simulations}},\ }\href
  {https://doi.org/https://doi.org/10.1016/0375-9601(90)90962-N} {\bibfield
  {journal} {\bibinfo  {journal} {Physics Letters A}\ }\textbf {\bibinfo
  {volume} {146}},\ \bibinfo {pages} {319 } (\bibinfo {year}
  {1990})}\BibitemShut {NoStop}%
\bibitem [{\citenamefont {{Yoshida}}(1990)}]{Yoshida1990}%
  \BibitemOpen
  \bibfield  {author} {\bibinfo {author} {\bibfnamefont {H.}~\bibnamefont
  {{Yoshida}}},\ }\bibfield  {title} {\bibinfo {title} {{Construction of higher
  order symplectic integrators}},\ }\href
  {https://doi.org/10.1016/0375-9601(90)90092-3} {\bibfield  {journal}
  {\bibinfo  {journal} {Physics Letters A}\ }\textbf {\bibinfo {volume}
  {150}},\ \bibinfo {pages} {262} (\bibinfo {year} {1990})}\BibitemShut
  {NoStop}%
\bibitem [{\citenamefont {{Suzuki}}(1991)}]{Suzuki1991JMP}%
  \BibitemOpen
  \bibfield  {author} {\bibinfo {author} {\bibfnamefont {M.}~\bibnamefont
  {{Suzuki}}},\ }\bibfield  {title} {\bibinfo {title} {{General theory of
  fractal path integrals with applications to many-body theories and
  statistical physics}},\ }\href {https://doi.org/10.1063/1.529425} {\bibfield
  {journal} {\bibinfo  {journal} {Journal of Mathematical Physics}\ }\textbf
  {\bibinfo {volume} {32}},\ \bibinfo {pages} {400} (\bibinfo {year}
  {1991})}\BibitemShut {NoStop}%
\bibitem [{\citenamefont {Hatano}\ and\ \citenamefont
  {Suzuki}(2005)}]{Hatano2005}%
  \BibitemOpen
  \bibfield  {author} {\bibinfo {author} {\bibfnamefont {N.}~\bibnamefont
  {Hatano}}\ and\ \bibinfo {author} {\bibfnamefont {M.}~\bibnamefont
  {Suzuki}},\ }\bibfield  {title} {\bibinfo {title} {{Finding Exponential
  Product Formulas of Higher Orders}},\ }\href
  {https://doi.org/10.1007/11526216_2} {\bibfield  {journal} {\bibinfo
  {journal} {Lecture Notes in Physics}\ }\textbf {\bibinfo {volume} {679}},\
  \bibinfo {pages} {37} (\bibinfo {year} {2005})}\BibitemShut {NoStop}%
\bibitem [{\citenamefont {Suzuki}(1992)}]{Suzuki1992}%
  \BibitemOpen
  \bibfield  {author} {\bibinfo {author} {\bibfnamefont {M.}~\bibnamefont
  {Suzuki}},\ }\bibfield  {title} {\bibinfo {title} {{General Nonsymmetric
  Higher-Order Decomposition of Exponential Operators and Symplectic
  Integrators}},\ }\href {https://doi.org/10.1143/JPSJ.61.3015} {\bibfield
  {journal} {\bibinfo  {journal} {Journal of the Physical Society of Japan}\
  }\textbf {\bibinfo {volume} {61}},\ \bibinfo {pages} {3015} (\bibinfo {year}
  {1992})}\BibitemShut {NoStop}%
\bibitem [{\citenamefont {Chatelin}(2012)}]{Chatelin}%
  \BibitemOpen
  \bibfield  {author} {\bibinfo {author} {\bibfnamefont {F.}~\bibnamefont
  {Chatelin}},\ }\href@noop {} {\emph {\bibinfo {title} {{Eigenvalues of
  Matrices}}}}\ (\bibinfo  {publisher} {SIAM},\ \bibinfo {address}
  {Philadelphia},\ \bibinfo {year} {2012})\BibitemShut {NoStop}%
\bibitem [{\citenamefont {Romero}\ \emph {et~al.}(2018)\citenamefont {Romero},
  \citenamefont {Babbush}, \citenamefont {McClean}, \citenamefont {Hempel},
  \citenamefont {Love},\ and\ \citenamefont {Aspuru-Guzik}}]{Romero2018}%
  \BibitemOpen
  \bibfield  {author} {\bibinfo {author} {\bibfnamefont {J.}~\bibnamefont
  {Romero}}, \bibinfo {author} {\bibfnamefont {R.}~\bibnamefont {Babbush}},
  \bibinfo {author} {\bibfnamefont {J.~R.}\ \bibnamefont {McClean}}, \bibinfo
  {author} {\bibfnamefont {C.}~\bibnamefont {Hempel}}, \bibinfo {author}
  {\bibfnamefont {P.~J.}\ \bibnamefont {Love}},\ and\ \bibinfo {author}
  {\bibfnamefont {A.}~\bibnamefont {Aspuru-Guzik}},\ }\bibfield  {title}
  {\bibinfo {title} {Strategies for quantum computing molecular energies using
  the unitary coupled cluster ansatz},\ }\href
  {https://doi.org/10.1088/2058-9565/aad3e4} {\bibfield  {journal} {\bibinfo
  {journal} {Quantum Science and Technology}\ }\textbf {\bibinfo {volume}
  {4}},\ \bibinfo {pages} {014008} (\bibinfo {year} {2018})}\BibitemShut
  {NoStop}%
\bibitem [{\citenamefont {Dallaire-Demers}\ \emph {et~al.}(2019)\citenamefont
  {Dallaire-Demers}, \citenamefont {Romero}, \citenamefont {Veis},
  \citenamefont {Sim},\ and\ \citenamefont
  {Aspuru-Guzik}}]{Dallaire-Demers2019}%
  \BibitemOpen
  \bibfield  {author} {\bibinfo {author} {\bibfnamefont {P.-L.}\ \bibnamefont
  {Dallaire-Demers}}, \bibinfo {author} {\bibfnamefont {J.}~\bibnamefont
  {Romero}}, \bibinfo {author} {\bibfnamefont {L.}~\bibnamefont {Veis}},
  \bibinfo {author} {\bibfnamefont {S.}~\bibnamefont {Sim}},\ and\ \bibinfo
  {author} {\bibfnamefont {A.}~\bibnamefont {Aspuru-Guzik}},\ }\bibfield
  {title} {\bibinfo {title} {Low-depth circuit ansatz for preparing correlated
  fermionic states on a quantum computer},\ }\href
  {https://doi.org/10.1088/2058-9565/ab3951} {\bibfield  {journal} {\bibinfo
  {journal} {Quantum Science and Technology}\ }\textbf {\bibinfo {volume}
  {4}},\ \bibinfo {pages} {045005} (\bibinfo {year} {2019})}\BibitemShut
  {NoStop}%
\bibitem [{\citenamefont {McArdle}\ \emph {et~al.}(2019)\citenamefont
  {McArdle}, \citenamefont {Jones}, \citenamefont {Endo}, \citenamefont {Li},
  \citenamefont {Benjamin},\ and\ \citenamefont {Yuan}}]{McArdle2019}%
  \BibitemOpen
  \bibfield  {author} {\bibinfo {author} {\bibfnamefont {S.}~\bibnamefont
  {McArdle}}, \bibinfo {author} {\bibfnamefont {T.}~\bibnamefont {Jones}},
  \bibinfo {author} {\bibfnamefont {S.}~\bibnamefont {Endo}}, \bibinfo {author}
  {\bibfnamefont {Y.}~\bibnamefont {Li}}, \bibinfo {author} {\bibfnamefont
  {S.~C.}\ \bibnamefont {Benjamin}},\ and\ \bibinfo {author} {\bibfnamefont
  {X.}~\bibnamefont {Yuan}},\ }\bibfield  {title} {\bibinfo {title}
  {Variational ansatz-based quantum simulation of imaginary time evolution},\
  }\href {https://doi.org/10.1038/s41534-019-0187-2} {\bibfield  {journal}
  {\bibinfo  {journal} {npj Quantum Information}\ }\textbf {\bibinfo {volume}
  {5}},\ \bibinfo {pages} {75} (\bibinfo {year} {2019})}\BibitemShut {NoStop}%
\bibitem [{\citenamefont {Loss}\ and\ \citenamefont
  {DiVincenzo}(1998)}]{Loss1998}%
  \BibitemOpen
  \bibfield  {author} {\bibinfo {author} {\bibfnamefont {D.}~\bibnamefont
  {Loss}}\ and\ \bibinfo {author} {\bibfnamefont {D.~P.}\ \bibnamefont
  {DiVincenzo}},\ }\bibfield  {title} {\bibinfo {title} {Quantum computation
  with quantum dots},\ }\href {https://doi.org/10.1103/PhysRevA.57.120}
  {\bibfield  {journal} {\bibinfo  {journal} {Phys. Rev. A}\ }\textbf {\bibinfo
  {volume} {57}},\ \bibinfo {pages} {120} (\bibinfo {year} {1998})}\BibitemShut
  {NoStop}%
\bibitem [{\citenamefont {DiVincenzo}\ \emph {et~al.}(2000)\citenamefont
  {DiVincenzo}, \citenamefont {Bacon}, \citenamefont {Kempe}, \citenamefont
  {Burkard},\ and\ \citenamefont {Whaley}}]{DiVincenzo2000}%
  \BibitemOpen
  \bibfield  {author} {\bibinfo {author} {\bibfnamefont {D.~P.}\ \bibnamefont
  {DiVincenzo}}, \bibinfo {author} {\bibfnamefont {D.}~\bibnamefont {Bacon}},
  \bibinfo {author} {\bibfnamefont {J.}~\bibnamefont {Kempe}}, \bibinfo
  {author} {\bibfnamefont {G.}~\bibnamefont {Burkard}},\ and\ \bibinfo {author}
  {\bibfnamefont {K.~B.}\ \bibnamefont {Whaley}},\ }\bibfield  {title}
  {\bibinfo {title} {Universal quantum computation with the exchange
  interaction},\ }\href {https://doi.org/10.1038/35042541} {\bibfield
  {journal} {\bibinfo  {journal} {Nature}\ }\textbf {\bibinfo {volume} {408}},\
  \bibinfo {pages} {339} (\bibinfo {year} {2000})}\BibitemShut {NoStop}%
\bibitem [{\citenamefont {Brunner}\ \emph {et~al.}(2011)\citenamefont
  {Brunner}, \citenamefont {Shin}, \citenamefont {Obata}, \citenamefont
  {Pioro-Ladri\`ere}, \citenamefont {Kubo}, \citenamefont {Yoshida},
  \citenamefont {Taniyama}, \citenamefont {Tokura},\ and\ \citenamefont
  {Tarucha}}]{Brunner2011}%
  \BibitemOpen
  \bibfield  {author} {\bibinfo {author} {\bibfnamefont {R.}~\bibnamefont
  {Brunner}}, \bibinfo {author} {\bibfnamefont {Y.-S.}\ \bibnamefont {Shin}},
  \bibinfo {author} {\bibfnamefont {T.}~\bibnamefont {Obata}}, \bibinfo
  {author} {\bibfnamefont {M.}~\bibnamefont {Pioro-Ladri\`ere}}, \bibinfo
  {author} {\bibfnamefont {T.}~\bibnamefont {Kubo}}, \bibinfo {author}
  {\bibfnamefont {K.}~\bibnamefont {Yoshida}}, \bibinfo {author} {\bibfnamefont
  {T.}~\bibnamefont {Taniyama}}, \bibinfo {author} {\bibfnamefont
  {Y.}~\bibnamefont {Tokura}},\ and\ \bibinfo {author} {\bibfnamefont
  {S.}~\bibnamefont {Tarucha}},\ }\bibfield  {title} {\bibinfo {title}
  {{Two-Qubit Gate of Combined Single-Spin Rotation and Interdot Spin Exchange
  in a Double Quantum Dot}},\ }\href
  {https://doi.org/10.1103/PhysRevLett.107.146801} {\bibfield  {journal}
  {\bibinfo  {journal} {Phys. Rev. Lett.}\ }\textbf {\bibinfo {volume} {107}},\
  \bibinfo {pages} {146801} (\bibinfo {year} {2011})}\BibitemShut {NoStop}%
\bibitem [{\citenamefont {Lloyd}\ \emph {et~al.}(2014)\citenamefont {Lloyd},
  \citenamefont {Mohseni},\ and\ \citenamefont {Rebentrost}}]{Lloyd2014}%
  \BibitemOpen
  \bibfield  {author} {\bibinfo {author} {\bibfnamefont {S.}~\bibnamefont
  {Lloyd}}, \bibinfo {author} {\bibfnamefont {M.}~\bibnamefont {Mohseni}},\
  and\ \bibinfo {author} {\bibfnamefont {P.}~\bibnamefont {Rebentrost}},\
  }\bibfield  {title} {\bibinfo {title} {Quantum principal component
  analysis},\ }\href {https://doi.org/10.1038/nphys3029} {\bibfield  {journal}
  {\bibinfo  {journal} {Nature Physics}\ }\textbf {\bibinfo {volume} {10}},\
  \bibinfo {pages} {631} (\bibinfo {year} {2014})}\BibitemShut {NoStop}%
\bibitem [{\citenamefont {Lau}\ and\ \citenamefont {Plenio}(2016)}]{Lau2016}%
  \BibitemOpen
  \bibfield  {author} {\bibinfo {author} {\bibfnamefont {H.-K.}\ \bibnamefont
  {Lau}}\ and\ \bibinfo {author} {\bibfnamefont {M.~B.}\ \bibnamefont
  {Plenio}},\ }\bibfield  {title} {\bibinfo {title} {{Universal Quantum
  Computing with Arbitrary Continuous-Variable Encoding}},\ }\href
  {https://doi.org/10.1103/PhysRevLett.117.100501} {\bibfield  {journal}
  {\bibinfo  {journal} {Phys. Rev. Lett.}\ }\textbf {\bibinfo {volume} {117}},\
  \bibinfo {pages} {100501} (\bibinfo {year} {2016})}\BibitemShut {NoStop}%
\bibitem [{\citenamefont {Fan}\ \emph {et~al.}(2005)\citenamefont {Fan},
  \citenamefont {Roychowdhury},\ and\ \citenamefont {Szkopek}}]{Fan2005}%
  \BibitemOpen
  \bibfield  {author} {\bibinfo {author} {\bibfnamefont {H.}~\bibnamefont
  {Fan}}, \bibinfo {author} {\bibfnamefont {V.}~\bibnamefont {Roychowdhury}},\
  and\ \bibinfo {author} {\bibfnamefont {T.}~\bibnamefont {Szkopek}},\
  }\bibfield  {title} {\bibinfo {title} {Optimal two-qubit quantum circuits
  using exchange interactions},\ }\href
  {https://doi.org/10.1103/PhysRevA.72.052323} {\bibfield  {journal} {\bibinfo
  {journal} {Phys. Rev. A}\ }\textbf {\bibinfo {volume} {72}},\ \bibinfo
  {pages} {052323} (\bibinfo {year} {2005})}\BibitemShut {NoStop}%
\bibitem [{\citenamefont {Balakrishnan}\ and\ \citenamefont
  {Sankaranarayanan}(2008)}]{Balakrishnan2008}%
  \BibitemOpen
  \bibfield  {author} {\bibinfo {author} {\bibfnamefont {S.}~\bibnamefont
  {Balakrishnan}}\ and\ \bibinfo {author} {\bibfnamefont {R.}~\bibnamefont
  {Sankaranarayanan}},\ }\bibfield  {title} {\bibinfo {title} {{Entangling
  characterization of ${\mathrm{SWAP}}^{1/m}$ and controlled unitary gates}},\
  }\href {https://doi.org/10.1103/PhysRevA.78.052305} {\bibfield  {journal}
  {\bibinfo  {journal} {Phys. Rev. A}\ }\textbf {\bibinfo {volume} {78}},\
  \bibinfo {pages} {052305} (\bibinfo {year} {2008})}\BibitemShut {NoStop}%
\bibitem [{\citenamefont {Gard}\ \emph {et~al.}(2020)\citenamefont {Gard},
  \citenamefont {Zhu}, \citenamefont {Barron}, \citenamefont {Mayhall},
  \citenamefont {Economou},\ and\ \citenamefont {Barnes}}]{Gard2019}%
  \BibitemOpen
  \bibfield  {author} {\bibinfo {author} {\bibfnamefont {B.~T.}\ \bibnamefont
  {Gard}}, \bibinfo {author} {\bibfnamefont {L.}~\bibnamefont {Zhu}}, \bibinfo
  {author} {\bibfnamefont {G.~S.}\ \bibnamefont {Barron}}, \bibinfo {author}
  {\bibfnamefont {N.~J.}\ \bibnamefont {Mayhall}}, \bibinfo {author}
  {\bibfnamefont {S.~E.}\ \bibnamefont {Economou}},\ and\ \bibinfo {author}
  {\bibfnamefont {E.}~\bibnamefont {Barnes}},\ }\bibfield  {title} {\bibinfo
  {title} {Efficient symmetry-preserving state preparation circuits for the
  variational quantum eigensolver algorithm},\ }\href
  {http://dx.doi.org/10.1038/s41534-019-0240-1} {\bibfield  {journal} {\bibinfo
   {journal} {npj Quantum Information}\ }\textbf {\bibinfo {volume} {6}},\
  \bibinfo {pages} {10} (\bibinfo {year} {2020})}\BibitemShut {NoStop}%
\bibitem [{\citenamefont {Vidal}\ and\ \citenamefont
  {Dawson}(2004)}]{Vidal2004}%
  \BibitemOpen
  \bibfield  {author} {\bibinfo {author} {\bibfnamefont {G.}~\bibnamefont
  {Vidal}}\ and\ \bibinfo {author} {\bibfnamefont {C.~M.}\ \bibnamefont
  {Dawson}},\ }\bibfield  {title} {\bibinfo {title} {{Universal quantum circuit
  for two-qubit transformations with three controlled-NOT gates}},\ }\href
  {https://doi.org/10.1103/PhysRevA.69.010301} {\bibfield  {journal} {\bibinfo
  {journal} {Phys. Rev. A}\ }\textbf {\bibinfo {volume} {69}},\ \bibinfo
  {pages} {010301} (\bibinfo {year} {2004})}\BibitemShut {NoStop}%
\bibitem [{\citenamefont {Chiesa}\ \emph {et~al.}(2019)\citenamefont {Chiesa},
  \citenamefont {Tacchino}, \citenamefont {Grossi}, \citenamefont {Santini},
  \citenamefont {Tavernelli}, \citenamefont {Gerace},\ and\ \citenamefont
  {Carretta}}]{Chiesa2019}%
  \BibitemOpen
  \bibfield  {author} {\bibinfo {author} {\bibfnamefont {A.}~\bibnamefont
  {Chiesa}}, \bibinfo {author} {\bibfnamefont {F.}~\bibnamefont {Tacchino}},
  \bibinfo {author} {\bibfnamefont {M.}~\bibnamefont {Grossi}}, \bibinfo
  {author} {\bibfnamefont {P.}~\bibnamefont {Santini}}, \bibinfo {author}
  {\bibfnamefont {I.}~\bibnamefont {Tavernelli}}, \bibinfo {author}
  {\bibfnamefont {D.}~\bibnamefont {Gerace}},\ and\ \bibinfo {author}
  {\bibfnamefont {S.}~\bibnamefont {Carretta}},\ }\bibfield  {title} {\bibinfo
  {title} {Quantum hardware simulating four-dimensional inelastic neutron
  scattering},\ }\href {https://doi.org/10.1038/s41567-019-0437-4} {\bibfield
  {journal} {\bibinfo  {journal} {Nature Physics}\ }\textbf {\bibinfo {volume}
  {15}},\ \bibinfo {pages} {455} (\bibinfo {year} {2019})}\BibitemShut
  {NoStop}%
\bibitem [{not()}]{note_eSWAP}%
  \BibitemOpen
  \href@noop {} {}\bibinfo {note} {Our decomposition of the e-{\sc swap} gate
  in Ref.~\cite{Seki2020vqe} was not optimal in terms of the number of {\sc
  cnot} gates.}\BibitemShut {Stop}%
\bibitem [{\citenamefont {Drabold}\ and\ \citenamefont
  {Sankey}(1993)}]{Drabold1993}%
  \BibitemOpen
  \bibfield  {author} {\bibinfo {author} {\bibfnamefont {D.~A.}\ \bibnamefont
  {Drabold}}\ and\ \bibinfo {author} {\bibfnamefont {O.~F.}\ \bibnamefont
  {Sankey}},\ }\bibfield  {title} {\bibinfo {title} {{Maximum entropy approach
  for linear scaling in the electronic structure problem}},\ }\href
  {https://doi.org/10.1103/PhysRevLett.70.3631} {\bibfield  {journal} {\bibinfo
   {journal} {Phys. Rev. Lett.}\ }\textbf {\bibinfo {volume} {70}},\ \bibinfo
  {pages} {3631} (\bibinfo {year} {1993})}\BibitemShut {NoStop}%
\bibitem [{\citenamefont {Hams}\ and\ \citenamefont
  {De~Raedt}(2000)}]{Hams2000}%
  \BibitemOpen
  \bibfield  {author} {\bibinfo {author} {\bibfnamefont {A.}~\bibnamefont
  {Hams}}\ and\ \bibinfo {author} {\bibfnamefont {H.}~\bibnamefont
  {De~Raedt}},\ }\bibfield  {title} {\bibinfo {title} {Fast algorithm for
  finding the eigenvalue distribution of very large matrices},\ }\href
  {https://doi.org/10.1103/PhysRevE.62.4365} {\bibfield  {journal} {\bibinfo
  {journal} {Phys. Rev. E}\ }\textbf {\bibinfo {volume} {62}},\ \bibinfo
  {pages} {4365} (\bibinfo {year} {2000})}\BibitemShut {NoStop}%
\bibitem [{\citenamefont {Iitaka}\ and\ \citenamefont
  {Ebisuzaki}(2004)}]{IItaka2004}%
  \BibitemOpen
  \bibfield  {author} {\bibinfo {author} {\bibfnamefont {T.}~\bibnamefont
  {Iitaka}}\ and\ \bibinfo {author} {\bibfnamefont {T.}~\bibnamefont
  {Ebisuzaki}},\ }\bibfield  {title} {\bibinfo {title} {Random phase vector for
  calculating the trace of a large matrix},\ }\href
  {https://doi.org/10.1103/PhysRevE.69.057701} {\bibfield  {journal} {\bibinfo
  {journal} {Phys. Rev. E}\ }\textbf {\bibinfo {volume} {69}},\ \bibinfo
  {pages} {057701} (\bibinfo {year} {2004})}\BibitemShut {NoStop}%
\bibitem [{\citenamefont {Wei\ss{}e}\ \emph {et~al.}(2006)\citenamefont
  {Wei\ss{}e}, \citenamefont {Wellein}, \citenamefont {Alvermann},\ and\
  \citenamefont {Fehske}}]{Weisse2006}%
  \BibitemOpen
  \bibfield  {author} {\bibinfo {author} {\bibfnamefont {A.}~\bibnamefont
  {Wei\ss{}e}}, \bibinfo {author} {\bibfnamefont {G.}~\bibnamefont {Wellein}},
  \bibinfo {author} {\bibfnamefont {A.}~\bibnamefont {Alvermann}},\ and\
  \bibinfo {author} {\bibfnamefont {H.}~\bibnamefont {Fehske}},\ }\bibfield
  {title} {\bibinfo {title} {The kernel polynomial method},\ }\href
  {https://doi.org/10.1103/RevModPhys.78.275} {\bibfield  {journal} {\bibinfo
  {journal} {Rev. Mod. Phys.}\ }\textbf {\bibinfo {volume} {78}},\ \bibinfo
  {pages} {275} (\bibinfo {year} {2006})}\BibitemShut {NoStop}%
\bibitem [{\citenamefont {Seki}\ and\ \citenamefont
  {Yunoki}(2020)}]{Seki2020ftlm}%
  \BibitemOpen
  \bibfield  {author} {\bibinfo {author} {\bibfnamefont {K.}~\bibnamefont
  {Seki}}\ and\ \bibinfo {author} {\bibfnamefont {S.}~\bibnamefont {Yunoki}},\
  }\bibfield  {title} {\bibinfo {title} {{Thermodynamic properties of an
  $S=\frac{1}{2}$ ring-exchange model on the triangular lattice}},\ }\href
  {https://doi.org/10.1103/PhysRevB.101.235115} {\bibfield  {journal} {\bibinfo
   {journal} {Phys. Rev. B}\ }\textbf {\bibinfo {volume} {101}},\ \bibinfo
  {pages} {235115} (\bibinfo {year} {2020})}\BibitemShut {NoStop}%
\bibitem [{\citenamefont {Seki}\ \emph {et~al.}(2020)\citenamefont {Seki},
  \citenamefont {Shirakawa},\ and\ \citenamefont {Yunoki}}]{Seki2020vqe}%
  \BibitemOpen
  \bibfield  {author} {\bibinfo {author} {\bibfnamefont {K.}~\bibnamefont
  {Seki}}, \bibinfo {author} {\bibfnamefont {T.}~\bibnamefont {Shirakawa}},\
  and\ \bibinfo {author} {\bibfnamefont {S.}~\bibnamefont {Yunoki}},\
  }\bibfield  {title} {\bibinfo {title} {Symmetry-adapted variational quantum
  eigensolver},\ }\href {https://doi.org/10.1103/PhysRevA.101.052340}
  {\bibfield  {journal} {\bibinfo  {journal} {Phys. Rev. A}\ }\textbf {\bibinfo
  {volume} {101}},\ \bibinfo {pages} {052340} (\bibinfo {year}
  {2020})}\BibitemShut {NoStop}%
\bibitem [{\citenamefont {Marshall}(1955)}]{Marshall1955}%
  \BibitemOpen
  \bibfield  {author} {\bibinfo {author} {\bibfnamefont {W.}~\bibnamefont
  {Marshall}},\ }\bibfield  {title} {\bibinfo {title} {Antiferromagnetism},\
  }\href {https://doi.org/10.1098/rspa.1955.0200} {\bibfield  {journal}
  {\bibinfo  {journal} {Proceedings of the Royal Society of London. Series A.
  Mathematical and Physical Sciences}\ }\textbf {\bibinfo {volume} {232}},\
  \bibinfo {pages} {48} (\bibinfo {year} {1955})}\BibitemShut {NoStop}%
\bibitem [{\citenamefont {Lieb}\ and\ \citenamefont {Mattis}(1962)}]{Lieb1962}%
  \BibitemOpen
  \bibfield  {author} {\bibinfo {author} {\bibfnamefont {E.}~\bibnamefont
  {Lieb}}\ and\ \bibinfo {author} {\bibfnamefont {D.}~\bibnamefont {Mattis}},\
  }\bibfield  {title} {\bibinfo {title} {{Ordering Energy Levels of Interacting
  Spin Systems}},\ }\href {https://doi.org/10.1063/1.1724276} {\bibfield
  {journal} {\bibinfo  {journal} {Journal of Mathematical Physics}\ }\textbf
  {\bibinfo {volume} {3}},\ \bibinfo {pages} {749} (\bibinfo {year}
  {1962})}\BibitemShut {NoStop}%
\bibitem [{\citenamefont {Li}\ and\ \citenamefont
  {Benjamin}(2017)}]{Li2017PRX}%
  \BibitemOpen
  \bibfield  {author} {\bibinfo {author} {\bibfnamefont {Y.}~\bibnamefont
  {Li}}\ and\ \bibinfo {author} {\bibfnamefont {S.~C.}\ \bibnamefont
  {Benjamin}},\ }\bibfield  {title} {\bibinfo {title} {{Efficient Variational
  Quantum Simulator Incorporating Active Error Minimization}},\ }\href
  {https://doi.org/10.1103/PhysRevX.7.021050} {\bibfield  {journal} {\bibinfo
  {journal} {Phys. Rev. X}\ }\textbf {\bibinfo {volume} {7}},\ \bibinfo {pages}
  {021050} (\bibinfo {year} {2017})}\BibitemShut {NoStop}%
\bibitem [{\citenamefont {Endo}\ \emph {et~al.}(2018)\citenamefont {Endo},
  \citenamefont {Benjamin},\ and\ \citenamefont {Li}}]{Endo2018PRX}%
  \BibitemOpen
  \bibfield  {author} {\bibinfo {author} {\bibfnamefont {S.}~\bibnamefont
  {Endo}}, \bibinfo {author} {\bibfnamefont {S.~C.}\ \bibnamefont {Benjamin}},\
  and\ \bibinfo {author} {\bibfnamefont {Y.}~\bibnamefont {Li}},\ }\bibfield
  {title} {\bibinfo {title} {Practical quantum error mitigation for near-future
  applications},\ }\href {https://doi.org/10.1103/PhysRevX.8.031027} {\bibfield
   {journal} {\bibinfo  {journal} {Phys. Rev. X}\ }\textbf {\bibinfo {volume}
  {8}},\ \bibinfo {pages} {031027} (\bibinfo {year} {2018})}\BibitemShut
  {NoStop}%
\bibitem [{\citenamefont {Seki}\ and\ \citenamefont
  {Sorella}(2019)}]{Seki2019}%
  \BibitemOpen
  \bibfield  {author} {\bibinfo {author} {\bibfnamefont {K.}~\bibnamefont
  {Seki}}\ and\ \bibinfo {author} {\bibfnamefont {S.}~\bibnamefont {Sorella}},\
  }\bibfield  {title} {\bibinfo {title} {{Benchmark study of an auxiliary-field
  quantum Monte Carlo technique for the Hubbard model with shifted-discrete
  Hubbard-Stratonovich transformations}},\ }\href
  {https://doi.org/10.1103/PhysRevB.99.144407} {\bibfield  {journal} {\bibinfo
  {journal} {Phys. Rev. B}\ }\textbf {\bibinfo {volume} {99}},\ \bibinfo
  {pages} {144407} (\bibinfo {year} {2019})}\BibitemShut {NoStop}%
\bibitem [{\citenamefont {Oitmaa}\ \emph {et~al.}(2006)\citenamefont {Oitmaa},
  \citenamefont {Hamer},\ and\ \citenamefont {Zheng}}]{Oitmaa}%
  \BibitemOpen
  \bibfield  {author} {\bibinfo {author} {\bibfnamefont {J.}~\bibnamefont
  {Oitmaa}}, \bibinfo {author} {\bibfnamefont {C.}~\bibnamefont {Hamer}},\ and\
  \bibinfo {author} {\bibfnamefont {W.}~\bibnamefont {Zheng}},\ }\href@noop {}
  {\emph {\bibinfo {title} {{Series Expansion Methods for Strongly Interacting
  Lattice Models}}}}\ (\bibinfo  {publisher} {Cambridge},\ \bibinfo {address}
  {New York},\ \bibinfo {year} {2006})\BibitemShut {NoStop}%
\bibitem [{\citenamefont {Bespalova}\ and\ \citenamefont
  {Kyriienko}(2020)}]{bespalova2020hamiltonian}%
  \BibitemOpen
  \bibfield  {author} {\bibinfo {author} {\bibfnamefont {T.~A.}\ \bibnamefont
  {Bespalova}}\ and\ \bibinfo {author} {\bibfnamefont {O.}~\bibnamefont
  {Kyriienko}},\ }\href@noop {} {\bibinfo {title} {{Hamiltonian operator
  approximation for energy measurement and ground state preparation}}}
  (\bibinfo {year} {2020}),\ \Eprint {https://arxiv.org/abs/2009.03351}
  {arXiv:2009.03351 [quant-ph]} \BibitemShut {NoStop}%
\bibitem [{\citenamefont {Reiner}\ \emph {et~al.}(2016)\citenamefont {Reiner},
  \citenamefont {Marthaler}, \citenamefont {Braum\"uller}, \citenamefont
  {Weides},\ and\ \citenamefont {Sch\"on}}]{Reiner2016}%
  \BibitemOpen
  \bibfield  {author} {\bibinfo {author} {\bibfnamefont {J.-M.}\ \bibnamefont
  {Reiner}}, \bibinfo {author} {\bibfnamefont {M.}~\bibnamefont {Marthaler}},
  \bibinfo {author} {\bibfnamefont {J.}~\bibnamefont {Braum\"uller}}, \bibinfo
  {author} {\bibfnamefont {M.}~\bibnamefont {Weides}},\ and\ \bibinfo {author}
  {\bibfnamefont {G.}~\bibnamefont {Sch\"on}},\ }\bibfield  {title} {\bibinfo
  {title} {{Emulating the one-dimensional Fermi-Hubbard model by a double chain
  of qubits}},\ }\href {https://doi.org/10.1103/PhysRevA.94.032338} {\bibfield
  {journal} {\bibinfo  {journal} {Phys. Rev. A}\ }\textbf {\bibinfo {volume}
  {94}},\ \bibinfo {pages} {032338} (\bibinfo {year} {2016})}\BibitemShut
  {NoStop}%
\bibitem [{\citenamefont {Rodriguez}(1959)}]{Rodriguez1959}%
  \BibitemOpen
  \bibfield  {author} {\bibinfo {author} {\bibfnamefont {S.}~\bibnamefont
  {Rodriguez}},\ }\bibfield  {title} {\bibinfo {title} {{Linear
  Antiferromagnetic Chain}},\ }\href {https://doi.org/10.1103/PhysRev.116.1474}
  {\bibfield  {journal} {\bibinfo  {journal} {Phys. Rev.}\ }\textbf {\bibinfo
  {volume} {116}},\ \bibinfo {pages} {1474} (\bibinfo {year}
  {1959})}\BibitemShut {NoStop}%
\bibitem [{\citenamefont {Reiner}\ \emph {et~al.}(2019)\citenamefont {Reiner},
  \citenamefont {Wilhelm-Mauch}, \citenamefont {Sch{\"o}n},\ and\ \citenamefont
  {Marthaler}}]{Reiner2019}%
  \BibitemOpen
  \bibfield  {author} {\bibinfo {author} {\bibfnamefont {J.-M.}\ \bibnamefont
  {Reiner}}, \bibinfo {author} {\bibfnamefont {F.}~\bibnamefont
  {Wilhelm-Mauch}}, \bibinfo {author} {\bibfnamefont {G.}~\bibnamefont
  {Sch{\"o}n}},\ and\ \bibinfo {author} {\bibfnamefont {M.}~\bibnamefont
  {Marthaler}},\ }\bibfield  {title} {\bibinfo {title} {{Finding the ground
  state of the Hubbard model by variational methods on a quantum computer with
  gate errors}},\ }\href {https://doi.org/10.1088/2058-9565/ab1e85} {\bibfield
  {journal} {\bibinfo  {journal} {Quantum Science and Technology}\ }\textbf
  {\bibinfo {volume} {4}},\ \bibinfo {pages} {035005} (\bibinfo {year}
  {2019})}\BibitemShut {NoStop}%
\bibitem [{\citenamefont {Dallaire-Demers}\ and\ \citenamefont
  {Wilhelm}(2016)}]{Dallaire-Demers2016gates}%
  \BibitemOpen
  \bibfield  {author} {\bibinfo {author} {\bibfnamefont {P.-L.}\ \bibnamefont
  {Dallaire-Demers}}\ and\ \bibinfo {author} {\bibfnamefont {F.~K.}\
  \bibnamefont {Wilhelm}},\ }\bibfield  {title} {\bibinfo {title} {{Quantum
  gates and architecture for the quantum simulation of the Fermi-Hubbard
  model}},\ }\href {https://doi.org/10.1103/PhysRevA.94.062304} {\bibfield
  {journal} {\bibinfo  {journal} {Phys. Rev. A}\ }\textbf {\bibinfo {volume}
  {94}},\ \bibinfo {pages} {062304} (\bibinfo {year} {2016})}\BibitemShut
  {NoStop}%
\bibitem [{\citenamefont {Reiner}\ \emph {et~al.}(2018)\citenamefont {Reiner},
  \citenamefont {Zanker}, \citenamefont {Schwenk}, \citenamefont
  {Lepp{\"a}kangas}, \citenamefont {Wilhelm-Mauch}, \citenamefont {Sch{\"o}n},\
  and\ \citenamefont {Marthaler}}]{Reiner2018}%
  \BibitemOpen
  \bibfield  {author} {\bibinfo {author} {\bibfnamefont {J.-M.}\ \bibnamefont
  {Reiner}}, \bibinfo {author} {\bibfnamefont {S.}~\bibnamefont {Zanker}},
  \bibinfo {author} {\bibfnamefont {I.}~\bibnamefont {Schwenk}}, \bibinfo
  {author} {\bibfnamefont {J.}~\bibnamefont {Lepp{\"a}kangas}}, \bibinfo
  {author} {\bibfnamefont {F.}~\bibnamefont {Wilhelm-Mauch}}, \bibinfo {author}
  {\bibfnamefont {G.}~\bibnamefont {Sch{\"o}n}},\ and\ \bibinfo {author}
  {\bibfnamefont {M.}~\bibnamefont {Marthaler}},\ }\bibfield  {title} {\bibinfo
  {title} {{Effects of gate errors in digital quantum simulations of fermionic
  systems}},\ }\href {https://doi.org/10.1088/2058-9565/aad5ba} {\bibfield
  {journal} {\bibinfo  {journal} {Quantum Science and Technology}\ }\textbf
  {\bibinfo {volume} {3}},\ \bibinfo {pages} {045008} (\bibinfo {year}
  {2018})}\BibitemShut {NoStop}%
\bibitem [{\citenamefont {Dallaire-Demers}\ \emph {et~al.}(2020)\citenamefont
  {Dallaire-Demers}, \citenamefont {Stechly}, \citenamefont {Gonthier},
  \citenamefont {Bashige}, \citenamefont {Romero},\ and\ \citenamefont
  {Cao}}]{dallairedemers2020application}%
  \BibitemOpen
  \bibfield  {author} {\bibinfo {author} {\bibfnamefont {P.-L.}\ \bibnamefont
  {Dallaire-Demers}}, \bibinfo {author} {\bibfnamefont {M.}~\bibnamefont
  {Stechly}}, \bibinfo {author} {\bibfnamefont {J.~F.}\ \bibnamefont
  {Gonthier}}, \bibinfo {author} {\bibfnamefont {N.~T.}\ \bibnamefont
  {Bashige}}, \bibinfo {author} {\bibfnamefont {J.}~\bibnamefont {Romero}},\
  and\ \bibinfo {author} {\bibfnamefont {Y.}~\bibnamefont {Cao}},\ }\href@noop
  {} {\bibinfo {title} {An application benchmark for fermionic quantum
  simulations}} (\bibinfo {year} {2020}),\ \Eprint
  {https://arxiv.org/abs/2003.01862} {arXiv:2003.01862 [quant-ph]} \BibitemShut
  {NoStop}%
\bibitem [{\citenamefont {Arute}\ \emph
  {et~al.}(2020{\natexlab{b}})\citenamefont {Arute}, \citenamefont {Arya},
  \citenamefont {Babbush}, \citenamefont {Bacon}, \citenamefont {Bardin},
  \citenamefont {Barends}, \citenamefont {Bengtsson}, \citenamefont {Boixo},
  \citenamefont {Broughton}, \citenamefont {Buckley}, \citenamefont {Buell},
  \citenamefont {Burkett}, \citenamefont {Bushnell}, \citenamefont {Chen},
  \citenamefont {Chen}, \citenamefont {Chen}, \citenamefont {Chiaro},
  \citenamefont {Collins}, \citenamefont {Cotton}, \citenamefont {Courtney},
  \citenamefont {Demura}, \citenamefont {Derk}, \citenamefont {Dunsworth},
  \citenamefont {Eppens}, \citenamefont {Eckl}, \citenamefont {Erickson},
  \citenamefont {Farhi}, \citenamefont {Fowler}, \citenamefont {Foxen},
  \citenamefont {Gidney}, \citenamefont {Giustina}, \citenamefont {Graff},
  \citenamefont {Gross}, \citenamefont {Habegger}, \citenamefont {Harrigan},
  \citenamefont {Ho}, \citenamefont {Hong}, \citenamefont {Huang},
  \citenamefont {Huggins}, \citenamefont {Ioffe}, \citenamefont {Isakov},
  \citenamefont {Jeffrey}, \citenamefont {Jiang}, \citenamefont {Jones},
  \citenamefont {Kafri}, \citenamefont {Kechedzhi}, \citenamefont {Kelly},
  \citenamefont {Kim}, \citenamefont {Klimov}, \citenamefont {Korotkov},
  \citenamefont {Kostritsa}, \citenamefont {Landhuis}, \citenamefont {Laptev},
  \citenamefont {Lindmark}, \citenamefont {Lucero}, \citenamefont {Marthaler},
  \citenamefont {Martin}, \citenamefont {Martinis}, \citenamefont {Marusczyk},
  \citenamefont {McArdle}, \citenamefont {McClean}, \citenamefont {McCourt},
  \citenamefont {McEwen}, \citenamefont {Megrant}, \citenamefont
  {Mejuto-Zaera}, \citenamefont {Mi}, \citenamefont {Mohseni}, \citenamefont
  {Mruczkiewicz}, \citenamefont {Mutus}, \citenamefont {Naaman}, \citenamefont
  {Neeley}, \citenamefont {Neill}, \citenamefont {Neven}, \citenamefont
  {Newman}, \citenamefont {Niu}, \citenamefont {O'Brien}, \citenamefont
  {Ostby}, \citenamefont {Pató}, \citenamefont {Petukhov}, \citenamefont
  {Putterman}, \citenamefont {Quintana}, \citenamefont {Reiner}, \citenamefont
  {Roushan}, \citenamefont {Rubin}, \citenamefont {Sank}, \citenamefont
  {Satzinger}, \citenamefont {Smelyanskiy}, \citenamefont {Strain},
  \citenamefont {Sung}, \citenamefont {Schmitteckert}, \citenamefont {Szalay},
  \citenamefont {Tubman}, \citenamefont {Vainsencher}, \citenamefont {White},
  \citenamefont {Vogt}, \citenamefont {Yao}, \citenamefont {Yeh}, \citenamefont
  {Zalcman},\ and\ \citenamefont {Zanker}}]{arute2020observation}%
  \BibitemOpen
  \bibfield  {author} {\bibinfo {author} {\bibfnamefont {F.}~\bibnamefont
  {Arute}}, \bibinfo {author} {\bibfnamefont {K.}~\bibnamefont {Arya}},
  \bibinfo {author} {\bibfnamefont {R.}~\bibnamefont {Babbush}}, \bibinfo
  {author} {\bibfnamefont {D.}~\bibnamefont {Bacon}}, \bibinfo {author}
  {\bibfnamefont {J.~C.}\ \bibnamefont {Bardin}}, \bibinfo {author}
  {\bibfnamefont {R.}~\bibnamefont {Barends}}, \bibinfo {author} {\bibfnamefont
  {A.}~\bibnamefont {Bengtsson}}, \bibinfo {author} {\bibfnamefont
  {S.}~\bibnamefont {Boixo}}, \bibinfo {author} {\bibfnamefont
  {M.}~\bibnamefont {Broughton}}, \bibinfo {author} {\bibfnamefont {B.~B.}\
  \bibnamefont {Buckley}}, \bibinfo {author} {\bibfnamefont {D.~A.}\
  \bibnamefont {Buell}}, \bibinfo {author} {\bibfnamefont {B.}~\bibnamefont
  {Burkett}}, \bibinfo {author} {\bibfnamefont {N.}~\bibnamefont {Bushnell}},
  \bibinfo {author} {\bibfnamefont {Y.}~\bibnamefont {Chen}}, \bibinfo {author}
  {\bibfnamefont {Z.}~\bibnamefont {Chen}}, \bibinfo {author} {\bibfnamefont
  {Y.-A.}\ \bibnamefont {Chen}}, \bibinfo {author} {\bibfnamefont
  {B.}~\bibnamefont {Chiaro}}, \bibinfo {author} {\bibfnamefont
  {R.}~\bibnamefont {Collins}}, \bibinfo {author} {\bibfnamefont {S.~J.}\
  \bibnamefont {Cotton}}, \bibinfo {author} {\bibfnamefont {W.}~\bibnamefont
  {Courtney}}, \bibinfo {author} {\bibfnamefont {S.}~\bibnamefont {Demura}},
  \bibinfo {author} {\bibfnamefont {A.}~\bibnamefont {Derk}}, \bibinfo {author}
  {\bibfnamefont {A.}~\bibnamefont {Dunsworth}}, \bibinfo {author}
  {\bibfnamefont {D.}~\bibnamefont {Eppens}}, \bibinfo {author} {\bibfnamefont
  {T.}~\bibnamefont {Eckl}}, \bibinfo {author} {\bibfnamefont {C.}~\bibnamefont
  {Erickson}}, \bibinfo {author} {\bibfnamefont {E.}~\bibnamefont {Farhi}},
  \bibinfo {author} {\bibfnamefont {A.}~\bibnamefont {Fowler}}, \bibinfo
  {author} {\bibfnamefont {B.}~\bibnamefont {Foxen}}, \bibinfo {author}
  {\bibfnamefont {C.}~\bibnamefont {Gidney}}, \bibinfo {author} {\bibfnamefont
  {M.}~\bibnamefont {Giustina}}, \bibinfo {author} {\bibfnamefont
  {R.}~\bibnamefont {Graff}}, \bibinfo {author} {\bibfnamefont {J.~A.}\
  \bibnamefont {Gross}}, \bibinfo {author} {\bibfnamefont {S.}~\bibnamefont
  {Habegger}}, \bibinfo {author} {\bibfnamefont {M.~P.}\ \bibnamefont
  {Harrigan}}, \bibinfo {author} {\bibfnamefont {A.}~\bibnamefont {Ho}},
  \bibinfo {author} {\bibfnamefont {S.}~\bibnamefont {Hong}}, \bibinfo {author}
  {\bibfnamefont {T.}~\bibnamefont {Huang}}, \bibinfo {author} {\bibfnamefont
  {W.}~\bibnamefont {Huggins}}, \bibinfo {author} {\bibfnamefont {L.~B.}\
  \bibnamefont {Ioffe}}, \bibinfo {author} {\bibfnamefont {S.~V.}\ \bibnamefont
  {Isakov}}, \bibinfo {author} {\bibfnamefont {E.}~\bibnamefont {Jeffrey}},
  \bibinfo {author} {\bibfnamefont {Z.}~\bibnamefont {Jiang}}, \bibinfo
  {author} {\bibfnamefont {C.}~\bibnamefont {Jones}}, \bibinfo {author}
  {\bibfnamefont {D.}~\bibnamefont {Kafri}}, \bibinfo {author} {\bibfnamefont
  {K.}~\bibnamefont {Kechedzhi}}, \bibinfo {author} {\bibfnamefont
  {J.}~\bibnamefont {Kelly}}, \bibinfo {author} {\bibfnamefont
  {S.}~\bibnamefont {Kim}}, \bibinfo {author} {\bibfnamefont {P.~V.}\
  \bibnamefont {Klimov}}, \bibinfo {author} {\bibfnamefont {A.~N.}\
  \bibnamefont {Korotkov}}, \bibinfo {author} {\bibfnamefont {F.}~\bibnamefont
  {Kostritsa}}, \bibinfo {author} {\bibfnamefont {D.}~\bibnamefont {Landhuis}},
  \bibinfo {author} {\bibfnamefont {P.}~\bibnamefont {Laptev}}, \bibinfo
  {author} {\bibfnamefont {M.}~\bibnamefont {Lindmark}}, \bibinfo {author}
  {\bibfnamefont {E.}~\bibnamefont {Lucero}}, \bibinfo {author} {\bibfnamefont
  {M.}~\bibnamefont {Marthaler}}, \bibinfo {author} {\bibfnamefont
  {O.}~\bibnamefont {Martin}}, \bibinfo {author} {\bibfnamefont {J.~M.}\
  \bibnamefont {Martinis}}, \bibinfo {author} {\bibfnamefont {A.}~\bibnamefont
  {Marusczyk}}, \bibinfo {author} {\bibfnamefont {S.}~\bibnamefont {McArdle}},
  \bibinfo {author} {\bibfnamefont {J.~R.}\ \bibnamefont {McClean}}, \bibinfo
  {author} {\bibfnamefont {T.}~\bibnamefont {McCourt}}, \bibinfo {author}
  {\bibfnamefont {M.}~\bibnamefont {McEwen}}, \bibinfo {author} {\bibfnamefont
  {A.}~\bibnamefont {Megrant}}, \bibinfo {author} {\bibfnamefont
  {C.}~\bibnamefont {Mejuto-Zaera}}, \bibinfo {author} {\bibfnamefont
  {X.}~\bibnamefont {Mi}}, \bibinfo {author} {\bibfnamefont {M.}~\bibnamefont
  {Mohseni}}, \bibinfo {author} {\bibfnamefont {W.}~\bibnamefont
  {Mruczkiewicz}}, \bibinfo {author} {\bibfnamefont {J.}~\bibnamefont {Mutus}},
  \bibinfo {author} {\bibfnamefont {O.}~\bibnamefont {Naaman}}, \bibinfo
  {author} {\bibfnamefont {M.}~\bibnamefont {Neeley}}, \bibinfo {author}
  {\bibfnamefont {C.}~\bibnamefont {Neill}}, \bibinfo {author} {\bibfnamefont
  {H.}~\bibnamefont {Neven}}, \bibinfo {author} {\bibfnamefont
  {M.}~\bibnamefont {Newman}}, \bibinfo {author} {\bibfnamefont {M.~Y.}\
  \bibnamefont {Niu}}, \bibinfo {author} {\bibfnamefont {T.~E.}\ \bibnamefont
  {O'Brien}}, \bibinfo {author} {\bibfnamefont {E.}~\bibnamefont {Ostby}},
  \bibinfo {author} {\bibfnamefont {B.}~\bibnamefont {Pató}}, \bibinfo
  {author} {\bibfnamefont {A.}~\bibnamefont {Petukhov}}, \bibinfo {author}
  {\bibfnamefont {H.}~\bibnamefont {Putterman}}, \bibinfo {author}
  {\bibfnamefont {C.}~\bibnamefont {Quintana}}, \bibinfo {author}
  {\bibfnamefont {J.-M.}\ \bibnamefont {Reiner}}, \bibinfo {author}
  {\bibfnamefont {P.}~\bibnamefont {Roushan}}, \bibinfo {author} {\bibfnamefont
  {N.~C.}\ \bibnamefont {Rubin}}, \bibinfo {author} {\bibfnamefont
  {D.}~\bibnamefont {Sank}}, \bibinfo {author} {\bibfnamefont {K.~J.}\
  \bibnamefont {Satzinger}}, \bibinfo {author} {\bibfnamefont {V.}~\bibnamefont
  {Smelyanskiy}}, \bibinfo {author} {\bibfnamefont {D.}~\bibnamefont {Strain}},
  \bibinfo {author} {\bibfnamefont {K.~J.}\ \bibnamefont {Sung}}, \bibinfo
  {author} {\bibfnamefont {P.}~\bibnamefont {Schmitteckert}}, \bibinfo {author}
  {\bibfnamefont {M.}~\bibnamefont {Szalay}}, \bibinfo {author} {\bibfnamefont
  {N.~M.}\ \bibnamefont {Tubman}}, \bibinfo {author} {\bibfnamefont
  {A.}~\bibnamefont {Vainsencher}}, \bibinfo {author} {\bibfnamefont
  {T.}~\bibnamefont {White}}, \bibinfo {author} {\bibfnamefont
  {N.}~\bibnamefont {Vogt}}, \bibinfo {author} {\bibfnamefont {Z.~J.}\
  \bibnamefont {Yao}}, \bibinfo {author} {\bibfnamefont {P.}~\bibnamefont
  {Yeh}}, \bibinfo {author} {\bibfnamefont {A.}~\bibnamefont {Zalcman}},\ and\
  \bibinfo {author} {\bibfnamefont {S.}~\bibnamefont {Zanker}},\ }\href@noop {}
  {\bibinfo {title} {{Observation of separated dynamics of charge and spin in
  the Fermi-Hubbard model}}} (\bibinfo {year} {2020}{\natexlab{b}}),\ \Eprint
  {https://arxiv.org/abs/2010.07965} {arXiv:2010.07965 [quant-ph]} \BibitemShut
  {NoStop}%
\bibitem [{\citenamefont {Kivlichan}\ \emph {et~al.}(2018)\citenamefont
  {Kivlichan}, \citenamefont {McClean}, \citenamefont {Wiebe}, \citenamefont
  {Gidney}, \citenamefont {Aspuru-Guzik}, \citenamefont {Chan},\ and\
  \citenamefont {Babbush}}]{Kivlichan2018}%
  \BibitemOpen
  \bibfield  {author} {\bibinfo {author} {\bibfnamefont {I.~D.}\ \bibnamefont
  {Kivlichan}}, \bibinfo {author} {\bibfnamefont {J.}~\bibnamefont {McClean}},
  \bibinfo {author} {\bibfnamefont {N.}~\bibnamefont {Wiebe}}, \bibinfo
  {author} {\bibfnamefont {C.}~\bibnamefont {Gidney}}, \bibinfo {author}
  {\bibfnamefont {A.}~\bibnamefont {Aspuru-Guzik}}, \bibinfo {author}
  {\bibfnamefont {G.~K.-L.}\ \bibnamefont {Chan}},\ and\ \bibinfo {author}
  {\bibfnamefont {R.}~\bibnamefont {Babbush}},\ }\bibfield  {title} {\bibinfo
  {title} {{Quantum Simulation of Electronic Structure with Linear Depth and
  Connectivity}},\ }\href {https://doi.org/10.1103/PhysRevLett.120.110501}
  {\bibfield  {journal} {\bibinfo  {journal} {Phys. Rev. Lett.}\ }\textbf
  {\bibinfo {volume} {120}},\ \bibinfo {pages} {110501} (\bibinfo {year}
  {2018})}\BibitemShut {NoStop}%
\bibitem [{\citenamefont {Jiang}\ \emph {et~al.}(2018)\citenamefont {Jiang},
  \citenamefont {Sung}, \citenamefont {Kechedzhi}, \citenamefont
  {Smelyanskiy},\ and\ \citenamefont {Boixo}}]{Jiang2018}%
  \BibitemOpen
  \bibfield  {author} {\bibinfo {author} {\bibfnamefont {Z.}~\bibnamefont
  {Jiang}}, \bibinfo {author} {\bibfnamefont {K.~J.}\ \bibnamefont {Sung}},
  \bibinfo {author} {\bibfnamefont {K.}~\bibnamefont {Kechedzhi}}, \bibinfo
  {author} {\bibfnamefont {V.~N.}\ \bibnamefont {Smelyanskiy}},\ and\ \bibinfo
  {author} {\bibfnamefont {S.}~\bibnamefont {Boixo}},\ }\bibfield  {title}
  {\bibinfo {title} {{Quantum Algorithms to Simulate Many-Body Physics of
  Correlated Fermions}},\ }\href
  {https://doi.org/10.1103/PhysRevApplied.9.044036} {\bibfield  {journal}
  {\bibinfo  {journal} {Phys. Rev. Applied}\ }\textbf {\bibinfo {volume} {9}},\
  \bibinfo {pages} {044036} (\bibinfo {year} {2018})}\BibitemShut {NoStop}%
\bibitem [{\citenamefont {Shirakawa}\ \emph {et~al.}(2021)\citenamefont
  {Shirakawa}, \citenamefont {Seki},\ and\ \citenamefont
  {Yunoki}}]{Shirakawa2020}%
  \BibitemOpen
  \bibfield  {author} {\bibinfo {author} {\bibfnamefont {T.}~\bibnamefont
  {Shirakawa}}, \bibinfo {author} {\bibfnamefont {K.}~\bibnamefont {Seki}},\
  and\ \bibinfo {author} {\bibfnamefont {S.}~\bibnamefont {Yunoki}},\
  }\bibfield  {title} {\bibinfo {title} {Discretized quantum adiabatic process
  for free fermions and comparison with the imaginary-time evolution},\ }\href
  {https://doi.org/10.1103/PhysRevResearch.3.013004} {\bibfield  {journal}
  {\bibinfo  {journal} {Phys. Rev. Research}\ }\textbf {\bibinfo {volume}
  {3}},\ \bibinfo {pages} {013004} (\bibinfo {year} {2021})}\BibitemShut
  {NoStop}%
\bibitem [{\citenamefont {Tranter}\ \emph {et~al.}(2018)\citenamefont
  {Tranter}, \citenamefont {Love}, \citenamefont {Mintert},\ and\ \citenamefont
  {Coveney}}]{Tranter2018}%
  \BibitemOpen
  \bibfield  {author} {\bibinfo {author} {\bibfnamefont {A.}~\bibnamefont
  {Tranter}}, \bibinfo {author} {\bibfnamefont {P.~J.}\ \bibnamefont {Love}},
  \bibinfo {author} {\bibfnamefont {F.}~\bibnamefont {Mintert}},\ and\ \bibinfo
  {author} {\bibfnamefont {P.~V.}\ \bibnamefont {Coveney}},\ }\bibfield
  {title} {\bibinfo {title} {{A Comparison of the Bravyi-Kitaev and
  Jordan-Wigner Transformations for the Quantum Simulation of Quantum
  Chemistry}},\ }\href {https://doi.org/10.1021/acs.jctc.8b00450} {\bibfield
  {journal} {\bibinfo  {journal} {Journal of Chemical Theory and Computation}\
  }\textbf {\bibinfo {volume} {14}},\ \bibinfo {pages} {5617} (\bibinfo {year}
  {2018})}\BibitemShut {NoStop}%
\bibitem [{\citenamefont {Cade}\ \emph {et~al.}(2020)\citenamefont {Cade},
  \citenamefont {Mineh}, \citenamefont {Montanaro},\ and\ \citenamefont
  {Stanisic}}]{cade2019strategies}%
  \BibitemOpen
  \bibfield  {author} {\bibinfo {author} {\bibfnamefont {C.}~\bibnamefont
  {Cade}}, \bibinfo {author} {\bibfnamefont {L.}~\bibnamefont {Mineh}},
  \bibinfo {author} {\bibfnamefont {A.}~\bibnamefont {Montanaro}},\ and\
  \bibinfo {author} {\bibfnamefont {S.}~\bibnamefont {Stanisic}},\ }\bibfield
  {title} {\bibinfo {title} {{Strategies for solving the Fermi-Hubbard model on
  near-term quantum computers}},\ }\href
  {https://doi.org/10.1103/PhysRevB.102.235122} {\bibfield  {journal} {\bibinfo
   {journal} {Phys. Rev. B}\ }\textbf {\bibinfo {volume} {102}},\ \bibinfo
  {pages} {235122} (\bibinfo {year} {2020})}\BibitemShut {NoStop}%
\bibitem [{\citenamefont {Trotter}(1959)}]{Trotter1959}%
  \BibitemOpen
  \bibfield  {author} {\bibinfo {author} {\bibfnamefont {H.~F.}\ \bibnamefont
  {Trotter}},\ }\bibfield  {title} {\bibinfo {title} {On the product of
  semi-groups of operators},\ }\href
  {https://doi.org/10.1090/S0002-9939-1959-0108732-6} {\bibfield  {journal}
  {\bibinfo  {journal} {Proc. Am. Math. Soc.}\ }\textbf {\bibinfo {volume}
  {10}},\ \bibinfo {pages} {545} (\bibinfo {year} {1959})}\BibitemShut
  {NoStop}%
\bibitem [{\citenamefont {Suzuki}(1976{\natexlab{a}})}]{Suzuki1976PTP}%
  \BibitemOpen
  \bibfield  {author} {\bibinfo {author} {\bibfnamefont {M.}~\bibnamefont
  {Suzuki}},\ }\bibfield  {title} {\bibinfo {title} {{Relationship between
  $d$-Dimensional Quantal Spin Systems and ($d+1$)-Dimensional Ising Systems:
  Equivalence, Critical Exponents and Systematic Approximants of the Partition
  Function and Spin Correlations}},\ }\href
  {https://doi.org/10.1143/PTP.56.1454} {\bibfield  {journal} {\bibinfo
  {journal} {Progress of Theoretical Physics}\ }\textbf {\bibinfo {volume}
  {56}},\ \bibinfo {pages} {1454} (\bibinfo {year}
  {1976}{\natexlab{a}})}\BibitemShut {NoStop}%
\bibitem [{\citenamefont {Suzuki}(1976{\natexlab{b}})}]{Suzuki1976}%
  \BibitemOpen
  \bibfield  {author} {\bibinfo {author} {\bibfnamefont {M.}~\bibnamefont
  {Suzuki}},\ }\bibfield  {title} {\bibinfo {title} {{Generalized Trotter's
  formula and systematic approximants of exponential operators and inner
  derivations with applications to many-body problems}},\ }\href
  {http://projecteuclid.org/euclid.cmp/1103900351} {\bibfield  {journal}
  {\bibinfo  {journal} {Comm. Math. Phys.}\ }\textbf {\bibinfo {volume} {51}},\
  \bibinfo {pages} {183} (\bibinfo {year} {1976}{\natexlab{b}})}\BibitemShut
  {NoStop}%
\bibitem [{\citenamefont {Omelyan}\ \emph {et~al.}(2003)\citenamefont
  {Omelyan}, \citenamefont {Mryglod},\ and\ \citenamefont
  {Folk}}]{Omelyan2003}%
  \BibitemOpen
  \bibfield  {author} {\bibinfo {author} {\bibfnamefont {I.}~\bibnamefont
  {Omelyan}}, \bibinfo {author} {\bibfnamefont {I.}~\bibnamefont {Mryglod}},\
  and\ \bibinfo {author} {\bibfnamefont {R.}~\bibnamefont {Folk}},\ }\bibfield
  {title} {\bibinfo {title} {Symplectic analytically integrable decomposition
  algorithms: classification, derivation, and application to molecular
  dynamics, quantum and celestial mechanics simulations},\ }\href
  {https://doi.org/https://doi.org/10.1016/S0010-4655(02)00754-3} {\bibfield
  {journal} {\bibinfo  {journal} {Computer Physics Communications}\ }\textbf
  {\bibinfo {volume} {151}},\ \bibinfo {pages} {272 } (\bibinfo {year}
  {2003})}\BibitemShut {NoStop}%
\bibitem [{\citenamefont {Papageorgiou}\ and\ \citenamefont
  {Zhang}(2012)}]{Papageorgiou2012}%
  \BibitemOpen
  \bibfield  {author} {\bibinfo {author} {\bibfnamefont {A.}~\bibnamefont
  {Papageorgiou}}\ and\ \bibinfo {author} {\bibfnamefont {C.}~\bibnamefont
  {Zhang}},\ }\bibfield  {title} {\bibinfo {title} {{On the efficiency of
  quantum algorithms for Hamiltonian simulation}},\ }\href
  {https://doi.org/10.1007/s11128-011-0263-9} {\bibfield  {journal} {\bibinfo
  {journal} {Quantum Information Processing}\ }\textbf {\bibinfo {volume}
  {11}},\ \bibinfo {pages} {541} (\bibinfo {year} {2012})}\BibitemShut
  {NoStop}%
\bibitem [{\citenamefont {Heyl}\ \emph {et~al.}(2019)\citenamefont {Heyl},
  \citenamefont {Hauke},\ and\ \citenamefont {Zoller}}]{Heyl2019}%
  \BibitemOpen
  \bibfield  {author} {\bibinfo {author} {\bibfnamefont {M.}~\bibnamefont
  {Heyl}}, \bibinfo {author} {\bibfnamefont {P.}~\bibnamefont {Hauke}},\ and\
  \bibinfo {author} {\bibfnamefont {P.}~\bibnamefont {Zoller}},\ }\bibfield
  {title} {\bibinfo {title} {{Quantum localization bounds Trotter errors in
  digital quantum simulation}},\ }\href
  {https://doi.org/10.1126/sciadv.aau8342} {\bibfield  {journal} {\bibinfo
  {journal} {Science Advances}\ }\textbf {\bibinfo {volume} {5}},\ \bibinfo
  {pages} {eaau8342} (\bibinfo {year} {2019})}\BibitemShut {NoStop}%
\bibitem [{\citenamefont {Childs}\ \emph {et~al.}(2021)\citenamefont {Childs},
  \citenamefont {Su}, \citenamefont {Tran}, \citenamefont {Wiebe},\ and\
  \citenamefont {Zhu}}]{childs2019theory}%
  \BibitemOpen
  \bibfield  {author} {\bibinfo {author} {\bibfnamefont {A.~M.}\ \bibnamefont
  {Childs}}, \bibinfo {author} {\bibfnamefont {Y.}~\bibnamefont {Su}}, \bibinfo
  {author} {\bibfnamefont {M.~C.}\ \bibnamefont {Tran}}, \bibinfo {author}
  {\bibfnamefont {N.}~\bibnamefont {Wiebe}},\ and\ \bibinfo {author}
  {\bibfnamefont {S.}~\bibnamefont {Zhu}},\ }\bibfield  {title} {\bibinfo
  {title} {{Theory of Trotter Error with Commutator Scaling}},\ }\href
  {https://doi.org/10.1103/PhysRevX.11.011020} {\bibfield  {journal} {\bibinfo
  {journal} {Phys. Rev. X}\ }\textbf {\bibinfo {volume} {11}},\ \bibinfo
  {pages} {011020} (\bibinfo {year} {2021})}\BibitemShut {NoStop}%
\bibitem [{\citenamefont {Becca}\ and\ \citenamefont
  {Sorella}(2017)}]{Becca_Sorella_book}%
  \BibitemOpen
  \bibfield  {author} {\bibinfo {author} {\bibfnamefont {F.}~\bibnamefont
  {Becca}}\ and\ \bibinfo {author} {\bibfnamefont {S.}~\bibnamefont
  {Sorella}},\ }\href@noop {} {\emph {\bibinfo {title} {Quantum Monte Carlo
  Approaches for Correlated Systems}}}\ (\bibinfo  {publisher} {Cambridge
  University Press},\ \bibinfo {address} {Cambridge},\ \bibinfo {year}
  {2017})\BibitemShut {NoStop}%
\bibitem [{gat()}]{gate4}%
  \BibitemOpen
  \href@noop {} {}\bibinfo {note} {This is because the circuit depth for
  $\hat{\mathcal{H}}_{\underline{\rm ST}}^n(\Delta_\tau)$ is given by
  $D_{2\lceil n/2 \rceil}^{(3)}=2(N_\Gamma-1)3^{\lceil n/2\rceil-1}+1$, while
  the largest circuit depth for $\hat{\mathcal{H}}_{{\rm ST}}^n(\Delta_\tau)$
  involving $[\hat{S}_{2}^{(p)}(\pm \Delta/2)]^n$ is
  $n(D_{2}^{(3)}-1)+1=2(N_\Gamma-1)n+1$, when $p=3$}\BibitemShut {NoStop}%
\bibitem [{\citenamefont {Somma}(2019)}]{rol2019quantum}%
  \BibitemOpen
  \bibfield  {author} {\bibinfo {author} {\bibfnamefont {R.~D.}\ \bibnamefont
  {Somma}},\ }\bibfield  {title} {\bibinfo {title} {Quantum eigenvalue
  estimation via time series analysis},\ }\href
  {https://doi.org/10.1088/1367-2630/ab5c60} {\bibfield  {journal} {\bibinfo
  {journal} {New Journal of Physics}\ }\textbf {\bibinfo {volume} {21}},\
  \bibinfo {pages} {123025} (\bibinfo {year} {2019})}\BibitemShut {NoStop}%
\bibitem [{\citenamefont {Horn}\ and\ \citenamefont
  {Weinstein}(1984)}]{Horn1984}%
  \BibitemOpen
  \bibfield  {author} {\bibinfo {author} {\bibfnamefont {D.}~\bibnamefont
  {Horn}}\ and\ \bibinfo {author} {\bibfnamefont {M.}~\bibnamefont
  {Weinstein}},\ }\bibfield  {title} {\bibinfo {title} {{The $t$ expansion: A
  nonperturbative analytic tool for Hamiltonian systems}},\ }\href
  {https://doi.org/10.1103/PhysRevD.30.1256} {\bibfield  {journal} {\bibinfo
  {journal} {Phys. Rev. D}\ }\textbf {\bibinfo {volume} {30}},\ \bibinfo
  {pages} {1256} (\bibinfo {year} {1984})}\BibitemShut {NoStop}%
\bibitem [{\citenamefont {Kubo}(1962)}]{Kubo1962}%
  \BibitemOpen
  \bibfield  {author} {\bibinfo {author} {\bibfnamefont {R.}~\bibnamefont
  {Kubo}},\ }\bibfield  {title} {\bibinfo {title} {{Generalized Cumulant
  Expansion Method}},\ }\href {http://dx.doi.org/10.1143/JPSJ.17.1100}
  {\bibfield  {journal} {\bibinfo  {journal} {J. Phys. Soc. Jpn.}\ }\textbf
  {\bibinfo {volume} {17}},\ \bibinfo {pages} {1100} (\bibinfo {year}
  {1962})}\BibitemShut {NoStop}%
\bibitem [{\citenamefont {Fanizza}\ \emph {et~al.}(2020)\citenamefont
  {Fanizza}, \citenamefont {Rosati}, \citenamefont {Skotiniotis}, \citenamefont
  {Calsamiglia},\ and\ \citenamefont {Giovannetti}}]{Fanizza2020}%
  \BibitemOpen
  \bibfield  {author} {\bibinfo {author} {\bibfnamefont {M.}~\bibnamefont
  {Fanizza}}, \bibinfo {author} {\bibfnamefont {M.}~\bibnamefont {Rosati}},
  \bibinfo {author} {\bibfnamefont {M.}~\bibnamefont {Skotiniotis}}, \bibinfo
  {author} {\bibfnamefont {J.}~\bibnamefont {Calsamiglia}},\ and\ \bibinfo
  {author} {\bibfnamefont {V.}~\bibnamefont {Giovannetti}},\ }\bibfield
  {title} {\bibinfo {title} {{Beyond the Swap Test: Optimal Estimation of
  Quantum State Overlap}},\ }\href
  {https://doi.org/10.1103/PhysRevLett.124.060503} {\bibfield  {journal}
  {\bibinfo  {journal} {Phys. Rev. Lett.}\ }\textbf {\bibinfo {volume} {124}},\
  \bibinfo {pages} {060503} (\bibinfo {year} {2020})}\BibitemShut {NoStop}%
\bibitem [{\citenamefont {Zhang}\ \emph {et~al.}(2020)\citenamefont {Zhang},
  \citenamefont {Yuan},\ and\ \citenamefont {Yin}}]{zhang2020variational}%
  \BibitemOpen
  \bibfield  {author} {\bibinfo {author} {\bibfnamefont {D.-B.}\ \bibnamefont
  {Zhang}}, \bibinfo {author} {\bibfnamefont {Z.-H.}\ \bibnamefont {Yuan}},\
  and\ \bibinfo {author} {\bibfnamefont {T.}~\bibnamefont {Yin}},\ }\href@noop
  {} {\bibinfo {title} {{Variational quantum eigensolvers by variance
  minimization}}} (\bibinfo {year} {2020}),\ \Eprint
  {https://arxiv.org/abs/2006.15781} {arXiv:2006.15781 [quant-ph]} \BibitemShut
  {NoStop}%
\bibitem [{\citenamefont {Kowalski}\ and\ \citenamefont
  {Peng}(2020)}]{kowalski2020quantum}%
  \BibitemOpen
  \bibfield  {author} {\bibinfo {author} {\bibfnamefont {K.}~\bibnamefont
  {Kowalski}}\ and\ \bibinfo {author} {\bibfnamefont {B.}~\bibnamefont
  {Peng}},\ }\bibfield  {title} {\bibinfo {title} {Quantum simulations
  employing connected moments expansions},\ }\href
  {https://doi.org/10.1063/5.0030688} {\bibfield  {journal} {\bibinfo
  {journal} {The Journal of Chemical Physics}\ }\textbf {\bibinfo {volume}
  {153}},\ \bibinfo {pages} {201102} (\bibinfo {year} {2020})}\BibitemShut
  {NoStop}%
\bibitem [{\citenamefont {Vallury}\ \emph {et~al.}(2020)\citenamefont
  {Vallury}, \citenamefont {Jones}, \citenamefont {Hill},\ and\ \citenamefont
  {Hollenberg}}]{vallury2020quantum}%
  \BibitemOpen
  \bibfield  {author} {\bibinfo {author} {\bibfnamefont {H.~J.}\ \bibnamefont
  {Vallury}}, \bibinfo {author} {\bibfnamefont {M.~A.}\ \bibnamefont {Jones}},
  \bibinfo {author} {\bibfnamefont {C.~D.}\ \bibnamefont {Hill}},\ and\
  \bibinfo {author} {\bibfnamefont {L.~C.~L.}\ \bibnamefont {Hollenberg}},\
  }\bibfield  {title} {\bibinfo {title} {Quantum computed moments correction to
  variational estimates},\ }\href {https://doi.org/10.22331/q-2020-12-15-373}
  {\bibfield  {journal} {\bibinfo  {journal} {{Quantum}}\ }\textbf {\bibinfo
  {volume} {4}},\ \bibinfo {pages} {373} (\bibinfo {year} {2020})}\BibitemShut
  {NoStop}%
\bibitem [{\citenamefont {Witte}\ and\ \citenamefont
  {Hollenberg}(1994)}]{Witte1994}%
  \BibitemOpen
  \bibfield  {author} {\bibinfo {author} {\bibfnamefont {N.~S.}\ \bibnamefont
  {Witte}}\ and\ \bibinfo {author} {\bibfnamefont {L.~C.~L.}\ \bibnamefont
  {Hollenberg}},\ }\bibfield  {title} {\bibinfo {title} {Plaquette expansion
  proof and interpretation},\ }\href@noop {} {\bibfield  {journal} {\bibinfo
  {journal} {Z. Phys. B}\ }\textbf {\bibinfo {volume} {95}},\ \bibinfo {pages}
  {531} (\bibinfo {year} {1994})}\BibitemShut {NoStop}%
\bibitem [{\citenamefont {Haxton}\ \emph {et~al.}(2005)\citenamefont {Haxton},
  \citenamefont {Nollett},\ and\ \citenamefont {Zurek}}]{Haxton2005}%
  \BibitemOpen
  \bibfield  {author} {\bibinfo {author} {\bibfnamefont {W.~C.}\ \bibnamefont
  {Haxton}}, \bibinfo {author} {\bibfnamefont {K.~M.}\ \bibnamefont
  {Nollett}},\ and\ \bibinfo {author} {\bibfnamefont {K.~M.}\ \bibnamefont
  {Zurek}},\ }\bibfield  {title} {\bibinfo {title} {{Piecewise moments method:
  Generalized Lanczos technique for nuclear response surfaces}},\ }\href
  {https://doi.org/10.1103/PhysRevC.72.065501} {\bibfield  {journal} {\bibinfo
  {journal} {Phys. Rev. C}\ }\textbf {\bibinfo {volume} {72}},\ \bibinfo
  {pages} {065501} (\bibinfo {year} {2005})}\BibitemShut {NoStop}%
\end{thebibliography}%
\end{document}